\documentclass[a4paper,twoside,11pt]{book}  
\newcommand{\be}{\begin{equation}}
\newcommand{\ee}{\end{equation}}
\newcommand{\bea}{\begin{eqnarray}}
\newcommand{\eea}{\end{eqnarray}}
\newcommand{\beas}{\begin{eqnarray*}}
\newcommand{\eeas}{\end{eqnarray*}}
\newcommand{\bd}{\begin{displaymath}}
\newcommand{\ed}{\end{displaymath}}

\usepackage[sort&compress]{natbib} %Pruebon

\usepackage[dvips]{graphicx}
\def\shiftdown#1{#1\llap{\lower.04ex\hbox{#1}}}

\makeatletter
\def\cleardoublepage{\clearpage\if@twoside \ifodd\c@page\else
 \hbox{}
 \vspace*{\fill}
 \begin{center}

 \end{center}
 \vspace{\fill}
 \thispagestyle{empty}
 \newpage
 \if@twocolumn\hbox{}\newpage\fi\fi\fi}
\makeatother

\bibpunct[,]{[}{]}{,}{n}{,}{,}

\usepackage{dcolumn}
\usepackage{multirow}

 \def\tstrut{\vrule height2.5ex depth0pt width0pt} % used in tables
\usepackage{slashed} %PARA HACER BIEN LAS COSAS SLASHED

\usepackage{amsmath} %OJO, para que pueda saltar \allowdisplaybreaks

\begin{document}
\author{Jos\'e Mar\'{\i}a Verde Velasco}
\title{Static and dynamic properties of hadronic systems with heavy
  quarks b and c}
\date{}

\maketitle

%\printindex
\tableofcontents
%\listoffigures
%\listoftables
%-----------------------------------------------------------------------------------------------------
\chapter{Introduction}
\label{cha:introduction}

Quantum Chromodynamics (QCD) is widely accepted as the theory of the
strong interaction.  QCD is a gauge theory for the strong interaction
of quarks and gluons based on the SU(3) color group.  Quarks have been
with us since the 60's when the first three flavors, up ($u$), down
($d$) and strange ($s$), were proposed as members of the fundamental
representation for the approximate SU(3) flavor symmetry observed in
the hadron spectra.  In the simplest picture baryons are made up of
three quarks, while most mesons can be explained as made up of a quark
and an antiquark. Deep inelastic scattering (DIS) on the proton
confirmed this picture of hadrons as bound states of more fundamental
particles. The data supported the idea of charged spin-1/2 components
or partons inside the nucleon. Those partons were identified with the
quarks.

Quarks carry a new quantum number that was named color. The color
quantum number can take on three different values and was introduced
to solve the statistics problem associated with ground state baryons
with three equal quarks.  Experimental evidence for the color quantum
number comes from the analysis of the
$\sigma(e^+e^-\rightarrow\mathrm{hadrons})/\sigma(e^+e^-\to
\mu^+\mu^-)$ cross sections ratio or the $\pi^0\to\gamma\gamma$ decay
width. The absence of color will render the theoretical predictions a
factor of 3 and 9 respectively below experimental data. Hadrons do not
carry the color quantum number, and thus quarks have to combine in
such a way that hadrons are ``colorless'' or color singlets.

Quarks have never been observed as free particles and the consensus is
that they are confined in the physical colorless hadrons. While
confinement has never been proven in a rigorous way in QCD, there are
hints, obtained in  Monte Carlo simulations performed in a discretized
space-time (see, for example Ref. \cite{Bali:2000gf}), that QCD leads
to confinement.

In the 70's 't Hooft, Politzer~\cite{Politzer:1973fx}, Gross and
Wilczek~\cite{Gross:1973id,Gross:1973ju} found that in nonabelian
gauge theories the effective coupling constant decreases at large
momentum transfers or short distances while it increases at low
momentum transfers or large distances. Both features, called asymptotic
freedom and infrared slavery respectively, are necessary to describe
quark dynamics. While DIS data, where large momentum transfers are
involved, shows the existence of almost free partons inside the
nucleon, the absence of free quarks suggests confinement at large
distances.  Fritzsch and
Gell-Mann~\cite{Fritzsch:1972jv,Fritzsch:1973pi} proposed that quark
dynamics was described by a nonabelian gauge theory based on the color
quantum number, QCD. The appropriate gauge group was color
SU(3). SU(3) has a fundamental representation of dimension 3 that can
accommodate the three values for the quark color quantum number. This
representation is complex and the antiquark color states belong to the
complex conjugate of the fundamental representation. The combination
of three quarks or of a quark and an antiquark can give rise to color
states which are SU(3) color singlets.

The QCD gauge boson  is called gluon. The gluon appears in 8
different color states corresponding to the adjoint representation of
color SU(3). Gluons are not observed as free particles and as in the
quark case, we think they are bound in the physical hadrons. Evidence for
gluons comes from DIS data both from the observation of three jets
events, and from the fact that only half of the proton momentum is
carried by charged partons. The rest must be attributed to non charged
partons, the gluons.

Another important feature of QCD with three quark flavors, and which
is essential to understand its low energy spectrum, is chiral
symmetry. Chiral symmetry is an approximate symmetry (it is explicitly
broken by the quark masses) of the QCD Lagrangian. This symmetry is
not shared by the ground state and one talks of a spontaneously broken
symmetry. This property manifest itself in the existence of an octet
of almost-massless pseudoscalar Goldstone bosons.

With only three quark flavors, the theory of electroweak interactions
involving quarks leads to the existence of weak neutral currents with
$\Delta S=1$. Those processes were not observed experimentally. The
easiest way to avoid this unwanted feature is through the GIM
mechanism (from Glashow, Illiopoulos and Maiani~\cite{Glashow:1970gm}
) that required a new quark flavor, the charm
($c$)\footnote{References on earlier works on charm can be found
in~\cite{DeRujula:1975ge}. }.  This new flavor implied the existence of
new types of hadrons containing the charm quark or antiquark. The
$J/\Psi $ meson discovered in
1974~\cite{Aubert:1974js,Augustin:1974xw} was interpreted as a bound
state of the charm quark and its antiquark. Two years later came the
first observation of mesons with open
charm~\cite{Goldhaber:1976xn,Peruzzi:1976sv}. Two more quark flavors,
later named bottom ($b$) and top ($t$), were predicted in 1973 by
Kobayashi and Maskawa~\cite{Kobayashi:1973fv} in order to get a
realistic model of CP violation in weak theories. The discovery of the
$\tau$ lepton in 1974-1977 led further theoretical support to the
existence of the bottom and top quarks as the cancellation of the
triangle anomaly in the electroweak theory required as many quarks as
there were leptons (also that the former appeared in three colors).
The $\Upsilon$ meson discovered in 1977~\cite{herb:1977ek} was
interpreted as being a $b\bar b$ state. Top discovery had to await for
another 17 years when it was finally seen in proton-antiproton
collisions~\cite{Abe:1994st}.

In the large energy regime QCD can be treated perturbatively and
predictions of the theory have shown a remarkable agreement with
experiment. On the other hand at low energies QCD becomes
nonperturbative and it is not known how to solve it exactly. In this
low energy regime there are approaches which make approximations to the
exact solutions like QCD sum rules and Lattice QCD.

Another approach to QCD in the low energy regime is the use of
phenomenological models which are not directly derived from QCD but
include some of its basic properties.  Constituent quark models (CQM) are
among these. In CQM models hadrons are simply modeled, as bound states
of three valence quarks (baryons) or a quark and an antiquark
(mesons).  Those quarks are quasiparticle degrees of freedom with the
same quantum numbers as QCD quarks, differing from the latter in their
masses and in the fact that they could have an internal structure.

Among CQM, nonrelativistic quark models (NRQM) have shown to be
phenomenologically very successful. In these models constituent quarks
are treated nonrelativistically and they interact through potentials
that mimic QCD asymptotic freedom and confinement. In fact it was this
simple nonrelativistic picture that led to the necessity for the color
quantum number. The vast literature on NRQM calculations forbids to
quote here but a few works. An early success of NRQM calculations was
the satisfactory account of the magnetic moments of the octet
baryons~\cite{Close:1979bt}.  Baryon spectra was studied within the
NRQM by Isgur and Karl in a series of
papers~\cite{Isgur:1977ef,Isgur:1978xi,Isgur:1978xj,Isgur:1978wd,Isgur:1978xb,Isgur:1979be}
where it was shown that a model with nonrelativistic point-like quarks
moving in a flavor independent confining potential, perturbed by color
hyperfine interactions, was able to explain the main features of the
spectra.  Meson spectra was equally well described within
 the NRQM~\cite{Lucha:1991vn}.  Color hyperfine interactions among
quarks are generated through the one gluon exchange potential first
introduced in~\cite{DeRujula:1975ge}. The spin-spin part of this
potential explains in a simple way the difference in mass between
octet and decuplet baryons and between pseudoscalar and vector
mesons. The NRQM was extended to the two nucleon system in the 80's
being able to explain, in a more fundamental picture, the short range
nucleon-nucleon repulsion in S-wave scattering as due to one gluon
exchange between quarks with a simultaneous quark exchange between the
two three-quark
clusters~\cite{Oka:1981ri,Oka:1981rj,Faessler:1983yd}. Chiral symmetry
and its spontaneously breaking was also incorporated into the
NRQM~\cite{Manohar:1983md} and inter-quark potentials coming from
one-Goldstone boson exchanges naturally
appeared~\cite{Shimizu:1985mg,Maltman:1985mc,Fernandez:1986zn}. In a
linear realization of chiral symmetry with two quark flavors not only
the Golstone bosons (pions) but also their chiral partner (sigma) was
considered. The inclusion of the latter improved the theoretical
prediction of $NN$ phase
shifts~\cite{Obukhovsky:1990tx,Fernandez:1993hx} and deuteron
properties \cite{Valcarce:1994nr}.  Some authors advocated that the
one-gluon exchange part of the potential be discarded altogether in
favor of only Goldstone boson exchanges~\cite{Glozman:1995fu}. This
exclusive view had big problems and was heavily
critizied~\cite{Isgur:1999jv}. Most versions of the NRQM use a mixture
of both one-gluon and one-Golstone boson exchanges. More recently the
study of dibaryons~\cite{Buchmann:1995ek,Wagner:1995id,Mota:2001ee},
tribaryons~\cite{Mota:2001ee,Fernandez-Carames:2006ra} and
tetraquarks~\cite{Vijande:2003ki,Vijande:2006jf,Barnea:1} systems have
also been addressed in the NRQM. For a recent review on few baryon
systems studies see~\cite{Valcarce:2005em}.
Electromagnetic~\cite{Nag:1987cv,Giannini:1990pc,Buchmann:1991cy,Buchmann:1994bt,Buchmann:1996bd,Buchmann:1998yj,
Meyer:2001js,Julia-Diaz:2004eg} and weak
\cite{Beyer:1985iq,Liu:1995bu,Barquilla-Cano:2002cm,
Julia-Diaz:2004qr,Barquilla-Cano:2006ws} form factors have also been
analyzed in the NRQM and exchange current contributions to these
observables have been studied with
detail~\cite{Buchmann:1991cy,Buchmann:1994bt,Buchmann:1996bd,Buchmann:1998yj,
Meyer:2001js, Barquilla-Cano:2002cm,Barquilla-Cano:2006ws}.

In this thesis our interest is in the analysis, within the framework
of a NRQM, of static and dynamic properties, the latter related to
weak decays, of hadrons with one and two heavy quarks $c$ and/or
$b$. In systems with a heavy quark with mass much larger than the QCD
scale ($\Lambda_{QCD}$) a new symmetry known as heavy quark symmetry
(HQS)
arises~\cite{Nussinov:1986hw,Shifman:1986sm,Politzer:1988wp,Politzer:1988bs,Isgur:1989vq,Isgur:1989ed}. HQS
is an approximate SU($N_F$) symmetry of QCD, being $N_F$ the number of
heavy flavors ($c$, $b$,\dots), that appears in systems containing
heavy quarks with masses much larger than the typical quantities
($\Lambda_{QCD}, m_u, m_d,m_s,\dots)$ which set up the energy scale of
the dynamics of the remaining degrees of freedom. In that limit the
dynamics of the light quark degrees of freedom becomes independent of
the heavy quark flavor and spin. This is similar to what happens in
atomic physics where the electron properties are approximately
independent of the spin and mass of the nucleus (for a fixed nuclear
charge).  HQS can be cast into the language of an effective theory
(HQET)\cite{Georgi:1990um} that allows a systematic, order by order,
evaluation of corrections to the infinite mass limit in inverse powers
of the heavy quark masses. HQS and HQET
 have proved very useful tools to understand bottom and charm physics
and they have been extensively used to describe the dynamics of
systems containing a heavy $c$ or $b$
quark~\cite{Neubert:1993mb,Korner:1994nh}. For instance, all lattice
QCD simulations rely on HQS to describe bottom
systems~\cite{Booth:1993zb,Burford:1995fc,Bowler:1994zr,Flynn:1995dc,DelDebbio:1997kr}.
In the case of systems with two heavy quarks one can not directly
apply HQS.  It is known that the kinetic energy term, needed in those
systems to regulate infrared divergences, breaks heavy flavor
symmetry~\cite{Thacker:1990bm}. Still, there is a symmetry that
survives: heavy quark spin symmetry (HQSS)~\cite{Jenkins:1992nb}. This
symmetry amounts to the decoupling of the two heavy quark spins since
the spin--spin interaction vanishes for infinite heavy quark masses.

The constraints imposed by HQS and HQSS have not been systematically
employed in the context of nonrelativistic quark models. We intend to
study hadrons with one and two heavy $c$ and/or $b$ quarks in a NRQM
making full use of those constraints. For instance, those constraints
will simplify notably the resolution of the three-body problem in
baryons allowing for the use of a simple variational ansatz. A NRQM
treatment of those systems will comply with the model independent
predictions of HQS and HQSS in the infinite heavy quark mass limit, as
in that limit all spin-spin interactions terms involving heavy quarks
vanish. Another question is to what extend the deviations from the
infinite heavy quark mass limit in a NRQM calculation agree with the
predictions of HQET.

This thesis is organized as follows. In chapter \ref{cha:potential} we
briefly introduce the quark-quark potentials that have been used
throughout this work.

In chapter \ref{cha:BD} we study leptonic decays of pseudoscalars
$B,\,D$ and vector $B^*,\,D^*$ mesons, and $B\to D$ and $B\to D^*$
semileptonic decays. In the latter case full form factors are computed
and used to get differential and total decay widths. When possible
these form factors are improved using corrections from HQET. We make a
prediction for the value of the $|V_{cb}|$ Cabbibo-Kobayashi-Maskawa
(CKM) matrix element which compares well with experimental
determinations.  \\ Also in this chapter we will use partial
conservation of the axial current (PCAC) to relate axial current
matrix elements with the pion emission amplitude and in this way make
a prediction for the strong coupling constants $g_{BB^*\pi}$ and
$g_{DD^*\pi}$.

Exclusive semileptonic and nonleptonic two-meson decays of the $B_c^-$
meson have been analyzed in chapter \ref{cha:Bc}, where a fairly
exhaustive study of these decays is done. In semileptonic decays we
have excluded processes that involve a $b\to u$ transition at the
quark level where the NRQM fails due to the large recoil momenta and
the ignorance of resonance exchanges. Similarly in nonleptonic
two-meson decays we shall only consider channels with at least a
$c\bar c$ or $B$ final meson.

In chapter \ref{cha:dp_masas} we study baryons with two heavy
quarks. A procedure to solve the three-body problem in an approximate
way using a variational ansatz is presented. We obtain the spectrum
and wave functions that are further used to compute different static
properties. The infinite heavy quark mass limit of the wave functions
is studied and as expected they comply with HQSS model independent
predictions.

In chapter \ref{cha:dp_sl} we use the wave functions computed in the
previous chapter %\ref{cha:dp_masas}
 to study semileptonic decays of  doubly heavy
baryons. Full form factors, decay widths,  and angular asymmetries are studied.

Finally chapter \ref{cha:lambdapi} is devoted to the study of the
strong one pion decay of baryons with a heavy quark. As in chapter
\ref{cha:BD} PCAC is used to relate weak axial current matrix elements
with the pion emission amplitude.

\chapter{Quark quark potentials used in this work}
\label{cha:potential}

\section{Introduction}

In this chapter we give a brief account of the quark-quark potentials
that we use in this work to get the hadron wave functions.  We shall
use five different two-body phenomenological
  potentials.  One of these potentials (BHAD) has been taken from Ref.
\cite{Bhaduri:1981pn} and four others (AL1, AL2, AP1, AP2) from
Refs. \cite{Semay:1994ht, Silvestre-Brac:1996bg}. The use of different
potentials will allow us to check the sensitivity of our results to
the inter-quark interaction.

All these potentials share a common structure that includes a
confining term plus Coulomb and hyperfine terms coming from the
one-gluon exchange potential. They differ in the form factor used in
the hyperfine and $ 1/r$ terms and/or in the power chosen for the
confining term. All of them neglect the tensor and spin-orbit pieces
also present in the one-gluon exchange
potential~\cite{DeRujula:1975ge} on the account that they are not
essential to describe the global features of hadron spectroscopy. As a
consequence both the total orbital angular momentum $L$ and the total spin $S$
are separately good quantum numbers. Chiral symmetry is not
incorporated in these potentials and thus no terms generated by
Goldstone boson exchange are included. Explicit expressions appear in
what follows.

\section{BHAD, AL1, AL2, AP1 and  AP2 potentials}
For the $q\overline q$ case they can be given in  the general form

\begin{eqnarray}
\hspace*{-1.cm}V_{ij}^{q\bar q}(r) = &&\hspace{-.5cm}-\frac{\kappa\left ( 1 - e^{-r/r_c}\right )}{r}
+\lambda r^p - \Lambda 
\nonumber\\  &&\hspace{-.5cm}
+ \left\{a_0\frac{\kappa}{m_im_j}
\frac{e^{-r/r_0}}{rr_0^2} + \frac{2\pi}{3m_im_j}\kappa^\prime \left (
1 - e^{-r/r_c}\right ) \frac{e^{-r^2/x_0^2}}{\pi^\frac32 x_0^3} \right
\}\vec{\sigma}_i\vec{\sigma}_j  \label{eq:phe}
\end{eqnarray}
where $i,j=(l(u.d),s,c,b)$, $\vec{\sigma}$ are the spin Pauli matrices, $m_i$ the quark masses and
\begin{equation}
\label{eq:x0}
x_0(m_i,m_j) = A \left ( \frac{2m_im_j}{m_i+m_j} \right )^{-B} 
\end{equation}
Parameters for the different potentials are summarized 
in Table~\ref{tab:param}. 
\begin{table}[h!!!]
\begin{center}
\begin{tabular}{c|lllll}
                          & BHAD     & AL1    & AL2    & AP1     & AP2  \\\hline
$\kappa$                  & 0.52   & 0.5069 & 0.5871 & 0.4242  & 0.5743  \\
$r_c$[GeV$^{-1}$]         & 0.     & 0.     & 0.1844 & 0.      & 0.3466 \\ 
$p$                       & 1      & 1      &   1    & 2/3     & 2/3   \\
$\lambda$ [GeV$^{(1+p)}$] & 0.186  & 0.1653 & 0.1673 & 0.3898  & 0.3978 \\ 
$\Lambda$ [GeV]           & 0.9135 & 0.8321 & 0.8182 & 1.1313  &
1.1146 \\
$a_0$                     & 1      &   0    &   0    &   0     &   0 \\
$r_0$ [GeV$^{-1}$]        & 2.305  &   \hspace{.2cm} ---  &   \hspace{.2cm} ---  &  \hspace{.2cm} ---    & \hspace{.2cm} --- \\  
$\kappa^\prime$           & 0.     & 1.8609 & 1.8475 & 1.8025  &
1.8993 \\
$m_u=m_d$ [GeV]           & 0.337  & 0.315  & 0.320  & 0.277   & 0.280
\\
$m_s$  [GeV]              & 0.600  & 0.577  & 0.587  & 0.553   & 0.569
\\
$m_c$  [GeV]              & 1.870  & 1.836  & 1.851  & 1.819   &
1.840\\
$m_b$ [GeV]               & 5.259  & 5.227  & 5.231  & 5.206   & 5.213\\
$B$                       & \hspace{.2cm} ---    & 0.2204 & 0.2132 & 0.3263  &
0.3478 \\
$A$ [GeV$^{B-1}$]         &\hspace{.2cm} ---    & 1.6553 & 1.6560 & 1.5296  &
1.5321\\\hline  
\end{tabular}
\end{center}
\caption{Parameters for the BHAD, AL1, AL2, AP1 and AP2 potentials  taken from
  Refs.~\cite{Bhaduri:1981pn,Silvestre-Brac:1996bg}} 
% en el de Semay
% la not es distinta 
\label{tab:param}
\end{table}
Note that the $a_0$ and $\kappa^\prime$ terms are mutually exclusive.  The
power $p$ chosen for the confining term is $p=1$ for the BHAD, AL1,
AL2 potentials. This linear confinement is suggested by lattice
results \cite{Gutbrod:1983yi}.  The AP1 and AP2 potentials use instead
$p=\frac{2}{3}$ which is needed, in a nonrelativistic approach, to get
the correct asymptotic Regge trajectories for large angular momentum
mesons~\cite{FabreDeLaRipelle:1988zr}. The hyperfine term comes from
the one-gluon exchange potential and the original $\delta(\vec r)$
dependence has been smeared out in order that nonperturbative
calculations are possible. The BHAD potential uses a Yukawa form
factor, while AL1, AL2, AP1 and AP2 potentials use a gaussian form
factor. In these latter cases the depth and width of the gaussian
depends on the constituent quark masses allowing for a better
reproduction of the hyperfine splitting of heavy mesons. The
phenomenological formula in Eq.~(\ref{eq:x0}), originally proposed in
\cite{Ono:1982ft}, seems to be well suited for that
purpose~\cite{Narodetsky:1992zy}. The AL2 and AP2 potentials include a
further form factor in the $1/r$ and hyperfine terms with a shape
given by $(1-\exp(-r/r_c))$. This factor simulates asymptotic freedom
and its inclusion seems to be more important for heavy mesons than for
light ones.  Finally note that the $\kappa$ and $\kappa^\prime$
parameters in the AL1, AL2, AP1 and AP2 potentials would be the same
if one simply replaced the original $\delta(\vec r)$ function in the hyperfine term 
by the gaussian
form factor. The fact that they are different allows for a great
improvement in the results~\cite{Narodetsky:1992zy}. This is also true
for the BHAD potential where the strength for the hyperfine term, where the $\delta(\vec r)$ function has been
replaced by a Yukawa form factor,  is six times larger than the one used in the $1/r$ term.
This, in effect, is allowing for  sources of hyperfine interaction other  than 
the one-gluon exchange.

All free parameters in the potentials were adjusted in the original
works to reproduce the light and heavy-light meson spectra
(AL1,AL2,AP1,AP2), or to fit the low-lying levels of charmonium
(BHAD). In this latter case $m_b$ and $m_s$ were adjusted to reproduce
the ground state of the $\Upsilon$ and $\phi$ mesons respectively
while the light quark mass $m_l$ was chosen from magnetic moments
considerations.

To obtain the quark-quark potential we shall use, as has been done
in Refs.~\cite{Bhaduri:1981pn,Semay:1994ht,
 Silvestre-Brac:1996bg}, the prescription
\bea
 V_{ij}^{qq}&=&\frac{V_{ij}^{q\bar{q}}}{2} 
\eea 
This relation assumes the potential has a global color dependence given by
 $\vec{\lambda}^c_i\vec{\lambda}^c_j$ , where $\vec\lambda^c$ stands
for the Gell-Mann matrices in the
 color space.

\chapter{Leptonic and semileptonic decays of mesons with a heavy
  quark}
\chaptermark{Decays of $B$,$D$,$B^*$ and $D^*$}
\label{cha:BD}
%EL CAPITULO DE LAS BD%

\section{Introduction}

This chapter is devoted to the calculation of different weak
observables of pseudoscalar and vector mesons with a heavy quark. These
weak observables are of great interest since they help to probe the
quark structure of the involved hadrons, and also provide information
to determinate the absolute value of elements of the
CKM  matrix.

In Ref.~\cite{Albertus:2004iw}, in which a similar model was
used to study  the  $\Lambda_b^0\to \Lambda^+_c\,l^-\bar{\nu}_l$ and 
$\Xi_b^0\to \Xi^+_c\,l^-\bar{\nu}_l$ reactions, it  was shown that a direct
nonrelativistic calculation does not meet HQET constraints.
This problem  was
solved imposing HQET relations among form factors. That led to a
determination  of the $|V_{cb}|$  CKM matrix element in good agreement
with experimental results. 

One would expect this problem will also show up in this calculation,
so that 
constraints coming from HQET will be included at some point in the
calculation in order  to improve the results.

In the case of mesons with a heavy quark, HQS leads to many model independent
predictions. For instance in the HQS limit the masses of  the lowest lying (s-wave) 
pseudoscalar and vector mesons with a heavy quark  are degenerate. 
Nonrelativistic quark models  satisfy this constraint:  the spin-spin 
terms, that distinguishes vector from pseudoscalar, are zero if the mass of
 the heavy quark goes to infinity. HQS also predicts that the
  masses and leptonic decay constants of pseudoscalars and vector mesons
 are related via 
\bea
f_P\,m_P\Bigg|_{HQS}=f_V\,m_V\Bigg|_{HQS}
\eea
 relation that is also satisfied in the quark model in the HQS limit.
 If one looks now at the form factors for the semileptonic $B\to D$ and $B\to
 D^*$  decays HQS predicts  relations among different form factors
 that are also met by the quark model in the HQS limit. The question is
 to what extent the deviations from the HQS limit evaluated in the quark model
 agree with the constraints deduced from HQET. In addition we will make use of
 these HQET constraints to improve the quark model results and thus come up with
 reliable predictions.

Apart from lattice QCD and QCD sum rules (QCDSR) calculations with which we
shall compare our results, and that will be quoted in the following, the different observables analyzed in this work 
have been studied in
the quark model starting with the pioneering work of
 Ref.~\cite{Isgur:1989qw,Isgur:1988gb,Scora:1995ty} within a non relativistic version, to continue with
 different versions of the relativistic quark model  applied to the
 determination of decay 
constants~\cite{Capstick:1989ra,Barik:1993aw,Hwang:1995uy,Hwang:1995vt,Micu:1996hj,AbdEl-Hady:1997rj,Morenas:1997rx,Melikhov:2000yu,Wang:2004xs,DeVito:2004zs}, 
form factors and differential decay
widths~\cite{Melikhov:2000yu,DeVito:2004zs,Jaus:1989au,Jaus:1991cy,Jaus:1996np,Faustov:1995xc,Barik:1996xf,Ishida:1997ft,Ebert:1997mg,Ebert:2002qa,Ivanov:1999tv}, 
Isgur-Wise
functions~\cite{DeVito:2004zs,Close:1993yk,Kiselev:1994ay,LeYaouanc:1995wv,Morenas:1996bn,Morenas:1997nk,Deandrea:1998uz,Choi:1999nu,Krutov:2000kt} 
or strong coupling 
constants~\cite{Melikhov:2000yu,Deandrea:1998uz,Miller:1988tz,O'Donnell:1994ey,Colangelo:1994jc,Becirevic:1999fr}.

For calculations in this chapter we have used experimental meson masses taken from Ref. \cite{Eidelman:2004wy}.

\section{Meson states}
For a meson $M$ we use the following expression for the wave function
in the Fock-space representation.
\begin{eqnarray}
\label{BD_wf}
&& \hspace{-2.2cm}
\left|{M,\lambda\,\vec{P}}\,\right\rangle_{NR}
\,=\,\int d^3p \sum_{\alpha_1,\alpha_2}\hat{\phi}^{(M,\lambda)}_{\alpha_1,\alpha_2}(\,\vec{p}\,)
\nonumber\\ && \hspace{-.5cm}
\frac{(-1)^{\frac{1}{2}-s_2}}{(2\pi)^{\frac{3}{2}}
\sqrt{2E_{f_1}(\vec{p}_1)2E_{f_2}(\vec{p}_2)}}\ 
%\ \nonumber\\
%&&\hspace{3cm}
\left|\ q,\ \alpha_1\
\vec{p}_1=\frac{m_{f_1}}{m_{f_1}+m_{f_2}}\vec{P}-\vec{p}\ \right\rangle
\nonumber \\&&\hspace{4.7cm}
\left|\ \bar{q},\ \alpha_2\ \vec{p}_2=\frac{m_{f_2}}{m_{f_1}+m_{f_2}}\vec{P}+\vec{p}\
 \right\rangle
\end{eqnarray}
where $\vec{P}$ stands for the meson three momentum and $\lambda$ represents
the spin projection in the meson center of mass. $\alpha_1$ and $\alpha_2$ represent
the quantum numbers of spin (s), flavor (f) and color (c)%, $\alpha\equiv
%(s,f,c)$, of the
\begin{equation}
\alpha\equiv(s,f,c)
\end{equation}
of the
quark and the antiquark, while $E_{f_1}(\vec{p}_1),\,\vec{p}_1$ and $E_{f_2}(\vec{p}_2),\,\vec{p}_2$ are their 
respective energies and three-momenta. $m_f$ is the mass of the quark or
antiquark with flavor $f$. The factor $(-1)^{\frac{1}{2}-s_2}$ 
is included in order that the antiquark spin states have the correct relative
phase\footnote{Note that under charge conjugation $({\cal C})$  quark and
antiquark creation operators are related via \hbox{${\cal
C}\,c^{\dagger}_\alpha(\,\vec{p}\,)\,{\cal C}^{\dagger}= (-1)^{\frac{1}{2}-s}\,d^{\dagger}_\alpha(\,\vec{p}\,)
$}. This means that the antiquark states with the correct spin relative phase 
are not $d^{\dagger}_\alpha(\,\vec{p}\,)\,|0\rangle=
|\,\bar{q},\ \alpha\ \vec{p}\ \rangle$ but are instead given by
$(-1)^{\frac{1}{2}-s}\,d^{\dagger}_\alpha(\,\vec{p}\,)\,|0\rangle=
(-1)^{\frac{1}{2}-s}\,|\,\bar{q},\ \alpha\ \vec{p}\ \rangle$.}.
The normalization of the quark and antiquark
states is 
\begin{eqnarray}
\left\langle\ \alpha^{\prime}\ \vec{p}^{\ \prime}\,|\,\alpha\ \vec{p}\,
\right\rangle=\delta_{\alpha^{\prime},\ \alpha}\, (2\pi)^3\, 2E_f(\vec{p})\,\delta(
\vec{p}^{\ \prime}-\vec{p}\,)
\end{eqnarray}
Furthermore, $\hat{\phi}^{\,(M,\lambda)}_{\alpha_1,\alpha_2}(\,\vec{p}\,)$ is the
momentum space wave
function  for the relative motion of the quark-antiquark
system. 
Its normalization is given  by
\begin{equation}
\int\, d^3p\ \sum_{\alpha_1\,\alpha_2} \left(
\hat{\phi}^{\,(M,\lambda')}_{\alpha_1,\,\alpha_2}(\,\vec{p}\,)
\right)^* \hat{\phi}^{\,(M,\lambda)}_{\alpha_1,\,\alpha_2}(\,\vec{p}\,)
=\delta_{\lambda',\,\lambda}
\end{equation}
and, thus,  the normalization of our meson states is 
\begin{equation}
\label{BD:eq01}
{}_{\stackrel{}{NR}}\left\langle\, {M,\lambda'\,\vec{P}^{\,\prime}}\,|\,{M,\lambda
\,\vec{P}}\,\right\rangle_{NR}
=\delta_{\lambda',\,\lambda}\,(2\pi)^3\,\delta(\vec{P}^{\,\prime}-\vec{P}\,)
\end{equation}
For the particular case of ground state pseudoscalar ($P$) and vector ($V$) mesons we can assume the
orbital angular momentum to be zero and then we will have
\begin{eqnarray}
\hat{\phi}^{\,(P)}_{\alpha_1,\,\alpha_2}(\,\vec{p}\,)
&=&\frac{1}{\sqrt{3}}\,\delta_{c_1,\,c_2}\
\hat{\phi}^{\,(P)}_{(s_1,\,f_1),\,(s_2,\,f_2)}(\,\vec{p}\,)\nonumber\\
&=&\frac{1}{\sqrt{3}}\,\delta_{c_1,\,c_2}\ (-i)
\ \hat{\phi}^{\,(P)}_{f_1,\,f_2}(\,|\vec{p}\,|)\ 
Y_{00}(\,\widehat{\vec{p}}\ )\  \left(\left.\frac{1}{2},\frac12,0\right|s_1,s_2,0\right)\nonumber\\
\hat{\phi}^{\,(V,\,\lambda)}_{\alpha_1,\,\alpha_2}(\,\vec{p}\,)
&=&\frac{1}{\sqrt{3}}\,\delta_{c_1,\,c_2}\
\hat{\phi}^{\,(V,\,\lambda)}_{(s_1,\,f_1),\,(s_2,\,f_2)}(\,\vec{p}\,)\nonumber\\
&=&\frac{1}{\sqrt{3}}\,\delta_{c_1,\,c_2}\ (-1)\ 
\hat{\phi}^{\,(V)}_{f_1,\,f_2}(\,|\vec{p}\,|)\
Y_{00}(\,\widehat{\vec{p}}\ )\
\left(\left.\frac12,\frac12,1\right|s_1,s_2,\lambda\right)
\end{eqnarray}
where $(j_1,j_2,j_3|m_1,m_2,m_3)$ is a Clebsch-Gordan coefficient,
$Y_{00}=1/\sqrt{4\pi}$ is the $l=m=0$ spherical harmonic, and
$\hat{\phi}^{\,(M)}_{f_1,\,f_2}(|\,\vec{p}\,|)$ is the Fourier
transform of the radial coordinate space wave function.  The phases
are introduced for later convenience.

To evaluate the coordinate space wave function
we shall use the  potentials presented in chapter \ref{cha:potential}.
This will provide us with a spread of
results that we will consider, and quote, as a theoretical error to
the averaged value that we will quote as our central result.
 Another source of theoretical uncertainty, that we can not account
for, is the use of 
nonrelativistic kinematics in the evaluation of the orbital wave
functions and the construction of our states defined  above.
While this is a very good approximation for mesons with two heavy
quarks (as the  $B_c$ that will be studied in chapter \ref{cha:Bc}), it is not that
good for mesons with a light quark. That notwithstanding note that any
nonrelativistic quark model has free parameters in the inter-quark interaction
that are fitted to experimental data. In that sense we think that at least 
part of the ignored relativistic effects are included in an effective way 
in their fitted values.

\section[Leptonic decay of pseudoscalar and vector $B$ and $D$
  mesons]{Leptonic decay of pseudoscalar and vector $B$ and $D$
  mesons\sectionmark{Leptonic  decay}}
\sectionmark{Leptonic  decay}
\label{sect:leptonic}
In this section we will consider the purely leptonic decay of
pseudoscalars 
($B$, $D$) and vector ($B^*$, $D^*$) mesons.  The charged
 weak current operator for a specific pair of quark flavors $f_1$ and
 $f_2$ reads
\begin{equation}
J^{f_1\,f_2}_\mu(0)=J^{f_1\,f_2}_{V\,\mu}(0)-J^{f_1\,f_2}_{A\,\mu}(0)
=\sum_{(c_1,\,s_1),\,(c_2,\,s_2)}\,\delta_{c_1,\,c_2}\ \overline{\Psi}_{\alpha_1}(0)\gamma_\mu
(1-\gamma_5)\Psi_{\alpha_2}(0)
\end{equation}
with $\Psi_{\alpha_1}$ a quark field of a definite spin, flavor and
color. The hadronic matrix elements involved in the processes can be
parametrized in terms of a unique pseudoscalar $f_P$ or vector $f_V$
decay constant as
\begin{eqnarray}
\label{fpv}
&&\hspace*{-.7cm}
\left\langle 0 \right|\, J_\mu^{f_1\,f_2}(0)\,\left|\, P, \vec{P}\,\right\rangle
=\left\langle 0 \right|\, -J_{A\,\mu}^{f_1\,f_2}(0)\,\left|\, P, \vec{P}\,
\right\rangle= -iP_\mu\,f_P \nonumber\\
%\end{eqnarray}
%\begin{eqnarray}
%\label{fv}
&&\hspace*{-.7cm}
\left\langle 0 \right|\, J_\mu^{f_1\,f_2}(0)\,\left|\, V, \lambda\,\vec{P}\,
\right\rangle
=\left\langle 0 \right|\, J_{V\,\mu}^{f_1\,f_2}(0)\,\left|\, V,\,\lambda\, 
\vec{P}\,
\right\rangle= \varepsilon_{(\lambda)\mu}(\,\vec{P}\,)m_Vf_V
\end{eqnarray}
where the meson states are normalized such that
\begin{equation}
\left\langle\, {M,\lambda'\,\vec{P}^{\,\prime}}\,|\,{M,\lambda
\,\vec{P}}\,\right\rangle
=\delta_{\lambda',\,\lambda}\,(2\pi)^3\,2E_M(\vec{P})\,\delta(\vec{P}^{\,\prime}-\vec{P}\,)
\end{equation}
with $E_M(\vec{P})$ the energy of the meson.

In the first of Eqs.(\ref{fpv}) $P_\mu$ is the four-momentum of the
meson, while in the second $m_V$ and
$\varepsilon_{(\lambda)}(\,\vec{P}\,)$ are the mass and the
polarization vector of the vector meson. 
In both cases $f_1$ and $f_2$
are the flavors of the quark and the antiquark that make up the meson.
Expressions for the different
polarization vectors used in this memory are presented in appendix \ref{app:epsilon}.

Concerns about the experimental determination of the pseudoscalars
decay constants have been raised in Ref.~\cite{Burdman:1994ip}. There
the effect of radiative decays was analyzed concluding that for $B$
mesons the decay constant determination could be greatly affected by
radiative corrections. In the vector sector, and as rightly pointed
out in Ref.~\cite{Lellouch:1994dy}, the vector decay constants are not
relevant from a phenomenological point of view since $B^*$ and $D^*$
will decay through the electromagnetic and/or strong interaction. They
are nevertheless interesting as a mean to test HQS relations.

%\noindent
For  mesons at rest we will obtain
\begin{eqnarray}
&&f_P=\frac{-i}{m_P}\,\left\langle 0 \right|\, J_{A\,0}^{f_1\,f_2}(0)
\,\left|\, P, \vec{0}\,\right\rangle\nonumber\\
&&f_V=\frac{-1}{m_V}\,\left\langle 0 \right|\, J_{V\,3}^{f_1\,f_2}(0)
\,\left|\, V,\,0\, 
\vec{0}\,
\right\rangle
%&&f_V=\frac{-M_V^2}{\,\left\langle 0 \right|\, J_{V\,3}^{f_1\,f_2}(0)
%\,\left|\, V,\,0\, 
%\vec{0}\,
%\right\rangle}
\end{eqnarray}
with $m_P$ the mass of the pseudoscalar meson.
 In our model, and due to the different normalization of our meson states, we shall 
 evaluate the decay constants as
\begin{eqnarray}
&&f_P=-i\,\sqrt{\frac{2}{m_P}}\,\left\langle 0 \right|\, J_{A\,0}^{f_1\,f_2}(0)
\,\left|\, P, \vec{0}\,\right\rangle_{NR}\nonumber\\
&&f_V=-\sqrt{\frac{2}{m_V}}\,\left\langle 0 \right|\, J_{V\,3}^{f_1\,f_2}(0)
\,\left|\, V,\,0\, 
\vec{0}\,
\right\rangle_{NR}
%&&f_V=\frac{-M_V^{3/2}}{\sqrt2\,\left\langle 0 \right|\, J_{V\,3}^{f_1\,f_2}(0)
%\,\left|\, V,\,0\, 
%\vec{0}\,
%\right\rangle_{NR}}
\end{eqnarray}
The  corresponding matrix elements are given in appendix~\ref{app:lep}.

The results that we obtain for the different decay constants appear
 in Tables~\ref{tab:fp} and ~\ref{tab:fv}. Starting with $f_D$ and $f_{D_s}$ our
 results are larger that the ones obtained in the lattice by the UKQCD
 Collaboration~\cite{Bowler:2000xw} or the ones evaluated
 using QCD spectral sum rules (QSSR)~\cite{Narison:2001pu,Narison:book}. Not only the independent values are larger but
 also the ratio $f_{D_s}/f_D$ is larger in our case. On the other hand our results
 are in better agreement with other  lattice
 determinations~\cite{Aubin:2005ar,Wingate:2003gm}. They also  compare very well
 with the experimental measurements of $f_D$ and $f_{D_s}$ in
  Refs.~\cite{Artuso:2005ym,Chadha:1997zh,Heister:2002fp,Abbiendi:2001nb,Alexandrov:2000ns,Ablikim:2004ry}.
% being our $f_{D_s}/f_D$
% ratio  in very good agreement with the value obtained using recent CLEO Collaboration
% data~\cite{Bonvicini:2004gv,Chadha:1997zh}.        %OBSOLETO 
As for $f_B$ and $f_{B_s}$, we find a very good
 agreement in the case of $f_{B_s}$ between  our results and 
 the ones obtained in the lattice or with the use of QSSR. For $f_B$ our result
 is smaller and then also our ratio
$f_{B_s}/f_B$ is larger. 
\begin{table}[p]
\begin{center}
{\footnotesize
\begin{tabular}{c|ccc}
 & $f_D$ [MeV] & $f_{D_s}$  [MeV]  & $f_{D_s}/f_D$ \\
\hline
& & &\\
This work & $243^{+21}_{-17}$  & $341^{+7}_{-5}$  & $1.41^{+0.08}_{-0.09}$ \\ \hline\\
%PDG~\protect\cite{pdg04} & $300^{+180+80}_{-150-40}$~\protect\cite{bai98}  &
%$266\pm32$ &  ---\\
Experimental data & & & \\
& & &\\
CLEO & $222.6\pm16.7^{+2.8}_{-3.4}$~\protect\cite{Artuso:2005ym} %CAMBIADO
&$280 \pm19\pm28\pm34$~\protect\cite{Chadha:1997zh} & ---\\
ALEPH~\cite{Heister:2002fp} & --- & $285\pm19\pm40$ & ---\\
OPAL~\cite{Abbiendi:2001nb} & --- & $286\pm44\pm41$ &---\\
BEATRICE~\cite{Alexandrov:2000ns} & --- &$323\pm44\pm 12\pm34$ &--- \\
E653~\cite{Kodama:1996xq} & ---&$194\pm35\pm20\pm14$ &---\\
BES~\cite{Ablikim:2004ry} & $371^{+129}_{-119}\pm 25$ & --- & ---\\ \hline\\
Lattice data & & & \\
& & & \\
UKQCD~\protect\cite{Bowler:2000xw} & $206(4)^{+17}_{-10}$ & $229(3)^{+23}_{-12}$ &
$1.11(1)^{+1}_{-1}$\\
Fermilab Lattice~\cite{Aubin:2005ar} & $201\pm3\pm6\pm9\pm13$ & 
$249\pm3\pm7\pm11\pm10$&---\\ %CAMBIADO
M. Wingate {\it et al.}~\cite{Wingate:2003gm} & ---& $290\pm20\pm29\pm29\pm6$   &---\\
\hline\\
QCD Spectral Sum Rules & & &\\
& &   &\\
S. Narison:~\protect\cite{Narison:2001pu,Narison:book} &     $203\pm23$ & $235\pm24$&$1.15\pm0.04$\\
\hline
& & &\\
& & &\\
 & $f_B$ [MeV] & $f_{B_s}$  [MeV]  & $f_{B_s}/f_B$ \\
\hline
& & &\\
This work & $155^{+15}_{-12}$  & $239^{+9}_{-7}$ &
$1.54^{+0.09}_{-0.08}$ \\ \hline\\
Experimental data \\ \\
BELLE~\protect\cite{Ikado:2006un} & $229^{+36}_{-31}\,^{+34}_{-37}$
&---&---
\\
\hline \\
Lattice data &  & &\\
& & & \\
UKQCD~\protect\cite{Bowler:2000xw} & $195(6)^{+24}_{-23}$& $220(6)^{+23}_{-18}$ &
$1.13(1)^{+1}_{-1}$\\
M. Wingate {\it et al.}~\cite{Wingate:2003gm} & ---& $260\pm7\pm26\pm8\pm5$   &---\\
Lattice world averages & $200\pm30$~\cite{Bernard:2000ki} & $230\pm30$
~\cite{Hashimoto:2004hn} &  $1.16\pm0.04$~\cite{Bernard:2000ki}\\ \hline\\
QCD Spectral Sum Rules & & &\\ %AKI
& &   &\\
S. Narison~\protect\cite{Narison:2001pu,Narison:book} & $207\pm21$ & $240\pm24$& $1.16\pm0.04$\\
\end{tabular}
}
\end{center}
\caption{ Pseudoscalar $f_P$ decay constants  for $B$ and $D$ mesons}\vspace{1cm}
\label{tab:fp}
\end{table}
\begin{table}[h!!!]
\begin{center}
\begin{tabular}{c|ccc}
 & $\tilde{f}_{D^*}$  & $\tilde{f}_{D^*_s}$    &
 $\tilde{f}_{D^*}/\tilde{f}_{D^*_s}$ \\
\hline
& & &\\
This work & $9.1^{+0.9}_{-0.9}$  & $6.5^{+0.3}_{-0.4}$ & $1.41^{+0.06}_{-0.05}$ \\
UKQCD~\cite{Bowler:2000xw} & $8.6(3)^{+5}_{-9}$ & $8.3(2)^{+5}_{-5}$&
$1.04(1)^{+2}_{-2}$\\
\hline
& & &\\
 & $\tilde{f}_{B^*}$ [MeV] & $\tilde{f}_{B^*_s}$   [MeV]  & $\tilde{f}_{B^*}
 /\tilde{f}_{B^*_s}$ \\
\hline
& & &\\
This work & $35.6^{+3.4}_{-3.4}$  & $23.0^{+1.0}_{-1.5}$ & $1.55^{+0.07}_{-0.06}$ \\
UKQCD~\cite{Bowler:2000xw} & $28(1)^{+3}_{-4}$ & $25(1)^{+2}_{-3}$&
$1.10(2)^{+2}_{-2}$\\
\end{tabular}
\end{center}
\caption{ $\tilde{f}_V=m_V/f_V$   for $B^*$ and $D^*$ mesons}
\label{tab:fv}
\end{table}
\noindent

For the vector meson decay constants we obtain the values
\begin{eqnarray}
f_{D^*}=223^{+23}_{-19}\, {\mathrm MeV} \hspace{1cm} f_{D_s^*}=326^{+21}_{-17}\, {\mathrm MeV}\nonumber\\
f_{B^*}=151^{+15}_{-13}\, {\mathrm MeV} \hspace{1cm} f_{B_s^*}=236^{+14}_{-11}\, {\mathrm MeV}
\end{eqnarray}
which are very much the same as the values obtained
for the decay constants of their pseudoscalar counterparts. This almost equality
of pseudoscalar and vector decay constants is expected in HQS  in the limit
where the heavy quark masses go to infinity where one would have~\cite{Isgur:1989vq,Isgur:1989ed}
\begin{eqnarray}
f_V\,m_V=f_P\,m_P \hspace{1cm},\hspace{1cm} m_V=m_P
\end{eqnarray}
Our decay constants  satisfy the above relation within  2\%. On the
other hand 
UKQCD lattice data show deviations as large as 20\% for D mesons~\cite{Bowler:2000xw}.
 
In order to compare  the values of the vector decay constants with lattice data
 from Ref.~\cite{Bowler:2000xw} we give in 
Table~\ref{tab:fv} the quantity 
 $\tilde{f}_V=m_V/f_V$. We find good agreement  for 
$\tilde{f}_{D^*}$ and
$\tilde{f}_{B_s^*}$ but not so much for the other two. Also our ratios
$\tilde{f}_{D^*}/\tilde{f}_{D_s^*}$ and $\tilde{f}_{B^*}/\tilde{f}_{B_s^*}$
are larger than the ones favored by lattice calculations. 

On the other hand the
ratio
\begin{eqnarray}
\frac{f_{B^*}\sqrt{m_B}}{f_{D^*}\sqrt{m_D}}=1.138^{+0.011}_{-0.008}
\end{eqnarray}
is in very good agreement with the expectation in Ref.~\cite{Burdman:1993es} where
they would get
$1.05\sim1.20$ for that ratio.

%--------------------------------------------------------------
%--------------------------------------------------------------

\section[Semileptonic decay of $B$ into $Dl\bar{\nu}$ and $D^*l\bar{\nu}$]
{Semileptonic decay of $B$ into $Dl\bar{\nu}$ and
$D^*l\bar{\nu}$\sectionmark{Semileptonic decay of $B$ into
$D\lowercase{l}\bar{\nu}$ and $D^*\lowercase{l}\bar{\nu}$}}
\sectionmark{Semileptonic decay of $B$ into $D\lowercase{l}\bar{\nu}$
and $D^*\lowercase{l}\bar{\nu}$}
\label{sect:semileptonic}
In this case the hadronic  matrix elements are parametrized as
%
%CAMBIADAS LAS EPSILONS 30-01
%
\begin{eqnarray}
\frac{\left\langle\, D,\,\vec{P}^\prime\,\left|\, J^{c\,b}_\mu(0)\,
\right| \, B,\,\vec{P}\right\rangle}{\sqrt{m_B
m_D}}&=&\frac{\left\langle\, D,\,\vec{P}^\prime\,\left|\, J^{c\,b}_{V\,\mu}(0)\,
\right| \, B,\,\vec{P}\right\rangle}{\sqrt{m_B
m_D}}\nonumber\\ &=&
\left(v+v^\prime\right)_\mu\,h_+(w)+\left(v-v^\prime\right)_\mu\,h_-(w)
\end{eqnarray}
\begin{eqnarray}
\label{bd*}
\frac{\left\langle\, D^*,\,\lambda\,\vec{P}^\prime\,\left|\, J^{c\,b}_\mu(0)\,
\right| \, B,\,\vec{P}\right\rangle}{\sqrt{m_B
m_{D^*}}}&=&-\varepsilon_{\mu\nu\alpha\beta}\,\left(\varepsilon_{(\lambda)}^\nu
(\,\vec{P}^\prime\,)\right)^*\,v^\alpha\,v^{\prime\,
\beta}\,h_V(w)\nonumber\\
&&\hspace{-.3cm}-\, i\, \left(\varepsilon_{(\lambda)\mu}(\,\vec{P}^\prime\,)\right)^*\,(w+1)
\,h_{A_1}(w)\nonumber\\
&&\hspace{-.3cm}+\, i\, \left(\varepsilon_{(\lambda)\,
}(\,\vec{P}^\prime\,)\right)^*\cdot v\,     \left( 
v_\mu\,h_{A_2}(w)+v^\prime_\mu\,h_{A_3}(w)\right)
\nonumber \\
\end{eqnarray}
where $v=P/m_B$ and $v^\prime=P^\prime/m_{D,\,D^*}$  are
the four velocities of the initial $B$ and final $D,\,D^*$ mesons, 
$w=v\cdot v^\prime\,$ \footnote{ $w$ is related to the four momentum transferred
 square
$q^2$ via  
\bea
q^2&=&m_B^2+m_{D,\,D^*}^2-2\ w\, m_B\,m_{D,\,D^*}
\eea
} and
$\varepsilon_{\mu\nu\alpha\beta}$ is the fully antisymmetric tensor with 
$\varepsilon^{0123}=+1$.

In the limit of infinite heavy quark masses $m_c,\,m_b\to \infty$ 
HQS reduces the six form factors to  a unique universal function
$\xi(w)$ known as the Isgur-Wise function~\cite{Isgur:1989vq,Isgur:1989ed}
\begin{eqnarray}
&&h_+(w)=h_{A_1}(w)=h_{A_3}(w)=h_{V}(w)=\xi(w)\\
&&h_{-}(w)=h_{A_2}(w)=0
\end{eqnarray}
Vector current conservation in the equal mass case implies the normalization
\begin{equation}
\xi(1)=1
\end{equation}
Away from the heavy quark limit those relations are modified by QCD corrections
so that one has
\begin{equation}
\label{hiw}
h_j(w)=\left( \alpha_j+\beta_j(w)+\gamma_j(w)+{\cal
O}(1/m_{c,b}^2)\right)\,\xi(w)
\end{equation}
The $\alpha_j$ are constants  fixed by the behavior of the form factor in the
heavy quark limit
\begin{eqnarray}
&&\alpha_+=\alpha_{A_1}=\alpha_{A_3}=\alpha_{V}=1\nonumber\\
&&\alpha_{-}=\alpha_{A_2}=0
\end{eqnarray}
The different $\beta_j$ account for perturbative radiative corrections~\cite{Neubert:1992hb}
while the $\gamma_j$ are non perturbative in nature and are proportional to the inverse
of the heavy quark masses~\cite{Neubert:1992tg}. At zero recoil ($w=1$) Luke's
theorem~\cite{Luke:1990eg} imposes the restriction. 
\begin{equation}
\gamma_+(1)=\gamma_{A_1}(1)=0
\end{equation}
so that power corrections to $h_+(1)$ and $h_{A_1}(1)$ are of order 
${\cal O}(1/m_{c,b}^2)$. In Tables~\ref{tab:beta}
and~\ref{tab:gamma} we collect the values for the different $\beta_j$ and $\gamma_j$
in the full interval of $w$ values allowed in the two decays. These two tables have
been taken from Ref.~\cite{Neubert:1992tg}.
\begin{table}[t]
\begin{center}
\begin{tabular}{c|r r r r r r}
w & $\beta_+$  & $\beta_{-}$    & $\beta_{V}$ & $\beta_{A_1}$ & $\beta_{A_2}$
& $\beta_{A_3}$\\
\hline
%& & & & &  &\\
1.0 & 2.6  & $-$5.4 & 11.9 & $-$1.5 & $-$11.0 & 2.2 \\
1.1 & $-$0.3 & $-$5.4 & 8.9 & $-$3.8 & $-$10.3 & $-$0.2 \\
1.2 & $-$3.1 & $-$5.3 & 6.1 & $-$5.9 & $-$9.8 & $-$2.5 \\
1.3 & $-$5.6 & $-$5.3 & 3.5 & $-$7.9 & $-$9.3 & $-$4.6 \\
1.4 & $-$8.0 & $-$5.2 & 1.1 & $-$9.7 & $-$8.8 & $-$6.6 \\
1.5 & $-$10.2 & $-$5.2 & $-$1.1 & $-$11.5 & $-$8.4 & $-$8.5 \\
1.59 & $-$12.1 & $-$5.1 & & & &\\
\end{tabular}
\end{center}
\caption{ QCD corrections $\beta_j(w)$ in \% as evaluated in
Ref.~\cite{Neubert:1992tg}}
\label{tab:beta}
\end{table}
\begin{table}[h!!]
\begin{center}
\begin{tabular}{c|r r r r r r}
w & $\gamma_+$  & $\gamma_{-}$    & $\gamma_{V}$ & $\gamma_{A_1}$ & $\gamma_{A_2}$
& $\gamma_{A_3}$\\
\hline
%& & & & &  &\\
1.0 & 0.0 & $-$4.1 & 19.1 & 0.0 & $-$23.1 &$-$4.1 \\
1.1 & 2.7 & $-$4.1 & 20.7 & 2.9 & $-$21.4 & $-$0.7  \\
1.2 & 6.2 & $-$4.1 & 23.1 & 6.5 & $-$19.8 & 3.4 \\
1.3 & 10.5 & $-$4.2 & 26.3 & 10.7 & $-$18.3 & 8.0\\
1.4 & 15.3 &  $-$4.4 & 30.0 & 15.4 & $-$17.0 &13.0 \\
1.5 & 20.6 & $-$4.5 & 34.3 & 20.5 & $-$15.8 & 18.5 \\
1.59 & 25.7 & $-$4.7\\
\end{tabular}
\end{center}
\caption{ Power corrections $\gamma_j(w)$ in \% as evaluated in
Ref.~\cite{Neubert:1992tg}}
\label{tab:gamma}
\end{table}
\subsection{$B\to D\, l\, \bar{\nu}$ decay}%and $B_s\to D_s\, l\, \bar{\nu}$ decays}
\label{subsect:btod}
Let us start with the $B\to D\, l\, \bar{\nu}$ case. In the center of mass of the $B$ meson and taking $\vec{P}^{\,\prime}=-\vec{q}=-|\vec{q}\,|\,
\vec{k}$ in the z 
direction\footnote{We use $\vec{k}$ for the unit vector in the z
  direction.}
  we will have for the form factors $h_+(w)$ and $h_-(w)$\footnote{In this case 
$w$ is related to $|\vec{q}\,|$ via $|\vec{q}\,|=m_D\,\sqrt{w^2-1}$.}
\begin{eqnarray}
\label{hpm}
h_+(w)=\frac{1}{\sqrt{2m_B2m_D}}\,\left(
V^0(|\vec{q}\,|)+\frac{V^3(|\vec{q}\,|)}{|\vec{q}\,|}\,\left(E_D(-\vec{q})-
m_D\right)\right)\nonumber\\
h_-(w)=\frac{1}{\sqrt{2m_B2m_D}}\,\left(
V^0(|\vec{q}\,|)+\frac{V^3(|\vec{q}\,|)}{|\vec{q}\,|}\,\left(
E_D(-\vec{q})+m_D
\right)\right)
\end{eqnarray}
where $E_D(-\vec{q})=\sqrt{m_D^2+\vec{q}\,^2}$ and $V^\mu(|\vec{q}\,|)$ ($\mu=0,3$)
is given by
\begin{eqnarray}
V^{\mu}(|\vec{q}\,|)=\left\langle\, D,\,-|\vec{q}\,|\,\vec{k}\,\left|\, (J^{c\,b}_V)^\mu(0)\,
\right| \, B,\,\vec{0}\right\rangle%\nonumber\\
%V^3=\left\langle\, D,\,-|\vec{q}\,|\,\vec{k}\,\left|\, (J^{c\,b}_V)^3(0)\,
%\right| \, B,\,\vec{0}\right\rangle
\end{eqnarray}
In our model $V^\mu(|\vec{q}\,|)$ is evaluated as
\begin{eqnarray}
\label{vmu}
V^\mu(|\vec{q}\,|)&=&\sqrt{2m_B2E_D(-\vec{q})}\ \ 
{}_{\stackrel{}{NR}}\left\langle\, D,\,
-|\vec{q}\,|\,{\vec{k}}\,\left|\, (J^{c\,b}_V)^\mu(0)\,
\right| \, B,\,\vec{0}\right\rangle_{NR}\nonumber\\
\end{eqnarray}
which expression is given in appendix~\ref{app:v}.

In the case of equal masses $m_b=m_c$ vector current conservation demands that
\begin{eqnarray}
h_+(1)=1\hspace{1cm};\hspace{1cm}
h_-(w)=0
\end{eqnarray}
In this limit we find that $h_+(1)=1$ %and for  $w=1$ ($|\vec{q}\,|=0$) is that
so that our value for $h_+(1)$ complies with vector current conservation. On the
other hand $h_-(w)\ne 0$  
violating vector current conservation.
\begin{figure}[p]
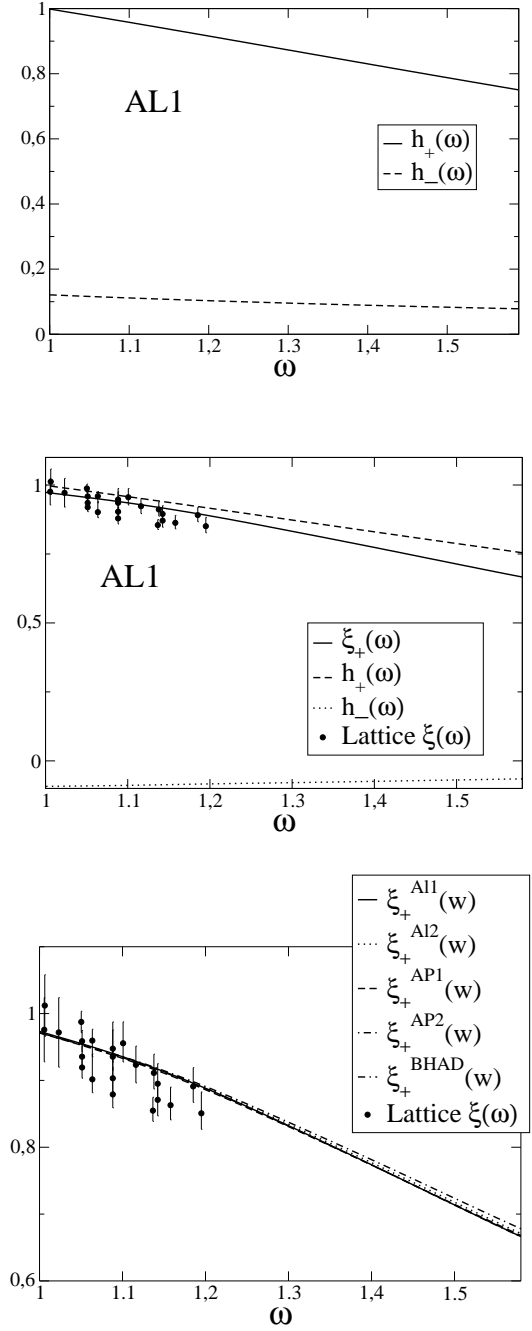

\vspace{1cm}
\centering
\resizebox{!}{5cm}{\includegraphics{BD_hpml.eps}}\vspace{1.cm}\\
\resizebox{!}{5cm}{\includegraphics{BD_hiw.eps}}\vspace{.5cm}\\
\resizebox{!}{6cm}{\includegraphics{BD_xiwpot.eps}}
%7,4
%\resizebox{10cm}{!}
% cambio el caption para la nueva posicion
\caption{Top  panel: $h_+(w)$ and $h_-(w)$  for the $B\to D$
transition as obtained from Eqs.~(\ref{hpm}, \ref{vmu})  using
the AL1 interquark potential. Middle  panel:
$h_+(w)$ as before, $\xi_+(w)$ obtained from the values of $h_+(w)$ using
Eq.~(\ref{hiw}), $h_-(w)$ obtained from $\xi_+(w)$ using Eq.~(\ref{hiw}). We
also show the UKQCD lattice data by Bowler {\it et al.}~\cite{Bowler:2002zh}. Lower
panel: Different $\xi_+(w)$ obtained with the above procedure for the different
interquark interactions. Lattice data is also shown.}
%\caption{Upper left panel: $h_+(w)$ and $h_-(w)$  for the $B\to D$
%transition as obtained from Eqs.~(\ref{hpm}, \ref{vmu})  using
%the AL1 interquark potential. Upper right panel:
%$h_+(w)$ as before, $\xi_+(w)$ obtained from the values of $h_+(w)$ using
%Eq.~(\ref{hiw}), $h_-(w)$ obtained from $\xi_+(w)$ using Eq.~(\ref{hiw}). We
%also show the UKQCD lattice data by Bowler {\it et al.}~\cite{Bowler:2002zh}. Lower
%panel: Different $\xi_+(w)$ obtained with the above procedure for the different
%interquark interactions. Lattice data is also shown.}
\label{fig:hpml}
\end{figure}

In the top  panel of Fig.~\ref{fig:hpml} we show the values of
$h_+(w)$ and $h_-(w)$ for the $B\to D$
transition as obtained from Eqs.~(\ref{hpm}, \ref{vmu}) with the use of the AL1
interquark potential. The values for $h_-(w)$ 
are not reliable. Actual calculation shows that they are of the same size as 
the deviations from zero that
one computes in the equal mass case. 
To improve on this what we shall do instead is to use the
form factor $h_+(w)$ and Eq.~(\ref{hiw}) to extract $\xi(w)$ (we shall call it
$\xi_+(w)$\,) and from there we can
re-evaluate $h_-(w)$  with the use of Eq.~(\ref{hiw}). The results appear 
in the middle  panel of Fig.~\ref{fig:hpml} where we also show the lattice results
for $\xi(w)$ obtained by the UKQCD Collaboration in Ref.~\cite{Bowler:2002zh}. 
We find good agreement with lattice data. Finally in the lower panel of Fig.~\ref{fig:hpml} 
we show the different $\xi_+(w)$ obtained with the use of the different interquark
potentials. As we see from the figure all $\xi_+(w)$ are very much the same in the
whole interval for $w$.

 The slope at the
origin of our Isgur-Wise function is given by
\begin{eqnarray}
\rho^2=-\frac{1}{\xi_+(w)}\left.\frac{d\,\xi_+(w)}{dw}\right|_{w=1}
=0.35 \pm 0.02
\end{eqnarray}
small  compared to the lattice value of $\rho^2=0.81^{+17}_{-11}$ extracted from
a best fit to data.

\subsubsection{Differential decay width}
Neglecting lepton masses the differential decay width for the process
$B\to D\,l\,\bar{\nu}$ is given by~\cite{Neubert:1991td}
\begin{eqnarray}
\frac{d\Gamma}{d w}=\frac{G_F^2}{48\pi^3}\,|V_{cb}|^2\,m_D^3\,
(w^2-1)^{3/2}(m_B+m_D)^2\ F_D^2(w)                                       
\end{eqnarray}
where $G_F=1.16637(1)\times
10^{-5}$\, GeV$^{-2}$~\cite{Eidelman:2004wy} is the Fermi decay constant, $V_{cb}$ is the
CKM matrix
element for the $b\to c$ weak transition, and $F_D(w)$ is given by
\begin{eqnarray}
F_D(w)=\bigg[h_+(w)-\frac{1-r}{1+r}\,h_-(w)\bigg] 
\end{eqnarray}
with $r=m_D/m_B$.

\begin{figure}[h!!!!]
\vspace{1cm}
\centering
\resizebox{!}{7cm}{\includegraphics{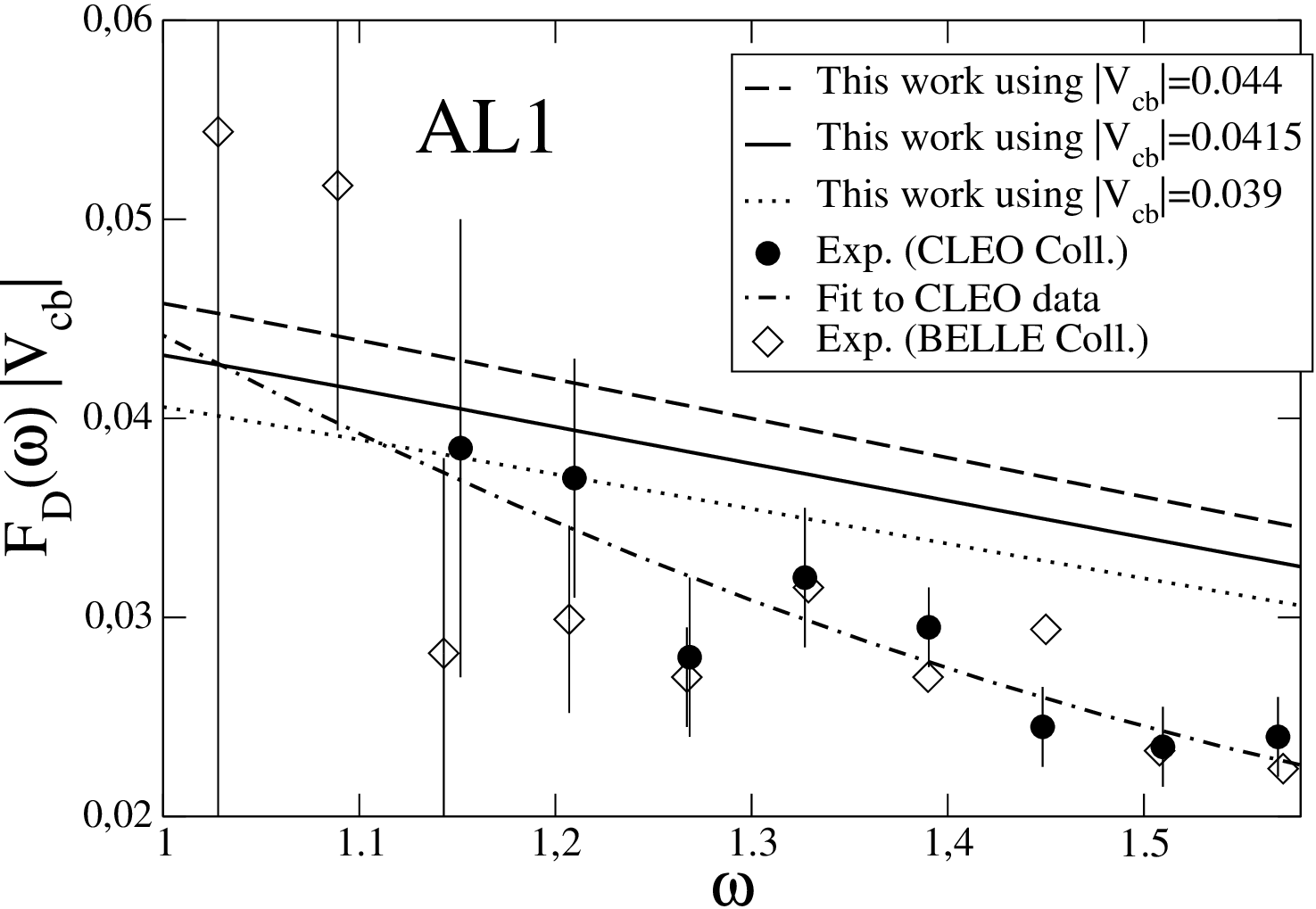}}
\caption{$F_D(r,w)\ |V_{cb}|$ obtained with the AL1 interquark potential.
 Solid line: our results
using $|V_{cb}|=0.0415$. Dashed line: our results
using \mbox{$|V_{cb}|=0.044$}. Dotted line: our results
using $|V_{cb}|=0.039$.
Circles: experimental data by the CLEO Collaboration~\cite{Bartelt:1998dq}.
Dashed-dotted line:
fit to CLEO data using the form factors of Boyd {\it et al.}~\cite{Boyd:1997kz}.
Diamonds: experimental data the by BELLE Collaboration~\cite{Abe:2001yf}.} 
\label{fig:fd}
\end{figure}

In Fig.~\ref{fig:fd} we show our calculation for $F_D(w)\ |V_{cb}|$ obtained
with the AL1 interquark potential and using three
different values of  $|V_{cb}|$ corresponding to the 
central and extreme values of the range for $|V_{cb}|$ favored by the
Particle Data Group (PDG), 
 $|V_{cb}|=0.039 \sim 0.044$~\cite{Eidelman:2004wy}. We also show the experimental data 
 for the decays $B^-\to D^0\, l\,\bar{\nu}$ and 
$\bar{B}^0\to D^+\,l\,\bar{\nu}$ obtained by the CLEO
Collaboration~\cite{Bartelt:1998dq}, a fit to CLEO data using the form factors of
 Boyd {\it et al.}~\cite{Boyd:1997kz}, and the experimental data for the decay 
 $\bar{B}^0\to D^+\,l\,\bar{\nu}$ obtained by the BELLE
Collaboration~\cite{Abe:2001yf}. Our results are larger than experimental
data for $w >1.2$. Our total integrated width will thus be larger than
the experimental one for any reasonable value of $|V_{cb}|$. 
From our data we
extract the slope at $w=1$ given by
\begin{eqnarray}
\rho_D^2=\left.-\frac{1}{F_D(w)}\frac{d F_D(w)}{d w}\right|_{w=1}=
0.38 \pm 0.02
\end{eqnarray}
which is  small compared to the values extracted from the experimental data: 
 \hbox{$\rho_D^2=0.76\pm0.16\pm0.08$}~\cite{Bartelt:1998dq} and
\hbox{$\rho_D^2=0.69\pm0.14$}~\cite{Abe:2001yf} obtained from a linear fit to the
data, or 
$\rho_D^2=1.30\pm0.27\pm0.14$~\cite{Bartelt:1998dq} and
$\rho_D^2=1.16\pm0.25$~\cite{Abe:2001yf} obtained using the form factors of Boyd
{\it et al.}~\cite{Boyd:1997kz}. Thus, only our
results close to $w=1$ seem to be reliable. We can use our prediction for
 $F_D(1)$ to extract the value of $|V_{cb}|$ from the experimental
determination of the quantity $|V_{cb}|F_D(1)$. Different values of that
quantity
appear in  Table~\ref{tab:vcbfd}.
 \begin{table}[h!!!!]
\begin{center}
\begin{tabular}{l|c  }
 & $|V_{cb}|\,F_{D}(1)$ \\
\hline
& \\
CLEO Collaboration\hfill ~\protect{\cite{Bartelt:1998dq}} & $\ 0.0416\pm0.0047\pm0.0037$\\
BELLE Collaboration\hfill~\protect{\cite{Abe:2001yf}} & $\ 0.0411\pm0.0044\pm0.0052$
\end{tabular}
\end{center}
\caption{ $|V_{cb}|\,F_{D}(1)$ values obtained by different experiments. }
\label{tab:vcbfd}
\end{table}

Our result for  $F_D(1)$ is given by (we do not show the theoretical
 error which
is or the order of $10^{-4}$)
\begin{eqnarray}
F_D(1)=1.04
\end{eqnarray}
which is in good agreement with other calculations
$F_D(1)=0.98\pm0.07$~\cite{Caprini:1995wq}, $F_D(1)=1.04$~\cite{Scora:1995ty} or
\hbox{$F_D(1)=1.069\pm0.008\pm0.002\pm0.025$~\cite{Hashimoto:1998ia}}.
From our value for $F_D(1)$ and the experimental values for $|V_{cb}|F_D(1)$
we can obtain  $|V_{cb}|$  in the range
\begin{eqnarray}
|V_{cb}|=0.040\pm0.006
\end{eqnarray}
%
%This result agrees with our recent determination based on the analysis of the 
%$\Lambda_b\to\Lambda_c\,l\,\bar{\nu}_l$ reaction from where we got 
%$|V_{cb}|=0.040\pm0.005$~\cite{Albertus:2004wj}.
% Lo cambio que no es muy mio
%
This result agrees with a recent determination based on the analysis of the 
\mbox{$\Lambda_b\to\Lambda_c\,l\,\bar{\nu}_l$} reaction using  the same
model, from where 
$|V_{cb}|=0.040\pm0.005$ was obtained~\cite{Albertus:2004wj}.

\subsection{$B\to D^*\,l\,\bar{\nu}$ decay}
\label{subsect:btod*}
Working again in the center of mass of  the $B$ meson and taking $\vec{P}^{\,\prime}=
-\vec{q}=-|\vec{q}\,|\,
{\vec{k}}$ in the z 
direction  we will have for the form factors $h_V(w)$, $h_{A_1}(w)$, 
$h_{A_2}(w)$ and $h_{A_3}(w)$ the expressions
\begin{eqnarray}
\label{hva123}
h_V(w)\hspace{0.0cm}&=&\sqrt2\,\sqrt{\frac{m_{D^*}}{m_B}}\,\frac{V^{(*)}_{-1,\,2}(|\vec{q}\,|)}{|\vec{q}\,|}\nonumber\\
h_{A_1}(w)&=&i\,\frac{\sqrt2}{w+1}\,\frac{1}{\sqrt{m_Bm_{D^*}}}\,
A^{(*)}_{-1,\, 1}(|\vec{q}\,|)\nonumber\\
h_{A_2}(w)&=&i\,\sqrt{\frac{m_{D^*}}{m_B}}\,
\left(-\frac{A^{(*)}_{0,\, 0}(|\vec{q}\,|)}{|\vec{q}\,|}+
\frac{E_{D^*}(-\vec{q})\,A^{(*)}_{0,\, 3}(|\vec{q}\,|)}{|\vec{q}\,|^2}
-\sqrt2\,m_{D^*}
\frac{A^{(*)}_{-1,\, 1}(|\vec{q}\,|)}{|\vec{q}\,|^2}
\right)\nonumber\\
h_{A_3}(w)&=&i\frac{m_{D^*}^2}{\sqrt{m_Bm_{D^*}}}
\left(-\frac{A^{(*)}_{0,\, 3}(|\vec{q}\,|)}{|\vec{q}\,|^2}+\,
\frac{\sqrt2}{m_{D^*}}
\frac{E_{D^*}(-\vec{q})\,A^{(*)}_{-1,\, 1}(|\vec{q}\,|)}{|\vec{q}\,|^2}\right)
\end{eqnarray}
with $E_{D^*}(-\vec{q})=\sqrt{m_{D^*}^2+\vec{q}\,^2}$, and
where $V^{(*)}_{\lambda,\,\mu}(|\vec{q}\,|)$ and 
$A^{(*)}_{\lambda,\,\mu}(|\vec{q}\,|)$
are given by
\begin{eqnarray}
V^{(*)}_{\lambda,\,\mu}(|\vec{q}\,|)=\left\langle\, D^*,\,\lambda\ -|\vec{q}\,
|\,\vec{k}\,\left|\, (J^{c\,b}_V)_\mu(0)\,
\right| \, B,\,\vec{0}\right\rangle\nonumber\\
A^{(*)}_{\lambda,\,\mu}(|\vec{q}\,|)=\left\langle\, D^*,\,\lambda\ -|\vec{q}\,
|\,\vec{k}\,\left|\, (J^{c\,b}_A)_\mu(0)\,
\right| \, B,\,\vec{0}\right\rangle
\end{eqnarray}
In our model $V^{(*)}_{\lambda,\,\mu}(|\vec{q}\,|)$ and 
$A^{(*)}_{\lambda,\,\mu}(|\vec{q}\,|)$ are evaluated as
\begin{eqnarray}
\label{va123}
V^{(*)}_{\lambda,\,\mu}(|\vec{q}\,|)&=&\sqrt{2m_B2E_{D^*}(-\vec{q})}
\ \ {}_{\stackrel{}{NR}}
\left\langle\, D^*,\,\lambda\ 
-|\vec{q}\,|\,{\vec{k}}\,\left|\, (J^{c\,b}_V)_\mu(0)\,
\right| \, B,\,\vec{0}\right\rangle_{NR}\nonumber\\
A^{(*)}_{\lambda,\,\mu}(|\vec{q}\,|)&=&\sqrt{2m_B2E_{D^*}(-\vec{q})}\ \ 
{}_{\stackrel{}{NR}}
\left\langle\, D^*,\,\lambda\ 
-|\vec{q}\,|\,{\vec{k}}\,\left|\, (J^{c\,b}_A)_\mu(0)\,
\right| \, B,\,\vec{0}\right\rangle_{NR}\nonumber\\
\end{eqnarray}
with expressions given in appendix~\ref{app:va}.

In the top panel of Fig.~\ref{fig:hva123r12} we show our results for the $h_V(w), h_{A_1}(w), 
h_{A_2}(w)$ and $h_{A_3}(w)$ form factors, obtained with the AL1 interquark
potential and the use of Eq.~(\ref{hva123}). In the lower panel of the same figure we
show the ratios
\begin{figure}[t]
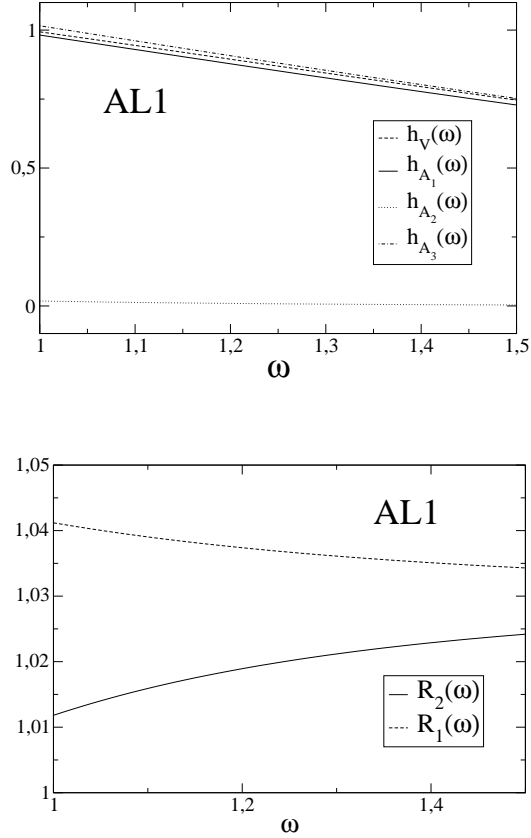

\vspace{1cm}
\centering
\resizebox{!}{5cm}{\includegraphics{BD_hva123.eps}}\vspace{1cm}
\\\resizebox{!}{5cm}{\includegraphics{BD_r12.eps}}
\caption{Top panel: $h_V(w)$, $h_{A_1}(w)$, $h_{A_2}(w)$ and $h_{A_3}(w)$ form
factors obtained  using Eq.~(\ref{hva123}). Lower panel: 
$R_1(w)$ and $R_2(w)$ ratios. In both panels the AL1 interquark potential has
been used. }
\label{fig:hva123r12}
\end{figure}
\begin{eqnarray}
&&R_1(w)=\frac{h_V(w)}{h_{A_1}(w)}\nonumber\\
&&R_2(w)=\frac{h_{A_3}(w)+r\,h_{A_2}(w)}{h_{A_1}(w)}
\end{eqnarray}
where now $r=m_{D^*}/m_B$. These ratios are expected to vary very weakly with
$w$.
We find indeed that this  is so in our case being our values of $R_1(w)$ 
and $R_2(w)$ 
within 4\% of  unity.
In Table~\ref{tab:r12} we give now our results for $R_1(1)$ and $R_2(1)$ and compare
them to different experimental\footnote{The experimental
results by the CLEO and BABAR Collaborations have been  obtained with the
assumption that $R_1(w)$ and $R_2(w)$ are constants.} and theoretical 
determinations. We find discrepancies of
the order of $15\sim 33\, \% 
$ 
for $R_1(1)$ and $13\sim 46\, \%
 $ for $R_2(1)$. 
 
\begin{figure}[p]
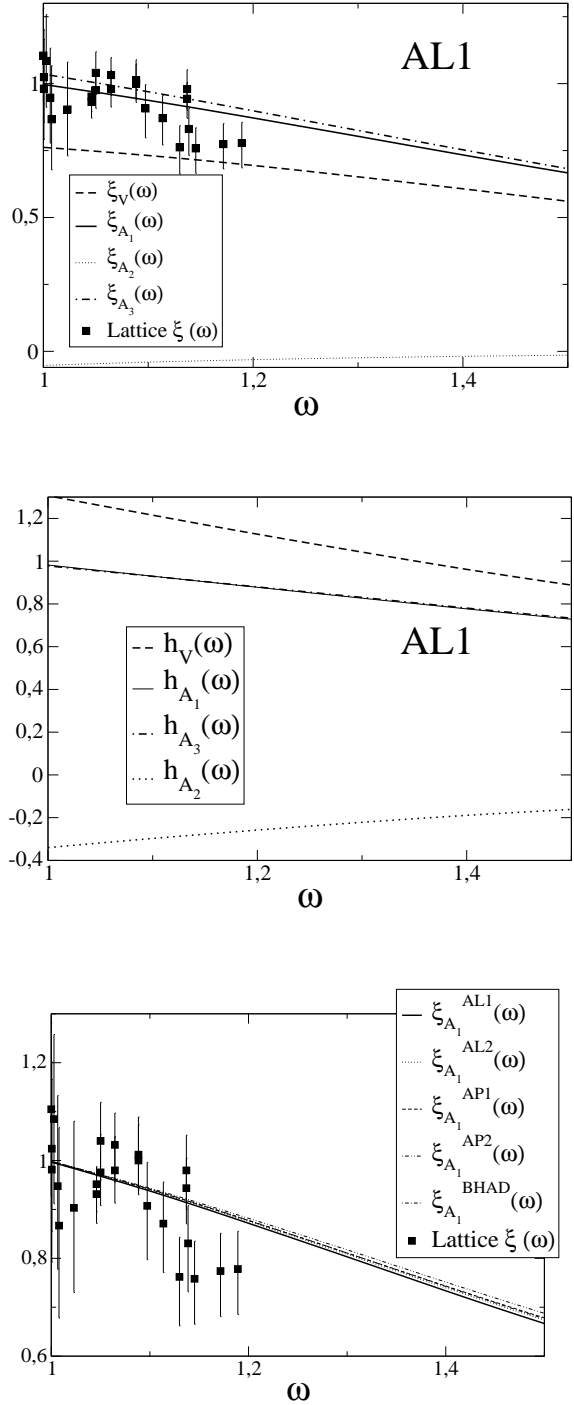

\vspace{.8cm}
\centering
\resizebox{!}{5.5cm}{\includegraphics{BD_xiva123.eps}}\vspace{1cm}\\
\resizebox{!}{5.5cm}{\includegraphics{BD_xia1hva123.eps}}\vspace{1cm}\\
\resizebox{!}{5.5cm}{\includegraphics{BD_xipoten.eps}}
\caption{Top panel: different $\xi(w)$ functions obtained from the $h_V(w)$, 
$h_{A_1}(w)$, $h_{A_2}(w)$ and $h_{A_3}(w)$ form
factors
using Eq.~(\ref{hiw}) and the AL1 interquark potential. Lattice data  by K. C. Bowler {\it et
al.} from
Ref.~\cite{Bowler:2002zh} are also shown. Middle panel: form factors obtained  from
$\xi_{A_1}(w)$ with the use of Eq.~(\ref{hiw}). Lower panel: Different 
$\xi_{A_1}(w)$ obtained with the different interquark potentials.}
\label{fig:xiva123}
\end{figure}

One can understand these discrepancies by evaluating the different $\xi(w)$
functions obtained from the  form factors with the use of 
Eq.~(\ref{hiw})
and the $\beta$ and $\gamma$ coefficients of Neubert given in
 Tables~\ref{tab:beta} and \ref{tab:gamma}.
The results appear in the top  panel of Fig.~\ref{fig:xiva123}. One can 
infer from the
figure that our results for $h_{A_2}(w)$ are not reliable. 
%In fact one should
%expect for $h_{A_2}(w)$ errors similar in size to the ones encountered in the
%evaluation of $h_-(w)$ is subsection~\ref{subsect:btod}. Those errors are of
%the same size as the values of $h_{A_2}(w)$ themselves. 
Also we somehow miss the
correct normalization for $h_{V}(1)$. On the other hand the values of 
$\xi_ {A_1}(w)$ and $\xi_{A_3}(w)$ are equal within 4\% and in reasonable
agreement with   lattice data from Ref.~\cite{Bowler:2002zh}. 

To improve the nonrelativistic quark model prediction,
 and similarly to what we did in subsection~\ref{subsect:btod},
we will take $\xi_{A_1}(w)$ as our model determination of the
Isgur-Wise function $\xi(w)$ and we will reevaluate the form factors with the
use of Eq.~(\ref{hiw}). What we obtain is now depicted in the middle  panel of
Fig.~\ref{fig:xiva123}. In the lower panel we give the different $\xi_{A_1}(w)$ obtained
with the different interquark potentials. They do not show any significant
difference.

The slope of the $\xi_{A_1}(w)$ function at the origin is  given by
\begin{eqnarray}
\rho^2=0.55 \pm 0.02
\end{eqnarray}
to be compared to the lattice result $\rho^2=0.93^{+47}_{-59}$~\cite{Bowler:2002zh}.
In this case we are within lattice errors, but one
can not be very conclusive due to the 
large value of the latter in this case.
\begin{table}[t]
\begin{center}
{\footnotesize
\begin{tabular}{c|c c }
 & $R_1(1)$  & $R_2(1)$\\
\hline
%& & & & &  &\\
This work &  $1.01\pm0.02$ & $1.04\pm0.01$\\
CLEO~\cite{Adam:2002uw} & $1.18\pm0.30\pm0.12$ & $0.71\pm0.22\pm0.07$ \\
BABAR (Preliminary)~\cite{Aubert:2004nz} &\hspace{.5cm}$1.328\pm0.055\pm0.025\pm0.025$\hspace{.5cm} &
$0.920\pm0.044\pm0.020\pm0.013$ \\ \hline
& &\\
Caprini {\it et al.}~\cite{Caprini:1997mu}& 1.27 & 0.80\\
Grinstein {\it et al.}~\cite{Grinstein:2001yg}& 1.25& 0.81\\
Close {\it et al.}~\cite{Close:1993yk}& 1.15& 0.91
\end{tabular}
}
\end{center}
\caption{ $R_1(1)$ and $R_2(1)$ }
\label{tab:r12}
\end{table}

\begin{figure}[h!!!]
\vspace{.8cm}
\centering
\resizebox{9cm}{7cm}{\includegraphics{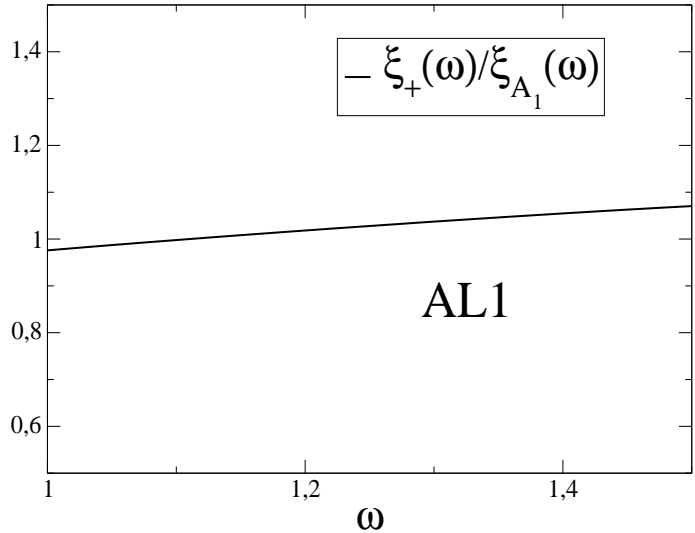}}
\caption{Ratio of our two Isgur-Wise functions calculated with the AL1
interquark potential.}
\label{fig:xipxia1}
\end{figure}

Finally in Fig.~\ref{fig:xipxia1} we give the ratio $\xi_+(w)/\xi_{A_1}(w)$
evaluated with the AL1 interquark potential. We
see  the differences between the two Isgur-Wise functions are at the level of 
3-7\%.

\subsubsection{Differential decay width}
Neglecting lepton masses the differential decay width for the process
$B\to D^*\,l\,\bar{\nu}$ is given by~\cite{Richman:1995wm}
\begin{eqnarray}
\hspace{-2cm}
\frac{d\Gamma}{d w}=\frac{G_F^2}{48\pi^3}\,|V_{cb}|^2\,(m_B-m_{D^*})^2\,m_{D^*}^3\,
\sqrt{(w^2-1)}\,(w+1)^2
\nonumber \\ \hspace{5cm} \times
\bigg[
1+\frac{4w}{w+1}\ \frac{1-2wr+r^2}{(1-r)^2}
\bigg]\ F_{D^*}^2(w)                                       
\end{eqnarray}
where  $F_{D^*}(w)$ is defined as
\begin{eqnarray}
F_{D^*}(w)=h_{A_1}(w)\sqrt{\frac{\tilde{H}_0^2(w)+\tilde{H}_+^2(w)+
\tilde{H}_-^2(w)}
{1+\frac{4w}{w+1}\ \frac{1-2wr+r^2}{(1-r)^2}}} 
\end{eqnarray}
The $\tilde{H}_j(w)$ are helicity form factors given in terms of the $R_1(w)$ 
and $R_2(w)$ ratios as
\begin{eqnarray}
\tilde{H}_0(w)&=&1+\frac{w-1}{1-r}\,[1-R_2(w)]\nonumber\\
\tilde{H}_\pm(w)&=&\frac{\sqrt{1-2wr+r^2}}{1-r}\left[
1\mp\sqrt{\frac{w-1}{w+1}}\ R_1(w)
\right]
\end{eqnarray}

\begin{figure}[h!!!]
\vspace{1cm}
\centering
\resizebox{!}{7cm}{\includegraphics{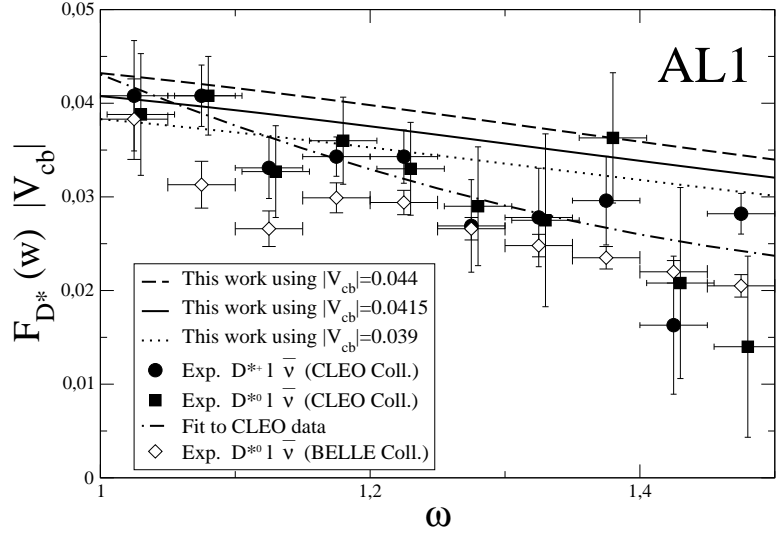}}
\caption{$F_{D^*}(r,w)\ |V_{cb}|$ obtained with the AL1 interquark potential.
 Solid line: our results
using $|V_{cb}|=0.0415$. Dashed line: our results
using $|V_{cb}|=0.044$. Dotted line: our results
using $|V_{cb}|=0.039$.
Circles and squares: experimental data by the CLEO Collaboration~\cite{Adam:2002uw}.
Dashed-dotted line: fit to CLEO Collaboration data. Diamonds: experimental data by the
BELLE Collaboration~\cite{Abe:2001yf}.}
\label{fig:fdstar}
\end{figure}

Similarly to Fig.~\ref{fig:fd}, in Fig.~\ref{fig:fdstar} we show our results 
for the quantity  $F_{D^*}(w)\
|V_{cb}|$ evaluated with the AL1 interquark potential and using the values of $|V_{cb}|$ corresponding to the central and
extreme values of the range for  $|V_{cb}|$  favored by the PDG. We also show the experimental data by the CLEO
 Collaboration~\cite{Adam:2002uw} for the $B^-\to D^{*0}\,l\,\bar{\nu}$ reaction
  (squares),
  and for the  $\bar{B}^0\to D^{*+}\,l\,\bar{\nu}$ reaction
 (circles) together with a best fit, and the
 experimental data by the BELLE Collaboration~\cite{Abe:2001cs} for the $\bar{B}^0\to D^{*+}\,l\,\bar{\nu}$ 
 reaction (diamonds). We find 
 good agreement
 with CLEO data for small $w$ values. Disagreement starts already
 at around $w=1.1$ where our results start to go above the experimental data.
   BELLE data are systematically below our results.
  
 Also our slope at the origin
 \begin{eqnarray}
 \rho^2_{D^*}=-\frac{1}{F_{D^*}(w)}\left.\frac{d F_{D^*}(w)}{d w}\right|_{w=1}
 =0.31 \pm 0.02
 \end{eqnarray}
 is  smaller  than the value
  obtained by the BELLE Collaboration
  \hbox{$\rho^2_{D^*}=0.81\pm0.12$} \cite{Abe:2001cs} using a linear fit to their
  data.
 All this means that our total width would be larger than the 
 experimental one for any reasonable value of $V_{cb}$. On the other hand
  experimentalists are able to extract the value of $|V_{cb}|\,F_{D^*}(1)$.
 Different experimental results for that quantity appear in
 Table~\ref{tab:vcbfds}. 
 \begin{table}[h!!!!]
\begin{center}
\begin{tabular}{l|c  }
 & $|V_{cb}|\,F_{D^*}(1)$ \\
\hline
& \\
(CLEO Coll.)\ \hfill \protect{\cite{Adam:2002uw}} & $\ 0.0431\pm0.0013\pm0.0018$\\
(DELPHI Coll.)\ \hfill\protect{\cite{Abdallah:2004rz}} & $\ 0.0392\pm0.0018\pm0.0023$\\
(BELLE Coll.)\ \hfill\protect{\cite{Abe:2001cs}}& $\ 0.0354\pm0.0019\pm0.0018$\\
(BABAR Coll.)\ \hfill\protect{\cite{Aubert:2004bw}} & $\ 0.0355\pm0.0003\pm0.0016$ 
\end{tabular}
\end{center}
\caption{ $|V_{cb}|\,F_{D^*}(1)$ values obtained by different experiments. }
\label{tab:vcbfds}
\end{table}

Our result  for  $F_{D^*}(1)$ is given by 
  \begin{eqnarray}
 F_{D^*}(1)=h_{A_1}(1)=0.983\pm 0.001
 \end{eqnarray}
 Comparison with the experimental data for $|V_{cb}|\,F_{D^*}(1)$ allows us to 
  extract
  values for $|V_{cb}|$ in the range
 \begin{equation}
 |V_{cb}|=0.0333\sim0.0461
 \end{equation}
 One can not be more conclusive due to  the dispersion in the experimental data
 for $|V_{cb}|\,F_{D^*}(1)$.  From DELPHI data alone we would obtain 
 $|V_{cb}|=0.040\pm0.003$ in perfect agreement with our determination using
 the $B\to D$ reaction data.  We
 should also say that our value for $F_{D^*}(1)$
is larger than the lattice determination 
$F_{D^*}(1)=0.919^{+0.030}_{-0.035}$ by S. Hashimoto {\it et al.}~\cite{Hashimoto:2001nb}
normally  used
by experimentalists to extract their $|V_{cb}|$ values. A new unquenched lattice
determination of this quantity by the Fermilab Lattice
Collaboration is in progress~\cite{Hashimoto:2004hn}.

\section[Strong coupling constants $g_{H^*H\pi}$]{Strong coupling constants $g_{H^*H\pi}$\sectionmark{Strong coupling constants $\lowercase{g}_{H^*H\pi}$}}
\sectionmark{Strong coupling constants $\lowercase{g}_{H^*H\pi}$}
\label{sect:gh*hpi}
In this section we will evaluate the strong coupling constants $g_{H^*H\pi}$ 
where $H$ stands for a $B$ or $D$ meson. To this end we shall make use of the
partial conservation of the axial current hypothesis (PCAC) which relates the
divergence of the axial current to the pion field as
\begin{eqnarray}
\label{pcac}
\partial^\mu\, J_{A\,\mu}^{d\, u}(x)=\,f_{\pi}\,m_{\pi}^2\, \Phi_{\pi^-}(x)
\end{eqnarray}
where $f_{\pi}=130.7\pm0.1\pm0.36 $\,MeV~\cite{Eidelman:2004wy} is the pion decay constant, \mbox{$m_{\pi}=
139.57$\,MeV} \cite{Eidelman:2004wy} is the pion mass, and $\Phi_{\pi^-}(x)$ is the charged pion
field that destroys a $\pi^-$ and creates a $\pi^+$. Using the Lehmann-Symanzik-Zimmermann reduction 
formula one can  relate the matrix element
of the divergence of the axial current to the pion emission amplitude as
\begin{eqnarray}
\label{pcac2}
\langle H,\ \vec{P}^\prime\,|\,q^\mu\, J_{A\,\mu}^{d\, u}(0)\,|
\,H^*,\lambda\,\vec{P}\,\rangle= -i\,f_{\pi}\frac{m_{\pi}^2}{q^2-m_{\pi}^2}\ 
{\cal A}^{(\lambda)}_{H^*H\pi}(P^\prime,P)
\end{eqnarray}
where $q=P-P^\prime$ and ${\cal A}^{(\lambda)}_{H^*H\pi}(P^\prime,P)$ is 
the pion emission amplitude for the process $H^*\to H\pi$  given
by\footnote{Corresponding to the emission  of a $\pi^+$.}
\begin{eqnarray}
{\cal A}^{(\lambda)}_{H^*H\pi}(P^\prime,P)=-\,g_{H^*H\pi}(q^2)\,
\left(q\cdot\varepsilon_{(\lambda)}(\vec{P}\,)\right)
\end{eqnarray}
The matrix element on the left hand side of Eq.(\ref{pcac2}) has a pion pole
contribution that can be easily evaluated to be
\begin{eqnarray}
\label{pp}
\langle H,\ \vec{P}^\prime\,|\,q^\mu\, J_{A\,\mu}^{d\, u}(0)\,|
\,H^*,\lambda\,\vec{P}\,\rangle_{pion-pole}= -i\,f_{\pi}\frac{q^2}{q^2-m_{\pi}^2}\ 
{\cal A}^{(\lambda)}_{H^*H\pi}(P^\prime,P)
\nonumber\\
\end{eqnarray}
so that we can extract a non-pole contribution
\begin{eqnarray}
\label{np}
\langle H,\ \vec{P}^\prime\,|\,q^\mu\, J_{A\,\mu}^{d\, u}(0)\,|
\,H^*,\lambda\,\vec{P}\,\rangle_{non-pole}&=& i\,f_{\pi}\ 
{\cal A}^{(\lambda)}_{H^*H\pi}(P^\prime,P)\nonumber\\
&=&-i\,f_{\pi}\ g_{H^*H\pi}(q^2)
\,\left(q^{\mu}\,\varepsilon_{(\lambda)\mu}(\vec{P}\,)\right)
\nonumber\\
\end{eqnarray}
which is the one we shall evaluate within the quark model.
For the matrix element on the left hand side of Eq.(\ref{np}) we can use a
parametrization similar to the one used in Eq.({\ref{bd*}})
\begin{eqnarray}
&&\hspace{-2cm}
\langle H,\ \vec{P}^\prime\,|\,q^\mu\, J_{A\,\mu}^{d\, u}(0)\,|
\,H^*,\lambda\,\vec{P}\,\rangle_{non-pole}=
\nonumber\\
&&\hspace{-1cm}
q^\mu\bigg\{-i\,
\varepsilon_{(\lambda)\mu}(\vec{P}\,)\ (w+1)\ h_{A_1}(w)\nonumber\\
&&+i \left(\varepsilon_{(\lambda)}(\vec{P}\,)\cdot v^\prime\right)
\left( v^\prime_\mu\, h_{A_2}(w)+v_\mu\,h_{A_3}(w) 
\right)\bigg\} \sqrt{m_H m_{H^*}}
\end{eqnarray}
%
%\begin{eqnarray}
%\langle H,\ \vec{P}^\prime\,|\,q^\mu\, J_{A\,\mu}^{d\, u}(0)\,|
%\,H^*,\lambda\,\vec{P}\,\rangle_{non-pole}=q^\mu\bigg\{\hspace{-.4cm}&&-i\,
%\varepsilon^{(\lambda)}_{\mu}(\vec{P}\,)\ (w+1)\ h_{A_1}(w)\nonumber\\
%&&+i \left(\varepsilon^{(\lambda)}(\vec{P}\,)\cdot v^\prime\right)
%\left( v^\prime_\mu\, h_{A_2}(w)+v_\mu\,h_{A_3}(w) 
%\right)\bigg\} \sqrt{M_H M_{H^*}}
%\end{eqnarray}
%
% Antes de cambiarla
%
with the result that
\begin{eqnarray}
g_{H^*H\pi}(q^2)=\frac{1}{f_{\pi}}\bigg\{
(w+1)\,h_{A_1}(w)+w\left(\frac{m_{H^*}}{m_H}\,h_{A_2}(w)-h_{A_3}(w)\right)\nonumber\\
+\left(\frac{m_{H^*}}{m_H}\,h_{A_3}(w)-h_{A_2}(w)\right)
\bigg\} \sqrt{m_H m_{H^*}}
\end{eqnarray}
The evaluation of the form factors is done is a similar way as the one
described in subsection~\ref{subsect:btod*}. The results that we get for
$q^2=0$ are
\begin{eqnarray}
g_{D^*D\pi}(0)=22.1\pm 0.4\hspace{.5cm},\hspace{.5cm}g_{B^*B\pi}(0)=60.5\pm1.1
\end{eqnarray}
to be compared to the experimental determination 
$g_{D^*D\pi}(m_\pi^2)=17.9\pm0.3\pm1.9$ by the CLEO Collaboration~\cite{Anastassov:2001cw},
the lattice results
$g_{D^*D\pi}(m_\pi^2)=18.8\pm2.3^{+1.1}_{-2.0}$~\cite{Abada:2002xe}
 and $g_{B^*B\pi}(0)=47\pm5\pm8$~\cite{Abada:2003un}, or  a recent
 determination using QCDSR for which
 $g_{D^*D\pi}(m_\pi^2)=14.0\pm1.5$ and
 $g_{B^*B\pi}(0)=42.5\pm2.6$~\cite{Navarra:2001ju}. Older QCDSR results give 
 smaller values for both coupling
 constants. For instance, the calculation within QCDSR on the light cone in 
 Ref.~\cite{Belyaev:1994zk} give
  \mbox{$g_{D^*D\pi}(m_\pi^2)=12.5\pm1$} and \mbox{$g_{B^*B\pi}(0)=29\pm3$\ }\footnote{
  Values for both coupling
  constants obtained prior to 1995 within different approaches  can be found
  in Ref.~\cite{Belyaev:1994zk} and references therein. }. The latter are
   small compared to lattice data or the experimental determination of
  $g_{D^*D\pi}(m_\pi^2)$ by the CLEO Collaboration.
 
From our results we obtain the ratio
\begin{eqnarray}
R=\frac{g_{B^*B\pi}(0)\ f_{B^*}\,\sqrt{m_D}}
{g_{D^*D\pi}(0)\ f_{D^*}\,\sqrt{m_B}}=1.105\pm 0.005
\label{R}
\end{eqnarray}
in good agreement with  HQS that predicts a value of 1
with
 $1/m_{c,b}$ corrections appearing  in next to leading
 order~\cite{Burdman:1993es}
 \footnote{Note  the  strong coupling constant used in 
 Ref.~\cite{Burdman:1993es}
 is  given in terms of ours as
 $(g_{H^*H\pi}\,f_{\pi})/(2 m_{H^*})$ with $H=B,D$.}. Our result in Eq.(\ref{R}) is also in agreement with the one
 obtained combining lattice data for $f_{B^*}$ and $f_{D^*}$ from 
 Ref.~\cite{Bowler:2000xw}, for $g_{B^*B\pi}(0)$ from Ref.~\cite{Abada:2002xe}, and the
 experimental CLEO
 Collaboration data for $g_{D^*D\pi}(m_{\pi}^2)$ from Ref.~\cite{Anastassov:2001cw}. In
 this case one gets 
 \begin{equation}
 \left.R\right|_{Exp.-Latt.}=1.26\pm0.36
 \end{equation}
  where we have added errors in quadratures. A calculation using
 light cone QCDSR gives~\cite{Belyaev:1994zk}
 \begin{equation}
 \left.R\right|_{QCDSR}=0.92
 \end{equation}

\chapter {Semileptonic and non-leptonic decays of the $B_c^{-}$ meson}
%\chaptermark{Semi and non-leptonic decays of the $B_{\lowercase{c}}$ meson}
\chaptermark{ Decays of the $B_{\lowercase{c}}$ meson}
\label{cha:Bc} 
% REVISADO 9-11
\section{Introduction}
\label{Bc_intro}
Since its discovery at Fermilab by the CDF
Collaboration~\cite{Abe:1998fb,Abe:1998wi} the $B_c$ meson has drawn a
lot of attention. Unlike other heavy mesons it is composed of two
heavy quarks of different flavor ($b,\,c$) and, being below the
$B$--$D$ threshold, it can only decay through weak interactions making
an ideal system to study weak decays of heavy quarks.

Using HQSS
Jenkins {\it et al.}~\cite{Jenkins:1992nb} were able to obtain, in the
infinite heavy quark mass limit, relations between different form
factors for semileptonic $B_c$ decays into pseudoscalar and vector
mesons. Contrary to the heavy-light meson case where standard HQS
applies, no determination of corrections in inverse powers of the
heavy quark masses has been worked out in this case. So one can only
test any model calculation against HQSS predictions in the infinite
heavy quark mass limit.

With both quarks being heavy, a nonrelativistic treatment of the $B_c$
meson should provide reliable results. Besides a nonrelativistic model
will comply with the constraints imposed by HQSS as the spin--spin
interaction vanish in the infinity heavy quark mass limit. In this
chapter we will study, within the framework of a nonrelativistic quark
model, exclusive semileptonic and nonleptonic decays of the $B_c^-$
meson driven by a $b\to c$ or $\bar c\to \bar d,\,\bar s$ transitions
at the quark level. We will not consider semileptonic processes driven
by the quark $b\to u$ transition.  Our experience with this kind of
processes, like the analogous $B\to\pi$ semileptonic decay
~\cite{Albertus:2005ud}, shows that the nonrelativistic model without
any improvements underestimates the decay width for two reasons: first
at high $q^2$ transfers one might need to include the exchange of a
$B^*$ meson, and second the model underestimates the form factors at
low $q^2$ or high three--momentum transfers.  We will concentrate thus
on semileptonic $B_c^-\to c\bar c$ and $B_c^-\to \overline B$
transitions. As for two--meson nonleptonic decay we will only consider
channels with a least a $c\bar c$ or $ B$ final meson. In the first
case we will include channels with final $D$ mesons for which there is
a contribution coming from an effective $b\to d,\,s $ transition. As
later explained this is not the main contribution to the decay
amplitude and besides the momentum transfer in those cases is neither
too high nor too low so that the problems mentioned above are avoided.

The observables studied here have been analyzed before in the context
of different models like the relativistic constituent quark
model~\cite{Ivanov:2006ni,Ivanov:2005fd,Ivanov:2000aj}, the
quasi-potential approach to the relativistic quark
model~\cite{Ebert:2003cn,Ebert:2003wc}, the instantaneous
nonrelativistic approach to the Bethe--Salpeter
equation~\cite{Chang:1992pt,Chang:2001pm,Chang:2001vq}, the
Bethe--Salpeter equation~\cite{AbdEl-Hady:1999xh,Liu:1997hr}, the
three point sum rules of QCD and nonrelativistic
QCD~\cite{Kiselev:1999sc,Kiselev:2000pp,Kiselev:2002vz,Kiselev:2001zb},
the QCD relativistic potential model~\cite{Colangelo:1999zn}, the
relativistic constituent quark model formulated on the light
front~\cite{Anisimov:1998xv}, the relativistic quark--meson
model~\cite{Nobes:2000pm} or in models that use the Isgur, Scora,
Grinstein and Wise wave functions~\cite{Isgur:1988gb,Scora:1995ty}
like the calculations in
Refs.~\cite{Sanchis-Lozano:1994vh,LopezCastro:2002ud,Lu:1995ug}. We
will compare our results with those obtained in these latter
references whenever is possible. Besides, we will perform an
exhaustive study for
exclusive semileptonic and nonleptonic $B_c^-$ decays, paying an
special attention to the theoretical uncertainties affecting our
predictions and providing reliable estimates for all of them.

In the present calculation we shall use physical masses taken from
Ref.~\cite{Eidelman:2004wy}.  For the $B_c$ meson mass and lifetime we shall use
the central values of the recent experimental determinations by the
CDF Collaboration of
$m_{B_c}=6285.7\pm5.3\pm1.2$\,MeV/$c^2$\ \cite{Abulencia:2005us} and
$\tau_{B_c}=(0.463^{+0.073}_{-0.065}\pm0.036)\times 10^{-12}$\,s
~\cite{Abulencia:2006zu}.  This new mass value is very close to the one we
obtain with the different quark-quark potentials that we use in this
memory  from where we get $m_{B_c}=6291.6^{+12}_{-33}$\,MeV.

We shall also need CKM matrix elements and
different meson decay constants.
For the former we shall use the ones quoted in Ref.~\cite{Ivanov:2006ni} that we
reproduce in Table~\ref{tab:ckm}. All of them are within the ranges quoted by
the  Particle Data Group (PDG)~\cite{Eidelman:2004wy}.
\begin{table}[h!]
\begin{center}
\begin{tabular}{c c c c c  c }
$|V_{ud}|$&$|V_{us}|$&$|V_{cd}|$&$|V_{cs}|$&$|V_{cb}|$\\%&$|V_{ub}|$\\
\hline
&&&&&\\
 0.975  &0.224   & 0.224  &  0.974 & 0.0413 \\%& 0.0037
 \end{tabular}
\caption{Values for Cabibbo-Kobayashi-Maskawa matrix elements used 
in
this work.}
\label{tab:ckm}
\end{center}
\end{table}

For the meson decay constants the values used in this work are
compiled in Table~\ref{tab:dc}. They correspond to central values of
experimental measurements or lattice determinations. 
The results for
$f_\rho $ and $f_{K^*}$ have been obtained by the authors in
Ref.~\cite{Ivanov:2006ni} using  $\tau $ lepton decay data.
  Our own theoretical calculation, obtained with the model described
  chapter \ref{cha:BD}, give $f_{\rho}=0.189\sim 0.227$\,GeV,
\mbox{$f_{K^*}=0.180\sim 0.220$\,GeV} depending on the inter-quark interaction
used, results  which agree with the determinations in Ref.~\cite{Ivanov:2006ni}. 
We shall nevertheless use
the latter for our calculations.  
For $f_{\eta_c}$ we have been unable to find an experimental
result or a lattice determination\footnote{There is a determination by
  the CLEO Colaboration~\cite{Edwards:2000bb} using factorization
  aproximation and thus model--dependent. }.
 There are at least two theoretical
determinations that predict $f_{\eta_c}=0.484$\,GeV~\cite{Ivanov:2006ni}
and $f_{\eta_c}=0.420\pm0.052$\,GeV~\cite{Hwang:1997ie}. Again our own
calculation gives values in the range $f_{\eta_c}=0.485\sim 0.500$\,GeV
depending on the inter-quark interaction used.  Here we will take
$f_{\eta_c}=0.490$\,GeV.
\begin{table}[h!]
\begin{center}
\begin{tabular}{c c c c c  }
$f_{\pi^-}$&$f_{\pi^0}$&$f_{\rho^-,\, \rho^0}$&$f_{K^-,\,K^0}$
&$f_{K^{*-},\,K^{*0}}$\\
\hline\\
 0.1307~\cite{Eidelman:2004wy}  &0.130~\cite{Eidelman:2004wy}&0.210~\cite{Ivanov:2006ni}  &
 0.1598~\cite{Eidelman:2004wy}  &  0.217~\cite{Ivanov:2006ni}\\
&&&\\
 \end{tabular}
 
\begin{tabular}{c c    }
$f_{\eta_c}$&$f_{J/\Psi}$\\
\hline\\
 0.490&0.405~\cite{Caso:1998tx}\\
 &\\
 \end{tabular}

\begin{tabular}{c c c c   }
$f_{D^-}$&$f_{D^{*-}}$&$f_{D_s^-}$&$f_{D_s^{*-}}$\\
\hline\\
 0.2226~\cite{Artuso:2005ym}  &0.245~\cite{Becirevic:1998ua}   & 0.294~\cite{Yao:2006px}  
 &  0.272~\cite{Becirevic:1998ua}\\
 \end{tabular}

\end{center}
 \caption{Meson decay constants in GeV used in this work.}
\label{tab:dc}
\end{table}%

%
%
%
% SECTION 1
%
%
\section{Meson states}
\label{Bcsect:wf}

The meson states were introduced earlier in chapter \ref{cha:BD}.
Here  we will need the ground state wave function for
scalar ($0^+$), pseudoscalar ($0^-$), vector ($1^-$), axial vector
($1^+$), tensor ($2^+$) and pseudotensor ($2^-$) mesons. Assuming
always the lowest possible value for the orbital angular momentum we
will have for a meson $M$ with scalar, pseudoscalar and vector quantum
numbers:

% DEJO COPIA POR SI LA ROMPO (0911)
%\begin{eqnarray}
%\label{eq:0pm}
%\hat{\phi}^{\,(M(0^+))}_{\alpha_1,\,\alpha_2}(\,\vec{p}\,)
%&=&\frac{1}{\sqrt{3}}\,\delta_{c_1,\,c_2}\
%\hat{\phi}^{\,(M(0^+))}_{(s_1,\,f_1),\,(s_2,\,f_2)}(\,\vec{p}\,)\nonumber\\
%&=&\frac{1}{\sqrt{3}}\,\delta_{c_1,\,c_2}\ i
%\ \hat{\phi}^{\,(M(0^+))}_{f_1,\,f_2}(\,|\vec{p}\,|)\sum_m\ (1/2,1/2,1\,;\,s_1,s_2,-m)
%\ (1,1,0\,;\,m,-m,0)\
%Y_{1m}(\,{\vec{p}}\ )\nonumber\\
%%
%%
%%
%%
%\hat{\phi}^{\,(M(0^-))}_{\alpha_1,\,\alpha_2}(\,\vec{p}\,)
%&=&\frac{1}{\sqrt{3}}\,\delta_{c_1,\,c_2}\
%\hat{\phi}^{\,(M(0^-))}_{(s_1,\,f_1),\,(s_2,\,f_2)}(\,\vec{p}\,)\nonumber\\
%&=&\frac{1}{\sqrt{3}}\,\delta_{c_1,\,c_2}\ (-i)
%\ \hat{\phi}^{\,(M(0^-))}_{f_1,\,f_2}(\,|\vec{p}\,|) 
%\ (1/2,1/2,0\,;\,s_1,s_2,0)\ Y_{00}(\,{\vec{p}}\ )\nonumber\\
%%
%%
%%
%%
%\hat{\phi}^{\,(M(1^-),\,\lambda)}_{\alpha_1,\,\alpha_2}(\,\vec{p}\,)
%&=&\frac{1}{\sqrt{3}}\,\delta_{c_1,\,c_2}\
%\hat{\phi}^{\,(M(1^-),\,\lambda)}_{(s_1,\,f_1),\,(s_2,\,f_2)}(\,\vec{p}\,)\nonumber\\
%&=&\frac{1}{\sqrt{3}}\,\delta_{c_1,\,c_2}\ (-1)\ 
%\hat{\phi}^{\,(M(1^-))}_{f_1,\,f_2}(\,|\vec{p}\,|)\
%(1/2,1/2,1\,;\,s_1,s_2,\lambda)\ Y_{00}(\,{\vec{p}}\ )
%\end{eqnarray}

\begin{eqnarray}
\label{eq:0pm}
\hat{\phi}^{\,(0^+)}_{\alpha_1,\,\alpha_2}(\,\vec{p}\,)
&=&\frac{1}{\sqrt{3}}\,\delta_{c_1,\,c_2}\
\hat{\phi}^{\,(0^+)}_{(s_1,\,f_1),\,(s_2,\,f_2)}(\,\vec{p}\,)\nonumber\\
&=&\frac{1}{\sqrt{3}}\,\delta_{c_1,\,c_2}\ i
\ \hat{\phi}^{\,(0^+)}_{f_1,\,f_2}(\,|\vec{p}\,|)
\nonumber \\ && \hspace*{0.7cm}
\sum_m\ \left(\left.\frac12,\frac12,1\right|s_1,s_2,-m\right)
\ \left(\left.1,1,0\right|m,-m,0\right)\
Y_{1m}(\,{\vec{p}}\ )\nonumber\\
\hat{\phi}^{\,(0^-)}_{\alpha_1,\,\alpha_2}(\,\vec{p}\,)
&=&\frac{1}{\sqrt{3}}\,\delta_{c_1,\,c_2}\
\hat{\phi}^{\,(0^-)}_{(s_1,\,f_1),\,(s_2,\,f_2)}(\,\vec{p}\,)\nonumber\\
&=&\frac{1}{\sqrt{3}}\,\delta_{c_1,\,c_2}\ (-i)
\ \hat{\phi}^{\,(0^-)}_{f_1,\,f_2}(\,|\vec{p}\,|) 
\ \left(\left.\frac12,\frac12,0\right|s_1,s_2,0\right)\ Y_{00}(\,{\vec{p}}\ )\nonumber\\
\hat{\phi}^{\,(1^-,\,\lambda)}_{\alpha_1,\,\alpha_2}(\,\vec{p}\,)
&=&\frac{1}{\sqrt{3}}\,\delta_{c_1,\,c_2}\
\hat{\phi}^{\,(1^-,\,\lambda)}_{(s_1,\,f_1),\,(s_2,\,f_2)}(\,\vec{p}\,)\nonumber\\
&=&\frac{1}{\sqrt{3}}\,\delta_{c_1,\,c_2}\ (-1)\ 
\hat{\phi}^{\,(1^-)}_{f_1,\,f_2}(\,|\vec{p}\,|)\
\left(\left.\frac12,\frac12,1\right|s_1,s_2,\lambda\right)\ Y_{00}(\,{\vec{p}}\ ) \nonumber\\
\end{eqnarray}

For axial mesons we need orbital angular momentum $L=1$. In this case two values
of the total quark--antiquark spin  $S_{q\bar{q}}=0,1$ are possible, giving
rise to the two states:
\begin{eqnarray}
\label{eq:1pm}
\hat{\phi}^{\,((1^+,S_{q\bar{q}}=0),\,\lambda)}_{\alpha_1,\,\alpha_2}(\,\vec{p}\,)
&=&\frac{1}{\sqrt{3}}\,\delta_{c_1,\,c_2}\
\hat{\phi}^{\,((1^+,S_{q\bar{q}}=0),\,\lambda)}_{(s_1,\,f_1),\,(s_2,\,f_2)}(\,\vec{p}\,)\nonumber\\
&=&\frac{1}{\sqrt{3}}\,\delta_{c_1,\,c_2}\ (-1)\ 
\hat{\phi}^{\,(1^+,S_{q\bar{q}}=0)}_{f_1,\,f_2}(\,|\vec{p}\,|)\
\left(\left.\frac12,\frac12,0\right|s_1,s_2,0\right)\
Y_{1\lambda}(\,{\vec{p}}\ )\nonumber\\
\hat{\phi}^{\,((1^+,S_{q\bar{q}}=1),\,\lambda)}_{\alpha_1,\,\alpha_2}(\,\vec{p}\,)
&=&\frac{1}{\sqrt{3}}\,\delta_{c_1,\,c_2}\
\hat{\phi}^{\,((1^+,S_{q\bar{q}}=1),\,\lambda)}_{(s_1,\,f_1),\,(s_2,\,f_2)}(\,\vec{p}\,)\nonumber\\
&=&\frac{1}{\sqrt{3}}\,\delta_{c_1,\,c_2}\ (-1)\ 
\hat{\phi}^{\,(1^+,S_{q\bar{q}}=1)}_{f_1,\,f_2}(\,|\vec{p}\,|)\nonumber\\
&&\times \sum_{m}\ 
\left(\left.\frac12,\frac12,1\right|s_1,s_2,\lambda-m\right)\ \left(\left.1,1,1\right|m,\lambda-m,\lambda\right)
Y_{1m}(\,{\vec{p}}\ ) \nonumber\\
\end{eqnarray}
Finally for  tensor and pseudotensor mesons we have the wave functions:
\begin{eqnarray}
\label{eq:2pm}
\hat{\phi}^{\,(2^+,\,\lambda)}_{\alpha_1,\,\alpha_2}(\,\vec{p}\,)
&=&\frac{1}{\sqrt{3}}\,\delta_{c_1,\,c_2}\
\hat{\phi}^{\,(2^+,\,\lambda)}_{(s_1,\,f_1),\,(s_2,\,f_2)}(\,\vec{p}\,)\nonumber\\
&=&\frac{1}{\sqrt{3}}\,\delta_{c_1,\,c_2}\ \ 
\hat{\phi}^{\,(2^+)}_{f_1,\,f_2}(\,|\vec{p}\,|)\
\nonumber\\ && \hspace*{.3cm} 
\sum_m\ \left(\left.\frac12,\frac12,1\right|s_1,s_2,\lambda-m\right)\ \left(\left.1,1,2\right|m,\lambda-m,\lambda\right)\
Y_{1m}(\,{\vec{p}}\ )\nonumber\\
\hat{\phi}^{\,(2^-,\,\lambda)}_{\alpha_1,\,\alpha_2}(\,\vec{p}\,)
&=&\frac{1}{\sqrt{3}}\,\delta_{c_1,\,c_2}\
\hat{\phi}^{\,(2^-,\,\lambda)}_{(s_1,\,f_1),\,(s_2,\,f_2)}(\,\vec{p}\,)\nonumber\\
&=&\frac{1}{\sqrt{3}}\,\delta_{c_1,\,c_2}\ (-1)\ 
\hat{\phi}^{\,(2^-)}_{f_1,\,f_2}(\,|\vec{p}\,|)\
\nonumber\\ && \hspace*{.3cm} 
\sum_{m}\ \left(\left.\frac12,\frac12,1\right|s_1,s_2,\lambda-m\right)\ \left(\left.2,1,2\right|m,\lambda-m,\lambda\right)
Y_{2m}(\,{\vec{p}}\ ) \hspace*{1cm}%\nonumber\\
\end{eqnarray}
All phases have been introduced for later convenience, and radial wave
function are computed using the potentials described in chapter \ref{cha:potential}. 
The use of different inter-quark interactions will provide us with a
spread in the results that we will consider, and quote, as a
theoretical error added to the value obtained with the AL1 potential, that we will use to get our central
results.
%
%
%
%      SECTION SEMILEPTONIC C\BAR{C} DECAYS
%
%
\section[Semileptonic $B_c^-\to c\bar{c}$ decays]{Semileptonic $B_c^-\to c\bar{c}$ decays\sectionmark{ Semileptonic $B_{\lowercase{c}}^-\to \lowercase{c}\bar{\lowercase{c}}$ decays}}
\sectionmark{ Semileptonic $B_{\lowercase{c}}^-\to
\lowercase{c}\bar{\lowercase{c}}$ decays}
\label{Bcsect:semileptonic}
In this section we will consider the semileptonic decay of the $B_c^-$
meson into different $c\bar{c}$ states with $0^+$, $0^-$, $1^+$,
$1^-$, $2^+$ and $2^-$ spin--parity quantum numbers.  Those decays
correspond to a $b\to c $ transition at the quark level which is
governed by the current
\begin{equation}
J^{c\,b}_\mu(0)=J^{c\,b}_{V\,\mu}(0)-J^{c\,b}_{A\,\mu}(0)
=
\overline{\Psi}_{c}(0)\gamma_\mu
(I-\gamma_5)\Psi_{b}(0)
\end{equation}
with $\Psi_{f}$  a quark field of a definite  flavor $f$.
%
%
%        SUBSECTION   FORM FACTOR DECOMPOSITION
%
\subsection{Form factor decomposition of hadronic matrix elements}
 The hadronic matrix elements involved in these processes can be
 parametrized in terms of a few form factors as
\begin{eqnarray}
\label{eq:ff}
\left\langle\, c\bar{c}\, (0^-),\,\vec{P}_{c\bar{c}}\,\left|\, J^{c\,b}_\mu(0)\,
\right| \, B_c,\,\vec{P}_{B_c}\right\rangle&=&
\left\langle\,  {c\bar{c}}\, (0^-),\,\vec{P}_{c\bar{c}}\,\left|\hspace{.3cm}J^{c\,b}_{V\,\mu}(0)\,
\right| \, B_c,\,\vec{P}_{B_c}\right\rangle
\nonumber \\
&=&P_\mu\,F_+(q^2)+q_\mu\,F_-(q^2)\nonumber\\
%\left\langle\, M (0^+),\,\vec{P}^\prime\,\left|\, J^{c\,b}_\mu(0)\,
%\right| \, B_c,\,\vec{P}\right\rangle&=&
%\left\langle\,M (0^+) ,\,\vec{P}^\prime\,\left|\, -J^{c\,b}_{A\,\mu}(0)\,
%\right| \, B_c,\,\vec{P}\right\rangle=P_\mu\,F_+(q^2)+q_\mu\,F_-(q^2)\nonumber\\
%
%
%
%
\left\langle\, {c\bar{c}}\, (1^-),\,\lambda\,\vec{P}_{c\bar{c}}\,\left|\, J^{c\,b}_\mu(0)\,
\right| \, B_c,\,\vec{P}_{B_c}\right\rangle&=&
\left\langle\, {c\bar{c}}\,(1^-),\,\lambda\,\vec{P}_{c\bar{c}}\,\left|\, J^{c\,b}_{V\mu}(0)-
J^{c\,b}_{A\mu}(0)\,
\right| \, B_c,\,\vec{P}_{B_c}\right\rangle\nonumber\\
&=&\frac{-1}{m_{B_c}+m_{c\bar{c}}} \varepsilon_{\mu\nu\alpha\beta}\ \varepsilon^{\,
\nu\,*}_{(\lambda)}(\,\vec{P
}_{c\bar{c}}\,)\,P^\alpha\,q^{\,
\beta}\,V(q^2)\nonumber\\
&&-\, i\, \bigg\{\ (m_{B_c}-m_{c\bar{c}})\ \varepsilon_{(\lambda)\,\mu}^*
(\,\vec{P}_{c\bar{c}}\,)\,A_0(q^2)\nonumber\\
&&\hspace{+.8cm}
-\frac{P\cdot\varepsilon^*_{(\lambda)}(\,\vec{P}_{c\bar{c}}\,)}
{m_{B_c}+m_{c\bar{c}}}
\left(P_\mu\,A_+(q^2)+q_\mu\,A_-(q^2)\right)\bigg\}\nonumber\\
\left\langle\, {c\bar{c}}\, (2^+),\,\lambda\,\vec{P}_{c\bar{c}}\,\left|\, J^{c\,b}_\mu(0)\,
\right| \, B_c,\,\vec{P}_{B_c}\right\rangle&=&
\left\langle\, {c\bar{c}}\, (2^+),\,\lambda\,\vec{P}_{c\bar{c}}\,\left|\, J^{c\,b}_{V\mu}(0)-
J^{c\,b}_{A\mu}(0)\,
\right| \, B_c,\,\vec{P}_{B_c}\right\rangle\nonumber\\
&=&\ \varepsilon_{\mu\nu\alpha\beta}\ \varepsilon^{\,
\nu\delta\,*}_{(\lambda)}(\,\vec{P
}_ {c\bar{c}}\,)P_\delta\,P^\alpha\,q^{\,
\beta}\,T_4(q^2)\nonumber\\
&&-\, i\, \bigg\{\ \varepsilon_{(\lambda)\,\mu\delta}^*(\,\vec{P}_{c\bar{c}}\,)
P^\delta\,T_1(q^2)\nonumber\\
&&\hspace{+.8cm}
+P^\nu P^\delta\varepsilon_{(\lambda)\nu\delta}^*(\,\vec{P}_{c\bar{c}},)
\left(P_\mu\, T_2(q^2)+q_\mu\,T_3(q^2)\right)\bigg\} 
\nonumber\\
\label{eq:9}
\end{eqnarray}
In the above expressions $P=P_{B_c}+P_{c\bar{c}}$,
$q=P_{B_c}-P_{c\bar{c}}$, being $P_{B_c}$ and $P_{c\bar{c}}$ the meson
four--momenta, $m_{B_c}$ and $m_{c\bar{c}}$ are the meson masses and
$\varepsilon_{(\lambda)\mu}(\,\vec{P}\,)$ and
$\varepsilon_{(\lambda)\mu\nu}(\,\vec{P}\,)$ are the polarization
vector and tensor of vector and tensor mesons
respectively.\footnote{Note we have taken $\lambda$ to be the
third component of the meson spin measured in the meson center of
mass.} The latter can be evaluated in terms of the former as 
\begin{eqnarray}
\varepsilon_{(\lambda)}^{\mu\nu}(\,\vec{P}\,)
=\sum_m \left(\left.1,1,2\right|m,\lambda-m,\lambda\right)\,
\varepsilon_{(m)}^{\mu}(\,\vec{P}\,)
\varepsilon_{(\lambda-m)}^{\nu}(\,\vec{P}\,)
\end{eqnarray}

Besides the meson states in the Lorenz decompositions of
Eq.~(\ref{eq:9}) are normalized such that
\begin{equation}
\left\langle\, {M,\lambda'\,\vec{P}^{\,\prime}}\,|\,{M,\lambda
\,\vec{P}}\,\right\rangle
=\delta_{\lambda',\,\lambda}\,(2\pi)^3\,2\,E_M(\vec{P})\delta(\vec{P}^{\,\prime}-\vec{P}\,)
\end{equation}
Note the factor $2E_M$ of difference with Eq.~(\ref{BD:eq01})

For the $0^+$, $1^+$ and $2^-$ cases the form factor decomposition is
the same as for the $0^-$, $1^-$ and $2^+$ cases respectively, but
with $-J^{c\,b}_{A\mu}(0)$ contributing where $ J^{c\,b}_{V\mu}(0)$
contributed before and vice versa.

The different form factors in Eq.(\ref{eq:ff}) are all relatively real
thanks to time--reversal invariance. $F_+$ , $F_-$, $V$, $A_0$, $A_+$,
$A_-$ and $T_1$ are dimensionless, whereas $T_2$, $T_3$ and $T_4$ have
dimension of $E^{-2}$.  They can be easily evaluated working in the
center of mass of the $B_c$ meson and taking $\vec{q}$ in the positive
$z$
direction, so that $\vec{P}_{c\bar{c}}=-\vec{q}=-|\vec{q}\,| \vec{k}$.
%
%
%
%     SUBSUBSECTION B_c\to \eta_c  \chi_{c0}
%
%
%
\subsubsection{\bf $\bullet$ $B_c^-\to \eta_c\, l^-\, \bar{\nu}_l,\ \chi_{c0}\, l^-\, 
\bar{\nu}_l $ decays}
Let us start with the $B_c^-$ decays into pseudoscalar $\eta_c$ and scalar 
$\chi_{c0}$ $c\bar{c}$ mesons. For  $B_c^-\to\eta_c$ transitions the form 
factors are given by:
\begin{eqnarray}
\label{eq:fpm0-}
F_+(q^2)=\frac{1}{2m_{B_c}}\,\left(
V^0(|\vec{q}\,|)+\frac{V^3(|\vec{q}\,|)}{|\vec{q}\,|}\,\left(E_{\eta_c}
(-\vec{q}\,)-m_{B_c}\right)\right)\nonumber\\
F_-(q^2)=\frac{1}{2m_{B_c}}\,\left(
V^0(|\vec{q}\,|)+\frac{V^3(|\vec{q}\,|)}{|\vec{q}\,|}\,\left(
E_{\eta_c}
(-\vec{q}\,)+m_{B_c}
\right)\right)
\end{eqnarray}
whereas for $B_c\to\chi_{c0}$ transitions we have:
\begin{eqnarray}
\label{eq:fpm0+}F_+(q^2)=\frac{-1}{2m_{B_c}}\,\left(
A^0(|\vec{q}\,|)+\frac{A^3(|\vec{q}\,|)}{|\vec{q}\,|}\,
\left(E_{\chi_{c0}}(-\vec{q}\,)-m_{B_{c}}\right)\right)\nonumber\\
F_-(q^2)=\frac{-1}{2m_{B_c}}\,\left(
A^0(|\vec{q}\,|)+\frac{A^3(|\vec{q}\,|)}{|\vec{q}\,|}\,\left(
E_{\chi_{c0}}(-\vec{q}\,)+m_{B_{c}}
\right)\right)
\end{eqnarray}%
with $V^\mu(|\vec{q}\,|)$  and $A^\mu(|\vec{q}\,|)$ ($\mu=0,3$)
calculated in our model as
\begin{eqnarray}
\hspace*{-.5cm}V^{\mu}(|\vec{q}\,|) &=& \left\langle\, \eta_c,\hspace{.2cm}-|\vec{q}\,|\,\vec{k}\,\left|\,
 J^{c\,b\ \mu}_V(0)\,
\right| \, B_c^-,\,\vec{0}\right\rangle
\nonumber \\ &=&
\sqrt{2m_{B_c}2E_{\eta_c}(-\vec{q}\,)}\ \ 
\hspace{.2cm}
{}_{\stackrel{}{\stackrel{}{NR}}}
\left\langle\, \eta_c,
\hspace{.2cm}
-|\vec{q}\,|\,{\vec{k}}\,\left|\, J^{c\,b\ \mu}_V(0)\,
\right| \, B_c^-,\,\vec{0}\right\rangle_{NR}\nonumber\\
\hspace*{-.5cm}A^{\mu}(|\vec{q}\,|)&=&\left\langle\, \chi_{c0},\,-|\vec{q}\,|\,\vec{k}\,\left|\,
 J^{c\,b\ \mu}_A(0)\,
\right| \, B_c^-,\,\vec{0}\right\rangle
\nonumber \\ &=&
\sqrt{2m_{B_c}2E_{\chi_{c0}}(-\vec{q}\,)}\ \ 
{}_{\stackrel{}{\stackrel{}{NR}}}\left\langle\, \chi_{c0},\,
-|\vec{q}\,|\,{\vec{k}}\,\left|\, J^{c\,b\ \mu}_A(0)\,
\right| \, B_c^-,\,\vec{0}\right\rangle_{NR}\end{eqnarray}
which expressions are given in appendix~\ref{Bcapp:va}.

\begin{figure}[t!!]
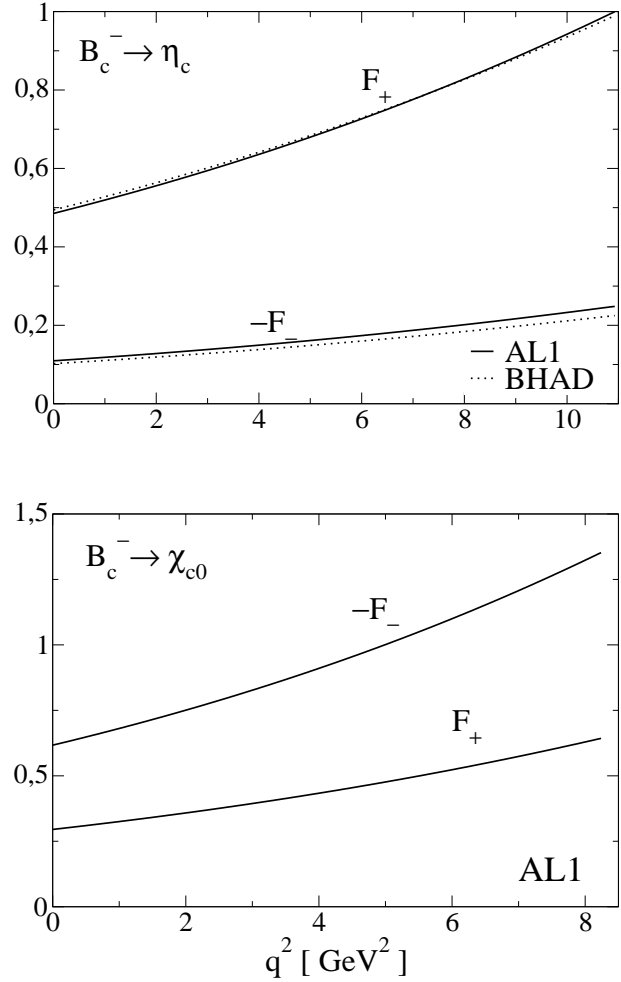

\vspace{.6cm}
\centering
\resizebox{8cm}{!}{\includegraphics{Bc_etacelectron.eps}}
\\
\vspace{.3cm}
\resizebox{8cm}{!}{\includegraphics{Bc_chic0electron.eps}}
\caption{$F_+$ and $F_-$ form factors for  $B_c^-\to\eta_c
$ and $B_c^-\to\chi_{c0}$ semileptonic decays evaluated with the AL1
 potential.  
In the first case, and for
comparison, we also show with dotted lines the results obtained with
the Bhaduri (BHAD) potential.} 
\label{fig:fpm}
\end{figure}
In Fig.~\ref{fig:fpm} we show our results for the $F_+$ and $F_-$ form
factors for the semileptonic $B_c^-\to \eta_c,\,\chi_{c0}$ transitions.
The minimum $q^2$ value depends on the actual final lepton and it is given, 
neglecting neutrino
masses, by the  lepton mass as
$q^2_{min}=m_l^2$. The form factors have
been evaluated using the AL1 potential. For decays into $\eta_c$, and for the
sake of comparison, we also show the results obtained using the BHAD
potential.  
As seen in the figures the differences
between the form factors evaluated with the two inter-quark
interactions are smaller than 10\%.

In Table~\ref{tab:fpm} we show  $F_+$ and $F_-$ evaluated at
$q^2_{\mathrm{min}}$ and $q^2_{\mathrm{max}}$ for a final light lepton
 ($l=e,\,\mu$)  and  compare them to  the ones
 obtained by Ivanov {\it et al.} in Ref.~\cite{Ivanov:2005fd}, and, when available, by Ebert {\it et al.} in
 Ref.~\cite{Ebert:2003cn}.
 For the $B_c\to \eta_c$ transition we also show the
 corresponding values for the $F_0$ form factor defined as
 \begin{equation}
\label{eq:f0}
 F_0(q^2)=F_+(q^2)+\frac{q^2}{m^2_{B_c}-m^2_{\eta_c}}\ F_-(q^2)
 \end{equation}
 Our results for the $\eta_c$ case are in excellent agreement with the ones obtained by  
 Ebert {\it et al.}. Compared to the results by Ivanov {\it et al.} we find
 large discrepancies  for $F_-$. 

%%%%%   TABLA 1

\begin{table}[h!]

%\begin{center}
\begin{tabular}{c| c c || c|  c c}
$B_c^-\to\eta_c\,l^-\bar{\nu}_l\ \ $ & $q^2_{\mathrm{min}}$ &
 $q^2_{\mathrm{max}}$ &
$B_c^-\to\chi_{c0}\,l^-\bar{\nu}_l\ \ $ & $q^2_{\mathrm{min}}$ &
 $q^2_{\mathrm{max}}$  \\
%\cline{1-3} \cline{5-7}
\hline
&&&&&\\
 $F_+$ &  &&$F_+$ & &\\
 This work & $0.49^{+0.01}$& $1.00_{-0.01}$  & This work & $0.30^{+0.01}$& 
 $0.64^{+0.01}_{-0.01}$\\
 \cite{Ivanov:2005fd} &0.61 &1.14&\cite{Ivanov:2005fd} &0.40 &0.65\\
 \cite{Ebert:2003cn} &0.47 &1.07  & &  &\\
&&&&&\\
$F_-$   & &&$F_-$  &&\\
 This work &$-0.11^{+0.01}_{-0.01}$ &$-0.25^{+0.03}$  &  This work
 &$-0.62^{+0.01}_{-0.03}$ &$-1.35^{+0.05}$\\
 \cite{Ivanov:2005fd} & $-0.32$&$-0.61$ & \cite{Ivanov:2005fd} & $-1.00$&$-1.63$\\
&&\\
$F_0$ & & \\
 This work & $0.49^{+0.01}$ &$ 0.91^{+0.01}$ \\
 \cite{Ebert:2003cn} & 0.47& 0.92 
 \end{tabular}\hspace{1cm}
%\end{center}
\caption{$F_+$ and $F_-$ evaluated at
$q^2_{\mathrm{min}}$ and $q^2_{\mathrm{max}}$ compared to  the ones obtained by 
Ivanov {\it et al.}~\cite{Ivanov:2005fd} and  Ebert {\it et al.}
Ref.~\cite{Ebert:2003cn}. Our central values have been obtained with the AL1
potential. For the $\eta_c$ channel we also show $F_0$ (see text for
definition).
Here $l$ stands for $l=e,\,\mu$.}
\label{tab:fpm}
\end{table}

%
%  SUBSUBSECTION DECAYS INTO VECTOR AND AXIAL VECTOR MESONS
%
%
\subsubsection{\bf $\bullet$ $B_c^-\to J/\Psi\, l\, \bar{\nu}_l,\  h_c\, l\, \bar{\nu}_l,
\ \chi_{c1}\, l\, \bar{\nu}_l$ decays}
Let us now see the form factors for the
semileptonic $B_c^-$ decays into vector $J/\Psi$ and axial vector
$h_c$ ($S_{q\bar{q}}=0$) and $\chi_{c1}$ ($S_{q\bar{q}}=1$) $c\bar{c}$
mesons. For the decay into $J/\Psi$ the form factors can be evaluated
in terms of matrix elements as:
\begin{eqnarray}
\label{eq:ffv}
V(q^2)&=&\frac{i}{\sqrt2}\,\frac{m_{B_c}+m_{J/\Psi}}{m_{B_c}\,|\vec{q}\,|}\
V^1_{\lambda=-1}(|\vec{q}\,|)\nonumber\\
A_+(q^2)&=&i\,\frac{m_{B_c}+m_{J/\Psi}}{2m_{B_c}}\,
\frac{m_{J/\Psi}}{|\vec{q}\,|\,m_{B_c}}\,
\bigg\{
-A^0_{\lambda=0}(|\vec{q}\,|)+\frac{m_{B_c}-E_{J/\Psi}(-\vec{q}\,)}{|\vec{q}\,|}\,
 A^3_{\lambda=0}(|\vec{q}\,|)\nonumber\\
&&\hspace{4cm}-\sqrt2\,\frac{m_{B_c}E_{J/\Psi}(-\vec{q}\,)-m^2_{J/\Psi}}{|\vec{q}\,|\,
m_{J/\Psi}}\,A^1_{\lambda=-1}(|\vec{q}\,|)\bigg\}\nonumber\\
A_-(q^2)&=&-i\,\frac{m_{B_c}+m_{J/\Psi}}{2m_{B_c}}\,
\frac{m_{J/\Psi}}{|\vec{q}\,|\,m_{B_c}}\,
\bigg\{
\ A^0_{\lambda=0}(|\vec{q}\,|)+\frac{m_{B_c}+E_{J/\Psi}(-\vec{q}\,)}{|\vec{q}\,|}\,
 A^3_{\lambda=0}(|\vec{q}\,|)\nonumber\\
&&\hspace{4cm}-\sqrt2\,\frac{m_{B_c}E_{J/\Psi}(-\vec{q}\,)+m^2_{J/\Psi}}{|\vec{q}\,|\,
m_{J/\Psi}}\,A^1_{\lambda=-1}(|\vec{q}\,|)\bigg\}\nonumber\\
A_0(q^2)&=&-i\sqrt2\,\frac{1}{m_{B_c}-m_{J/\Psi}}\, A^1_{\lambda=-1}(|\vec{q}\,|)
\end{eqnarray}
with $V^\mu_\lambda(|\vec{q}\,|)$  and $A^\mu_\lambda(|\vec{q}\,|)$ 
calculated in our model as
\begin{eqnarray}
\label{eq:vavector}
V^{\mu}_\lambda(|\vec{q}\,|)&=&\left\langle\, J/\Psi,\hspace{.2cm}\lambda\ -|\vec{q}\,|\,\vec{k}\,\left|\,
 J^{c\,b\ \mu}_V(0)\,
\right| \, B_c^-,\,\vec{0}\right\rangle\nonumber\\
&=&\sqrt{2m_{B_c}2E_{J/\Psi}(-\vec{q}\,)}\ 
\ 
%\hspace{.2cm}
{}_{\stackrel{}{\stackrel{}{NR}}}
\left\langle\, J/\Psi,
\hspace{.2cm}\lambda\,
-|\vec{q}\,|\,{\vec{k}}\,\left|\, J^{c\,b\ \mu}_V(0)\,
\right| \, B_c^-,\,\vec{0}\right\rangle_{NR}\nonumber\\
A^{\mu}_\lambda(|\vec{q}\,|)&=&\left\langle\, J/\Psi,\hspace{.2cm}\lambda\ -|\vec{q}\,|\,\vec{k}\,\left|\,
 J^{c\,b\ \mu}_A(0)\,
\right| \, B_c^-,\,\vec{0}\right\rangle\nonumber\\
&=&\sqrt{2m_{B_c}2E_{J/\Psi}(-\vec{q}\,)}\ 
\ 
%\hspace{.2cm}
{}_{\stackrel{}{\stackrel{}{NR}}}
\left\langle\, J/\Psi,
\hspace{.2cm}\lambda\,
-|\vec{q}\,|\,{\vec{k}}\,\left|\, J^{c\,b\ \mu}_A(0)\,
\right| \, B_c^-,\,\vec{0}\right\rangle_{NR}
\nonumber\\
\end{eqnarray}
which expressions are given in appendix~\ref{Bcapp:va}.

The  form factors corresponding to  transitions to the $\chi_{c1}$ 
and $h_c$ axial vector mesons are obtained from the expressions in 
Eq.(\ref{eq:ffv}) by just  changing
\begin{eqnarray}
V^\mu_\lambda(|\vec{q}\,|)\longleftrightarrow -A^\mu_\lambda(|\vec{q}\,|)
\end{eqnarray} 
and using the appropriate mass for the final meson. Obviously in 
Eq.~(\ref{eq:vavector}) $J/\Psi$ has to be replaced by $\chi_{c_1}$ or $h_c$.

In Table~\ref{tab:fva0pm} we show the results for the different form
factors evaluated at $q^2_{\mathrm{min}}$ and $q^2_{\mathrm{max}}$ for
the case where the final lepton is light ($l=e,\,\mu$). For the decay
into $J/\Psi$ we also show the combination of form
factors\footnote{This combination is called $A_0$ by the authors of
Ref.~\cite{Ebert:2003cn}.}:
\begin{eqnarray}
\label{eq:a0tilde}
\hspace*{-.4cm}
\widetilde{A}_0(q^2)=\frac{m_{B_c}-m_{J/\Psi}}{2m_{J/\Psi}}\left(
A_0(q^2)-A_+(q^2)\right)-\frac{q^2}{2m_{J/\Psi}\,(m_{B_c}+m_{J/\Psi})}\ A_-(q^2)
\end{eqnarray}
   Our results for the $B_c\to J/\Psi$ decay channel are in  agreement with the ones obtained by  
 Ebert {\it et al.}. They also agree reasonably well, with the exception of
 $A_-$, with the ones obtained by Ivanov 
 {\it et al.}. For the other two cases the discrepancies are in general large.
%
%      TABLA II
%
%
\begin{table}[p!!]
%\begin{center}
%Esta la rehago a lo barbaro
%\hspace*{-1cm}
\vspace{.5cm}
{\small
\begin{tabular}{c| c c||  c|  c c}
$B_c^-\to J/\Psi\,l^-\bar{\nu}_l\ \ $ & $q^2_{\mathrm{min}}$ &
 $q^2_{\mathrm{max}}$ &
$B_c^-\to h_{c}\,l^-\bar{\nu}_l\ \ $ & $q^2_{\mathrm{min}}$ &
 $q^2_{\mathrm{max}}$\\ 
\hline  
&&&&\\
 $V$ & & & $V$ & &  \\
 This work & $-0.61_{-0.03}$ & $-1.26^{+0.01}$  & This work &
 $-0.040_{-0.003}$& $-0.078_{-0.003}$ 
\\
 \cite{Ivanov:2005fd} &$-0.83$$^*$ &$-1.53$$^*$&\cite{Ivanov:2005fd} &$-0.25$
 $^*$ &$-0.365$$^*$
\\
 \cite{Ebert:2003cn} &$-0.49$ &$-1.34$ & & &  \\
&&&&&\\
 $A_+$ & & & $A_+$ & & 
\\
 This work & $0.56^{+0.03}$& $1.13^{+0.01}$  & This work &
 $-0.85^{+0.01}_{-0.05}$& $-1.90^{+0.06}$ 
\\
 \cite{Ivanov:2005fd} &0.54 &0.97&\cite{Ivanov:2005fd} &$-1.08$
 &$-1.80$
\\
 \cite{Ebert:2003cn} &0.73 &1.33 & & &  \\ 
&&&&\\
 $A_-$ & & & $A_-$ & & 
\\
 This work & $-0.60_{-0.03}$& $-1.24_{-0.01}$  & This work &
 $0.12^{+0.01}_{-0.02}$& $0.36_{-0.06}$
\\
 \cite{Ivanov:2005fd} &$-0.95$ &$-1.76$&\cite{Ivanov:2005fd} &0.52
 &0.89
\\
 &&&&\\
$A_0$ & & & $A_0$ & & 
\\
 This work & $1.44^{+0.08}$& $2.58^{+0.01}_{-0.02}$  & This work &
 $0.28^{+0.02}$& $0.52^{+0.02}$
\\
 \cite{Ivanov:2005fd} &1.64 &2.50&\cite{Ivanov:2005fd} &0.44 &0.54
\\
 \cite{Ebert:2003cn} &1.47 &2.59 & & &  \\  
&&&&\\
 $\widetilde{A}_0$ & & & & & \\
 This work & $0.45^{+0.03}$& $0.96_{-0.01}$ & &    &  \\
 \cite{Ebert:2003cn} &0.40 &1.06 & & &  \\  
&&&&\\
\hline
\hline
&&\\
$B_c^-\to\chi_{c1}\,l^-\bar{\nu}_l\ \ $ & $q^2_{\mathrm{min}}$ &
 $q^2_{\mathrm{max}}$ \\
&&\\
$V$ & &\\
  This work & 
 $0.92^{+0.04}_{-0.02}$&
 $1.86_{-0.12}$\\
\cite{Ivanov:2005fd} 
 &1.18$^*$ &1.81$^*$\\
&&\\
 $A_+$ & &\\
 This work & 
 $-0.44_{-0.03}$& $-0.78^{+0.04}$\\
\cite{Ivanov:2005fd} 
 &$-0.39$ &$-0.50$\\
&&\\
 $A_-$ & &\\
  This work &
 $ 0.96^{+0.04}_{-0.01}$& $1.97_{-0.13}$\\
\cite{Ivanov:2005fd} 
 &1.52 &2.36\\
&&\\
 $A_0$ & &\\
  This work &
 $ -0.50_{-0.02}$&$-0.32_{-0.02}$ \\
\cite{Ivanov:2005fd} 
 &$-0.064$ &0.46\\
 \end{tabular}
}
%\end{center}
\caption{$V$, $A_+$, $A_-$ and $A_0$ form factors evaluated at
$q^2_{\mathrm{min}}$ and $q^2_{\mathrm{max}}$ compared to  the ones obtained by 
Ivanov {\it et al.}~\cite{Ivanov:2005fd} and  Ebert {\it et al.} 
Ref.~\cite{Ebert:2003cn}. Our central values have been evaluated with the AL1
potential. For the $J/\Psi$ channel we also show $\widetilde{A}_0$ (see text
for definition). Here
$l$ stands for $l=e,\,\mu$. The asterisk to the right of a number means we 
have changed its sign to account
 for the different choice of $\varepsilon^{\,0\,1\,2\,3}$
in Ref.~\protect{\cite{Ivanov:2005fd}}.}
\label{tab:fva0pm}
\end{table}

\begin{figure}[h!!!]
\vspace{1cm}
\centering
\resizebox{8.cm}{!}{\includegraphics{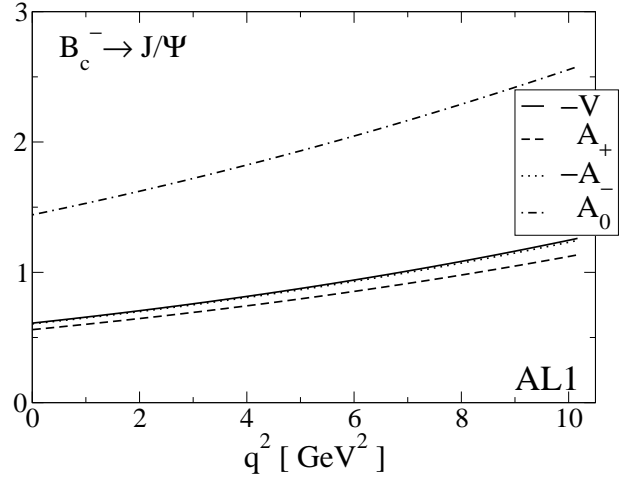}}
\caption{$V$ (solid line), $A_+$ (dashed line), $A_-$ (dotted line)
and $A_0$ (dashed-dotted line) form factors for the $B_c^-\to J/\Psi$
semileptonic decays evaluated with the AL1 potential. }
\label{fig:fva0pmjpsi}
\end{figure}

\begin{figure}[h!!!]
\vspace{1cm}
\centering
\resizebox{8.cm}{!}{\includegraphics{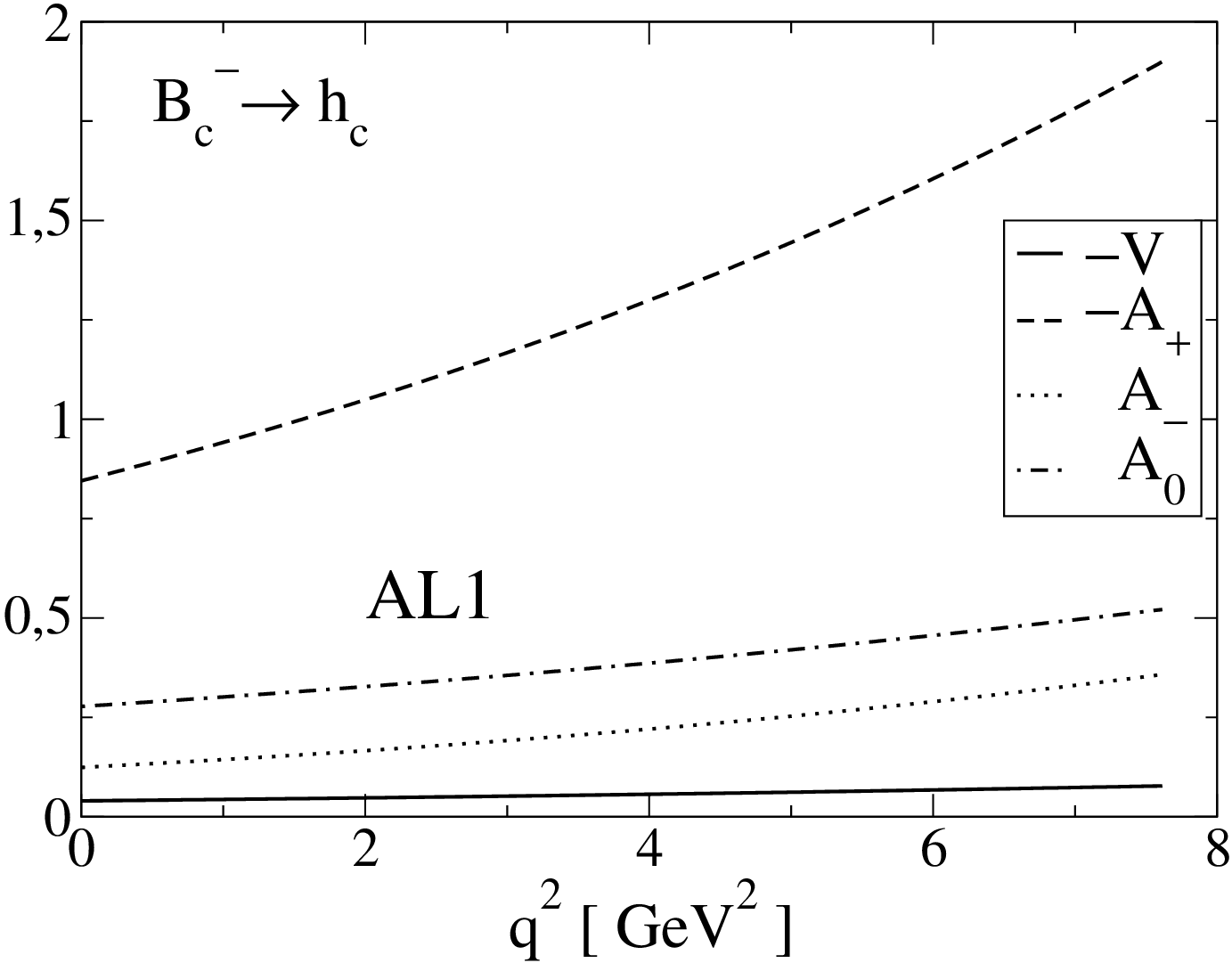}}
\\\vspace{1cm}
\resizebox{8.cm}{!}{\includegraphics{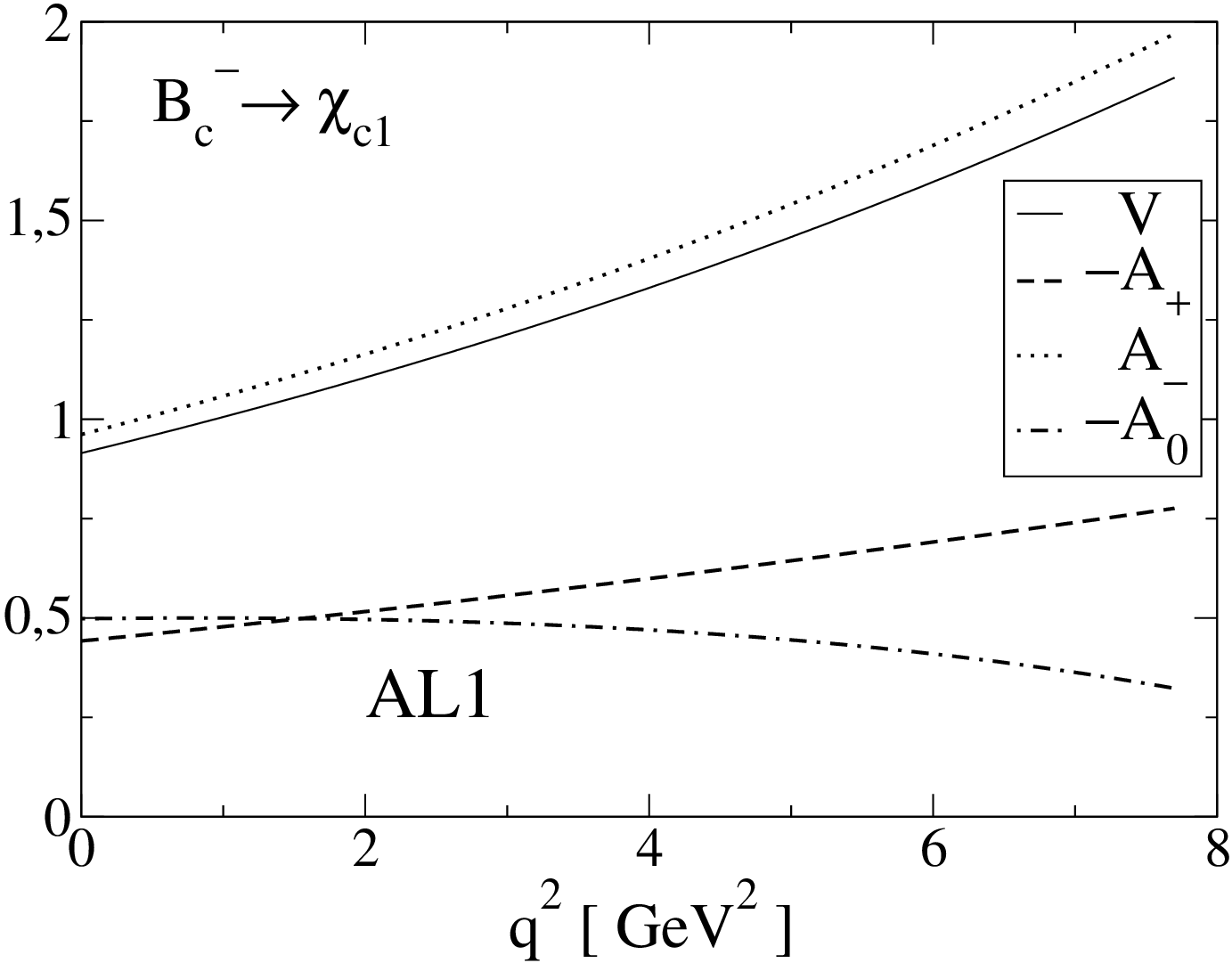}}
\caption{$V$ (solid line), $A_+$ (dashed line), $A_-$ (dotted line) 
and $A_0$ (dashed-dotted line) form factors for  
 $B_c^-\to h_c$ and $B_c^-\to\chi_{c1}$
semileptonic decays evaluated with the AL1 potential.}
\label{fig:fva0pm}
\end{figure}

All the  form factors are depicted in Figs.~\ref{fig:fva0pmjpsi} and \ref{fig:fva0pm}

%
%
%
%
%   SUBSUBSECTION DECAYS INTO TENSOR AND PSEUDOTENSOR MESONS
% 
%
%
\subsubsection{\bf $\bullet$$B_c^-\to \Psi(3836)\, l\, \bar{\nu}_l,\  \chi_{c2}\, l\,
 \bar{\nu}_l$, decays}
Finally let us  see the form factors for the $B_c^-$ decays into tensor
$\chi_{c2}$ and pseudotensor $\Psi(3836)$~\footnote{Note that while 
the $\Psi(3836)$ was still quoted in the particle listings of the former
Review of
Particle Physics~\cite{Eidelman:2004wy}, it has been excluded from the  more recent 
one~\cite{Yao:2006px}. We shall nevertheless keep it in our study to illustrate
the results to be expected for a ground state pseudotensor particle.}
 mesons. For the decay into $\chi_{c2}$ the form factors can be evaluated in 
 terms of matrix elements as:
\begin{eqnarray}
\label{eq:fft}
T_1(q^2)&=&-i\,\frac{2m_{\chi_{c2}}}{m_{B_c}|\vec{q}\,|}\
A^1_{T\lambda=+1}(|\vec{q}\,|)\nonumber\\
T_2(q^2)&=&i\,\frac{1}{2m^3_{B_c}}\,
\bigg\{
-\sqrt{\frac{3}{2}}\,\frac{m^2_{\chi_{c2}}}{|\vec{q}\,|^2}\,
A^0_{T\lambda=0}(|\vec{q}\,|)
-\sqrt{\frac{3}{2}}\,\frac{m^2_{\chi_{c2}}}
{|\vec{q}\,|^3}
\, \left(E_{\chi_{c2}}(-\vec{q}\,)-m_{B_c}\right)
 A^3_{T\lambda=0}(|\vec{q}\,|)\nonumber\\
&&\hspace{1.55cm}+\frac{2m_{\chi_{c2}}}{|\vec{q}\,|}
\left(1-\frac{E_{\chi_{c2}}(-\vec{q}\,)}{|\vec{q}\,|^2}
\left(E_{\chi_{c2}}(-\vec{q}\,)-m_{B_c}
\right)
\right)
\,A^1_{T\lambda=+1}(|\vec{q}\,|)\bigg\}\nonumber\\
T_3(q^2)&=&i\,\frac{1}{2m^3_{B_c}}\,
\bigg\{
-\sqrt{\frac{3}{2}}\,\frac{m^2_{\chi_{c2}}}{|\vec{q}\,|^2}\,
A^0_{T\lambda=0}(|\vec{q}\,|)
-\sqrt{\frac{3}{2}}\,\frac{m^2_{\chi_{c2}}}
{|\vec{q}\,|^3}
\, \left(E_{\chi_{c2}}(-\vec{q}\,)+m_{B_c}\right)
 A^3_{T\lambda=0}(|\vec{q}\,|)\nonumber\\
&&\hspace{1.45cm}+\frac{2m_{\chi_{c2}}}{|\vec{q}\,|}
\left(1-\frac{E_{\chi_{c2}}(-\vec{q}\,)}{|\vec{q}\,|^2}
\left(E_{\chi_{c2}}(-\vec{q}\,)+m_{B_c}
\right)
\right)
\,A^1_{T\lambda=+1}(|\vec{q}\,|)\bigg\}\nonumber\\
T_4(q^2)&=&i\,\frac{m_{\chi_{c2}}}{m^2_{B_c}|\vec{q}\,|^2}\, 
V^1_{T\lambda=+1}(|\vec{q}\,|)
\end{eqnarray}
with $V^\mu_{T\lambda}(|\vec{q}\,|)$  and $A^\mu_{T \lambda}(|\vec{q}\,|)$ 
calculated in our model as
\begin{eqnarray}
\label{eq:vatensor}
V^{\mu}_{T \lambda}(|\vec{q}\,|)&=&\left\langle\, \chi_{c2},\hspace{.2cm}\lambda\ -|\vec{q}\,|\,\vec{k}\,\left|\,
 J^{c\,b\ \mu}_V(0)\,
\right| \, B_c^-,\,\vec{0}\right\rangle\nonumber\\
&=&\sqrt{2m_{B_c}2E_{\chi_{c2}}(-\vec{q}\,)}\ 
\ 
%\hspace{.2cm}
{}_{\stackrel{}{\stackrel{}{NR}}}
\left\langle\, \chi_{c2},
\hspace{.2cm}\lambda\,
-|\vec{q}\,|\,{\vec{k}}\,\left|\, J^{c\,b\ \mu}_V(0)\,
\right| \, B_c^-,\,\vec{0}\right\rangle_{NR}\nonumber\\
A^{\mu}_{T\lambda}(|\vec{q}\,|)&=&\left\langle\, \chi_{c2},\hspace{.2cm}\lambda\ -|\vec{q}\,|\,\vec{k}\,\left|\,
 J^{c\,b\ \mu}_A(0)\,
\right| \, B_c^-,\,\vec{0}\right\rangle\nonumber\\
&=&\sqrt{2m_{B_c}2E_{\chi_{c2}}(-\vec{q}\,)}\ 
\ 
%\hspace{.2cm}
{}_{\stackrel{}{\stackrel{}{NR}}}
\left\langle\, \chi_{c2},
\hspace{.2cm}\lambda\,
-|\vec{q}\,|\,{\vec{k}}\,\left|\, J^{c\,b\ \mu}_A(0)\,
\right| \, B_c^-,\,\vec{0}\right\rangle_{NR}
\nonumber\\
\end{eqnarray}
which expressions are given in appendix~\ref{Bcapp:va}.

The  form factors corresponding to  transitions to  $\Psi(3836)$ 
 are obtained from the expressions in 
Eq.(\ref{eq:fft}) by just  changing
\begin{eqnarray}
V^\mu_{T\lambda}(|\vec{q}\,|)\longleftrightarrow -A^\mu_{T\lambda}(|\vec{q}\,|)
\end{eqnarray} 
and using the appropriate mass for the final meson. Besides in 
Eq.~(\ref{eq:vatensor}) $\chi_{c_2}$ has to be replaced by $\Psi(3836)$.

\begin{figure}[h!!!]
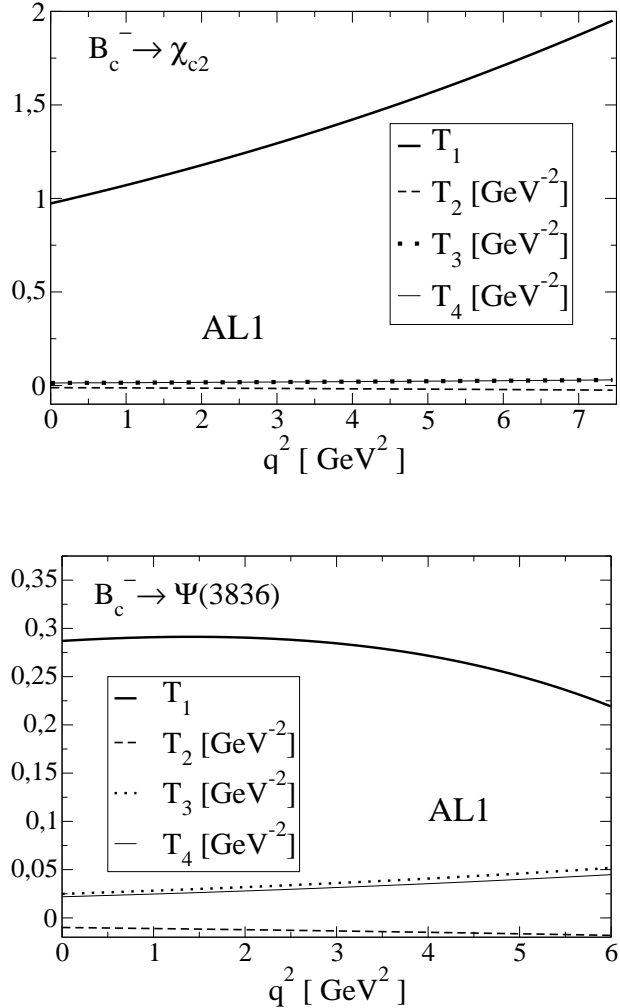

\vspace{1cm}
\centering
\resizebox{8.cm}{!}{\includegraphics{Bc_chic2electron.eps}}
\\ \vspace{1cm}
\resizebox{8.cm}{!}{\includegraphics{Bc_psi3836electron.eps}}
\caption{$T_1$ (bold solid line), $T_2$ (dashed line), $T_3$ (dotted line) 
and $T_4$ (thin solid line) form factors for  
 $B_c^-\to \chi_{c2}$ and
$B_c^-\to \Psi(3836)$
semileptonic decays evaluated with the AL1 potential.}
\label{fig:t1234}
\end{figure}

The results for the different form factors appear in  Fig.~\ref{fig:t1234}.
%
%
%%%%%%                         TABLA III
%
%
In Table~\ref{tab:t1234} we show $T_1$, $T_2$, $T_3$ and $T_4$ evaluated at
$q^2_{min}$ and $q^2_{max}$ for the case of a final light lepton, and compare
them to the values obtained by Ivanov {\it et al.}~\cite{Ivanov:2005fd}. For the 
$B_c^-\to \chi_{c2}$ transition
we find a reasonable agreement between the two calculations. For 
$B_c^-\to \Psi (3836)$ there is also a reasonable agreement for the absolute
values of the form factors but we disagree  for some of the signs.

\begin{table}[h!]
{\footnotesize
\setlength{\tabcolsep}{3pt}
\begin{center}
\begin{tabular}{c| c c || c|  c c}
$B_c^-\to\chi_{c2}\,l^-\bar{\nu}_l\ \ $ & $q^2_{\mathrm{min}}$ &
 $q^2_{\mathrm{max}}$ & 
$B_c^-\to\Psi(3836)\,l^-\bar{\nu}_l\ \ $ & $q^2_{\mathrm{min}}$ &
 $q^2_{\mathrm{max}}$  \\
\hline
&&&&&\\
 $T_1$ & & & $T_1$ & &\\
 This work & $
 0.97^{+0.08}_{-0.01}$&
 $1.95_{-0.06}$ &  This work &$0.29_{-0.02}$& $0.22_{-0.01}$\\
 \cite{Ivanov:2005fd} &1.22 &1.69&\cite{Ivanov:2005fd} &0.052 &0.35\\
&&&&&\\
$T_2\ [GeV^{-2}]$ & & & $T_2\ [GeV^{-2}]$ & &\\
 This work & $-0.012_{-0.001}$&$-0.025^{+0.001}$ &  This work &
 $-0.010^{+0.006}$ &$-0.018^{+0.001}$\\
 \cite{Ivanov:2005fd} & $-0.011$&$-0.018$&  \cite{Ivanov:2005fd} & 0.0071&0.0090\\
&&&&&\\
$T_3\ [GeV^{-2}]$ &  & &$T_3\ [GeV^{-2}]$ & &\\
 This work &$0.013^{+0.001}$& $0.030_{-0.001}$  & 
 This work & $ 0.025_{-0.002}$ & $0.052_{-0.004}$ \\
 \cite{Ivanov:2005fd} & 0.025&0.040&  \cite{Ivanov:2005fd} & $-0.036$&$ -0.052$ \\
&&&&&\\
$T_4\ [GeV^{-2}]$ & &  &$T_4\ [GeV^{-2}]$ & &\\
 This work & $0.013^{+0.001}$& $0.030_{-0.001}$&  
 This work &  $0.022_{-0.001}$ & $0.045_{-0.003}$ \\
 \cite{Ivanov:2005fd} & 0.021$^*$&0.033$^*$ & \cite{Ivanov:2005fd} 
 &  $-0.026$$^*$&$-0.038$$^*$
 \end{tabular}
\end{center}
}
 \caption{Values for $T_1$, $T_2$, $T_3$  and $T_4$ evaluated at
$q^2_{\mathrm{min}}$ and $q^2_{\mathrm{max}}$ compared to  the ones obtained by 
Ivanov {\it et al.}~\cite{Ivanov:2005fd}. Here
$l$ stands for $l=e,\,\mu$. Asterisk as in Table~\ref{tab:fva0pm}.}
\label{tab:t1234}
\end{table}
%
%
%              DECAY WIDTH
%
%
%
\subsection{Decay width}
For a $B_c$ at rest the double differential decay width with respect
to $q^2$ and $x_l$, being $x_l$ the cosine of the angle between the
final meson momentum and the momentum of the final charged lepton
measured in the lepton--neutrino center of mass frame (CMF), is given
by\footnote{We shall neglect neutrino masses in the calculation.}
\begin{eqnarray}
\frac{d^2\Gamma}{dq^2dx_l}=\frac{G_F^2}{64m^2_{B_c}}\,
\frac{|V_{bc}|^2}{8\pi^3}\,\frac{\lambda^{1/2}(q^2,m^2_{B_c},m^2_{c\bar{c}})}
{2m_{B_c}}\, \frac{q^2-m^2_{l}}{q^2}
 \ {\cal H}_{\alpha\,\beta} ({P}_{B_c},{P}_{c\bar{c}})
\ {\cal L}^{\alpha\,\beta}({p}_l,{p}_{{\nu}})
\nonumber\\
\end{eqnarray}
where  $\lambda(a,b,c)=(a+b-c)^2-4ab$, $m_l$ is the
mass of the charged lepton, ${\cal H}_{\alpha\,\beta}$ and ${\cal
L}^{\alpha\,\beta}$ are the hadron and lepton tensors, and
${P}_{B_c},\ {P}_{c\bar{c}},\ {p}_l,\ {p}_{{\nu}}$ are the meson and
lepton four--momenta.  The lepton tensor is\footnote{The $\mp$ signs
correspond respectively to decays into $l^-\bar{\nu}_l$ (for $B_c^-$
decays) and $l^+{\nu}_l$ (for $B_c^+$ decays).}
 \begin{eqnarray}
 {\cal L}^{\alpha\,\beta}({p}_l,{p}_{{\nu}})=
 8\left(
 {p}_l^\alpha{p}_{{\nu}}^\beta+
 {p}_l^\beta{p}_{{\nu}}^\alpha
 -g^{\alpha\beta}{p}_l\cdot{p}_{{\nu}}
 \mp i\,\varepsilon^{\alpha\,\beta\,\sigma\,\rho} p_{l\,\sigma}
 p_{\nu\,\rho}\right)
 \end{eqnarray}
As for the hadron tensor it is given by
 \begin{eqnarray}
 {\cal H}_{\alpha\,\beta}({P}_{B_c},{P}_{c\bar{c}})&=&
 \sum_{\lambda}\ h_{(\lambda)\,\alpha}({P}_{B_c},{P}_{c\bar{c}})\,
  h^*_{(\lambda)\,\beta}({P}_{B_c},{P}_{c\bar{c}})
 \end{eqnarray}
 with
 
 \begin{eqnarray}
 h_{(\lambda)\,\alpha}({P}_{B_c},{P}_{c\bar{c}})=
  \left\langle c\bar{c}, \, \lambda\,
 \vec P_{c\bar{c}}|J_{\alpha}^{cb}(0)|
 B_c,\,\vec P_{B_c}
 \right\rangle
 \end{eqnarray}

 The quantity 
\begin{eqnarray}
 {\cal H}_{\alpha\,\beta}({P}_{B_c},{P}_{c\bar{c}})
\ {\cal L}^{\alpha\,\beta}({p}_{l},{p}_{\nu})%
\end{eqnarray}
is a scalar and to evaluate it we can choose $\vec{P}_{c\bar c}$ along
the negative $z$-axis. This implies also that the CMF of the final
leptons moves in the positive $z$-direction.  Furthermore we shall
follow Ref.~\cite{Ivanov:2005fd} and introduce helicity components for the
hadron and lepton tensors. For that purpose we rewrite
\begin{eqnarray}
 {\cal H}_{\alpha\,\beta}({P}_{B_c},{P}_{c\bar{c}})
\ {\cal L}^{\alpha\,\beta}({p}_{l},{p}_{\nu})
={\cal H}^{\sigma\,\rho}({P}_{B_c},{P}_{c\bar{c}})
\ g_{\sigma\alpha}\,g_{\rho\beta}
\ {\cal L}^{\alpha\,\beta}({p}_{l},{p}_{\nu})
\end{eqnarray}
and use~\cite{Korner:1989qb}
\begin{eqnarray}
g_{\mu\nu}=\sum_{r=t,\pm 1,0} \ g_{rr}\ \varepsilon_{(r)\,\mu}({q})
\,\varepsilon_{(r)\,\nu}^*({q})
\hspace{0.5cm};\hspace{0.5cm} g_{tt}=1,g_{\pm 1,0}=-1
\end{eqnarray}
with $\varepsilon^\mu_{(t)}({q})=q^\mu/q^2$ and where the
  $\varepsilon_{(r)}({q}),\, r=\pm1,\,0$ are the polarization vectors
  for an on--shell vector particle with four--momentum ${q}$ and
  helicity $r$.  Defining helicity components for the hadron and
  lepton tensors as
\begin{eqnarray}
 {\cal H}_{r\,s}({P}_{B_c},{P}_{c\bar{c}})&=&
\varepsilon_{(r)\,\sigma}^*({q})\,
{\cal H}^{\sigma\,\rho}({P}_{B_c},{P}_{c\bar{c}})
\,\varepsilon_{(s)\,\rho}({q})\nonumber\\
 {\cal L}_{r\,s}({p}_{l},{p}_{\nu})&=&
\varepsilon_{(r)\,\alpha}({q})\,
{\cal L}^{\alpha\,\beta}({p}_{l},{p}_{\nu})
\,\varepsilon_{(s)\,\beta}^*({q})
\end{eqnarray}
we have that
\begin{eqnarray}
 {\cal H}_{\alpha\,\beta}({P}_{B_c},{P}_{c\bar{c}})
\ {\cal L}^{\alpha\,\beta}({p}_{l},{p}_{\nu})
=\sum_{r=t,\pm 1,0}\sum_{s=t,\pm 1,0}\ g_{rr}\,g_{ss}\
 {\cal H}_{r\,s}({P}_{B_c},{P}_{c\bar{c}})\
 {\cal L}_{r\,s}({p}_l,{p}_{\nu})
\nonumber \\
 \end{eqnarray}

Let us start with the lepton tensor. We can take advantage of the fact
that the Wigner rotation relating the original frame and the CMF of
the final leptons is the identity to evaluate the lepton tensor
helicity components in this latter reference system
\begin{eqnarray}
 {\cal L}_{r\,s}({p}_l,{p}_{\nu})&=&
\varepsilon_{(r)\,\alpha}({q})\,
  {\cal L}^{\alpha\,\beta}({p}_{l},{p}_{\nu})
\,\varepsilon_{(s)\,\beta}^*({q})=
\varepsilon_{(r)\,\alpha}(\widetilde{q})\,
{\cal L}^{\alpha,\beta}(\widetilde{p}_{l},\widetilde{p}_{\nu})
\,\varepsilon_{(s)\,\beta}^*(\widetilde{q})
\nonumber\\
\end{eqnarray}
were the tilde stands for momenta measured in the final leptons CMF. 
For the purpose of evaluation we can use\footnote{Note this is in accordance
with the definition of $x_l$ and the fact that we have taken $\vec{P}_{c\bar c}$
in the negative $z$ direction. Furthermore there can be no dependence on the
$\varphi_l$ azimuthal angle so that we  can take $\widetilde{\vec{p}}_{l}$, and
then $\widetilde{\vec{p}}_{\nu}$,
 in the $OXZ$ plane.}
\begin{eqnarray}
\label{eq:lm}
\widetilde{p}_l^\alpha&=&(E_l(|\widetilde p_l|),-|\widetilde p_l|
\,\sqrt{1-x_l^2},0,-|\widetilde p_l|\,x_l)\nonumber\\
\widetilde{p}_\nu^\alpha&=&(|\widetilde p_l|,|\widetilde p_l|
\,\sqrt{1-x_l^2},0,|\widetilde p_l|\,x_l)
\end{eqnarray}
with $|\widetilde p_l|$ the modulus of the lepton three-momentum measured in the leptons CMF.

The only  helicity components that we shall need
are the following.
\begin{eqnarray}
{\cal L}_{t\,t}({p}_l,{p}_{\nu})&=&
4 \frac{m_l^2 (q^2-m_l^2)}{q^2}\nonumber\\
{\cal L}_{t\,0}({p}_l,{p}_{\nu})&=&
{\cal L}_{0\,t}({p}_l,{p}_{\nu})=
-4x_l\frac{m_l^2 (q^2-m_l^2)}{q^2}\nonumber\\
{\cal L}_{+1\,+1}({p}_l,{p}_{\nu})&=&
(q^2-m_l^2)\,\left(
4(1\pm x_l)-2(1-x_l^2)\frac{q^2-m_l^2}{q^2}
\right)
\nonumber\\
{\cal L}_{-1\,-1}({p}_l,{p}_{\nu})&=&
(q^2-m_l^2)\,\left(
4(1\mp x_l)-2(1-x_l^2)\frac{q^2-m_l^2}{q^2}
\right)
\nonumber\\
{\cal L}_{0\,0}({p}_l,{p}_{\nu})&=&
4(q^2-m_l^2)\,\left(
1-x_l^2\frac{q^2-m_l^2}{q^2}
\right)
\nonumber\\
\end{eqnarray}
As for the hadron tensor  it is convenient to introduce 
helicity amplitudes defined as
 \begin{eqnarray}
 h_{(\lambda)\,r}({P}_{B_c},{P}_{c\bar{c}})=
\varepsilon_{(r)\,\alpha}^*({q})\ h_{(\lambda)}^\alpha({P}_{B_c},{P}_{c\bar{c}}) 
 \hspace{0.5cm},\hspace{0.5cm} r=t,\,\pm 1,\,0 \end{eqnarray}
  in terms of which
 \begin{eqnarray}
 {\cal H}_{r\,s}({P}_{B_c},{P}_{c\bar{c}})=\sum_\lambda
h_{(\lambda)\,r}({P}_{B_c},{P}_{c\bar{c}})\,
h_{(\lambda)\,s}^*({P}_{B_c},{P}_{c\bar{c}})
\label{eq:hcht}
 \end{eqnarray} 
We now give the expressions for the helicity amplitudes 
evaluated in the original frame. 

\begin{itemize}
\item[$\bullet$] Case $0^-\to 0^-,0^+$.
\begin{eqnarray}
h_t({P}_{B_c},{P}_{c\bar{c}})&=&\frac{m^2_{B_c}-m^2_{c\bar c}}{\sqrt{q^2}}\,F_+(q^2)+\sqrt{q^2}\,
F_-(q^2)\nonumber\\
h_0({P}_{B_c},{P}_{c\bar{c}})&=&\frac{\lambda^{1/2}(q^2,m^2_{B_c},
m^2_{c\bar c})}{\sqrt{q^2}}\, F_+(q^2)\nonumber\\
h_{+1}({P}_{B_c},{P}_{c\bar{c}})&=&h_{-1}({P}_{B_c},{P}_{c\bar{c}})=0
\end{eqnarray}
\item[$\bullet$] Case $0^-\to 1^-,1^+$.
\begin{eqnarray}
h_{(\lambda)\,t}({P}_{B_c},{P}_{c\bar{c}})&=&i\delta_{\lambda\,0}\,
\frac{\lambda^{1/2}(q^2,m^2_{B_c},m^2_{c\bar c})}{2m_{c\bar c}\sqrt{q^2}}\,
\nonumber \\ &&\hspace{.1cm}
\left((m_{B_c}-m_{c\bar c})\left(A_0(q^2)-A_+(q^2)\right)
-\frac{q^2}{m_{B_c}+m_{c\bar c}}A_-(q^2)\right)\nonumber\\
h_{(\lambda)\,+1}({P}_{B_c},{P}_{c\bar{c}})&=&-i\delta_{\lambda\,-1}\,
\left(\ \frac{\lambda^{1/2}(q^2,m^2_{B_c},m^2_{c\bar c})}{m_{B_c}+m_{c\bar c}}\,
V(q^2)+(m_{B_c}-m_{c\bar c})\,A_0(q^2)\right)
\nonumber\\
h_{(\lambda)\,-1} ({P}_{B_c},{P}_{c\bar{c}})&=&
-i\delta_{\lambda\,+1}\,
\left(-\frac{\lambda^{1/2}(q^2,m^2_{B_c},m^2_{c\bar c})}{m_{B_c}+m_{c\bar c}}\,
V(q^2)+(m_{B_c}-m_{c\bar c})\,A_0(q^2)\right)
\nonumber\\
h_{(\lambda)\,0}({P}_{B_c},{P}_{c\bar{c}})&=&i\delta_{\lambda\,0}\,
\left( (m_{B_c}-m_{c\bar c})\frac{m^2_{B_c}-q^2-m^2_{c\bar c}}{2m_{c\bar c}\sqrt{q^2}}\,
A_0(q^2)
\right. \nonumber \\ && \hspace*{2cm}\left.
-\frac{\lambda(q^2,m^2_{B_c},m^2_{c\bar c})}{2m_{c\bar
c}\sqrt{q^2}}\,\frac{A_+(q^2)}{m_{B_c}+m_{c\bar c}}
\right)
\end{eqnarray}
\item[$\bullet$] Case $0^-\to 2^-,2^+$.
\begin{eqnarray}
 h_{(\lambda)\,t}({P}_{B_c},{P}_{c\bar{c}})&=&-i\delta_{\lambda\,0}\,\sqrt{\frac{2}{3}}
\ \frac{\lambda(q^2,m^2_{B_c},m^2_{c\bar c})}{4\,m^2_{c\bar c}\,\sqrt{q^2}}\,
\nonumber\\ &&\hspace*{1cm}
\left(\, T_1(q^2)+(m^2_{B_c}-m^2_{c\bar c})\,T_2(q^2)+q^2\,T_3(q^2)\,
\right)\nonumber\\
h_{(\lambda)\,+1}({P}_{B_c},{P}_{c\bar{c}})&=&i\delta_{\lambda\,-1}\,
\frac{1}{\sqrt2}\,
\frac{\lambda^{1/2}(q^2,m^2_{B_c},m^2_{c\bar c})}{2\,m_{c\bar c}}\,
\nonumber\\ &&\hspace*{1cm}
\left(\,T_1(q^2)-\lambda^{1/2}(q^2,m^2_{B_c},m^2_{c\bar c})\,T_4(q^2)\,
\right)
\nonumber\\
h_{(\lambda)\,-1}({P}_{B_c},{P}_{c\bar{c}})&=&i\delta_{\lambda\,+1}\,
\frac{1}{\sqrt2}\,
\frac{\lambda^{1/2}(q^2,m^2_{B_c},m^2_{c\bar c})}{2\,m_{c\bar c}}\,
\nonumber\\ &&\hspace*{1cm}
\left(\,T_1(q^2)+\lambda^{1/2}(q^2,m^2_{B_c},m^2_{c\bar c})\,T_4(q^2)\,
\right)
\nonumber\\
h_{(\lambda)\,0}({P}_{B_c},{P}_{c\bar{c}})&=&
-i\delta_{\lambda\,0}\,\sqrt{\frac{2}{3}}
\ \frac{\lambda^{1/2}(q^2,m^2_{B_c},m^2_{c\bar c})}{4\,m^2_{c\bar c}\,\sqrt{q^2}}\,
\nonumber\\ &&\hspace*{1cm}
\left(\,(m^2_{B_c}-q^2-m^2_{c\bar c})\,T_1(q^2)+\lambda(q^2,m^2_{B_c},m^2_{c\bar c})\,T_2(q^2)\,
\right)\nonumber\\
\end{eqnarray}

\end{itemize}
We see that the helicity amplitudes, and thus the  helicity components of the
hadron tensor,  only depend on $q^2$. The expressions for the latter are
collected in appendix~\ref{Bcapp:hcht}.

We can now define the combinations~\cite{Ivanov:2005fd}
\begin{eqnarray}
H_U&=&{\cal H}_{+1\,+1}+{\cal H}_{-1\,-1}\nonumber\\
H_L&=&{\cal H}_{0\,0}\nonumber\\
H_P&=&{\cal H}_{+1\,+1}-{\cal H}_{-1\,-1}\nonumber\\
H_S&=&3\,{\cal H}_{t\,t}\nonumber\\
H_{SL}&=&{\cal H}_{t\,0}\nonumber\\
\widetilde{H}_J&=&\frac{m_l^2}{2\,q^2}\,H_J\hspace{0.5cm};\hspace{0.5cm}
J=U,L,S,SL
\label{eq:combinaciones}
\end{eqnarray}
with $U,\,L,\,P,\,S,\,SL$ representing respectively unpolarized--transverse,
longitudinal, parity--odd, scalar and scalar--longitudinal
interference.

Finally the double differential decay width is written in terms of the above
defined combinations 
as
\begin{eqnarray}
\frac{d^2\Gamma}{dq^2dx_l}&=&\frac{G_F^2}{8\pi^3}\,|V_{bc}|^2\,
\frac{(q^2-m^2_{l})^2}{12m^2_{B_c}q^2}\,
\frac{\lambda^{1/2}(q^2,m^2_{B_c},m^2_{c\bar{c}})}
{2m_{B_c}}\,
\nonumber \\ &&  \hspace{.5cm}
\bigg\{\
\frac{3}{8}(1+x_l^2)\,H_U+\frac{3}{4}(1-x_l^2)\,H_L\pm\frac{3}{4}x_lH_P
\nonumber \\ && \hspace{2cm}
+\frac{3}{4}(1-x_l^2)\widetilde{H}_U+\frac{3}{2}x^2_l
\widetilde{H}_L
+\frac{1}{2}\widetilde{H}_S+3x_l\widetilde{H}_{SL}
\bigg\}
\end{eqnarray}
Note that  for antiparticle decay $H_P$  has the opposite sign to the case of
particle decay while all other hadron tensor helicity components combinations defined in  Eq.(\ref{eq:combinaciones})
do not change (See
appendix~\ref{Bcapp:sign} for details). The sign change  of $H_P$ compensates
 the extra sign  coming from the lepton tensor.  This means that in fact the double
differential decay width is the same for $B_c^-$ or $B_c^+$ decay.

Integrating over $x_l$ we obtain the differential decay width
\begin{eqnarray}
\frac{d\Gamma}{dq^2}&=&\frac{G_F^2}{8\pi^3}\,|V_{bc}|^2\,
\frac{(q^2-m^2_{l})^2}{12m^2_{B_c}q^2}\,
\frac{\lambda^{1/2}(q^2,m^2_{B_c},m^2_{c\bar{c}})}
{2m_{B_c}}\, 
\bigg\{\
H_U+H_L+\widetilde{H}_U+\widetilde{H}_L
+\widetilde{H}_S
\bigg\}
\nonumber\\
\end{eqnarray}
from where, integrating over $q^2$, we obtain the total decay width that we
write, following Ref.~\cite{Ivanov:2005fd}, as
\begin{eqnarray}
\Gamma&=&\Gamma_U+\Gamma_L+\widetilde{\Gamma}_U+\widetilde{\Gamma}_L
+\widetilde{\Gamma}_S
\end{eqnarray}
with $\Gamma_J$ and $\widetilde{\Gamma}_J$ partial helicity widths defined as
\begin{eqnarray}
\Gamma_J=\int dq^2 \frac{G_F^2}{8\pi^3}\,|V_{bc}|^2\,
\frac{(q^2-m^2_{l})^2}{12m^2_{B_c}q^2}\,
\frac{\lambda^{1/2}(q^2,m^2_{B_c},m^2_{c\bar{c}})}
{2m_{B_c}}\, H_J
\end{eqnarray}
and similarly for $\widetilde{\Gamma}_J$ in terms of
$\widetilde{H}_J$. 

\begin{table}[!h]
\begin{center}
\hspace*{-1.7cm}
{\scriptsize \setlength{\tabcolsep}{2pt}
\begin{tabular}{l|c c c c c c c}
$B^-_c\to$ & \hspace*{-.4cm}$\Gamma_U$\hspace*{-.4cm} & \hspace*{-.4cm}$\widetilde{\Gamma}_U$\hspace*{-.4cm}
 &\hspace*{-.4cm} $\Gamma_L$\hspace*{-.4cm} & $\widetilde{\Gamma}_L$ & \hspace*{-.4cm}$\Gamma_P$
 \hspace*{-.4cm}& $\widetilde{\Gamma}_S $ & $\widetilde{\Gamma}_{SL}$\\ 
\hline
&&&&&&&\\
$\eta_c\,e^-\bar{\nu}_e$& 0 & 0 &$ 6.95^{+0.31}$&
$0.13^{+0.01}\ 10^{-5}$ & 0 &
 $ 0.44^{+0.03}\ 10^{-5}$  & $ 0.14^{+0.01}\ 10^{-5}$\\
%\hline
%&&&&&&&\\
$\eta_c\,\mu^-\bar{\nu}_\mu$& 0 & 0 &$ 6.80^{+0.31}$&
$0.28^{+0.02}\ 10^{-1}$ & 0 &
 $ 0.10^{+0.01}$  & $ 0.31^{+0.01}\ 10^{-1}$\\
%\hline
%&&&&&&&\\
$\eta_c\,\tau^-\bar{\nu}_\tau$
 & 0& 0&$0.71^{+0.02}$ &$0.17^{+0.01}$ &0 &$1.58^{+0.04}$
 &$0.29^{+0.01}$ \\
&&&&&&&\\

$\chi_{c0}\,e^-\bar{\nu}_e$& 0 & 0 &$ 1.55^{+0.14}_{-0.02}$&
$0.37^{+0.05}\ 10^{-6}$ & 0 &
 $ 0.11^{+0.01}\ 10^{-5}$  & $ 0.37^{+0.05}\ 10^{-6}$\\
%&&&&&&&\\
$\chi_{c0}\,\mu^-\bar{\nu}_\mu$& 0 & 0 &$ 1.51^{+0.13}_{-0.02}$&
$0.75^{+0.09}\ 10^{-2}$ & 0 &
 $ 0.23^{+0.02}\ 10^{-1}$  & $ 0.75^{+0.09}\ 10^{-2}$\\
%&&&&&&&\\
$ \chi_{c0}\,\tau^-\bar{\nu}_\tau$
 &0&0&$0.80^{+0.04}_{-0.02}\ 10^{-1}$
&$0.23^{+0.01}_{-0.01}\ 10^{-1}$
 &0 &
$0.84^{+0.07}\ 10^{-1} $ &$0.25^{+0.02}\ 10^{-1}$ \\
&&&&&&&\\

$J/\Psi\,e^-\bar{\nu}_e$
 & $11.5^{+0.6}$&$0.32^{+0.02}\ 10^{-6}$&$10.4^{+0.6}$ &$0.12^{+0.01}\ 10^{-5}$
 &$-5.48_{-0.24}$ &
$0.32^{+0.03}\ 10^{-5}$ &$0.11^{+0.01}\ 10^{-5}$ \\
%&&&&&&&\\
$J/\Psi\,\mu^-\bar{\nu}_\mu$
 & $11.4^{+0.6}$&$0.13^{+0.01}\ 10^{-1}$&$10.2^{+0.7}$ &$0.28^{+0.03}\ 10^{-1}$
 &$-5.45_{-0.24}$ &
$0.68^{+0.07}\ 10^{-1}$ &$0.25^{+0.02}\ 10^{-1}$ \\
%&&&&&&&\\
$J/\Psi\,\tau^-\bar{\nu}_\tau$
 & $2.78^{+0.10}_{-0.01}$&$0.59^{+0.02}$&$1.74^{+0.07}_{-0.01}$ &
 $0.39^{+0.02}$
 &$-1.10_{-0.03}$ &
$0.36^{+0.02}$ &$0.21^{+0.01}$ \\
&&&&&&&\\

$\chi_{c1}\,e^-\bar{\nu}_e$
 &$0.90^{+0.05}_{-0.03}$&$0.43^{+0.03}_{-0.01}\ 10^{-7}$
&$0.35^{+0.03}\ 10^{-1}$&$0.28^{+0.03}\
 10^{-8}$&$-0.75_{-0.04}^{+0.02}$
&$0.57^{+0.07}\ 10^{-8}$&$0.22^{+0.02}\ 10^{-8}$\\
%&&&&&&&\\
$\chi_{c1}\,\mu^-\bar{\nu}_\mu$
 &$0.89^{+0.05}_{-0.03}$&$0.18^{+0.01}_{-0.01}\ 10^{-2}$
&$0.35^{+0.03}\ 10^{-1}$&$0.77^{+0.08}\
 10^{-4}$&$-0.75_{-0.04}^{+0.02}$
&$0.11^{+0.01}\ 10^{-3}$&$0.49^{+0.05}\ 10^{-4}$\\
%&&&&&&&\\

$\chi_{c1}\,\tau^-\bar{\nu}_\tau$
 &$0.75^{+0.02}\ 10^{-1}$&$0.21^{+0.01}\ 10^{-1}$&$0.46^{+0.04}\ 10^{-2}$&
 $0.12^{+0.01}\ 10^{-2}$&
$-0.64_{-0.03}\ 10^{-1}$&$0.23^{+0.02}\ 10^{-3}$&$0.28^{+0.02}\ 10^{-3}$\\
&&&&&&&\\

$h_c\,e^-\bar{\nu}_e$ 
&$0.16^{+0.02}$&$0.57^{+0.07}\ 10^{-8}$&$2.23^{+0.12}$&
$0.72^{+0.09}\ 10^{-6}$&$-0.26_{-0.03}\ 10^{-1}$&
$0.23^{+0.03}\ 10^{-5}$&$0.74^{+0.09}\ 10^{-6}$\\
%& & & & & & &\\
$h_c\,\mu^-\bar{\nu}_\mu$ 
&$0.16^{+0.02}$&$0.24^{+0.03}\ 10^{-3}$&$2.16^{+0.21}_{-0.01}$&
$0.14^{+0.01}\ 10^{-1}$&$-0.26_{-0.03}\ 10^{-1}$&
$0.45^{+0.05}_{-0.01}\ 10^{-1}$&$0.14^{+0.02}\ 10^{-1}$\\
%& & & & & & &\\

$h_c \,\tau^-\bar{\nu}_\tau$ 
& $0.23^{+0.02} \ 10^{-1}$ & $0.60^{+0.06}\ 10^{-2}$ &
$0.67^{+0.05}\ 10^{-1}$&
$0.20^{+0.01}\ 10^{-1}$&
$-0.27_{-0.03}\ 10^{-2}$ &$0.97^{+0.05}_{-0.03}\ 10^{-1}$&
$0.26_{-0.01}^{+0.02}\ 10^{-1}$\\
& & & & & & &\\

$\chi_{c2}\,e^-\bar{\nu}_e$ 
 &$0.71^{+0.03}_{-0.03}$&$0.37^{+0.02}_{-0.02}\ 10^{-7}$
&$1.17^{+0.08}_{-0.05}$& $0.31^{+0.03}_{-0.01}\ 10^{-6}$
&$-0.35_{-0.02}^{+0.01}$&$0.88^{+0.09}_{-0.03}\ 10^{-6}$
&$0.30^{+0.03}_{-0.01}\ 10^{-6}$\\
%& & & & & & &\\
$\chi_{c2}\,\mu^-\bar{\nu}_\mu$ 
 &$0.71^{+0.02}_{-0.03}$&$0.15^{+0.01}\ 10^{-2}$
&$1.14^{+0.07}_{-0.05}$& $0.62^{+0.06}_{-0.02}\ 10^{-2}$
&$-0.34_{-0.02}^{+0.01}$&$0.16^{+0.02}\ 10^{-1}$
&$0.57^{+0.06}_{-0.02}\ 10^{-2}$\\
%& & & & & & &\\
$\chi_{c2}\,\tau^-\bar{\nu}_\tau$ 
 &$0.49^{+0.01}_{-0.03}\ 10^{-1}$ &$0.15_{-0.01}\ 10^{-1}$
 &$0.43^{+0.01}_{-0.02}\ 10^{-1}$ &$0.13_{-0.01}\ 10^{-1}$
 &$-0.18^{+0.01}\ 10^{-1}$ &$0.12_{-0.01}\ 10^{-1}$
 &$0.70^{+0.02}_{-0.04}\ 10^{-2}$\\
& & & & & & &\\

$\Psi(3836)\,e^-\bar{\nu}_e$ 
&$0.58_{-0.07} \ 10^{-1}$&$0.47_{-0.06}\ 10^{-8}$&$0.33^{+0.01}_{-0.02}\ 10^{-2}$
&$0.68^{+0.04}_{-0.04}\ 10^{-9}$&$-0.48^{+0.05}\ 10^{-1}$
&$0.17^{+0.01}_{-0.01}\ 10^{-8}$&$0.60^{+0.03}_{-0.04}\ 10^{-9}$\\
%& & & & & & &\\
$\Psi(3836)\,\mu^-\bar{\nu}_\mu$ 
&$0.57_{-0.06} \ 10^{-1}$&$0.19_{-0.02}\ 10^{-3}$&$0.32^{+0.01}_{-0.02}\ 10^{-2}$
&$0.15_{-0.01}\ 10^{-4}$&$-0.48^{+0.06}\ 10^{-1}$
&$0.28^{+0.02}_{-0.02}\ 10^{-4}$&$0.11^{+0.01}\ 10^{-4}$\\
%& & & & & & &\\
$\Psi(3836) \,\tau^-\bar{\nu}_\tau$
&$0.78_{-0.09}\ 10^{-3}$& $0.28_{-0.03}\ 10^{-3}$&$0.74_{-0.06}\ 10^{-4}$
&$0.25_{-0.02}\ 10^{-4}$& $-0.69^{+0.08}\ 10^{-3}$&$0.54_{-0.04}\ 10^{-5}$
&$0.65_{-0.05}\ 10^{-5}$\\
\end{tabular}
}
\end{center}
\caption{ Partial helicity widths in units of $10^{-15}$\,GeV for $B_c^-$ decays. Central values
have been evaluated with the AL1 potential. } 
\label{tab:ha}
\end{table}

Another quantity of interest is the forward-backward asymmetry of the charged
lepton measured in the leptons CMF. This asymmetry is defined as\footnote{The
forward direction is determined by the momentum of the final meson that we have
chosen in the negative $z$-direction.  }

\begin{eqnarray}
A_{FB}=\frac{\Gamma_{x_l>0}-\Gamma_{x_l<0}}{\Gamma_{x_l>0}+\Gamma_{x_l<0}}
\end{eqnarray}
and it is given in terms or partial helicity widths as
\begin{eqnarray}
=\frac{3}{4}\,\frac{\pm\Gamma_P+4\,\widetilde{\Gamma}_{SL}}
{\Gamma_U+\Gamma_L+\widetilde{\Gamma}_U+\widetilde{\Gamma}_L
+\widetilde{\Gamma}_S}
\end{eqnarray}
being the same for a negative charged lepton $l^-$ ($B_c^-$ decay) than for a
positive charged one $l^+$ ($B_c^+$ decay), as $\Gamma_P$ for antiparticle
decay has the opposite sign as for particle decay.

Finally for the decay channel $B_c\to J/\Psi\ l\bar{\nu}_l$ with the 
$J/\Psi$ decaying into $\mu^-\mu^+$
we can
evaluate the differential cross section

% COPIA POR SI LO ROMPO
%\begin{eqnarray}
%\frac{d\Gamma_{B_c\to\mu^-\mu^+ (J/\Psi)\,l\bar{\nu}_l}}{dx_\mu}&=&\left(
%1+\frac{\sqrt{1-4m_\mu^2/m_{J/\Psi}^2}\,\bigg(
%\Gamma_U+\widetilde{\Gamma}_U-2(\Gamma_L+\widetilde{\Gamma}_L
%+\widetilde{\Gamma}_S)\bigg)}{
%\left(2-\sqrt{1-4m_\mu^2/m_{J/\Psi}^2}\,
%\right)\,\left(\Gamma_U+\widetilde{\Gamma}_U\right)
%+2\left(\Gamma_L+\widetilde{\Gamma}_L
%+\widetilde{\Gamma}_S\right)
%}\ x_\mu^2\right)\nonumber\\
%&&\times\frac{\Gamma_{J/\Psi\to\mu^-\mu^+}}{\Gamma_{J/\Psi}}\,
%\frac{1}{1+2m_\mu^2/m_{J/\Psi}^2}
%\bigg[\ \frac{3}{4}\,\left(\Gamma_L+\widetilde{\Gamma}_L
%+\widetilde{\Gamma}_S\right)\nonumber\\
%&&\hspace{4.75cm}+\frac{3}{8}\,
%\left(2-\sqrt{1-4m_\mu^2/m_{J/\Psi}^2}\ \right)
%\left(\Gamma_U+\widetilde{\Gamma}_U\right)
%\bigg]
%\end{eqnarray}
%
\begin{eqnarray}
\lefteqn{\frac{d\Gamma_{B_c\to\mu^-\mu^+
      (J/\Psi)\,l\bar{\nu}_l}}{dx_\mu}\,=}
\nonumber\\&& \nonumber \left(
1+\frac{\sqrt{1-4m_\mu^2/m_{J/\Psi}^2}\,\bigg(
\Gamma_U+\widetilde{\Gamma}_U-2(\Gamma_L+\widetilde{\Gamma}_L
+\widetilde{\Gamma}_S)\bigg)}{
\left(2-\sqrt{1-4m_\mu^2/m_{J/\Psi}^2}\,
\right)\,\left(\Gamma_U+\widetilde{\Gamma}_U\right)
+2\left(\Gamma_L+\widetilde{\Gamma}_L
+\widetilde{\Gamma}_S\right)
}\ x_\mu^2\right)\nonumber\\&&
\times\frac{\Gamma_{J/\Psi\to\mu^-\mu^+}}{\Gamma_{J/\Psi}}\,
\frac{1}{1+2m_\mu^2/m_{J/\Psi}^2}
\bigg[\ \frac{3}{4}\,\left(\Gamma_L+\widetilde{\Gamma}_L
+\widetilde{\Gamma}_S\right)\nonumber\\ &&
\hspace{4.cm}+\frac{3}{8}\,
\left(2-\sqrt{1-4m_\mu^2/m_{J/\Psi}^2}\ \right)
\left(\Gamma_U+\widetilde{\Gamma}_U\right)
\bigg]
\end{eqnarray}
where $x_\mu$ is the cosine of the polar angle for the final
$\mu^-\mu^+$ pair, relative to the momentum of the decaying $J/\Psi$,
measured in the $\mu^-\mu^+$ CMF, $\Gamma_{J/\Psi\to\mu^-\mu^+}$ is
the $J/\Psi$ decay width into the $\mu^-\mu^+$ channel, and
$\Gamma_{J/\Psi}$ is the total $J/\Psi$ decay width. The asymmetry
parameter
\begin{eqnarray}
\alpha^*=\frac{\sqrt{1-4m_\mu^2/m_{J/\Psi}^2}\,\bigg(
\Gamma_U+\widetilde{\Gamma}_U-2(\Gamma_L+\widetilde{\Gamma}_L
+\widetilde{\Gamma}_S)\bigg)}{
\left(2-\sqrt{1-4m_\mu^2/m_{J/\Psi}^2}\ 
\right)\,\left(\Gamma_U+\widetilde{\Gamma}_U\right)
+2\left(\Gamma_L+\widetilde{\Gamma}_L
+\widetilde{\Gamma}_S\right)
}\
\end{eqnarray}
governs the muons angular distribution in their CMF.

\subsubsection{Results}

\begin{table}[!ht]
{\scriptsize
\setlength{\tabcolsep}{3pt}
%\begin{center}
\hspace*{-.4cm}
\begin{tabular}{lc|c c c c c c}
&&$A_{FB} (e)$
&$A_{FB} (\mu)$&$\ \ A_{FB} (\tau)\ \ $& $\  \ \alpha^* (e)$&
 $\  \ \alpha^* (\mu)$&$\  \
\alpha^* (\tau)$\\
\hline
&&&\\
$B_c\to \eta_c$&\\
 & This work &$0.60^{+0.01}\ 10^{-6}$&$0.13^{+0.01}\ 10^{-1}$ &$0.35$&&&\\
&\cite{Ivanov:2005fd}& $0.953\ 10^{-6}$&&0.36\\
$B_c\to \chi_{c0}$ &\\
& This work&$0.72^{+0.02}\ 10^{-6}$&$0.15\ 10^{-1}$&$0.40$&&&\\
&\cite{Ivanov:2005fd}& $1.31\ 10^{-6}$&&0.39\\
$B_c\to J/\Psi$ &\\
& This work&$-0.19$&$-0.18_{-0.01}$ &$-0.35^{+0.02}\ 10^{-1}$&$-0.29_{-0.01}$
&
$-0.29$&
$-0.19$\\

&\cite{Ivanov:2005fd}&$-0.21$&&$-0.48\ 10^{-1}$&$-0.34$&&$-0.24$\\
$B_c\to \chi_{c1}$ &\\
&This work &$-0.60_{-0.01}$&$-0.60_{-0.01}$&$-0.46$&&&\\
&\cite{Ivanov:2005fd}&0.19&&0.34\\
$B_c\to h_c$ &\\
&This work&$-0.83_{-0.05} \ 10^{-2}$&$0.97_{-0.05}^{+0.01}\ 10^{-2}$&
$0.35$&&&\\
&\cite{Ivanov:2005fd}&$-3.6\ 10^{-2}$&&0.31\\

$B_c\to \chi_{c2}$ &\\
& This work&$-0.14$&$-0.13$&$0.55^{+0.02}\ 10^{-1}$&&&\\
&\cite{Ivanov:2005fd}&$-0.16$&&$0.44\ 10 ^{-1}$&\\
$B_c\to \Psi (3836)$&\\
&This work &$-0.59$&$-0.59$&$-0.42$&&&\\
&\cite{Ivanov:2005fd}&0.21&&0.41\\
\end{tabular}
}
%\end{center}
\caption{Asymmetry parameters in semileptonic $B_c\to c\bar c$ decays. Our central values have been evaluated with the
AL1 potential. We also show the results obtained by Ivanov {\it et
al.}~\cite{Ivanov:2005fd}. } 
\label{tab:asy}
\end{table}

\begin{figure}[!p]
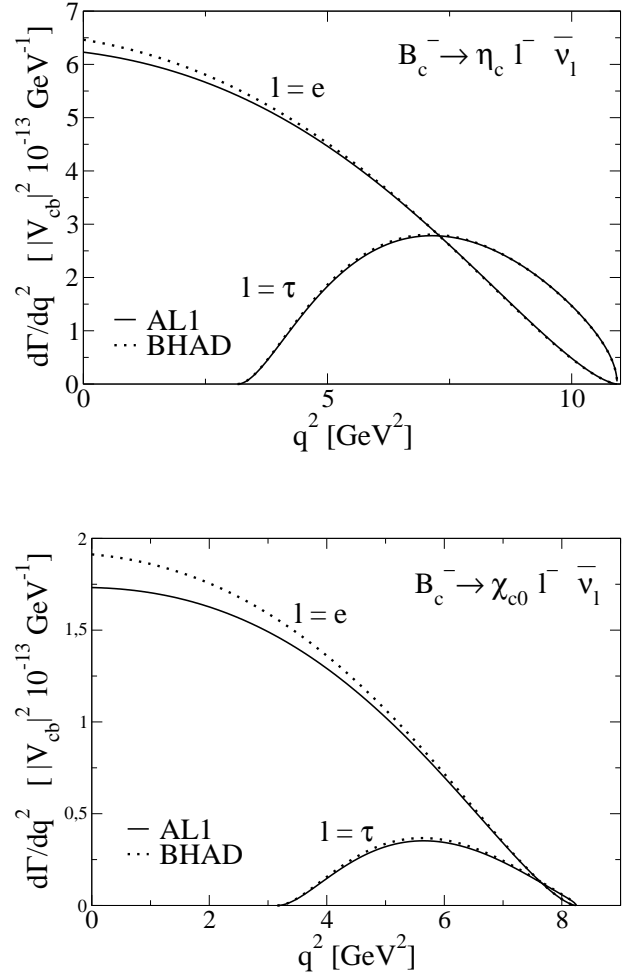

\vspace{1cm}
\centering
\resizebox{8.cm}{!}{\includegraphics{Bc_etacdgdq2.eps}}
\\ \vspace{1cm}
\resizebox{8.cm}{!}{\includegraphics{Bc_chic0dgdq2.eps}}
%\resizebox{10cm}{!}
\caption{Differential decay width for the $B_c^-\to \eta_c
l^-\bar{\nu}_l$ and $B_c^-\to \chi_{c0} l^-\bar{\nu}_l$ processes
obtained with the AL1 potential.  For comparison, we also show with
dotted lines the results obtained with the Bhaduri (BHAD) potential.
 The distribution for a final muon is not explicitly shown. It
differs appreciably from the corresponding to a final electron only
for $q^2$ around $m_\mu^2$.}
\label{fig:0pmdgdq2}
\end{figure}

In Table~\ref{tab:ha} we give our results for the partial helicity
widths corresponding to $B_c^-$ decays. For $B_c^+$ decay the ``$P$''
column changes sign while all others remain the same. The central values have been evaluated with the AL1
potential and the theoretical errors quoted reflect the dependence of the
results on the inter-quark potential.

In Table~\ref{tab:asy} we show the asymmetry parameters. Our values
for $\alpha^*$ compare well with the results obtained in
Ref.~\cite{Ivanov:2005fd}. The same is true for the forward--backward asymmetry
with some exceptions: most notably we get opposite signs for $B_c\to\chi_{c1}$ and
$B_c\to\Psi(3836)$.

In Fig.~\ref{fig:0pmdgdq2} we show the differential decay width
$d\Gamma/dq^2$ for the decay channels $\eta_c l^-\bar{\nu}_l$ and
$\chi_{c0} l^-\bar{\nu}_l$ for the case where the final lepton is a
light one $l=e,\mu$\footnote{We show the distribution corresponding to
a final electron. The distribution for a final muon differs from the
former only for $q^2$ around $m_\mu^2$.} or a heavy one $l=\tau$. We show
the results obtained with the AL1 and BHAD potentials, finding no
significant difference for the $\tau$ case, while for the light final
lepton case the differences are around 10\% at low $q^2$.

\begin{figure}[p!!!]
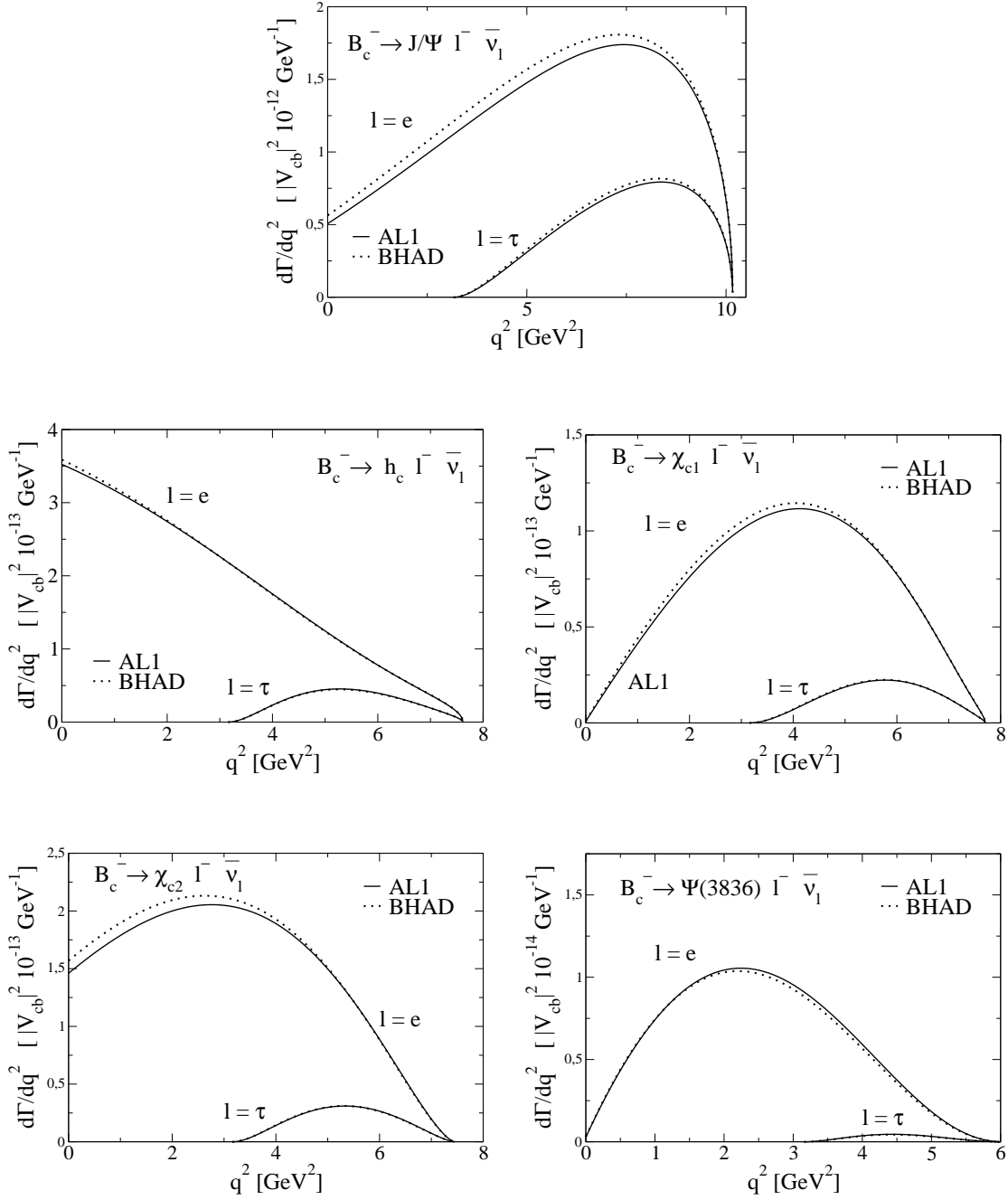

\vspace{1cm}
\centering
\hspace*{-1.cm}
\resizebox{7.cm}{!}{\includegraphics{Bc_jpsidgdq2.eps}}\vspace{1cm}\\
\hspace*{-1.cm}
\resizebox{7.cm}{!}{\includegraphics{Bc_hcdgdq2.eps}}\hspace{.4cm}
\resizebox{7.cm}{!}{\includegraphics{Bc_chic1dgdq2.eps}}\vspace{1cm}\\
\hspace*{-1.cm}
\resizebox{7.cm}{!}{\includegraphics{Bc_chic2dgdq2.eps}}\hspace{.4cm}
\resizebox{7.cm}{!}{\includegraphics{Bc_psi3836dgdq2.eps}}
%\resizebox{10cm}{!}
\caption{Differential decay width for the $B_c^-\to J/\Psi
  l^-\bar{\nu}_l$,
 $B_c^-\to h_{c} l^-\bar{\nu}_l$,
$B_c^-\to \chi_{c1} l^-\bar{\nu}_l$,
$B_c^-\to \chi_{c2} l^-\bar{\nu}_l$  and
$B_c^-\to \Psi (3836) l^-\bar{\nu}_l$
 decay channels obtained with
 the AL1 potential (solid lines) and  the Bhaduri (BHAD)
  potential (dotted lines)  . The distribution
for a final muon is not explicitly shown. It differs appreciably from the corresponding to a
final electron only for $q^2$ around $m_\mu^2$.}
\label{fig:12pmdgdq2}
\end{figure}

In Fig.~\ref{fig:12pmdgdq2} we show now the
results for vector and tensor mesons. As before only for the case
where the final lepton is light we see  up to 10\% differences 
  between the calculation with the AL1 and the BHAD potentials.

Finally in Tables~\ref{tab:dw},~\ref{tab:br} we give the total decay
widths and corresponding branching ratios for the different
transitions. The branching ratios evaluated by Ivanov {\it et
al.}~\cite{Ivanov:2006ni}, where they have used the new $B_c$ mass determination by the CDF 
Collaboration~\cite{Abulencia:2005us}, are in reasonable agreement with our results. Discrepancies are
larger for the decay widths in Table~\ref{tab:dw} where they use the larger 
mass value $m_{B_c}=6400$\,MeV quoted by the PDG~\cite{Eidelman:2004wy}.
\begin{table}[!h!!!!!]
%\begin{center}
\hspace*{-.8cm}
{\footnotesize \setlength{\tabcolsep}{2pt}
\begin{tabular}{l| c c c c c c c c c c}
\multicolumn{1}{l}{}&\multicolumn{10}{c}
{$\Gamma\ [10^{-15}\ \mathrm{GeV}]$}\\
&&&&\\
&This work&\cite{Ivanov:2005fd}&\cite{Ebert:2003cn}&\cite{Chang:1992pt,Chang:2001pm}&
\cite{AbdEl-Hady:1999xh}
&\cite{Liu:1997hr}&\cite{Kiselev:1999sc}&\cite{Colangelo:1999zn}&
\cite{Anisimov:1998xv}&\cite{Sanchis-Lozano:1994vh}\\
\hline
&&&&\\
$B^-_c\to \eta_c\, l^-\,\bar{\nu}_l$ &$6.95^{+0.29}$ 
& 10.7 &5.9&14.2&11.1&8.31&$11\pm 1$& 2.1 (6.9)&8.6&10\\
%$B^-_c\to \eta_c\, \mu\,\bar{\nu}_\mu$ &$6.93^{+0.31}$\\ 
$B^-_c\to \eta_c\, \tau^-\,\bar{\nu}_\tau$ &$2.46^{+0.07}$ 
&3.52&&&&&&&$3.3\pm0.9$&\\
&&&&\\
$B^-_c\to \chi_{c0}\, l^-\,\bar{\nu}_l$ &$1.55_{-0.02}^{+0.14}$
&2.52&&1.69&&&&&&\\
%$B^-_c\to \chi_{c0}\, \mu\,\bar{\nu}_\mu$ &$1.54_{-0.02}^{+0.14}$\\
$B^-_c\to \chi_{c0}\,\tau^-\,\bar{\nu}_\tau$ &$0.19^{+0.01}$
& 0.26&&0.25&&&&&&\\
&&&&\\

$B_c^-\to J/\Psi\, l^-\,\bar{\nu}_l$ &$21.9^{+1.2}$
&28.2&17.7& 34.4&30.2&20.3&$28\pm 5$& 21.6 (48.3)&17.2&42\\
%$B_c\to J/\Psi\, \mu\,\bar{\nu}_\mu$ &$21.8^{+1.2}_{-0.1}$\\
$B_c^-\to J/\Psi\,\tau^-\,\bar{\nu}_\tau$&$5.86^{+0.23}_{-0.03}$
&7.82&&&&&&&$7\pm 2$&\\
&&&&\\

$B_c^-\to \chi_{c1}\, l^-\,\bar{\nu}_l$ &$0.94^{+0.05}_{-0.03}$
&1.40&&2.21&&&&&&\\
%$B_c\to \chi_{c1}\, \mu\,\bar{\nu}_\mu$ &$0.93^{+0.05}_{-0.03}$\\
$B_c^-\to \chi_{c1}\, \tau^-\,\bar{\nu}_\tau$ &$0.10$
&0.17&&0.35&&&&&&\\
&&&&\\

$B_c^-\to h_c\, l^-\,\bar{\nu}_l$ &$2.40^{+0.23}_{-0.01}$
&4.42&&2.51&&&&&&\\
%$B_c\to h_c\, \mu\,\bar{\nu}_\mu$ &$2.38^{+0.23}_{-0.01}$\\
$B_c^-\to h_c\, \tau^-\,\bar{\nu}_\tau$ &$0.21^{+0.01}$
&0.38&&0.36&&&&&&\\
&&&&\\

$B_c^-\to \chi_{c2}\, l^-\,\bar{\nu}_l$&$1.89^{+0.11}_{-0.08}$
& 2.92&&2.73&&&&&\\
%$B_c\to \chi_{c2}\, \mu\,\bar{\nu}_\mu$&$1.87^{+0.11}_{-0.08}$\\
$B_c^-\to \chi_{c2}\, \tau^-\,\bar{\nu}_\tau$&$0.13_{-0.01}^{+0.01}$
& 0.20&&0.42&&&&&\\
&&&&\\

$B_c^-\to \Psi (3836)\, l^-\,\bar{\nu}_l$&$0.062_{-0.008}$ 
& 0.13&&&&&\\
%$B_c\to \Psi (3836)\, \mu\,\bar{\nu}_\mu$&$0.061_{-0.007}$ \\
$B_c^-\to \Psi (3836)\, \tau^-\,\bar{\nu}_\tau$ &$0.0012_{-0.0002}$
&0.0031&&&&&\\

\end{tabular}
%\end{center}
}

\caption{Decay widths in units of $10^{-15}$\,GeV for semileptonic $B_c^-\to c\bar c$ decays. Our central values have been evaluated with the
AL1 potential. Here $l$ stands for $l=e,\,\mu$.} 
\label{tab:dw}
%\vspace*{3cm}
\phantom{hola}
\end{table}

\begin{table}[t!!!!!]
%\begin{center}
{\footnotesize \setlength{\tabcolsep}{3pt}
\hspace*{-1.cm}\begin{tabular}{l| c c c c c c c c c}
\multicolumn{1}{l}{}&\multicolumn{6}{c}
{B.R. (\%)}\\
&&&&\\
&This work&\cite{Ivanov:2006ni}&\cite{Ebert:2003cn}&\cite{Chang:1992pt,Chang:2001pm,Chang:2001vq}&
\cite{AbdEl-Hady:1999xh}&\cite{Kiselev:2000pp,Kiselev:2002vz}&\cite{Colangelo:1999zn}&
\cite{Anisimov:1998xv}&\cite{Nobes:2000pm}\\
\hline
&&&&\\
$B^-_c\to \eta_c\, l^-\,\bar{\nu}_l$ &$0.48^{+0.02}$ 
& 0.81 &0.42&0.97&0.76&0.75&0.15&0.59&0.51\\
%$B^-_c\to \eta_c\, \mu\,\bar{\nu}_\mu$ &$0.49^{+0.02}$ \\
$B^-_c\to \eta_c\, \tau^-\,\bar{\nu}_\tau$ &$0.17^{+0.01}$ &0.22&& &&0.23&&0.20&\\
&&&&\\
$B^-_c\to \chi_{c0}\, l^-\,\bar{\nu}_l$ &$0.11^{+0.01}$&0.17&&0.12&&&
&&\\
%$B^-_c\to \chi_{c0}\, \mu\,\bar{\nu}_\mu$ &$0.11^{+0.01}$\\
$B^-_c\to \chi_{c0}\,\tau^-\,\bar{\nu}_\tau$ &$0.013^{+0.001}$&
 0.013&&0.017&&&&&\\
&&&&\\

$B_c^-\to J/\Psi\, l^-\,\bar{\nu}_l$ &$1.54^{+0.06}$&2.07&1.23& 2.35 &2.01&1.9&1.47
&1.20&1.44\\
%$B_c\to J/\Psi\, \mu\,\bar{\nu}_\mu$ &$1.53^{+0.09}$\\
$B_c^-\to J/\Psi\,\tau^-\,\bar{\nu}_\tau$&$0.41^{+0.02}$&0.49&&&&0.48&&0.34&
\\
&&&&\\

$B_c^-\to \chi_{c1}\, l^-\,\bar{\nu}_l$ &$0.066^{+0.003}_{-0.002}$&0.092&&0.15&
&&&&\\
%$B_c\to \chi_{c1}\, \mu\,\bar{\nu}_\mu$ &$0.065^{+0.004}_{-0.002}$\\
$B_c^-\to \chi_{c1}\, \tau^-\,\bar{\nu}_\tau$
&$0.0072^{+0.0002}_{-0.0003}$&0.0089&&0.024&&&&&\\
&&&&\\

$B_c^-\to h_c\, l^-\,\bar{\nu}_l$ &$0.17^{+0.02}$&0.27&&0.17&&&&&\\
%$B_c\to h_c\, \mu\,\bar{\nu}_\mu$ &$0.17^{+0.01}$\\
$B_c^-\to h_c\, \tau^-\,\bar{\nu}_\tau$ &$0.015^{+0.001}$&0.017&&0.024
&&&&&\\
&&&&\\

$B_c^-\to \chi_{c2}\, l^-\,\bar{\nu}_l$&$0.13^{+0.01}$& 0.17&&0.19&&&&&\\
%$B_c\to \chi_{c2}\, \mu\,\bar{\nu}_\mu$&$0.13^{+0.01}$\\
$B_c^-\to \chi_{c2}\, \tau^-\,\bar{\nu}_\tau$&$0.0093^{+0.0002}_{-0.0005}$&
 0.0082&&0.029&&&&&\\
&&&&\\

$B_c^-\to \Psi (3836)\, l^-\,\bar{\nu}_l$&$0.0043_{-0.0005}$ & 0.0066&&&&&&\\
%$B_c\to \Psi (3836)\, \mu\,\bar{\nu}_\mu$&$0.0043_{-0.0005}$\\
$B_c^-\to \Psi (3836)\, \tau^-\,\bar{\nu}_\tau$ &$0.000083_{-0.000010}$ 
&0.000099&&&&&&\\

\end{tabular}
}
%\end{center}
\caption{Branching ratios in \% for semileptonic $B_c^-\to c\bar c$ decays. Our central values have been evaluated with the
AL1 potential. Here $l$ stands for $l=e,\,\mu$.} 
\label{tab:br}
\end{table}

%\newpage
\subsection{Heavy quark spin symmetry}
As mentioned in the introduction one can not apply HQS to systems with
two heavy quarks due to flavor symmetry breaking by the kinetic energy
terms. The symmetry that survives for such systems is HQSS amounting
to the decoupling of the two heavy quark spins. Using HQSS Jenkins
{\it et al.}~\cite{Jenkins:1992nb} obtained relations between different
form factors for semileptonic $B_c$ decays into ground state vector
and pseudoscalar mesons. Let us check the agreement of our
calculations with their results. For that purpose let us re-write the
general form factor decompositions in Eq.~(\ref{eq:ff}) introducing
the four vectors $v$ and $k$ such that
\begin{eqnarray}
P_{B_c}=m_{B_c}\,v\hspace{.5cm};\hspace{.5cm}P_{c\bar c}=m_{c\bar c}\,v+k
\end{eqnarray}
$v$ is the four-velocity of the initial $B_c$ meson whereas $k$ is a residual
momentum. In terms of those we have
\begin{eqnarray}
P=P_{B_c}+P_{c\bar c}=(m_{B_c}+m_{c\bar c})\,v+k 
\\
q=P_{B_c}-P_{c\bar c}=(m_{B_c}-m_{c\bar c})\,v-k
\end{eqnarray}
and we can write for the $\eta_c\, (c\bar c\,(0^-))$ final state case

\begin{eqnarray}
\hspace*{-.5cm}
\left\langle\, \eta_c,\,\vec{P}_{\eta_c}\,\left|\, J^{c\,b}_\mu(0)\,
\right| \, B_c^-,\,\vec{P}_{B_c}\right\rangle&=&
\left\langle\,  \eta_c,\,\vec{P}_{\eta_c}\,\left|\hspace{.3cm}J^{c\,b}_{V\,\mu}(0)\,
\right| \, B_c^-,\,\vec{P}_{B_c}\right\rangle\nonumber\\
&=&\left((m_{B_c}+m_{\eta_c})\,F_+(q^2)
+(m_{B_c}-m_{\eta_c})\,F_-(q^2)\right)\,v_\mu
\nonumber\\ &&\hspace{.5cm}+(F_+(q^2)-F_-(q^2))\,k_\mu
\nonumber\\
&=&\sqrt{2m_{B_c}2m_{\eta_c}}\left(\Sigma_1^{(0^-)}(q^2)\, v_\mu
+\overline{\Sigma}_2^{(0^-)}(q^2)\,k_\mu\right)
\end{eqnarray}
where we have introduced the new form factors

\begin{eqnarray}
\Sigma_1^{(0^-)}(q^2)&=&\frac{1}{\sqrt{2m_{B_c}2m_{\eta_c}}}\,
\left((m_{B_c}+m_{\eta_c})\,F_+(q^2)
+(m_{B_c}-m_{\eta_c})\,F_-(q^2)\right)\nonumber\\
\overline{\Sigma}_2^{(0^-)}(q^2)&=&\frac{1}{\sqrt{2m_{B_c}2m_{\eta_c}}}\,
(F_+(q^2)-F_-(q^2))
\end{eqnarray}
Similarly for the  $J/\Psi\,(c\bar c\,(1^-))$ final state case we have
\begin{eqnarray}\hspace*{-.5cm}
\lefteqn{\left\langle\,J/\Psi ,\,\lambda\,\vec{P}_{J/\Psi}\,\left|\, J^{c\,b}_\mu(0)\,
\right| \, B_c^-,\,\vec{P}_{B_c}\right\rangle\,=}
\nonumber\\ &=&
\left\langle\, J/\Psi,\,\lambda\,\vec{P}_{J/\Psi}\,\left|\, J^{c\,b}_{V\mu}(0)-
J^{c\,b}_{A\mu}(0)\,
\right| \, B_c^-,\,\vec{P}_{B_c}\right\rangle\nonumber\\
&=&\frac{2m_{B_c}}{m_{B_c}+m_{J/\Psi}} \varepsilon_{\mu\nu\alpha\beta}\ \varepsilon^{\,
\nu\,*}_{(\lambda)}(\,\vec{P
}_{J/\Psi}\,)\,v^\alpha\,k^{\,
\beta}\,V(q^2)\nonumber\\
&&-\, i\, \bigg\{\ (m_{B_c}-m_{J/\Psi})\ \varepsilon_{(\lambda)\,\mu}^*
(\,\vec{P}_{J/\Psi}\,)\,A_0(q^2)\nonumber\\
&&\hspace{+.8cm}
+\frac{m_{B_c}}{m_{J/\Psi}}{k\cdot\varepsilon^*_{(\lambda)}
(\,\vec{P}_{J/\Psi}\,)}
\bigg(\ \frac{A_+(q^2)}{m_{B_c}+m_{J/\Psi}}\left(
(m_{B_c}+m_{J/\Psi})\,v_\mu+k_\mu\right)\nonumber\\
&&\hspace{4cm} +\frac{A_-(q^2)}{m_{B_c}+m_{J/\Psi}}\left(
(m_{B_c}-m_{J/\Psi})\,v_\mu-k_\mu\right) \bigg)\bigg\}\nonumber\\
&=&-\sqrt{2m_{B_c}2m_{J/\Psi}}\,
\varepsilon_{\mu\nu\alpha\beta}\ \varepsilon^{\,
\nu\,*}_{(\lambda)}(\,\vec{P
}_{J/\Psi}\,)\,v^\alpha\,k^{\,\beta}\,\overline{\Sigma}'^{(1^-)}_2(q^2)\nonumber\\
&&-i\sqrt{2m_{B_c}2m_{J/\Psi}}\bigg\{\
\varepsilon^*_{(\lambda)\,\mu}
(\,\vec{P}_{J/\Psi}\,)\,\Sigma_1^{(1^-)}(q^2)\nonumber\\
&&\hspace{2.55cm}+\left(k\cdot\varepsilon^*_{(\lambda)}
(\,\vec{P}_{J/\Psi}\,)\right)\,\overline{\Sigma}_2^{(1^-)}(q^2)\,v_\mu\nonumber\\
&&\hspace{2.55cm}+\left(k\cdot\varepsilon^*_{(\lambda)}
(\,\vec{P}_{J/\Psi}\,)\right)\,
\overline{\Sigma}_3^{(1^-)}(q^2)\,k_\mu
\bigg\}
\end{eqnarray}
with
\begin{eqnarray}
\Sigma_1^{(1^-)}(q^2)&=&\frac{1}{\sqrt{2m_{B_c}2m_{J/\Psi}}}\,
\left(m_{B_c}-m_{J/\Psi}\right)\,A_0(q^2)
\nonumber\\
\overline{\Sigma}_2^{(1^-)}(q^2)&=&\frac{1}{\sqrt{2m_{B_c}2m_{J/\Psi}}}\,
\frac{m_{B_c}}{m_{J/\Psi}}\,\left(\,A_+(q^2)
+\frac{m_{B_c}-m_{J/\Psi}}{m_{B_c}+m_{J/\Psi}}\,A_-(q^2)
\right)\nonumber\\
\overline{\Sigma}_2'^{(1^-)}(q^2)&=&\frac{1}{\sqrt{2m_{B_c}2m_{J/\Psi}}}\,
\frac{-2m_{B_c}}{m_{B_c}+m_{J/\Psi}}\,V(q^2)
\nonumber\\
\overline{\Sigma}_3^{(1^-)}(q^2)&=&\frac{1}{\sqrt{2m_{B_c}2m_{J/\Psi}}}\,
\frac{m_{B_c}}{m_{J/\Psi}}\,
\frac{1}{m_{B_c}+m_{J/\Psi}}\left(A_+(q^2)-A_-(q^2)\right)
\end{eqnarray}
$\Sigma_1^{(0^-)}$ and $\Sigma_1^{(1^-)}$ are dimensionless,
 $\overline{\Sigma}_2^{(0^-)}$, $\overline{\Sigma}_2^{(1^-)}$, 
$\overline{\Sigma}_2'^{(1^-)}$ 
have dimensions of $E^{-1}$, and $\overline{\Sigma}_3^{(1^-)}$ 
has dimensions of $E^{-2}$.

\begin{figure}[p!!]
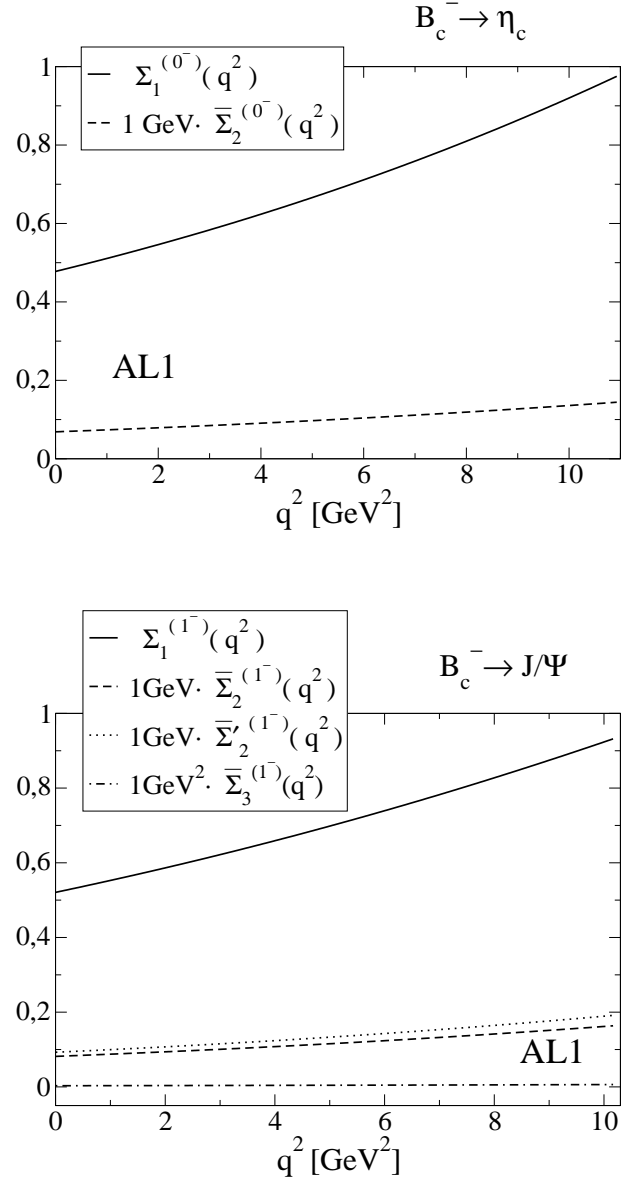

%\vspace{1cm}
\centering
\resizebox{8.cm}{!}{\includegraphics{Bc_sigescetac.eps}}
\\ \vspace{1cm}
\resizebox{8.cm}{!}{\includegraphics{Bc_sigvecjpsi.eps}}\\
%\resizebox{10cm}{!}
\caption{$\Sigma_1^{(0^-)}$ (solid line) and $\overline{\Sigma}_2^{(0^-)}$
(dashed line) of the $B_c^-\to \eta_c$, and
${\Sigma}_1^{(1^-)}$ (solid line), $\overline{\Sigma}_2^{(1^-)}$ (dashed line), 
$\overline{\Sigma}'^{\,(1^-)}_2$ (dotted line) and 
$\overline{\Sigma}_3^{(1^-)}$ (dashed--dotted line) of the
$B_c^-\to J/\Psi$ semileptonic decays evaluated 
 with the AL1 potential.}
\label{fig:sig01m}
\end{figure}
We can take the infinite heavy quark mass limit $m_b\gg m_c\gg \Lambda_{QCD}$
with the result that near zero recoil
\begin{eqnarray}
&&\Sigma_1^{(0^-)}=\Sigma_1^{(1^-)}\nonumber\\
&&\overline{\Sigma}_2^{(0^-)}=\overline{\Sigma}_2^{(1^-)}
=\overline{\Sigma}_2'^{(1^-)}= 0\nonumber\\
&&\overline{\Sigma}_3^{(1^-)}= 0
\end{eqnarray}
This agrees perfectly with the result obtained in
Ref.~\cite{Jenkins:1992nb} using HQSS\footnote{Note, however, the different
global phases and notation used in Ref.~\cite{Jenkins:1992nb}.}.

\begin{figure}[h!!!!]
\vspace{.2cm}
\centering
\resizebox{8cm}{!}{\includegraphics{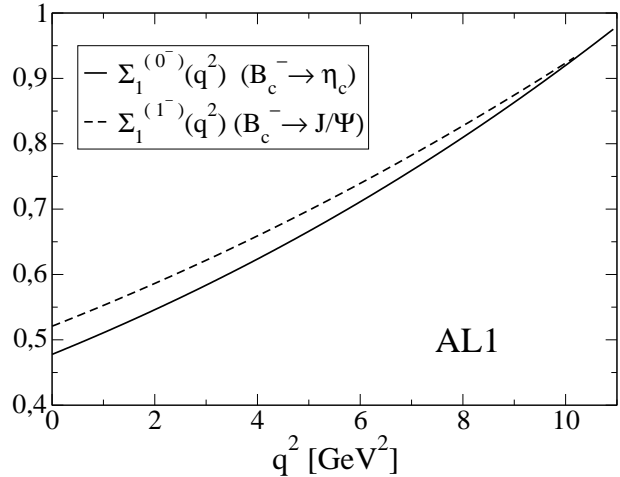}}
%\resizebox{10cm}{!}
\caption{$\Sigma_1^{(0^-)}$ (solid line) of the $B_c^-\to \eta_c$, and
${\Sigma}_1^{(1^-)}$ (dashed line) of the
$B_c^-\to J/\Psi$ semileptonic decays evaluated 
 with the AL1 potential.}
\label{fig:sig101m}
\end{figure}

In Fig.~\ref{fig:sig01m} we give our results for the above quantities
for the semileptonic $B_c^-\to \eta_c$ and $B_c^-\to J/\Psi$ decays
for the actual heavy quark masses. Even though we are not in the
infinite heavy quark mass limit we find that $\Sigma_1^{(0^-)}$ and
$\Sigma_1^{(1^-)}$ dominate over the whole $q^2$ interval.  This
dominant behavior would be more so near the zero--recoil point where
$k\approx 0$ and thus the contributions from the terms in
$\overline{\Sigma}_2^{(0^-)}$, $\overline{\Sigma}_2^{(1^-)}$,
$\overline{\Sigma}_2'^{(1^-)}$ and $\overline{\Sigma}_3^{(1^-)}$ are
even more suppressed. Thus, even for the actual heavy quark masses we
find that near zero recoil
\begin{eqnarray}
\left\langle\, \eta_c,\,\vec{P}_{\eta_c}
\,\left|\, J^{c\,b}_\mu(0)\,
\right| \, B_c^-,\,\vec{P}_{B_c}\right\rangle &\approx&
\sqrt{2m_{B_c}2m_{\eta_c}}\ \Sigma_1^{(0^-)}(q^2)\, v_\mu\nonumber\\
\left\langle\, J/\Psi,\,\lambda\,\vec{P}_{ J/\Psi}\,\left|\, J^{c\,b}_\mu(0)\,
\right| \, B_c^-,\,\vec{P}_{B_c}\right\rangle &\approx&
-i\sqrt{2m_{B_c}2m_{J/\Psi}}\ \varepsilon^*_{(\lambda)\,\mu}
(\,\vec{P}_{J/\Psi}\,)\,\Sigma_1^{(1^-)}(q^2)
\nonumber\\
\end{eqnarray}
Besides, as seen in In Fig.~\ref{fig:sig101m}, $\Sigma_1^{(0^-)}$ of
the $B_c^-\to \eta_c$, and ${\Sigma}_1^{(1^-)}$ of the $B_c^-\to
J/\Psi$ semileptonic decays are very close to each other over the
whole $q^2$ interval. This implies that the result obtained in
Ref.~\cite{Jenkins:1992nb} near zero recoil using HQSS seems to be valid,
to a very good approximation, outside the infinite heavy quark mass
limit.
%
%
%
%
%     NONLEPTONIC DECAYS !!!!!!!!!!!!!!!!!!!!!!!!!!
%
%

\newpage
\section[Nonleptonic $B_c^-\to c\bar c\ M_F^-$ two--meson decays.]%
{Nonleptonic $B_c^-\to c\bar c\ M_F^-$ two--meson
decays.\sectionmark{Nonleptonic $B_{\lowercase{c}}^-\to
\lowercase{c\bar c}\ M_F^-$ two--meson decays}}
\sectionmark{Nonleptonic $B_{\lowercase{c}}^-\to \lowercase{c\bar c}\
M_F^-$ two--meson decays}
\label{Bcsect:nonlepcc}
In this section we will evaluate decay widths for nonleptonic
 $B_c^-\to c\bar c\ M_F^-$ two--meson decays where $M_F^-$ is a
 pseudoscalar or vector meson.  These decay modes involve a $b\to c$
 transition at the quark level and they are governed, neglecting
 penguin operators, by the effective Hamiltonian
 ~\cite{Ebert:2003cn,Colangelo:1999zn,Ivanov:2006ni}
\begin{eqnarray}
H_{eff.}=\frac{G_F}{\sqrt2}\,\left\{ V_{cb}\left[
c_1(\mu)\,Q_1^{cb}+c_2(\mu)\,Q_2^{cb}
\right]
+H.c.\right\}
\end{eqnarray}
where $c_1,\,c_2$ are scale--dependent Wilson coefficients, and
$Q_1^{cb},\,Q_2^{cb}
%,\,Q_1^{ub},\,Q_2^{ub}
$ are local four--quark 
operators given by
\begin{eqnarray}
Q_1^{cb}&=&\overline{\Psi}_c(0)\gamma_\mu(I-\gamma_5)\Psi_b(0)\,\bigg[
\ V_{ud}^*\, \overline{\Psi}_d(0)\gamma^\mu(I-\gamma_5)\Psi_u(0)+
V_{us}^*\, \overline{\Psi}_s(0)\gamma^\mu(I-\gamma_5)\Psi_u(0)\nonumber\\
&&\hspace{3.5cm}+
V_{cd}^*\, \overline{\Psi}_d(0)\gamma^\mu(I-\gamma_5)\Psi_c(0)+
V_{cs}^*\, \overline{\Psi}_s(0)\gamma^\mu(I-\gamma_5)\Psi_c(0)\,
\bigg]\nonumber\\
Q_2^{cb}&=&\overline{\Psi}_d(0)\gamma_\mu(I-\gamma_5)\Psi_b(0)\,\bigg[
\ V_{ud}^*\, 
\overline{\Psi}_c(0)\gamma^\mu(I-\gamma_5)\Psi_u(0)
+ V_{cd}^*\, \overline{\Psi}_c(0)\gamma^\mu(I-\gamma_5)\Psi_c(0)\,
\bigg]\nonumber\\
&&\hspace{-.25cm}+\overline{\Psi}_s(0)\gamma_\mu(I-\gamma_5)\Psi_b(0)\, \bigg[
V_{us}^*\, \overline{\Psi}_c(0)\gamma^\mu(I-\gamma_5)\Psi_u(0)\,
+V_{cs}^*\,\overline{\Psi}_c(0)\gamma^\mu(I-\gamma_5)\Psi_c(0)
\bigg]\nonumber\\
\label{eq:q12cb}
\end{eqnarray}
where the different $V_{jk}$ are  CKM matrix elements. 

\begin{figure}[h]
%\vspace{1cm}
\centering
\resizebox{9.cm}{!}{\includegraphics{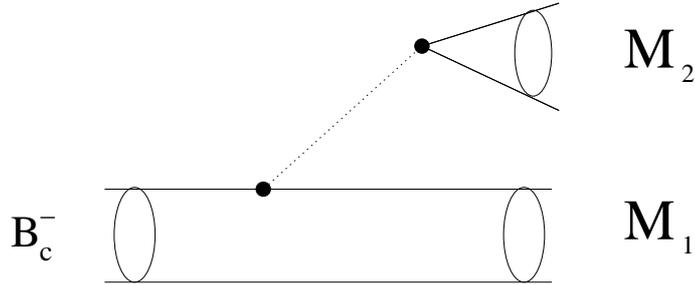}}
\caption{Diagrammatic representation of   $B_c^-$ two--meson decay in the factorization
approximation.}
\label{fig:feynman}
\end{figure}

We shall work in the
factorization approximation which amounts to  evaluate the hadron matrix
elements of the effective Hamiltonian as a product of quark--current matrix
elements: one of these is the matrix element for the $B_c^-$ transition to one
of the final mesons, while the other matrix element corresponds to the transition
from the vacuum to the other final meson. The latter  is given by the corresponding
meson decay constant.
This factorization approximation is schematically represented in Fig.~\ref{fig:feynman}.

When writing the factorization amplitude one has to take into account
the Fierz reordered contribution so that the relevant coefficients are
not $c_1$ and $c_2$ but the combinations
\begin{eqnarray}
a_1(\mu)=c_1(\mu)+\frac{1}{N_C}\,c_2(\mu)\hspace{0.5cm};\hspace{0.5cm}
a_2(\mu)=c_2(\mu)+\frac{1}{N_C}\,c_1(\mu)
\end{eqnarray}
with $N_C=3$ the number of colors. The energy scale $\mu$ appropriate in this
case is $\mu\simeq m_b$ and the values for $a_1$ and $a_2$ that we use
are~\cite{Ivanov:2006ni}
\begin{eqnarray}
a_1=1.14\hspace{0.5cm};\hspace{0.5cm}
a_2=-0.20
\end{eqnarray}
\subsubsection{\bf $\bullet$ $M_F^-=\pi^-,\,\rho^-,\,K^-,\,K^{*-}$ }
\begin{table}[!t]
\begin{center}
\begin{tabular}{l|r}
& $\Gamma\ [10^{-15}\ \mathrm{GeV}]$\\
\\
 \hline
$B_c^-\to\eta_{c}\,\pi^-$ & $1.02^{+0.07}\, a_1^2$\\
$B_c^-\to\eta_{c}\,\rho^-$ & $2.60^{+0.16}\, a_1^2$\\
$B_c^-\to\eta_{c}\,K^-$ & $0.082^{+0.004}\, a_1^2$\\
$B_c^-\to\eta_{c}\,K^{*-}$ & $0.15^{+0.01}\,  a_1^2$\\
\hline
$B_c^-\to J/\Psi\,\pi^-$ &$0.83^{+0.09}\, a_1^2$ \\
$B_c^-\to J/\Psi\,\rho^-$ &$2.61^{+0.27}\,a_1^2$\\
$B_c^-\to J/\Psi\,K^-$ &$0.065^{+0.007}\,a_1^2$\\
$B_c^-\to J/\Psi\,K^{*-}$ &$0.16^{+0.01} \,a_1^2$\\
\hline
$B_c^-\to \chi_{c0}\,\pi^-$  &$0.28^{+0.03}\, a_1^2$\\
$B_c^-\to \chi_{c0}\,\rho^-$ &$0.73^{+0.07}\, a_1^2$\\
$B_c^-\to \chi_{c0}\,K^-$    &$0.022^{+0.003} \,a_1^2$\\
$B_c^-\to \chi_{c0}\,K^{*-}$ &$0.041^{+0.005}\, a_1^2$\\
\hline
$B_c^-\to \chi_{c1}\,\pi^-$  &$0.0015^{+0.0002}\, a_1^2$\\
$B_c^-\to \chi_{c1}\,\rho^-$ &$0.11^{+0.01}\, a_1^2$\\
$B_c^-\to \chi_{c1}\,K^-$    &$0.00012^{+0.00001}\, a_1^2$\\
$B_c^-\to \chi_{c1}\,K^{*-}$ &$0.0080^{+0.0007}_{-0.0002}\, a_1^2$\\
\hline
$B_c^-\to h_c\,\pi^-$  &$0.58^{+0.07}\, a_1^2$\\
$B_c^-\to h_c\,\rho^-$ &$1.41^{+0.17}\, a_1^2$\\
$B_c^-\to h_c\,K^-$    &$0.045^{+0.006}\, a_1^2$\\
$B_c^-\to h_c\,K^{*-}$ &$0.078^{+0.009}\,a_1^2$\\
\hline
$B_c^-\to \chi_{c2}\,\pi^-$  &$0.24^{+0.02}\, a_1^2$\\
$B_c^-\to \chi_{c2}\,\rho^-$ &$0.71^{+0.07}_{-0.03} \,a_1^2$\\
$B_c^-\to \chi_{c2}\,K^-$    &$0.018^{+0.002} \,a_1^2$\\
$B_c^-\to \chi_{c2}\,K^{*-}$ &$0.041^{+0.004}_{-0.001}\, a_1^2$\\
\hline
$B_c^-\to \Psi(3836)\,\pi^-$  &$0.00045^{+0.00003}_{-0.00003}\, a_1^2$\\
$B_c^-\to \Psi(3836)\,\rho^-$ &$0.021^{+0.001}_{-0.002}\,a_1^2$\\
$B_c^-\to\Psi(3836) \,K^-$    &$0.000034^{+0.000002}_{-0.000002}\,a_1^2$\\
$B_c^-\to \Psi(3836)\,K^{*-}$ &$0.0015_{-0.0002} \,a_1^2$\\

\end{tabular}
\end{center}
\caption{Decay widths in units of
$10^{-15}$\,GeV, and for general values of the Wilson coefficient $a_1$, 
for exclusive nonleptonic decays of the $B_c^-$ meson. Our central values have
been obtained with the AL1 potential. }
\label{tab:dwccm}
\end{table}

This is the simplest case. 
The decay width is given by
\begin{eqnarray}
&&\Gamma=\frac{G_F^2}{16\pi m^2_{B_c}}\, |V_{cb}|^2\,|V_{F}|^2\ a_1^2\,
\frac{\lambda^{1/2}(m^2_{B_c},\,m^2_{c\bar c},\,m^2_F)}{2m_{B_c}}\,
{\cal H}_{\alpha\beta}(P_{B_c},P_{c\bar c})\, \widehat{\cal H}^{{\alpha\beta}}(P_F)
\nonumber\\
\end{eqnarray}
with $m_F$ the mass of the $M_F^-$ final meson, and $V_{F}=V_{ud}$ or
$V_{F}=V_{us}$ depending on whether $M_F^-=\pi^-,\,\rho^-$ or
$M_F^-=K^-,\,K ^{*-}$. ${\cal H}_{\alpha\beta}(P_{B_c},P_{c\bar c})$
is the hadron tensor for the $B_c^-\to c\bar c$ transition and
$\widehat{\cal H}^{{\alpha\beta}}(P_F)$ is the hadron tensor for the
$\mathrm{vacuum}\to M_F^-$ transition. The latter is
\begin{eqnarray}
\widehat{\cal H}^{{\alpha\beta}}(P_F)&=&P_F^\alpha\,P_F^\beta\,
 f_F^2\hspace{4.2cm}  M_F^-\equiv0^-\
\mathrm{case}\nonumber\\
\widehat{\cal
H}^{{\alpha\beta}}(P_F)&=&( P_F^\alpha\, P_F^\beta\,-m_F^2\,g^{\alpha\,\beta} )
f_F^2\hspace{2.15cm} M_F^-\equiv1^-\
\mathrm{case}
\end{eqnarray}
with $f_F$ being the $M_F^-$  decay constant.

Similarly to the semileptonic case, the product 
${\cal H}_{\alpha\beta}(P_{B_c},P_{c\bar c})\
\widehat{\cal H}^{{\alpha\beta}}(P_F)$ can now be easily written
in terms of helicity amplitudes for the
$B_c^-\to c\bar c$ transition so that the width is given as~\cite{Ivanov:2006ni}
\begin{eqnarray}
&&\hspace*{-1cm}\nonumber M_F^-\equiv 0^-\ \mathrm{case}\\
&&\Gamma=\frac{G_F^2}{16\pi m^2_{B_c}}\, |V_{cb}|^2\,|V_{F}|^2\ a_1^2\,
\frac{\lambda^{1/2}(m^2_{B_c},\,m^2_{c\bar c},\,m^2_F)}{2m_{B_c}}\,
m_F^2\,f^2_F\, {\cal H}^{B_c^-\to c\bar c}_{tt}(m_F^2)\nonumber\\
&&\hspace*{-1cm}M_F^-\equiv1^-\ \mathrm{case}\nonumber\\
&&\Gamma=\frac{G_F^2}{16\pi m^2_{B_c}}\, |V_{cb}|^2\,|V_{F}|^2\ a_1^2\
\frac{\lambda^{1/2}(m^2_{B_c},\,m^2_{c\bar c},\,m^2_F)}{2m_{B_c}}\,
m_F^2\,f^2_F\nonumber\\
 &&\hspace{2cm}\times\left({\cal H}^{B_c^-\to c\bar c}_{+1+1}(m_F^2)
 +{\cal H}^{B_c^-\to c\bar c}_{-1-1}(m_F^2)
 +{\cal H}_{00}^{B_c^-\to c\bar c}(m_F^2)
\right)
\label{eq:dwccm}
\end{eqnarray}
with the  different ${\cal H}_{rr}$  evaluated at $q^2=m_F^2$.
In Table~\ref{tab:dwccm} we show the decay widths for a general value of the
Wilson coefficient $a_1$, whereas in
Table~\ref{tab:brccm} we give the corresponding branching ratios evaluated
with $a_1=1.14$. Our results for a final $\eta_c$ or $J/\psi$ are in  good
agreement with the ones obtained by Ebert {\it et al.}~\cite{Ebert:2003cn}, El-Hady
{\it et al.}~\cite{AbdEl-Hady:1999xh} and Anisimov {\it et al.}~\cite{Anisimov:1998xv},
 but they are a
factor 2 smaller than the results by Ivanov {\it et al.}~\cite{Ivanov:2006ni}
and Kiselev~\cite{Kiselev:2002vz}. Large
discrepancies with Ivanov's results show up for the other transitions.\\

\begin{table}[!t]
\begin{center}
{\scriptsize \setlength{\tabcolsep}{3pt}
\hspace*{-1cm}
\begin{tabular}{l|r c c c c c c c c}
\multicolumn{1}{l}{}&\multicolumn{9}{c}{B.R. (\%)}\\
\multicolumn{1}{l}{}&\multicolumn{9}{c}{}\\
 & This work&\cite{Ivanov:2006ni}&\cite{Ebert:2003cn}&\cite{Chang:1992pt,Chang:2001pm,Chang:2001vq}
&\cite{AbdEl-Hady:1999xh}&\cite{Kiselev:2002vz,Kiselev:2001zb}
 &\cite{Colangelo:1999zn}&\cite{Anisimov:1998xv}&\cite{LopezCastro:2002ud}\\
\hline
$B_c^-\to\eta_{c}\,\pi^-$ &$0.094^{+0.006}$
&0.19&0.083&0.18&0.14&0.20&0.025&0.13&\\
$B_c^-\to\eta_{c}\,\rho^-$ & $0.24^{+0.01}$
&0.45&0.20&0.49&0.33&0.42&0.067&0.30&\\
$B_c^-\to\eta_{c}\,K^-$ &$0.0075^{+0.0005}$
&0.015&0.006&0.014&0.011&0.013&0.002&0.013&\\
$B_c^-\to\eta_{c}\,K^{*-}$ &$0.013^{+0.001}$
&0.025&0.011&0.025&0.018&0.020&0.004&0.021&\\
\hline
$B_c^-\to J/\Psi\,\pi^-$ &$0.076^{+0.008}$
&0.17 &0.060&0.18&0.11&0.13&0.13&0.073&\\
$B_c^-\to J/\Psi\,\rho^-$ &$0.24^{+0.02}$
&0.49&0.16&0.53&0.31&0.40&0.37&0.21&\\
$B_c^-\to J/\Psi\,K^-$&$0.0060^{+0.0006}$
&0.013&0.005&0.014&0.008&0.011&0.007&0.007&\\
$B_c^-\to J/\Psi\,K^{*-}$ &$0.014^{+0.002}$
&0.028&0.010&0.029&0.018&0.022&0.020&0.016&\\
\hline
$B_c^-\to \chi_{c0}\,\pi^-$  &$0.026^{+0.003}$
&0.055&&0.028&&0.98&&&\\
$B_c^-\to \chi_{c0}\,\rho^-$ &$0.067^{+0.006}_{-0.001}$
&0.13&&0.072&&3.29&&&\\
$B_c^-\to \chi_{c0}\,K^-$    &$0.0020^{+0.0002} $
&0.0042&&0.00021&&&&&\\
$B_c^-\to \chi_{c0}\,K^{*-}$ &$0.0037^{+0.0005}$
&0.0070&&0.00039&&&&&\\
\hline
$B_c^-\to \chi_{c1}\,\pi^-$  &$0.00014^{+0.00001}$
&0.0068&&0.007&&0.0089&&&\\
$B_c^-\to \chi_{c1}\,\rho^-$ &$0.010^{+0.001}_{-0.001}$
&0.029&&0.029&&0.46&&&\\
$B_c^-\to \chi_{c1}\,K^-$    &$1.1^{+0.1}\,10^{-5}$
&5.1\,$10^{-4}$&&5.2\,$10^{-5}$&&&&&\\
$B_c^-\to \chi_{c1}\,K^{*-}$ &$0.00073^{+0.00007}_{-0.00002}$
&0.0018&&0.00018&&&&&\\
\hline
$B_c^-\to h_c\,\pi^-$  &$0.053^{+0.007}$
&0.11&&0.05&&1.60&&&\\
$B_c^-\to h_c\,\rho^-$ &$0.13^{+0.01}$
&0.25&&0.12&&5.33&&&\\
$B_c^-\to h_c\,K^-$    &$0.0041^{+0.0006}$
&0.0083&&0.00038&&&&&\\
$B_c^-\to h_c\,K^{*-}$ &$0.0071^{+0.0008}$
&0.013&&0.00068&&&&&\\
\hline
$B_c^-\to \chi_{c2}\,\pi^-$  &$0.022^{+0.002}$
&0.046&&0.025&&0.79&&&0.0076\\
$B_c^-\to \chi_{c2}\,\rho^-$ &$0.065^{+0.006}_{-0.002}$
&0.12&&0.051&&3.20&&&0.023\\
$B_c^-\to \chi_{c2}\,K^-$    &$0.0017^{+0.0001} $
&0.0034&&0.00018&&&&&0.00056\\
$B_c^-\to \chi_{c2}\,K^{*-}$
&$0.0038^{+0.0003}_{-0.0002}$
&0.0065&&0.00031&&&&&0.0013\\
\hline
$B_c^-\to \Psi(3836)\,\pi^-$  &$4.1^{+0.03}_{-0.02}\,10^{-5}$
&0.0017&&&&0.030&&&\\
$B_c^-\to \Psi(3836)\,\rho^-$ &$0.0020_{-0.0003}$
&0.0055&&&&0.98&&&\\
$B_c^-\to\Psi(3836) \,K^-$    &$3.1^{+0.2}_{-0.2}\,10^{-6}$
&0.00012&&&&&&&\\
$B_c^-\to \Psi(3836)\,K^{*-}$ &$0.00014_{-0.00002} $
&0.00032&&&&&&&\\

\end{tabular}
}
\end{center}
\caption{ Branching ratios in \% for exclusive nonleptonic decays 
of the $B_c^-$ meson. Our central values have been obtained with the AL1
potential.  }
\label{tab:brccm}
\end{table}

\subsubsection{\bf $\bullet$ $M_F^-=D^-,\,D^{*-},\,D_s^-,\,D_s^{*-}$}
In this subsection we shall evaluate the nonleptonic two--meson
$B_c^-\to \eta_cD^-,$ $\eta_c D^{*-},\,
J/\Psi D^-,\,J/\Psi\,D^{*-}$ 
and $B_c^-\to \eta_cD_s^-,\,\eta_c D_s^{*-},\,
J/\Psi D_s^-,\, J/\Psi\,D_s^{*-}$ 
decay widths. 
In these cases there are two different contributions in the factorization
approximation. Following the same steps that lead to Eq.(\ref{eq:dwccm}) we
shall get
\begin{eqnarray}
&&\Gamma=\frac{G_F^2}{16\pi m^2_{B_c}}\, |V_{cb}|^2\,|V_{F}|^2\,
\frac{\lambda^{1/2}(m^2_{B_c},\,m^2_{c\bar c},\,m^2_F)}{2m_{B_c}}\ {\cal HH}
\end{eqnarray}
where now $V_F=V_{cd}$ for $M_F^-=D^-,D^{*-}$ and
$V_F=V_{cs}$ for $M_F^-=D_s^-,D_s^{*-}$. The quantity ${\cal HH}$ incorporates
all information on the hadron matrix elements and depends on the
transition as~\cite{Ivanov:2006ni}\footnote{Note the different phases used in
Ref.~\cite{Ivanov:2006ni}.}
\begin{eqnarray}
\label{eq:hh}
{\cal HH}^{B_c^-\to \eta_c D^-}&=&\left|
a_1\,h_t^{B_c^-\to\eta_c}(m^2_{D^-})\,m_{D^-}\,f_{D^-}
+a_2\,h_t^{B_c^-\to D^-}(m_{\eta_c}^2)\,m_{\eta_c}\,f_{\eta_c}
\right|^2\nonumber\\
{\cal HH}^{B_c^-\to \eta_c D^{*-}}&=&\left|
-a_1\,h_0^{B_c^-\to\eta_c}(m^2_{D^{*-}})\,m_{D^{*-}}\,f_{D^{*-}}
+a_2\,i\,h_{(0)\,t}^{B_c^-\to D^{*-}}(m_{\eta_c}^2)\,m_{\eta_c}\,f_{\eta_c}
\right|^2\nonumber\\
{\cal HH}^{B_c^-\to J/\Psi D^{-}}&=&\left|
a_1\, i\,h_{(0)\,t}^{B_c^-\to J/\Psi}(m_{D^{-}}^2)\,m_{D^{-}}\,f_{D^{-}}
-a_2\,\,h_{0}^{B_c^-\to D^{-}}(m_{J/\Psi}^2)\,m_{J/\Psi}\,f_{J/\Psi}
\right|^2\nonumber\\
{\cal HH}^{B_c^-\to J/\Psi D^{*-}}&=&\sum_{r=+1,\,-1,\,0}\,\left|
a_1\, \,h_{(r)\,r}^{B_c^-\to J/\Psi}(m_{D^{*-}}^2)\,m_{D^{*-}}\,f_{D^{*-}}
\right. \nonumber\\
&& \hspace{2cm}\left.
+a_2\,\,h_{(r)\,r}^{B_c^-\to D^{*-}}(m_{J/\Psi}^2)\,m_{J/\Psi}\,f_{J/\Psi}
\right|^2
\end{eqnarray}
and similarly for $D^-_s,\,D^{*-}_s$. Note  that the helicity amplitudes
corresponding to $B_c^-\to D^-,\,D^{*-},\,D_s^-,\,D_s^{*-}$ have been evaluated
from the matrix elements for  the  effective current operators
$\overline{\Psi}_{d,s}(0)\gamma^\mu(I-\gamma_5)\Psi_b(0)$  in
Eq.~(\ref{eq:q12cb}). While in practice this is a $b\to d,\,s$ transition, the
momentum transfer ($m_{\eta_c^2}$ or $m_{J/\Psi}^2$) is neither too high,
so that one has to include a $B_c^*$ resonance, nor too low, so as to have too high
three-momentum transfers\footnote{Our experience with the $B\to \pi$ 
decay~\cite{Albertus:2005ud}, where we have a similar $b\to u$ quark transition, shows that the naive nonrelativistic quark model
gives reliable results 
for $q^2\approx 9$\,GeV$^2$.}. Besides the contribution is weighed by the much smaller $a_2$ Wilson
coefficient.
 In Table~\ref{tab:dwcccm} we
give the decay widths for general values of the Wilson coefficients
$a_1$ and $a_2$, and in  Table~\ref{tab:brcccm} we show the branching ratios.
 We are in reasonable agreement with the
results by Ivanov {\it et al.}~\cite{Ivanov:2006ni}, El-Hady {\it et
al.}~\cite{AbdEl-Hady:1999xh} and Kiselev~\cite{Kiselev:2002vz}. For decays with a final
$D_s^-,D_s^{*-}$ the agreement is also reasonable with the results by
Colangelo {\it et
al.}~\cite{Colangelo:1999zn} and Anisimov {\it et
al.}~\cite{Anisimov:1998xv}.
\begin{table}[h]
\begin{center}
\begin{tabular}{l|c }

 & $\Gamma\ [10^{-15}\ \mathrm{GeV}]$\\
 &\\
\hline
$B_c^-\to\eta_{c}\,D^-$ & $(0.438^{+0.010} a_1+ 0.236^{+0.030}_{-0.023} a_2)^2$
\\
$B_c^-\to\eta_{c}\,D^{*-}$ & $(-0.390_{-0.009} a_1- 0.136_{-0.022}^{+0.015}a_2)^2$
\\
$B_c^-\to J/\Psi\,D^-$ & $( -0.328_{-0.012}a_1-0.156_{-0.019}^{+0.016} a_2)^2$
\\
$B_c^-\to J/\Psi\,D^{*-}$ & \ $(-0.195_{-0.008} a_1-0.066^{+0.006}_{-0.011} a_2)^2$\\
                          &  +$(-0.390_{-0.018} a_1-0.209^{+0.019}_{-0.032} a_2)^2$\\
                          & +$( 0.447_{-0.016} a_1+0.167_{-0.027}^{+0.016} a_2)^2$
		   \\
\hline
$B_c^-\to\eta_{c}\,D_s^-$ & $(2.54^{+0.05} a_1+ 1.93^{+0.10}a_2)^2$
\\
$B_c^-\to\eta_{c}\,D_s^{*-}$ & $(-1.84_{-0.04} a_1- 1.17_{-0.14}^{+0.02}a_2)^2$
\\
$B_c^-\to J/\Psi\,D_s^-$ & $(-1.85_{-0.06}^{+0.01} a_1-1.23_{-0.06} a_2)^2$
\\
$B_c^-\to J/\Psi\,D_s^{*-}$
& $(-1.01_{-0.04} a_1-0.60^{+0.02}_{-0.07} a_2)^2$\\ 
& $+(-2.00_{-0.06} a_1-1.71^{+0.03}_{-0.18} a_2)^2$\\
& $+( 2.17_{-0.08} a_1+1.42_{-0.16}^{+0.02} a_2)^2$
\\
\end{tabular}
\end{center}
\caption{Decay widths in units of $10^{-15}$\,GeV, and for general
values of the Wilson coefficients $a_1$ and $a_2$, for exclusive
nonleptonic decays of the $B_c^-$ meson. Our central values have been
obtained with the AL1 potential.  For vector--vector final state we
show the three different contributions corresponding to
$r=+1,\,-1,\,0$ (see Eq.(\ref{eq:hh})).}
\label{tab:dwcccm}
\end{table}
\begin{table}[h]
\begin{center}
\begin{tabular}{l|c c c c c c c}
\multicolumn{1}{l}{}&\multicolumn{7}{c}{B.R. (\%)}\\
\multicolumn{1}{l}{}&\multicolumn{7}{c}{}\\
%\hline
 & This work&\cite{Ivanov:2006ni}&\cite{Chang:1992pt}&\cite{AbdEl-Hady:1999xh}&\cite{Kiselev:2002vz}&\cite{Colangelo:1999zn}
& \cite{Anisimov:1998xv}\\
\hline
$B_c^-\to\eta_{c}\,D^-$ & $0.014^{+0.001}$
&0.019&0.0012&0.014&0.015&0.005&0.010\\
$B_c^-\to\eta_{c}\,D^{*-}$ &
$0.012^{+0.001}$
&0.019&0.0010&0.013&0.010&0.002&0.0055\\
$B_c^-\to J/\Psi\,D^-$ &
$0.0083^{+0.0005}$
&0.015&0.0009&0.009&0.009&0.013&0.0044\\
$B_c^-\to J/\Psi\,D^{*-}$ & $0.031^{+0.001}$
&0.045&&0.028&0.028&0.019&0.010\\
\hline
$B_c^-\to\eta_{c}\,D_s^-$ & $0.44^{+0.02}$
&0.44&0.054&0.26&0.28&0.50&0.35\\
$B_c^-\to\eta_{c}\,D_s^{*-}$ & $0.24^{+0.02}$
&0.37&0.044&0.24&0.27&0.038&0.36\\
$B_c^-\to J/\Psi\,D_s^-$ & $0.24^{+0.02}$
&0.34&0.041&0.15&0.17&0.34&0.12\\
$B_c^-\to J/\Psi\,D_s^{*-}$ & $0.68^{+0.03}$
&0.97&&0.55&0.67&0.59&0.62\\
\end{tabular}
\end{center}
\caption{Branching ratios in \% 
for exclusive nonleptonic decays of the $B_c^-$ meson. Our central values have
been obtained with the AL1 potential. }
\label{tab:brcccm}
\end{table}
%
%
%
%
%\newpage
\section[Semileptonic $B_c^-\to \overline{B}^0,\, \overline{B}^{*0},\,
 \overline{B}_s^0,\, \overline{B}_s^{*0}$  decays]{Semileptonic $B_c^-\to \overline{B}^0,\, \overline{B}^{*0},\,
 \overline{B}_s^0,\, \overline{B}_s^{*0}$  decays\sectionmark{Semileptonic $B_{\lowercase{c}}^-\to \overline{B}^0,\, \overline{B}^{*0},\,
 \overline{B}_{\lowercase{s}}^0,\, \overline{B}_{\lowercase{s}}^{*0}$} }
\sectionmark{semileptonic $B_{\lowercase{c}}^-\to \overline{B}^0,\, \overline{B}^{*0},\,
 \overline{B}_{\lowercase{s}}^0,\, \overline{B}_{\lowercase{s}}^{*0}$} 
\label{Bcsect:semilepb}
In this section we shall study the semileptonic $B_c^-\to
 \overline{B}^0,\, \overline{B}^{*0},\, \overline{B}_s^0,\,
 \overline{B}_s^{*0}$ decays. With obvious changes the calculations
 are done as before, with the only novel thing that now it is the
 antiquark that suffers the transition (we have $\bar c\to \bar d,\,
 \bar s$ ), and thus we have to take into account the changes in
 the form factors according to the results in appendix~\ref{Bcapp:sign}.

\subsection{Form factors}
\begin{figure}[!th]
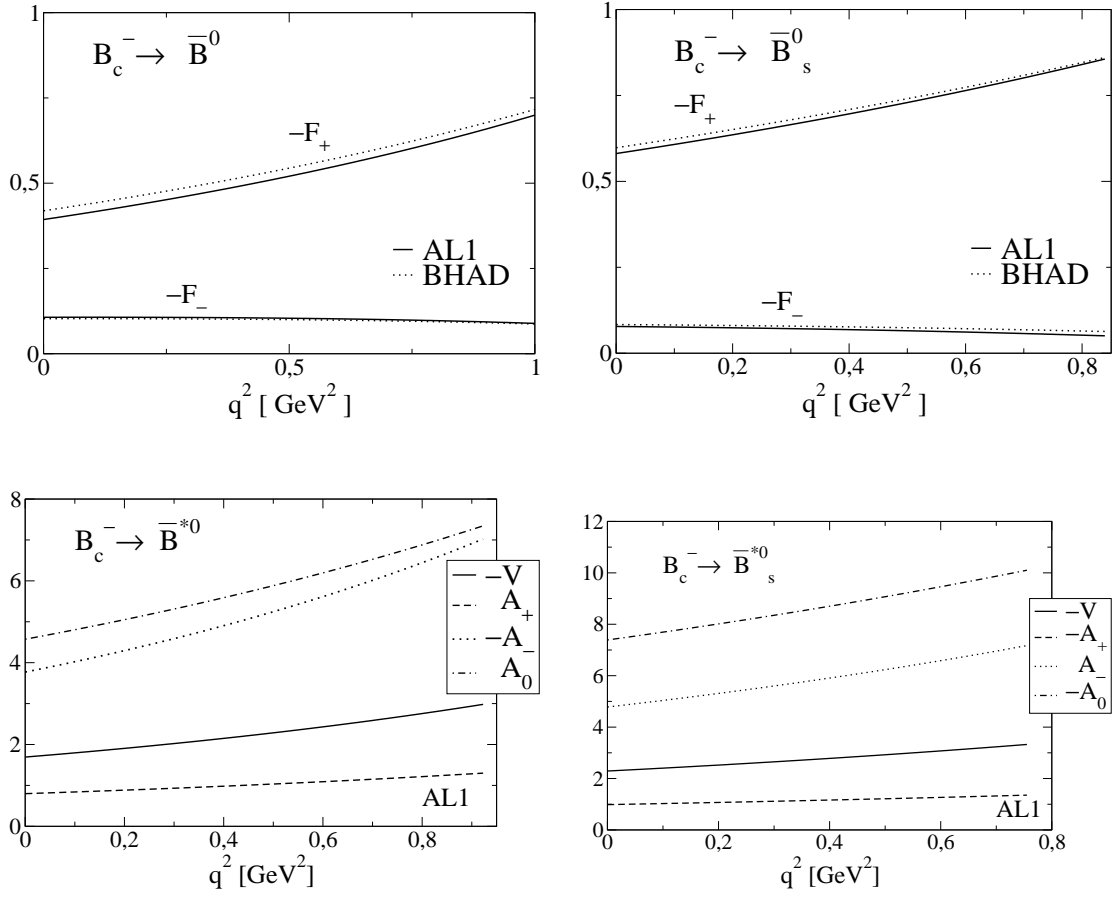

\vspace{1cm}
\centering
\hspace*{-1cm}
\resizebox{7.cm}{!}{\includegraphics{Bc_bctobelectron.eps}}\hspace{.4cm}
\resizebox{7.cm}{!}{\includegraphics{Bc_bctobselectron.eps}}\vspace{.9cm}\\
\hspace*{-1cm}
\resizebox{7.cm}{!}{\includegraphics{Bc_bctobstarelectron.eps}}\hspace{.4cm}
\resizebox{7.cm}{!}{\includegraphics{Bc_bctobsstarelectron.eps}}
%\resizebox{10cm}{!}
\caption{$F_+$ and $F_-$ form factors (solid lines) for 
$B_c^-\to\overline{B}^0$ and
$B_c^-\to\overline{B}^0_s$ semileptonic decay, and $V$ (solid line),
$A_+$ (dashed line), $A_-$ (dotted line) and $A_0$ (dashed--dotted line) form factors for
$B_c^-\to\overline{B}^{*0}$ and
$B_c^-\to\overline{B}^{*0}_s$ semileptonic decays evaluated
with the AL1 potential.  For the first two cases, and for comparison,
we also show with dotted lines the results obtained with the Bhaduri (BHAD)
potential.}
\label{fig:fpmvapm0}
\end{figure}

In Fig.~\ref{fig:fpmvapm0} we show the form factors for the above
transitions evaluated with the AL1 potential.  For the
$\overline{B}^0$ and $\overline{B}^0_s$ cases we also show the results
obtained with the BHAD potential. Although they are less visible in
the figures, the larger differences,  up  to 25\%, occur for the
$F_-$ form factor.
    
In Table~\ref{tab:fpmb} we show $F_+$, $F_-$ and $F_0$ (defined as in
Eq.~(\ref{eq:f0}) changing  the mass of the final meson) of the $B_c^-\to
\overline{B}^0,\,\overline{B}^0_s$ transitions evaluated at
$q^2_{min}$ and $q^2_{max}$ and compare them with the results by
Ivanov {\it et al.}~\cite{Ivanov:2000aj} and Ebert {\it et
al.}~\cite{Ebert:2003wc}.  Notice that, to favor comparison, we have
changed the signs of the form factors by Ebert {\it et al.} (they
evaluate $B_c^+$ decay) in accordance with the results in
appendix~\ref{Bcapp:sign}.  The agreement with the results by Ebert {\it
et al.}  is good. We also agree with Ivanov {\it et al.}  for $F_+$,
but get very different results for $F_-$.  As fermion masses are very
small the disagreement in the $F_-$ form factor will have a negligible
effect on the decay width.
 \begin{table}[h!!!!!!!]

%\begin{center}
{\small
\begin{tabular}{c| c c  ||c|  c c}
$B_c^-\to\overline{B}^0\,l^-\bar{\nu}_l\ \ $ & $q^2_{\mathrm{min}}$ &
 $q^2_{\mathrm{max}}$ & 
$B_c^-\to\overline{B}^0_s\,l^-\bar{\nu}_l\ \ $ & $q^2_{\mathrm{min}}$ &
 $q^2_{\mathrm{max}}$  \\
\hline
&&&&&\\
 $F_+$ & & & $F_+$ & &\\
 This work & $-0.39^{+0.03}_{-0.03}$& $-0.70^{+0.004}_{-0.02}$ &  
 This work & $-0.58^{+0.01}_{-0.02}$& 
 $-0.86^{+0.01}_{-0.01}$\\
 \cite{Ivanov:2000aj} &$-0.58$ &$-0.96$&\cite{Ivanov:2000aj} &$-0.61$ &$-0.92$\\
 \cite{Ebert:2003wc} &$-0.39$&$-0.96$&\cite{Ebert:2003wc}&$-0.50$&$-0.99$\\
&&&&\\
$F_-$ & & & $F_-$ & &\\
 This work &$-0.11^{+0.02}_{-0.03}$ &$-0.09^{+0.01}_{-0.06}$ &   This work
 &$-0.08^{+0.01}_{-0.02}$ &$-0.05^{+0.01}_{-0.03}$\\
 \cite{Ivanov:2000aj} & 2.14&2.98&  \cite{Ivanov:2000aj} & 1.83&2.35\\
 &&&&&\\
$F_0$ & & & $F_0$ & &\\
 This work &$-0.39^{+0.03}_{-0.03}$ &$-0.71^{+0.05}_{-0.02}$  &  This work
 &$-0.58_{-0.02}$ &$-0.86^{+0.01}_{-0.01}$\\
 \cite{Ebert:2003wc} &$-0.39$&$-0.80$&\cite{Ebert:2003wc}&$-0.50$ &$-0.86$\\
 \end{tabular}\hspace{1cm}
%\end{center}
}
\caption{$F_+$, $F_-$ and $F_0$ evaluated at
$q^2_{\mathrm{min}}$ and $q^2_{\mathrm{max}}$ compared to  the ones obtained by 
Ivanov {\it et al.}~\cite{Ivanov:2000aj} and Ebert {\it et al.}~\cite{Ebert:2003wc}. 
Our central values have been obtained with the AL1
potential. Here $l$ stands for $l=e,\,\mu$.}
\label{tab:fpmb}
\end{table}

In Table~\ref{tab:fva0pmb} we show $V$, $A_+$, $A_-$, $A_0$ and
$\widetilde{A}_0$ (defined as in
Eq.~(\ref{eq:a0tilde}) changing  the mass of the final meson) 
 of the
$B_c^-\to \overline{B}^{*0},\,\overline{B}^{*0}_s$ evaluated at
$q^2_{min}$ and $q^2_{max}$ and compare them with the results by
Ivanov {\it et al.}~\cite{Ivanov:2000aj} and Ebert {\it et
al.}~\cite{Ebert:2003wc}.  With some exceptions the agreement with Ebert's results 
is bad in this case. We are also in clear disagreement with Ivanov's results.

\begin{table}[h!]
{\small
\begin{tabular}{c| c c || c|  c c}
$B_c^-\to\overline{B}^{*0}\,l^-\bar{\nu}_l\ \ $ & $q^2_{\mathrm{min}}$ &
 $q^2_{\mathrm{max}}$ & 
$B_c^-\to\overline{B}^{*0}_s\,l^-\bar{\nu}_l\ \ $ & $q^2_{\mathrm{min}}$ &
 $q^2_{\mathrm{max}}$  \\
\hline
&&&&&\\
 $V$ &  & &$V$ & &\\
 This work & $-1.69^{+0.11}_{-0.09}$& $-2.98^{+0.17}_{-0.03}$ &  
 This work & $-2.29^{+0.02}_{-0.09}$& 
 $-3.32^{+0.04}_{-0.01}$\\
 \cite{Ivanov:2000aj} & &$-5.32^*$&\cite{Ivanov:2000aj} & &$-4.91^*$\\
 \cite{Ebert:2003wc} &$-3.94$&$-8.91$&\cite{Ebert:2003wc}&$-3.44$&$-6.25$\\
&&&&&\\
$A_+$ & &  &$A_+$ & &\\
 This work &$-0.80^{+0.02}_{-0.02}$ &$-1.30^{+0.02}$  &  This work
 &$-0.98^{+0.01}_{-0.03}$ &$-1.35^{+0.03}_{-0.02}$\\
 \cite{Ivanov:2000aj} & &0.49 & \cite{Ivanov:2000aj} & &0.21\\
 \cite{Ebert:2003wc} &$-2.89$ &$-2.83$ & \cite{Ebert:2003wc} &$-2.19$ &$-2.62$\\
 
 &&&&&\\
$A_-$ & & & $A_-$ & &\\
 This work &$3.77^{+0.15}_{-0.17}$ &$7.02_{-0.35}$ &  This work
 &$4.78^{+0.23}_{-0.01}$ &$7.17^{+0.06}_{-0.18}$\\
 \cite{Ivanov:2000aj} & &18.0&  \cite{Ivanov:2000aj} & &15.9\\
 &&&&&\\
$A_0$ & & & $A_0$ & &\\
 This work &$-4.57^{+0.27}_{-0.33}$ &$-7.34^{+0.28}_{-0.49}$  &  This work
 &$-7.39^{+0.14}_{-0.30}$ &$-10.10^{+0.22}_{-0.16}$\\
 \cite{Ivanov:2000aj} & &$-5.07$&  \cite{Ivanov:2000aj} & &$-6.60$\\
 \cite{Ebert:2003wc} & $-5.08$&$-8.70$ & \cite{Ebert:2003wc} & $-6.60$&$-10.23$\\
 
 &&&&&\\
$\widetilde{A}_0$ & &  &$\widetilde{A}_0$ & &\\
 This work &$-0.34^{+0.03}_{-0.03}$ &$-0.60^{+0.05}_{-0.02}$ &   This work
 &$-0.51^{+0.01}_{-0.03}$ &$-0.74^{+0.01}_{-0.01}$\\
 \cite{Ebert:2003wc} &$-0.20$&$-1.06$&\cite{Ebert:2003wc}&$-0.35$ &$-0.91$\\
 \end{tabular}\hspace{1cm}
%\end{center}
}
\caption{$V$, $A_+$, $A_-$, $A_0$ and $\widetilde A_0$ evaluated at
$q^2_{\mathrm{min}}$ and $q^2_{\mathrm{max}}$ compared to the ones
obtained by Ivanov {\it et al.}~\cite{Ivanov:2000aj} and Ebert {\it et
al.}~\cite{Ebert:2003wc}. Our central  values have been obtained with
the AL1 potential. Here $l$ stands for $l=e,\,\mu$. Asterisk as in
Table~\ref{tab:fva0pm}.}
\label{tab:fva0pmb}
\end{table}
%\vspace*{5cm}

\subsection{Decay width}
In Tables~\ref{tab:habtob}, \ref{tab:asybtob} we give respectively 
our results for the partial helicity widths and forward-backward asymmetries.

\begin{table}[h!!!]
\begin{center}
%\hspace*{-.75cm}
{\scriptsize
\hspace*{-1cm}
\begin{tabular}{l|c c c c c c c}
$B^-_c\to$ & \hspace{.2cm}$\Gamma_U$\hspace{.2cm} & \hspace{.2cm}$\widetilde{\Gamma}_U$\hspace{.2cm}
 &\hspace{.2cm} $\Gamma_L$\hspace{.2cm} & $\widetilde{\Gamma}_L$ & \hspace{.2cm}$\Gamma_P$
 \hspace{.2cm}& $\widetilde{\Gamma}_S $ & $\widetilde{\Gamma}_{SL}$\\ 
\hline
&&&&&&&\\
$\overline{B}^0\,e^-\bar{\nu}_e$& 0 & 0 &$ 0.65^{+0.07}_{-0.09}$&
$0.14^{+0.02}_{-0.02}\ 10^{-5}$ & 0 &
 $ 0.47^{+0.05}_{-0.07}\ 10^{-5}$  & $ 0.15^{+0.01}_{-0.02}\ 10^{-5}$\\
$\overline{B}^0\,\mu^-\bar{\nu}_\mu$& 0 & 0 &$ 0.55^{+0.06}_{-0.07}$&
$0.15^{+0.01}_{-0.02}\ 10^{-1}$ & 0 &
 $ 0.64^{+0.16}_{-0.09}\ 10^{-1}$  & $ 0.17^{+0.02}_{-0.02}\ 10^{-1}$\\
&&&&&&&\\
$\overline{B}_s^0\,e^-\bar{\nu}_e$& 0 & 0 &$ 15.1^{+0.7}_{-0.3}$&
$0.43^{+0.02}_{-0.01}\ 10^{-4}$ & 0 &
 $ 0.14^{+0.01}\ 10^{-3}$  & $ 0.45^{+0.03}_{-0.01}\ 10^{-4}$\\
$\overline{B}_s^0\,\mu^-\bar{\nu}_\mu$& 0 & 0 &$ 12.4^{+0.5}_{-0.3}$&
$0.40^{+0.02}_{-0.01}$ & 0 &
 $ 1.69^{+0.08}_{-0.03}$  & $ 0.47^{+0.02}_{-0.01}$\\
&&&&&&&\\

$\overline{B}^{*0}\,e^-\bar{\nu}_e$& $0.83^{+0.08}_{-0.11}$ 
& $0.26^{+0.03}_{-0.04}\ 10^{-6}$ &$ 0.76^{+0.09}_{-0.11}$&
$0.10^{+0.02}_{-0.01}\ 10^{-5}$ & $0.36_{-0.05}^{+0.03}$ &
 $ 0.27^{+0.04}_{-0.05}\ 10^{-5}$  & $ 0.96^{+0.15}_{-0.15}\ 10^{-6}$\\
 $\overline{B}^{*0}\,\mu^-\bar{\nu}_\mu$& $0.79^{+0.10}_{-0.11}$ 
& $0.97^{+0.11}_{-0.13}\ 10^{-2}$ &$ 0.68^{+0.08}_{-0.10}$&
$0.14^{+0.01}_{-0.03}\ 10^{-1}$ & $0.34_{-0.05}^{+0.02}$ &
 $ 0.28^{+0.04}_{-0.05}\ 10^{-1}$  & $ 0.11^{+0.02}_{-0.02}\ 10^{-1}$\\

&&&&&&&\\
$\overline{B}_s^{*0}\,e^-\bar{\nu}_e$& $16.7^{+0.8}_{-0.7}$ 
& $0.66^{+0.04}_{-0.03}\ 10^{-5}$ &$ 16.8^{+1.1}_{-0.8}$&
$0.33^{+0.02}_{-0.02}\ 10^{-4}$ & $6.11_{-0.20}^{+0.21}$ &
 $ 0.88^{+0.08}_{-0.05}\ 10^{-4}$  & $ 0.30^{+0.03}_{-0.01}\ 10^{-4}$\\
 $\overline{B}_s^{*0}\,\mu^-\bar{\nu}_\mu$& $15.6^{+0.8}_{-0.6}$ 
& $0.24^{+0.02}_{-0.01}$ &$ 14.5^{+1.0}_{-0.6}$&
$0.37^{+0.02}_{-0.02}$ & $5.66_{-0.19}^{+0.18}$ &
 $ 0.77^{+0.06}_{-0.04}$  & $ 0.30^{+0.02}_{-0.01}$\\

\end{tabular}
}
\end{center}

\caption{ Partial helicity widths in units of $10^{-15}$\,GeV for $B_c^-$ decay. Central values
have been evaluated with the AL1 potential. } 
\label{tab:habtob}
\end{table}

\begin{table}[h]
\begin{center}
\begin{tabular}{l|c c }
&$A_{FB} (e)$  &$A_{FB} (\mu)$  \\
\hline
&\\
$B_c^-\to \overline{B}^0$ &$0.67^{+0.02}\ 10^{-5}$&$0.82^{+0.01}\ 10^{-1}$\\
$B_c^-\to \overline{B}_s^0$ &$0.89^{+0.01}\ 10^{-5}$  &$0.96^{+0.01}\ 10^{-1}$\\
$B_c^-\to \overline{B}^{*0}$ &$0.17_{-0.01}$ &$0.19_{-0.01}$ \\
$B_c^-\to \overline{B}_s^{*0}$ &$0.14_{-0.01}$&$0.16^{+0.01}$\\

\end{tabular}
\end{center}
\caption{Forward-backward asymmetry. Our central values have been
evaluated with the AL1 potential. We would obtain the same results for
$B_c^+\to B^0$ decays. }
\label{tab:asybtob}
\end{table}

\begin{figure}[h!!]
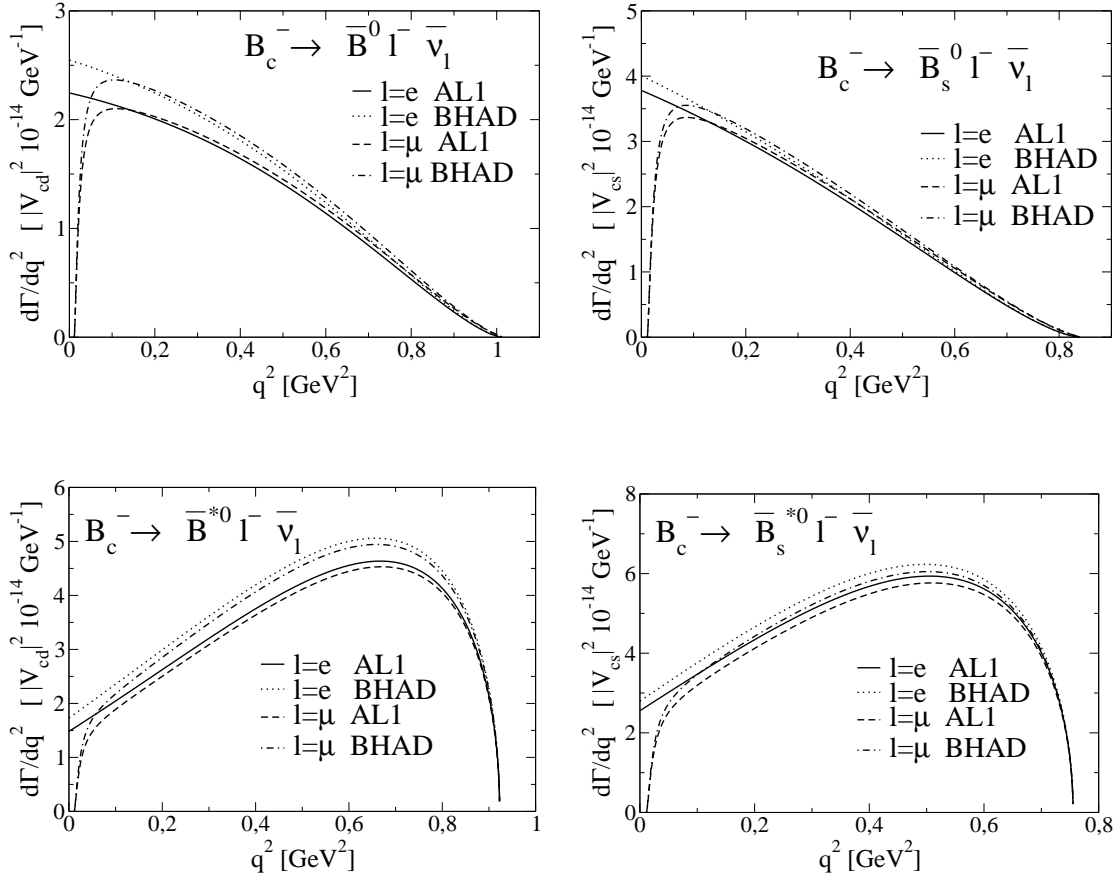

\vspace{1cm}
\centering
\hspace*{-1cm}
\resizebox{7cm}{!}{\includegraphics{Bc_dgdq2bctob.eps}}\hspace{.4cm}
\resizebox{7cm}{!}{\includegraphics{Bc_dgdq2bctobs.eps}}\vspace{1cm}\\
\hspace*{-1cm}
\resizebox{7cm}{!}{\includegraphics{Bc_dgdq2bctobstar.eps}}\hspace{.4cm}
\resizebox{7cm}{!}{\includegraphics{Bc_dgdq2bctobsstar.eps}}
%\resizebox{10cm}{!}
\caption{Differential decay width for the for the $B_c^-\to\overline{B}^0\, 
l^-\bar{\nu}_l$ and $B_c^-\to\overline{B}^0_s\, 
l^-\bar{\nu}_l$,
$B_c^-\to\overline{B}^{*0}\, 
l^-\bar{\nu}_l$ and $B_c^-\to\overline{B}^{*0}_s\, 
l^-\bar{\nu}_l$ ($l=e,\,\mu$)
transitions. Solid line: results for a final $e$ evaluated with the
 AL1 potential; dotted line: results for a final $e$ evaluated with the
Bhaduri (BHAD)  potential; dashed line: results for a final 
$\mu$ evaluated with the
 AL1 potential; dashed--dotted line: results for a final $\mu$ evaluated with the
Bhaduri (BHAD)  potential.}
\label{fig:dgdq2btob}
\end{figure}

In Fig.~\ref{fig:dgdq2btob} we show the differential decay width  for the $B_c^-\to\overline{B}^0\, 
l^-\bar{\nu}_l$, \mbox{$B_c^-\to\overline{B}^0_s\, 
l^-\bar{\nu}_l$},
$B_c^-\to\overline{B}^{*0}\, 
l^-\bar{\nu}_l$ and $B_c^-\to\overline{B}^{*0}_s\, 
l^-\bar{\nu}_l$
transitions ($l=e,\,\mu$). In Tables~\ref{tab:dwbtob}, \ref{tab:brbtob} we give the total decay 
widths and branching ratios and
compare them with determinations by other groups. Our results are in better
agreement with the ones obtained by Ebert {\it et al.}~\cite{Ebert:2003wc}, 
Colangelo {\it et al.}~\cite{Colangelo:1999zn},
Anisimov {\it et al.}~\cite{Anisimov:1998xv} 
and Lu {\it et al.}~\cite{Lu:1995ug}.

\begin{table}[h!]
{\footnotesize \setlength{\tabcolsep}{3pt}
\begin{center}
\hspace*{-.5cm}\begin{tabular}{l| c c c c c c c c c c c}
\multicolumn{1}{l}{}&\multicolumn{11}{c}
{$\Gamma\ [10^{-15}\ \mathrm{GeV}]$}\\
\multicolumn{1}{l}{}&\multicolumn{11}{c}{}\\
&This work&\cite{Ivanov:2000aj}&\cite{Ebert:2003wc}
&\cite{Chang:1992pt}&\cite{AbdEl-Hady:1999xh}&\cite{Liu:1997hr}&\cite{Kiselev:2002vz}&\cite{Colangelo:1999zn}&
\cite{Anisimov:1998xv}&\cite{Nobes:2000pm}&
\cite{Lu:1995ug} \\
\hline
&\\
$B^-_c\to \overline{B}^0\, e\,\bar{\nu}_e$&$0.65^{+0.07}_{-0.09}$
&2.1&0.6&2.30&1.14&1.90&4.9&0.9(1.0)&&&0.59\\
$B^-_c\to \overline{B}^0\, \mu\,\bar{\nu}_\mu$&$0.63^{+0.07}_{-0.09}$\\
$B^-_c\to \overline{B}_s^0\, e\,\bar{\nu}_e$&$15.1^{+0.7}_{-0.3}$
&29&12&26.6&14.3&26.8&59&11.1(12.9)&15&12.3&11.75\\
$B^-_c\to \overline{B}_s^0\, \mu\,\bar{\nu}_\mu$&$14.5^{+0.6}_{-0.3}$\\
$B^-_c\to \overline{B}^{*0}\, e\,\bar{\nu}_e$&$1.59^{+0.17}_{-0.22}$
&2.3&1.7&3.32&3.53&2.34&8.5&2.8(3.2)&&&2.44\\
$B^-_c\to \overline{B}^{*0}\, \mu\,\bar{\nu}_\mu$&$1.52^{+0.17}_{-0.21}$\\
$B^-_c\to \overline{B}_s^{*0}\, e\,\bar{\nu}_e$&$33.5^{+1.9}_{-1.5}$
&37&25&44.0&50.4&34.6&65&33.5(37.0)&34&19.0&32.56\\
$B^-_c\to \overline{B}_s^{*0}\, \mu\,\bar{\nu}_\mu$&$31.5^{+1.8}_{-1.4}$
\end{tabular}
\end{center}
}
\caption{Decay widths in units of $10^{-15}$\,GeV.
 Our central values have been evaluated with the
AL1 potential.} 
\label{tab:dwbtob}
\end{table}

\begin{table}[h!!!!]
{\footnotesize
\begin{center}

\begin{tabular}{l| c c c c c c c c  c}
\multicolumn{1}{l}{}&\multicolumn{8}{c}
{B.R. (\%)}\\
\multicolumn{1}{l}{}&\multicolumn{8}{c}{}\\
&This work&\cite{Ivanov:2006ni}&\cite{Ebert:2003wc}&\cite{Chang:1992pt}&\cite{AbdEl-Hady:1999xh}
&\cite{Kiselev:2002vz}&\cite{Colangelo:1999zn}&
\cite{Anisimov:1998xv}&
\cite{Nobes:2000pm}\\
\hline
&\\
$B^-_c\to \overline{B}^0\, e\,\bar{\nu}_e$&$0.046^{+0.004}_{-0.007}$
&0.071&0.042&0.16&0.078&0.34&0.06&&0.048\\
$B^-_c\to \overline{B}^0\, \mu\,\bar{\nu}_\mu$&$0.044^{+0.005}_{-0.006}$\\
$B^-_c\to \overline{B}_s^0\, e\,\bar{\nu}_e$&$1.06^{+0.05}_{-0.02}$
&1.10&0.84&1.82&0.98&4.03&0.8&0.99&0.92\\
$B^-_c\to \overline{B}_s^0\, \mu\,\bar{\nu}_\mu$&$1.02^{+0.04}_{-0.02}$\\
$B^-_c\to \overline{B}^{*0}\, e\,\bar{\nu}_e$&$0.11^{+0.01}_{-0.01}$
&0.063&0.12&0.23&0.24&0.58&0.19&&0.051\\
$B^-_c\to \overline{B}^{*0}\, \mu\,\bar{\nu}_\mu$&$0.11^{+0.01}_{-0.02}$\\
$B^-_c\to \overline{B}_s^{*0}\, e\,\bar{\nu}_e$&$2.35^{+0.14}_{-0.10}$
&2.37&1.75&3.01&3.45&5.06&2.3&2.30&1.41\\
$B^-_c\to \overline{B}_s^{*0}\, \mu\,\bar{\nu}_\mu$&$2.22^{+0.12}_{-0.10}$\\

\end{tabular}
\end{center}
}
\caption{Branching ratios in \% .
 Our central values have been evaluated with the
AL1 potential.} 
\label{tab:brbtob}
\end{table}

%\vspace*{5cm}

%\newpage
\subsection{Heavy quark spin symmetry}

\begin{figure}[!t]
\vspace{1cm}
\centering
\hspace*{-1cm}
\resizebox{7.cm}{!}{\includegraphics{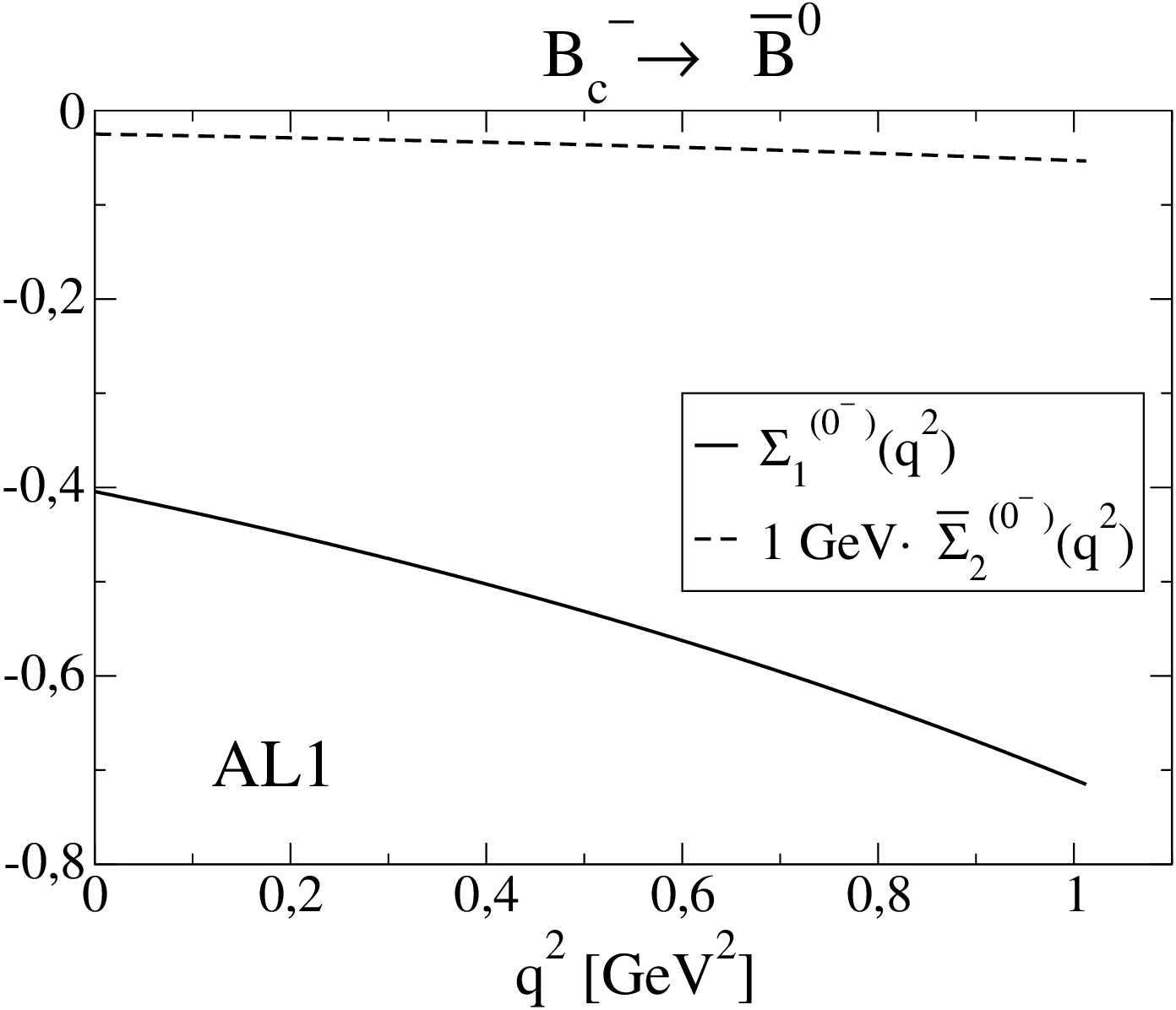}}\hspace{.4cm}
\resizebox{7.cm}{!}{\includegraphics{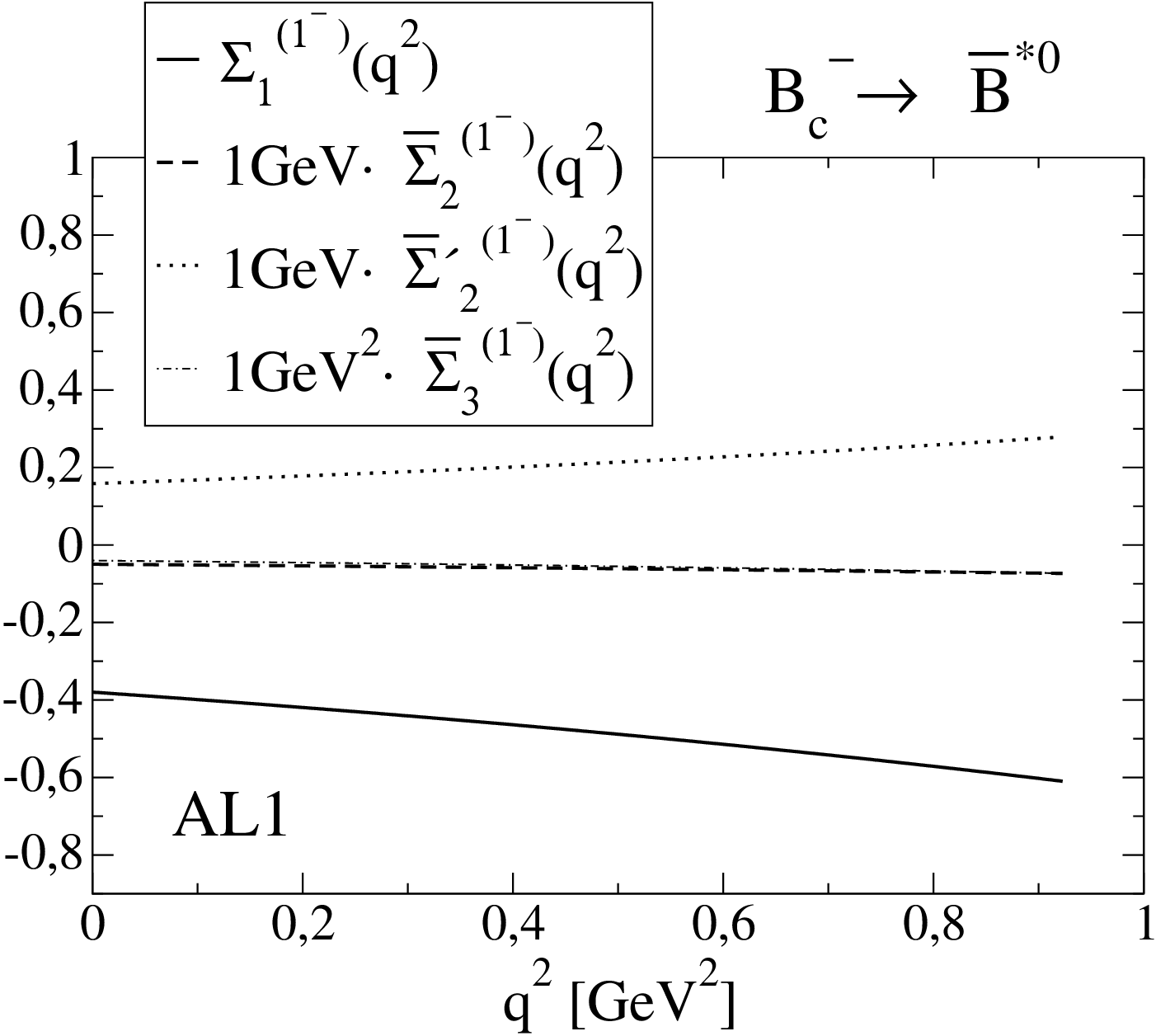}}\vspace{.9cm}\\
\hspace*{-1cm}
\resizebox{7.cm}{!}{\includegraphics{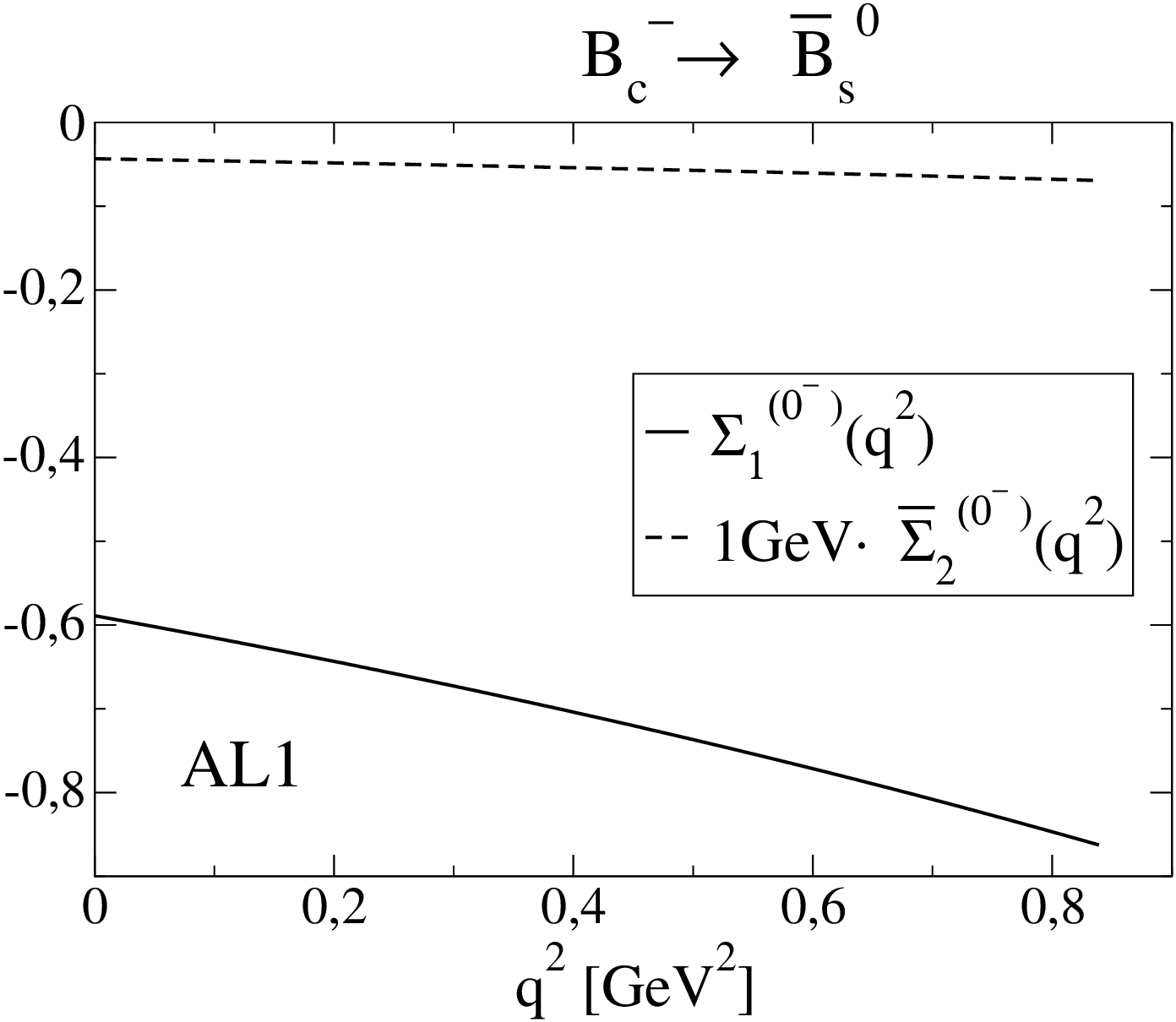}}\hspace{.4cm}
\resizebox{7.cm}{!}{\includegraphics{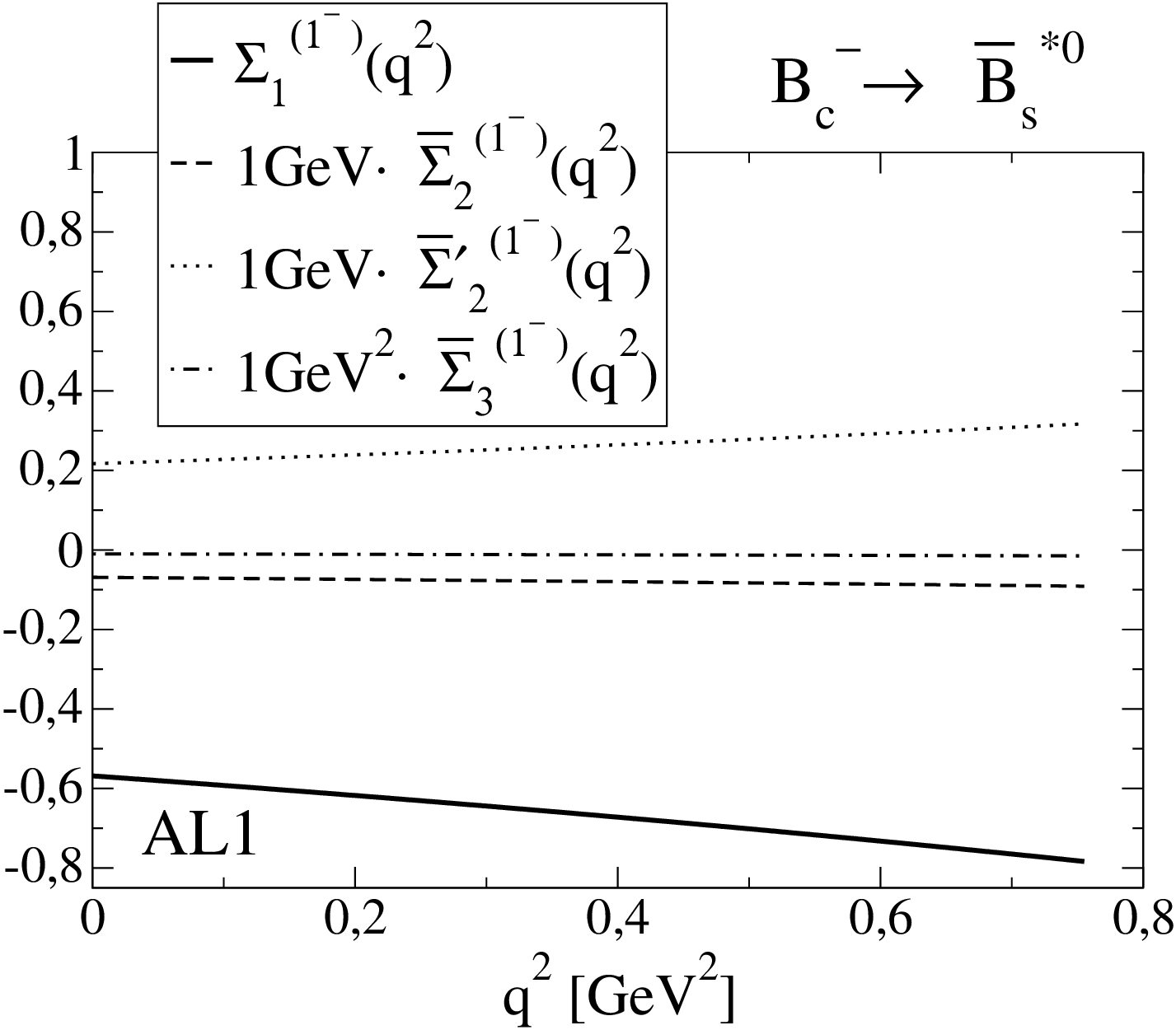}}
%\resizebox{10cm}{!}
\caption{$\Sigma_1^{(0^-)}$ (solid line) and $\Sigma_2^{(0^-)}$ (dashed line) of the $B_c^-\to
\overline{B}^0$ and $B_c^-\to \overline{B}^0_s$ transitions, and
$\Sigma_1^{(1^-)}$ (solid line), $\overline{\Sigma}_2^{(1^-)}$ (dashed line),
$\overline{\Sigma}_2'^{(1^-)}$ (dotted line) and 
$\overline{\Sigma}_3^{(1^-)}$ (dashed dotted line) of
the $B_c^-\to \overline{B}^{*0}$ and $B_c^-\to \overline{B}^{*0}_s$
transitions  evaluated with the AL1 potential.  }
\label{fig:sigbtob}
\end{figure} 
In Fig.~\ref{fig:sigbtob} we give $\Sigma_1^{(0^-)}$ and
$\Sigma_2^{(0^-)}$ of the $B_c^-\to \overline{B}^0$ and $B_c^-\to
\overline{B}^0_s$ transitions, and $\Sigma_1^{(1^-)}$,
$\overline{\Sigma}_2^{(1^-)}$, $\overline{\Sigma}_2'^{(1^-)}$ and
$\overline{\Sigma}_3^{(1^-)}$ of the $B_c^-\to \overline{B}^{*0}$ and
$B_c^-\to \overline{B}^{*0}_s$ transitions.

We can take the infinite heavy quark mass limit on our analytic expressions
with the result that near zero recoil 
\begin{eqnarray}
\label{eq:sigbtob}
\Sigma_1^{(1^-)}&=&\Sigma_1^{(0^-)}\nonumber\\
\overline{\Sigma}_2^{(1^-)}&=&\overline{\Sigma}_2'^{(1^-)}
=-\overline{\Sigma}_2^{(0^-)}\nonumber\\
\overline{\Sigma}_3^{(1^-)}&=&0
\end{eqnarray}
When compared to the results of HQSS by Jenkins {\it et
 al.}~\cite{Jenkins:1992nb} we see differences. In Ref.~\cite{Jenkins:1992nb} we
 find\footnote{Note the different notation and global phases used.}
 $\overline{\Sigma}_2^{(1^-)}=\overline{\Sigma}_2^{(0^-)}$
 instead. This is wrong as there is a misprint in
 Ref.~\cite{Jenkins:1992nb} that has not been noted before: the sign of the
 term in $v^\mu$ in the last expression of Eqs.~(2.9) and~(2.10) in
 Ref.~\cite{Jenkins:1992nb} should be a minus~\cite{Jenkins:PC}. Also from
 Ref.~\cite{Jenkins:1992nb} one would expect
\footnote{One would have to look at Eq.~(2.10) in
Ref.~\cite{Jenkins:1992nb}, even though it refers to $B_c^+$ decay into
$D^0,\,D^{*0}$, because that is the reaction where you have antiquark
decay in their case.}  $\overline{\Sigma}_2
'^{(1^-)}=\overline{\Sigma}_2^{(0^-)}$ contradicting our result in
Eq.~(\ref{eq:sigbtob}) were we find $\overline{\Sigma}_2
'^{(1^-)}=-\overline{\Sigma}_2^{(0^-)}$.  Our result is a clear
prediction of the quark model and comes from the extra signs that
appear due to the fact that it is the antiquark that decays (See
appendix~\ref{Bcapp:sign}). This difference between quark
and antiquark decay was not properly reflected in their published work~\cite{Jenkins:PC}.
\begin{figure}[h!!!!]
%\vspace{1cm}
\centering
\resizebox{8.cm}{7.cm}{\includegraphics{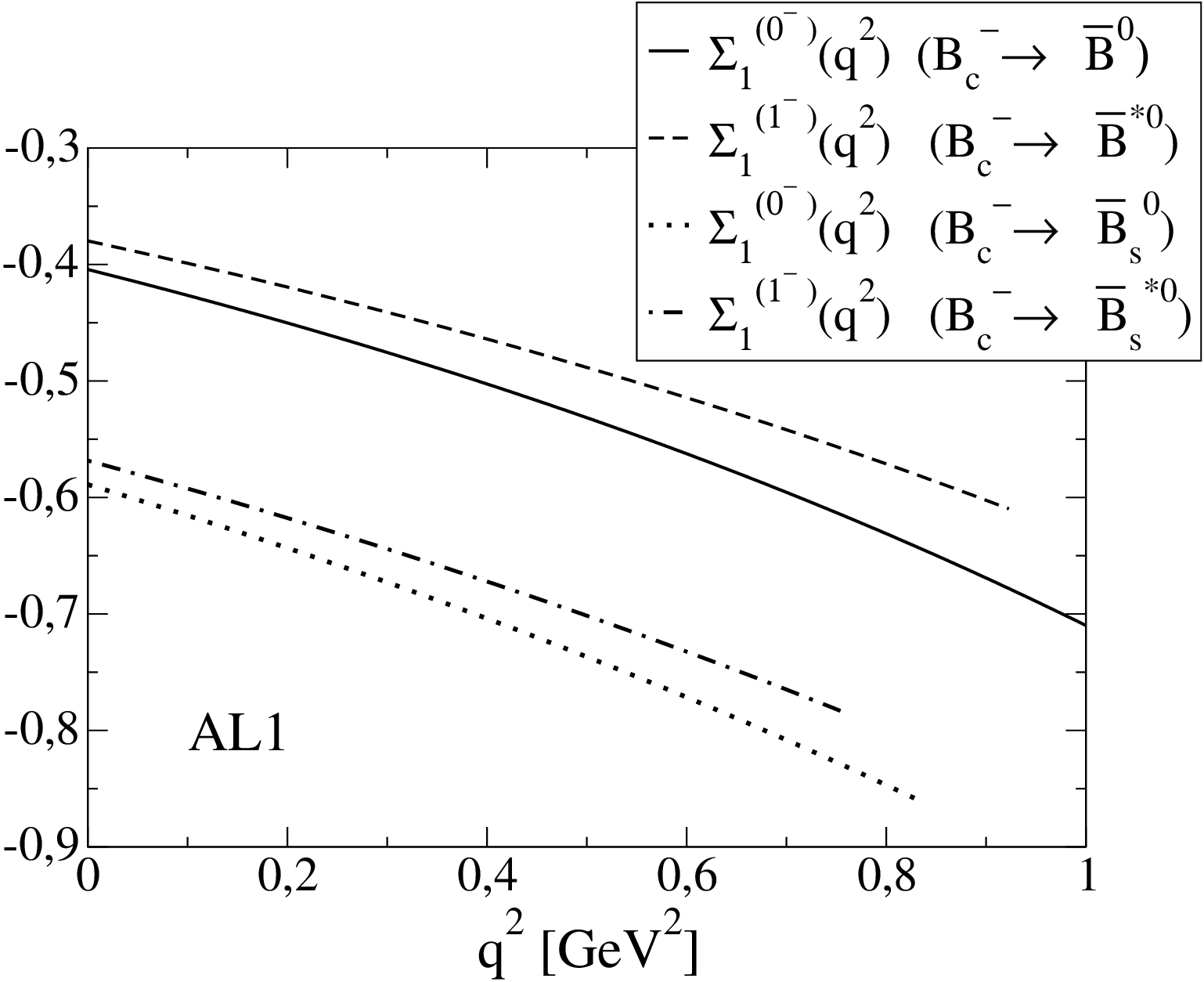}}
%\resizebox{10cm}{!}
\caption{$\Sigma_1^{(0^-)}$ of the semileptonic $B_c^-\to \overline{B}^0$
(solid line) and
$B_c^-\to \overline{B}_s^0$ (dotted line) transitions, and
$\Sigma_1^{(1^-)}$ of the semileptonic 
$B_c^-\to \overline{B}^{*0}$ (dashed line) and
$B_c^-\to \overline{B}_s^{*0}$ (dashed--dotted line) transitions
evaluated with the AL1 potential.}
\label{fig:sig1btob}
\end{figure} 

\begin{figure}[h!!]
%\vspace{1cm}
\centering
\resizebox{8.cm}{7.cm}{\includegraphics{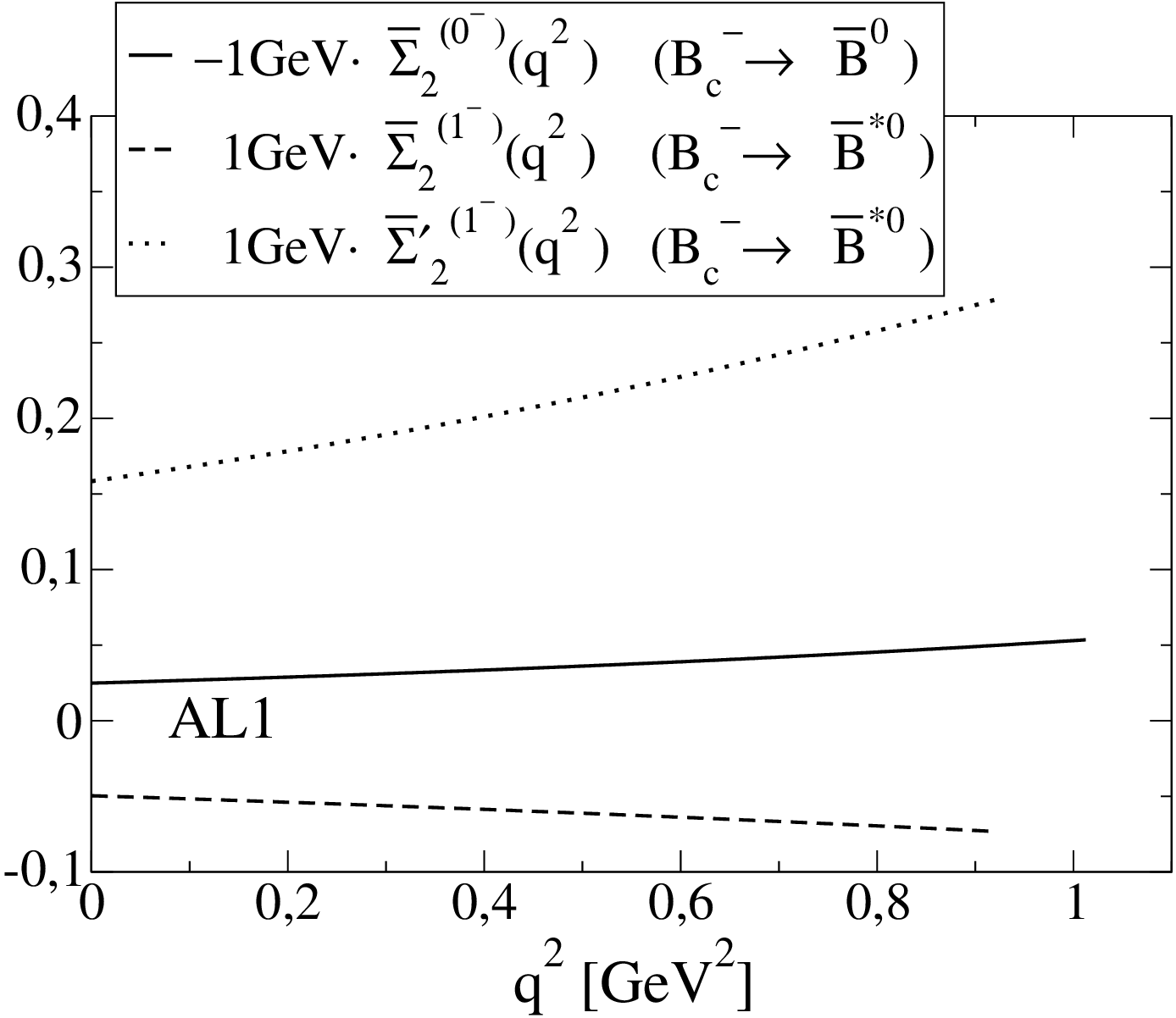}}\hspace{1cm}
\resizebox{8.cm}{7.cm}{\includegraphics{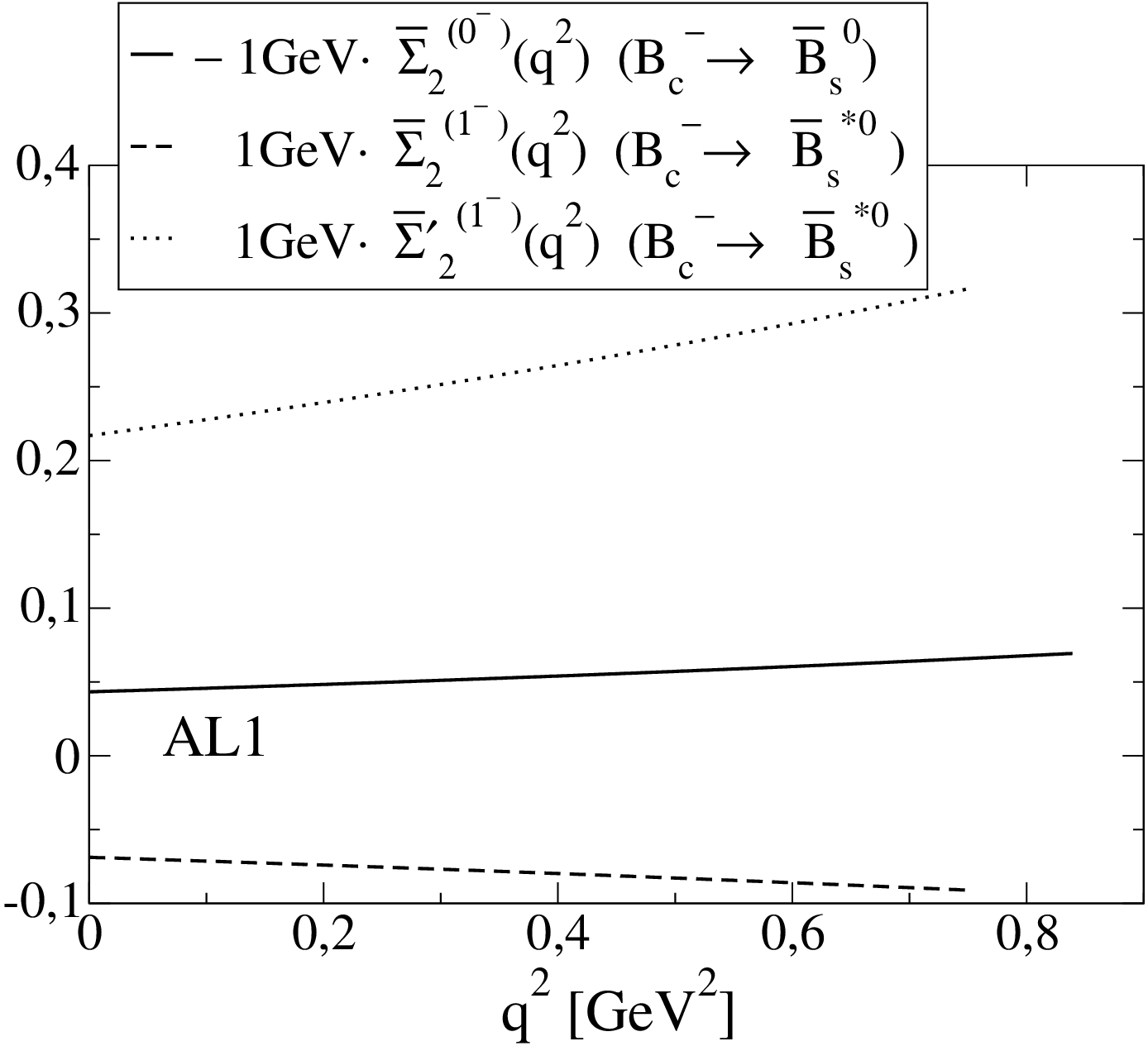}}\hspace{1cm}
%\resizebox{10cm}{!}
\caption{$-\overline{\Sigma}_2^{(0^-)}$ (solid line) of the semileptonic $B_c^-\to \overline{B}^0$ and
\mbox{$B_c^-\to \overline{B}_s^0$} transitions, and
$\overline{\Sigma}_2^{(1^-)}$ (dashed line), $\overline{\Sigma}_2'^{(1^-)}$
(dotted line) of the semileptonic \mbox{$B_c^-\to \overline{B}^{*0}$} and
$B_c^-\to \overline{B}_s^{*0}$ transitions
evaluated with the AL1 potential.}
\label{fig:sig2btob}
\end{figure}%
How far are we from the infinite heavy quark mass limit? In
Fig.~\ref{fig:sig1btob} we show $\Sigma_1^{(0^-)}$ of the semileptonic
$B_c^-\to \overline{B}^0$ and $B_c^-\to \overline{B}_s^0$ transitions,
and $\Sigma_1^{(1^-)}$ of the semileptonic $B_c^-\to
\overline{B}^{*0}$ and $B_c^-\to \overline{B}_s^{*0}$ transitions. The
differences between the corresponding $\Sigma_1^{(0^-)}$ and
$\Sigma_1^{(1^-)}$ are at the level of 10\%. The differences are much
more significant for $\overline{\Sigma}_2^{(0^-)}$,
$\overline{\Sigma}_2^{(1^-)}$ and $\overline{\Sigma}_2'^{(1^-)}$ that
we show in Fig.~\ref{fig:sig2btob}. In each case the three curves
shown would be the same in the infinite heavy quark mass limit.
Clearly in this case corrections on the inverse of the heavy quark
masses seem to be important.

\section{Nonleptonic $B_{\lowercase{c}}^-\to {B}\, M_F$ two-meson decays}
\label{Bcsect:nonlepb}
In this section we will evaluate decay widths for nonleptonic
 $B_c^-\to {B}\ M_F$ two--meson decays where $M_F$ is a pseudoscalar
 or vector meson with no $b$ quark content, and, at this point, ${B}$
 represents a meson with a $b$ quark.  These decay modes involve a
 $\bar c\to \bar d$ or $\bar c \to \bar s$ transition at the quark
 level and they are governed, neglecting penguin operators, by the
 effective Hamiltonian ~\cite{Ivanov:2006ni,Ebert:2003wc}
\begin{eqnarray}
H_{eff.}=\frac{G_F}{\sqrt2}\,\left\{ V_{cd}\left[
c_1(\mu)\,Q_1^{cd}+c_2(\mu)\,Q_2^{cd}
\right]
+V_{cs}\left[
c_1(\mu)\,Q_1^{cs}+c_2(\mu)\,Q_2^{cs}
\right]
+H.c.\right\}
\nonumber\\
\end{eqnarray}
where $c_1,\,c_2$ are scale--dependent Wilson coefficients, and
$Q_1^{cd},\,Q_2^{cd},\,Q_1^{cs},\,Q_2^{cs}$ are local four--quark 
operators given by
\begin{eqnarray}
Q_1^{cd}&=&\overline{\Psi}_c(0)\gamma_\mu(I-\gamma_5)\Psi_d(0)\,\bigg[
\ V_{ud}^*\, \overline{\Psi}_d(0)\gamma^\mu(I-\gamma_5)\Psi_u(0)+
V_{us}^*\, \overline{\Psi}_s(0)\gamma^\mu(I-\gamma_5)\Psi_u(0)
\bigg]\nonumber\\
Q_2^{cd}&=&\overline{\Psi}_c(0)\gamma_\mu(I-\gamma_5)\Psi_u(0)\,\bigg[
\ V_{ud}^*\, 
\overline{\Psi}_d(0)\gamma^\mu(I-\gamma_5)\Psi_d(0)
+V_{us}^*\, \overline{\Psi}_s(0)\gamma^\mu(I-\gamma_5)\Psi_d(0)
\bigg]\nonumber\\
Q_1^{cs}&=&\overline{\Psi}_c(0)\gamma_\mu(I-\gamma_5)\Psi_s(0)\,\bigg[
\ V_{ud}^*\, \overline{\Psi}_d(0)\gamma^\mu(I-\gamma_5)\Psi_u(0)+
V_{us}^*\, \overline{\Psi}_s(0)\gamma^\mu(I-\gamma_5)\Psi_u(0)
\bigg]\nonumber\\
Q_2^{cs}&=&\overline{\Psi}_c(0)\gamma_\mu(I-\gamma_5)\Psi_u(0)\,\bigg[
\ V_{ud}^*\, 
\overline{\Psi}_d(0)\gamma^\mu(I-\gamma_5)\Psi_s(0)
+V_{us}^*\, \overline{\Psi}_s(0)\gamma^\mu(I-\gamma_5)\Psi_s(0)\,
\bigg]
\nonumber\\
\end{eqnarray}

We shall work again in the factorization approximation taking into
account the Fierz reordered contribution so that the relevant
coefficients are not $c_1$ and $c_2$ but the combinations
\begin{eqnarray}
a_1(\mu)=c_1(\mu)+\frac{1}{N_C}\,c_2(\mu)\hspace{0.5cm};\hspace{0.5cm}
a_2(\mu)=c_2(\mu)+\frac{1}{N_C}\,c_1(\mu)
\end{eqnarray}
The energy scale $\mu$ appropriate in this
case is $\mu\simeq m_c$ and the values for $a_1$ and $a_2$ that we use
are~\cite{Ivanov:2006ni}
\begin{eqnarray}
a_1=1.20\hspace{0.5cm};\hspace{0.5cm}
a_2=-0.317
\end{eqnarray}
\subsubsection{\bf$\bullet$ $M_F=\pi^-,\,\rho^-,\,K^-,\,K^{*-}$ }

\begin{table}[h]
{\footnotesize
\centering
\begin{tabular}{l|r}
\multicolumn{1}{l}{}&\multicolumn{1}{c}{$\Gamma$ [$10^{-15}$\,GeV]}\\
\multicolumn{1}{l}{}&\multicolumn{1}{c}{}\\
 & This work\\
\hline
$B_c^-\to\overline{B}^0\,\pi^-$ & $1.10^{+0.14}_{-0.16}\, a_1^2$\\
$B_c^-\to\overline{B}^0\,\rho^-$ & $1.41^{+0.12}_{-0.19}\, a_1^2$\\
$B_c^-\to\overline{B}^0\,K^-$ & $0.098^{+0.012}_{-0.012}\, a_1^2$\\
$B_c^-\to\overline{B}^0\,K^{*-}$ & $0.038^{+0.003}_{-0.005}\,  a_1^2$\\
\hline
$B_c^-\to \overline{B}^{*0}\,\pi^-$ &$0.71^{+0.12}_{-0.11}\, a_1^2$ \\
$B_c^-\to \overline{B}^{*0}\,\rho^-$ &$5.68^{+0.55}_{-0.77}\,1a_1^2$\\
$B_c^-\to \overline{B}^{*0}\,K^-$ &$0.047^{+0.007}_{-0.007}\,a_1^2$\\
$B_c^-\to \overline{B}^{*0}\,K^{*-}$ &$0.29^{+0.03}_{-0.04} \,a_1^2$\\
\hline
$B_c^-\to\overline{B}_s^0\,\pi^-$ & $34.7^{+2.0}_{-0.6}\, a_1^2$\\
$B_c^-\to\overline{B}_s^0\,\rho^-$ & $23.1^{+0.5}_{-0.6}\, a_1^2$\\
$B_c^-\to\overline{B}_s^0\,K^-$ & $2.87^{+0.13}_{-0.06}\, a_1^2$\\
$B_c^-\to\overline{B}_s^0\,K^{*-}$ & $0.13_{-0.01}\,  a_1^2$\\
\hline
$B_c^-\to \overline{B}_s^{*0}\,\pi^-$ &$22.8^{+2.2}_{-1.0}\, a_1^2$ \\
$B_c^-\to \overline{B}_s^{*0}\,\rho^-$ &$132^{+5}_{-6}\,a_1^2$\\
$B_c^-\to \overline{B}_s^{*0}\,K^-$ &$1.29^{+0.10}_{-0.06}\,a_1^2$\\
\end{tabular}
\vspace{2cm}

\begin{tabular}{l|r c c c c c c c}
\multicolumn{1}{l}{}&\multicolumn{8}{c}{B.R. in \%}\\
\multicolumn{1}{l}{}&\multicolumn{8}{c}{}\\
 & This work&\cite{Ivanov:2006ni}&\cite{Ebert:2003wc}&\cite{Chang:1992pt}&\cite{AbdEl-Hady:1999xh}
&\cite{Kiselev:2002vz} &\cite{Colangelo:1999zn}
 &\cite{Anisimov:1998xv}\\
\hline
$B_c^-\to\overline{B}^0\,\pi^-$ & $0.11^{+0.01}_{-0.01}$
&0.20&0.10&0.32&0.10&1.06&0.19&0.15\\
$B_c^-\to\overline{B}^0\,\rho^-$ & $0.14^{+0.02}_{-0.02}$
&0.20&0.13&0.59&0.28&0.96&0.15&0.19\\
$B_c^-\to\overline{B}^0\,K^-$ & $0.010^{+0.001}_{-0.001}$
&0.015&0.009&0.025&0.010&0.07&0.014&\\
$B_c^-\to\overline{B}^0\,K^{*-}$ &
$0.0039^{+0.0003}_{-0.0005}$
&0.0048&0.004&0.018&0.012&0.015&0.003&\\
\hline
$B_c^-\to \overline{B}^{*0}\,\pi^-$ &$0.072^{+0.012}_{-0.012}$
&0.057&0.026&0.29&0.076&0.95&0.24&0.077 \\
$B_c^-\to \overline{B}^{*0}\,\rho^-$ &$0.58^{+0.05}_{-0.08}$
&0.30&0.67&1.17&0.89&2.57&0.85&0.67\\
$B_c^-\to \overline{B}^{*0}\,K^-$
&$0.0048^{+0.0007}_{-0.0008}$
&0.0036&0.004&0.019&0.006&0.055&0.012&\\
$B_c^-\to \overline{B}^{*0}\,K^{*-}$
&$0.030^{+0.002}_{-0.004}$
&0.013&0.032&0.037&0.065&0.058&0.033&\\
\hline
$B_c^-\to\overline{B}_s^0\,\pi^-$ & $3.51^{+0.19}_{-0.06}$
&3.9&2.46&5.75&1.56&16.4&3.01&3.42\\
$B_c^-\to\overline{B}_s^0\,\rho^-$ & $2.34^{+0.05}_{-0.06}$
&2.3&1.38&4.41&3.86&7.2&1.34&2.33\\
$B_c^-\to\overline{B}_s^0\,K^-$ & $0.29^{+0.01}_{-0.01}$
&0.29&0.21&0.41&0.17&1.06&0.21&\\
$B_c^-\to\overline{B}_s^0\,K^{*-}$ & $0.013_{-0.001}$
&0.011&0.0030&&0.10&&0.0043&\\
\hline
$B_c^-\to \overline{B}_s^{*0}\,\pi^-$ &$2.34^{+0.19}_{-0.14}$
&2.1&1.58&5.08&1.23&6.5&3.50&1.95 \\
$B_c^-\to \overline{B}_s^{*0}\,\rho^-$ &$13.4^{+0.5}_{-0.6}$
&11&10.8&14.8&16.8&20.2&10.8&12.1\\
$B_c^-\to \overline{B}_s^{*0}\,K^-$ &$0.13^{+0.01}_{-0.01}$
&0.13&0.11&0.29&0.13&0.37&0.16&\\

\end{tabular}
}
\caption{Decay widths in units of
$10^{-15}$\,GeV, and for general values of the Wilson coefficient $a_1$, and
branching ratios in \%
for exclusive nonleptonic decays of the $B_c^-$ meson. Our central values have
been obtained with the AL1 potential. }
\label{tab:dwbrbtobpik}
\end{table}

In this case $B$ denotes one of the 
$\overline{B}^{0},\,\overline{B}^{*0},\,\overline{B}_s^{0},\,\overline{B}_s^{*0}$.
The decay widths  are
\begin{eqnarray}
&&\hspace* {-.5cm}M_F\equiv 0^-\ \mathrm{case} \nonumber \\
&&\Gamma=\frac{G_F^2}{16\pi m^2_{B_c}}\, |V_{cd}|^2\,|V_{F}|^2\ a_1^2\,
\frac{\lambda^{1/2}(m^2_{B_c},\,m^2_{\overline{B}^{0},\,\overline{B}^{*0}},
\,m^2_F)}{2m_{B_c}}\,
m_F^2\,f^2_F\, {\cal H}_{tt}^{B_c^-\to\overline{B}^{0},\,\overline{B}^{*0}}
(m_F^2)\nonumber\\
&&\hspace*{-.5cm}M_F\equiv1^-\ \mathrm{case} \nonumber\\
&&\Gamma=\frac{G_F^2}{16\pi m^2_{B_c}}\, |V_{cd}|^2\,|V_{F}|^2\ a_1^2\
\frac{\lambda^{1/2}(m^2_{B_c},\,m^2_{\overline{B}^{0},\,
\overline{B}^{*0}},\,m^2_F)}{2m_{B_c}}\,
m_F^2\,f^2_F
\nonumber\\
 &&\hspace{2cm}\times\left({\cal H}_{+1+1}^{B_c^-\to\overline{B}^{0},\,\overline{B}^{*0}}(m_F^2)
 +{\cal H}_{-1-1}^{B_c^-\to\overline{B}^{0},\,\overline{B}^{*0}}(m_F^2)
 +{\cal H}_{00}^{B_c^-\to\overline{B}^{0},\,\overline{B}^{*0}}(m_F^2)
\right)
\nonumber\\
\label{eq:dwb}
\end{eqnarray}
and similarly for $\overline{B}^{0}_s,\,\overline{B}^{*0}_s$ with
$|V_{cd}|\to|V_{cs}|;\,\overline{B}^{0},\,\overline{B}^{*0}\to
\overline{B}^{0}_s,\,\overline{B}^{*0}_s$.  $V_F=V_{ud}$ or
$V_F=V_{us}$ depending on whether $M_F=\pi^-,\,\rho^-$ or
$M_F=K^-,\,K^{*-}$, $f_F$ is the decay constant of the $M_F$ meson, and
 the different ${\cal H}_{rr}$ have been evaluated at $q^2=m_F^2$.  In
Table~\ref{tab:dwbrbtobpik} we show the decay widths for a general
value of the Wilson coefficient $a_1$, and the corresponding branching
ratios evaluated with $a_1=1.20$. The transition
$B_c^-\to\overline{B}^{*0}_s K^{*-}$ is not allowed with the new $B_c$
mass value from Ref.~\cite{Abulencia:2005us}. Our branching ratios
for a final $\overline{B}^0_s$
or $\overline{B}^{*0}_s$ are in very good agreement with the results by
Ivanov {\it et al.}~\cite{Ivanov:2006ni}, while for a final $\overline{B}^0$ or
$\overline{B}^{*0}$
we are in very good agreement with the results by Ebert {\it et
al.}~\cite{Ebert:2003wc} (with the exception of the $B_c\to
\overline{B}^{*0}\pi^-$ decay).

\subsubsection{\bf $\bullet$$M_F=\pi^0,\,\rho^0,\,K^0,\,K^{*0}$}
Here the generic name ${B}$ stands for a ${B}^-$ or a ${B}^{*-}$
meson. The different decay widths are given by
\begin{eqnarray}
&&\hspace*{-.5cm}M_F\equiv 0^-\ \mathrm{case} \nonumber \\
&&\Gamma=\frac{G_F^2}{16\pi m^2_{B_c}}\, |V_{ud}|^2\,|V_{F}|^2\ a_2^2\,
\frac{\lambda^{1/2}(m^2_{B_c},\,m^2_{ B^{-},\, B^{*-}},
\,m^2_F)}{2m_{B_c}}\,
m_F^2\,\widetilde{f}^2_F\, {\cal H}_{tt}^{B_c^-\to {B}^{-},\,{B}^{*-}}
(m_F^2)\nonumber\\
&&\hspace*{-.5cm}M_F\equiv1^-\ \mathrm{case}\nonumber\\ 
&&\Gamma=\frac{G_F^2}{16\pi m^2_{B_c}}\, |V_{ud}|^2\,|V_{F}|^2\ a_2^2\
\frac{\lambda^{1/2}(m^2_{B_c},\,m^2_{{B}^{-},\,{B}^{*-}},
\,m^2_F)}{2m_{B_c}}\,
m_F^2\,\widetilde{f}^2_F
\nonumber\\
 &&\hspace{2cm}\times\left({\cal H}_{+1+1}^{B_c^-\to {B}^{-},\,{B}^{*-}}(m_F^2)
 +{\cal H}_{-1-1}^{B_c^-\to {B}^{-},\,{B}^{*-}}(m_F^2)
 +{\cal H}_{00}^{B_c^-\to {B}^{-},\,{B}^{*-}}(m_F^2)
\right) \nonumber\\
\label{eq:dwcc2m}
\end{eqnarray}
where $V_F=V_{cd}$ or $V_F=V_{cs}$ depending on whether
$M_F=\pi^0,\,\rho^0$ or \mbox{$M_F=K^0,\,K^{*0}$}, $\widetilde{f}_F=f_F$ for
$M_F=K^0,\,K^{*0}$ whereas $\widetilde{f}_F=f_F/\sqrt2$ for
$M_F=\pi^0,\,\rho^0$, with $f_F$ the $M_F$ meson decay constant, and
the different ${\cal H}_{rr}$  evaluated at $q^2=m_F^2$. The latter
 have been obtained
from the matrix elements for  the  effective current operator
$\overline{\Psi}_{c}(0)\gamma^\mu(I-\gamma_5)\Psi_u(0)$.
The decay
widths, for a general value of the Wilson coefficient $a_2$, and the
corresponding branching ratios are shown in
Table~\ref{tab:dwbrbtobm}. With the exception of the $B_c^-\to B^{*-}\pi^0$
case, our results are in a global  good agreement with
the ones by Ebert {\it et al.}~\cite{Ebert:2003wc}.

\begin{table}[h]
%\begin{center}
{\small
\centering
\begin{tabular}{l|r}
\multicolumn{1}{l}{}&\multicolumn{1}{c}{$\Gamma$ [$10^{-15}$\,GeV]}\\
\multicolumn{1}{l}{}&\multicolumn{1}{c}{}\\
 & This work\\
\hline
$B_c^-\to{B}^-\,\pi^0$ & $0.54^{+0.07}_{-0.07}\, a_2^2$\\
$B_c^-\to{B}^-\,\rho^0$ & $0.71^{+0.06}_{-0.10}\, a_2^2$\\
$B_c^-\to{B}^-\,K^0$ & $35.3^{+4.0}_{-4.9}\, a_2^2$\\
$B_c^-\to{B}^-\,K^{*0}$ & $13.1^{+0.9}_{-0.7}\,  a_2^2$\\
\hline
$B_c\to {B}^{*-}\,\pi^0$ &$0.35^{+0.06}_{-0.05}\, a_2^2$ \\
$B_c\to {B}^{*-}\,\rho^0$ &$2.84^{+0.27}_{-0.39}\,a_2^2$\\
$B_c\to {B}^{*-}\,K^0$ &$16.9^{+2.4}_{-2.7}\,a_2^2$\\
$B_c\to {B}^{*-}\,K^{*0}$ &$103^{+8}_{-13} \,a_2^2$\\

\end{tabular}
\vspace{2.cm}

\begin{tabular}{l|r c c c  c c c}
\multicolumn{1}{l}{}&\multicolumn{7}{c}{B.R. in \%}\\
\multicolumn{1}{l}{}&\multicolumn{7}{c}{}\\

 & This work&\cite{Ivanov:2006ni}&\cite{Ebert:2003wc}&\cite{Chang:1992pt}&\cite{AbdEl-Hady:1999xh}
&\cite{Kiselev:2002vz} &\cite{Anisimov:1998xv}\\
 \hline
$B_c^-\to{B}^-\,\pi^0$ & $0.0038^{+0.0005}_{-0.0006}$
&0.007&0.004&0.011&0.004&0.037&0.007\\
$B_c^-\to{B}^-\,\rho^0$ & $0.0050^{+0.0004}_{-0.0007}$
&0.0071&0.005&0.020&0.010&0.034&0.009\\
$B_c^-\to{B}^-\,K^0$ & $0.25^{+0.03}_{-0.04}$
&0.38  &0.24&0.66 &0.27&1.98 &0.17 \\
$B_c^-\to{B}^-\,K^{*0}$ &$0.093^{+0.006}_{-0.013}$
&0.11 &0.09 &0.47&0.32 &0.43 &0.095\\
\hline
$B_c\to {B}^{*-}\,\pi^0$ &$0.0025^{+0.0004}_{-0.0005}$
&0.0020&0.001&0.010&0.003&0.033&0.004 \\
$B_c\to {B}^{*-}\,\rho^0$ &$0.020^{+0.002}_{-0.003}$
&0.011 &0.024&0.041&0.031&0.09&0.031\\
$B_c\to {B}^{*-}\,K^0$&$0.12^{+0.02}_{-0.02}$
&0.088 &0.11&0.50 &0.16 &1.60&0.061\\
$B_c\to {B}^{*-}\,K^{*0}$&$0.73^{+0.06}_{-0.10}$
&0.32 &0.84 &0.97 &1.70 &1.67&0.57\\
\end{tabular}
\caption{Decay widths in units of
$10^{-15}$\,GeV, and for general values of the Wilson coefficient $a_2$, and
branching ratios in \%
for exclusive nonleptonic decays of the $B_c^-$ meson. Our central values have
been obtained with the AL1 potential. }
\label{tab:dwbrbtobm}
}
\end{table}

\chapter{Doubly heavy baryons spectroscopy and static properties}
\chaptermark{Doubly heavy baryons 1}
\label{cha:dp_masas}
\section{Introduction}
%************************

%
\begin{table}[t]

\centering
\begin{tabular}{cccccccc}\hline
Baryon &~~~~$S$~~~~&~~~~$J^P$~~~~&~~~~ $I$~~~~&~~~~$S_{ h}^\pi$~~~~& 
Quark content 

\\
       &       &         &   &          &                 
\\\hline
$\Xi_{cc}$& 0 &$\frac12^+$& $\frac12$ &$1^+$&$ccl$
\\
$\Xi^*_{cc}$ & 0 &$\frac32^+$&$\frac12$  &$1^+$&$ccl$
\\
$\Omega_{cc}$ & $-1$ &$\frac12^+$& 0 &$1^+$&$ccs$
\\
$\Omega^*_{cc}$ & $-1$ &$\frac32^+$&$0$&$1^+$&$ccs$
\\\\
$\Xi_{bb}$& 0 &$\frac12^+$& $\frac12$ &$1^+$&$bbl$
\\
$\Xi^*_{bb}$ & 0 &$\frac32^+$&$\frac12$  &$1^+$&$bbl$
\\
$\Omega_{bb}$ & $-1$ &$\frac12^+$& 0 &$1^+$&$bbs$
\\
$\Omega^*_{bb}$ & $-1$ &$\frac32^+$&$0$&$1^+$&$bbs$
\\\\
$\Xi'_{bc}$& 0 &$\frac12^+$& $\frac12$ &$0^+$&$bcl$
\\
$\Xi_{bc}$ & 0 &$\frac12^+$&$\frac12$  &$1^+$&$bcl$
\\
$\Xi^*_{bc}$& 0 &$\frac32^+$& $\frac12$ &$1^+$&$bcl$
\\
$\Omega'_{bc}$ & $-1$ &$\frac12^+$& 0 &$0^+$&$bcs$
\\
$\Omega_{bc}$ & $-1$ &$\frac12^+$& 0 &$1^+$&$bcs$
\\
$\Omega^*_{bc}$ & $-1$ &$\frac32^+$&$0$&$1^+$&$bcs$
\\
\hline
\end{tabular}
\caption{Quantum numbers of doubly heavy baryons analyzed in this study. $S$, $J^P$ are strangeness and the spin parity
of the baryon, $I$ is the isospin, and
$S_{h}^\pi$ is the spin parity of the heavy
degrees of freedom. $l$ denotes a light  $u$ or
$d$ quark .}
\label{tab:summ}
\end{table}

%=========================================

The subject of doubly heavy baryons has been
attracting attention for a long time. Magnetic moments of doubly
charmed baryons were evaluated back in the 70's by
Lichtenberg~\cite{Lichtenberg:1976fi} within a nonrelativistic approach.
The infinite heavy quark mass limit was already used in the 90's to
relate the spectrum of doubly heavy baryons to the one of mesons with
a single heavy quark~\cite{Savage:1990di}, or to analyze their semileptonic
decay~\cite{White:1991hz}. A factor of two error in the hyperfine splittings of Ref.~\cite{Savage:1990di} 
 have been recently noticed 
by the potential nonrelativistic QCD (pNRQCD) calculation of Ref.~\cite{Brambilla:2005yk}.

On the experimental side the SELEX Collaboration claimed
evidence for  the $\Xi^+_{cc}$ baryon, in the $\Lambda_c^+K^-\pi^+$ and 
$pD^+K^-$ decay modes, with a mass of
$M_{\Xi^+_{cc}}=3519\pm 1\ \mathrm{MeV/c^2}$~\cite{Mattson:2002vu}. Those results
were challenged by a theoretical analysis~\cite{Kiselev:2002an} which claimed 
 the observed events
by the SELEX Collaboration could be explained without the involvement
of doubly charmed baryons. Other experimental collaborations like 
FOCUS~\cite{Ratti:2003ez}, {\sl BABAR}  ~\cite{Aubert:2006qw} and BELLE~\cite{Lesiak:2006sk}
have found no evidence for doubly charmed baryons so far.   At present
 the    $\Xi^+_{cc}$ has only a one star status and it  is not listed in  the
particle summary table~\cite{Yao:2006px}.

As discussed before, for hadrons with two heavy quarks one can apply 
 HQSS~\cite{Jenkins:1992nb}, which
amounts to the decoupling of the heavy quark spins in the infinite
heavy quark mass limit. In that limit one can consider the total spin
of the two heavy quark  subsystem ($S_h$) to be well defined.  In this
work we shall assume this is a good approximation for the actual heavy
quark masses. This approximation, which is the only one related to the
infinite heavy quark mass limit that we shall use, will certainly
simplify the solution of the baryon three--quark problem.
Recently the authors of
Ref.~\cite{Brambilla:2005yk} have developed and effective theory (pNRQCD) suitable to
describe baryons with two and three heavy quarks.

%
% Parrafo anterior version antigua
%
%In hadrons with a heavy quark and working in the
%infinite heavy quark mass limit the dynamics of the light degrees of
%freedom becomes independent of the heavy quark flavor and spin. This
%is known as heavy quark symmetry (HQS)~\cite{Nussinov:1986hw,Shifman:1986sm,Politzer:1988wp,Politzer:1988bs,Isgur:1989vq,Isgur:1989ed}.
%This symmetry was developed into an effective theory
%(HQET)~\cite{Georgi:1990um} that allowed a systematic, order by order,
%evaluation of corrections in inverse powers of the heavy quark masses.
%Unfortunately ordinary HQS can not be applied directly to hadrons
%containing two heavy quarks as the kinetic energy term needed in those
%systems to regulate infrared divergences breaks heavy flavor
%symmetry~\cite{Thacker:1990bm}.  For those hadrons the symmetry that
%survives is heavy quark spin symmetry (HQSS)~\cite{Jenkins:1992nb}, which
%amounts to the decoupling of the heavy quark spins in the infinite
%heavy quark mass limit. In that limit one can consider the total spin
%of the two heavy quark  subsystem ($S_h$) to be well defined.  In this
%work we shall assume this is a good approximation for the actual heavy
%quark masses. This approximation, which is the only one related to the
%infinite heavy quark mass limit that we shall use, will certainly
%simplify the solution of the baryon three--quark problem.

Solving the three--body problem is not an easy task and here we shall
do it by means of a variational approach. The approach, with obvious
changes, was already applied with good results in the study of baryons
with one heavy quark~\cite{Albertus:2003sx}. This method, that leads to
simple and manageable wave functions, is made possible by the
simplifications introduced in the problem by the fact that we can
consider $S_h$ to be well defined.  

Our simple variational calculation reproduces the results for static
properties obtained in Ref.~\cite{Silvestre-Brac:1996bg} by solving more involved
Faddeev type equations. Our method has the  advantage that we provide explicit and
manageable wave
functions that can be used to evaluate further observables. Static properties like masses and magnetic
moments of doubly heavy baryons have also been studied in other
models.  Masses have been calculated in the relativistic quark model
assuming a light quark heavy diquark structure~\cite{Ebert:2002ig}, the
potential approach and sum rules of QCD~\cite{Kiselev:2001fw}, the
nonperturbative QCD approach~\cite{Narodetskii:2001bq}, the Bethe--Salpeter
equation applied to the light quark heavy diquark~\cite{Tong:1999qs}, the
nonrelativistic quark model with harmonic oscillator
potential~\cite{Itoh:2000um} or with the use of QCD derived
potentials~\cite{Vijande:2004at,Gershtein:2000nx}, the relativistic quasi--potential quark
model~\cite{Ebert:1996ec}, with the use of the Feynman-Hellman theorem and
semi-empirical mass formulas within the framework of a nonrelativistic
constituent quark model~\cite{Roncaglia:1995az,Roncaglia:1994ex},  in
effective field theories~\cite{Korner:1994nh,Brambilla:2005yk}, or in lattice nonrelativistic
QCD~\cite{Mathur:2002ce}. There are also lattice QCD determinations
~\cite{AliKhan:1999yb,Lewis:2001iz,Flynn:2003vz}. Similarly, magnetic moments have
been evaluated in a nonrelativistic approach~\cite{Lichtenberg:1976fi}, in
the relativistic three--quark model~\cite{Faessler:2006ft}, the
relativistic quark model using different forms of the relativistic
kinematics~\cite{Julia-Diaz:2004vh}, in the skyrmion model~\cite{Oh:1991ws}, in the
Dirac equation formalism~\cite{Jena:1986xs}, or using the MIT bag
model~\cite{Bose:1980vy}.

In Table~\ref{tab:summ} we summarize the quantum numbers of the
doubly heavy baryons considered in this study.

\section{Three Body Problem}
\label{sec:3bp}
\subsection{Intrinsic Hamiltonian}
\label{subsec:h}
\begin{figure}[!t]
%\vspace{-3cm}
\centering
\resizebox{10.cm}{!}{\includegraphics{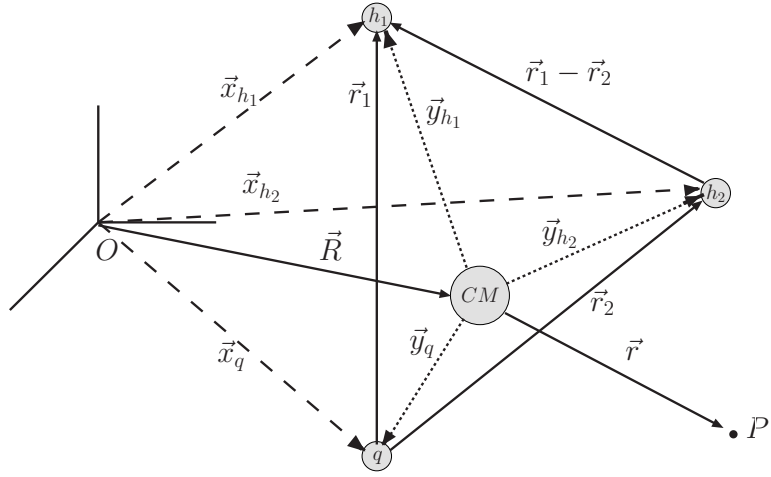}}
%\vspace{-15cm}
\caption{ Definition of different coordinates used through this
work.}\label{fig:coor}
\end{figure}

In the Laboratory (LAB) frame (see Fig.~\ref{fig:coor}), the
Hamiltonian ($H$) of the three quark ($h_1,\,h_2,\,q$, where
$h_1,\,h_2=c,b$ and $q=l(u,\,d),\,s$) system reads:
\begin{eqnarray}
H&=&  \sum_{j=h_1,\,h_2,\,q} \left (m_j
-\frac{\stackrel{\rightarrow}{\nabla}\stackrel{}{^2}_{\hspace{-.1cm}\vec x_j}}{2m_j}\right ) +
V_{h_1h_2} + V_{h_1q}+V_{h_2 q }
\end{eqnarray}
where $m_{h_1},\,m_{h_2},\,m_q$  are the quark masses and the
quark-quark interaction terms $V_{jk}$ depend on the quark
spin-flavor quantum numbers and the quark coordinates ($\vec{x}_{h_1},
\vec{x}_{h_2},\, \vec{x}_q$ for the $h_1,\,h_2,\,q$ quarks
respectively). To separate the Center of Mass (CM) free motion,
we go to the light quark frame ($\vec{R},\vec{r}_1,\vec{r}_2$),
\begin{eqnarray}
\vec{R}&=&\frac{m_{h_1}\vec{x}_{h_1} + m_{h_2}\vec{x}_{h_2} + m_q \vec{x}_q}
{m_{h_1}+m_{h_2}+m_q} \nonumber \\
\vec{r}_1 & = & \vec{x}_{h_1} - \vec{x}_q \nonumber \\
\vec{r}_2 & = & \vec{x}_{h_2} - \vec{x}_q 
\end{eqnarray}
where $\vec{R}$ and $\vec{r}_1,\, \vec{r}_2$ are the CM position in
the LAB frame and the relative positions of the $h_1,\,h_2$ heavy quarks 
with respect to the light quark $q$.   The Hamiltonian now reads
\begin{eqnarray}
H&=&
-\frac{\stackrel{\rightarrow}{\nabla}
\stackrel{}{^2}_{\hspace{-.1cm}\vec{R}}}{2 \overline M} +
H^{\rm int} 
\end{eqnarray}
\begin{eqnarray}
 H^{\rm
int}&=&\sum_{j=1,2}\,H_j^{sp}+V_{h_1h_2}(\vec r_1-\vec r_2,\, spin)-
\frac{\stackrel{\rightarrow}{\nabla}_1\cdot\stackrel{\rightarrow}{\nabla}_2}
{m_q}+\overline M
\nonumber\\
H_j^{sp}&=&-\frac{\stackrel{\rightarrow}{\nabla}\stackrel{}{^2}_{\hspace{-0.2cm}j}
}{2\mu_j}+V_{h_jq}(\vec r_j,\, spin),\ j=1,2
\label{eq:hint}
\end{eqnarray}
where $\overline M = m_{h_1}+m_{h_2}+m_{q}$, $\mu_{j} = \left (
1/m_{h_j} + 1/m_q\right)^{-1}$ and $\stackrel{\rightarrow}{\nabla}_{j}
= \partial/\partial_{\vec{r}_j},\ j=1,2$. The intrinsic Hamiltonian
$H^{\rm int}$ describes the dynamics of the baryon. 
Apart from the sum of the
 quark masses $\overline M$, it consists of the
sum of two single particle Hamiltonian $H^{sp}_j$, each of them
describing the dynamics of a heavy-light quark system, plus the
heavy--heavy interaction term, including the Hughes-Eckart term
($\stackrel{\rightarrow}{\nabla}_1\cdot
\stackrel{\rightarrow}{\nabla}_2$).  We will use a variational approach to solve it.

\section{Variational Wave Functions}\label{sec:vwf}

For the interactions presented in chapter \ref{cha:potential},
  we have that both the total spin  and
 the internal orbital angular momentum given  as
\begin{eqnarray}
{\vec
S} &=& \left ( {\vec \sigma}_{h_1} + {\vec \sigma}_{h_2} +
{\vec \sigma}_q \right )/2\nonumber\\
{\vec L} &=& {\vec l}_1 + {\vec l}_2, \qquad {\rm with}~~ {\vec
l}_j=-{\rm i}~~{\vec r}_j \times {\stackrel{\rightarrow}{\nabla}}_j, \quad j=1,2
\end{eqnarray}
commute with the intrinsic Hamiltonian and are thus well defined.  We
are interested in the ground state of baryons with total angular
momentum $J=1/2,\,3/2$\ so that we can assume the orbital angular
momentum of the baryons to be $L=0$.  This implies that the spatial
wave function can only depend on the relative distances $r_1$, $r_2$
and $r_{12}=|\vec{r}_1-\vec{r}_2|$. Furthermore when the heavy quark
mass is infinity ($m_h \to \infty$), the total spin of the heavy
degrees of freedom, $\vec{S}_{\rm heavy} = \left ( {\vec \sigma}_{h_1}
+ {\vec \sigma}_{h_2} \right )/2$, commutes with the intrinsic
Hamiltonian, since the spin--spin terms in the potentials vanish in
this limit. We can then assume the spin of the heavy degrees of
freedom to be well defined.

 With these simplifications we have used the following intrinsic wave
functions in our variational approach\footnote{We omit the
antisymmetric color wave function and the plane wave for the center of
mass motion which are common to all cases.}
\begin{itemize}
\item {\it $\Xi_{h_1h_2},\,\Omega_{h_1h_2}-$type baryons: }
\begin{eqnarray}
%\lefteqn
&&\hspace*{-2.4cm}{|\Xi_{h_1h_2},\,\Omega_{h_1h_2}; J=\frac12, M_J  \rangle \,=}
\nonumber \\ %
&& \hspace*{-2.1cm} 
\sum_{M_{S_h}M_{S_q}} \left(\left. 1,\frac12,\frac12\right|M_{S_h,}M_{S_q,}M_J\right)\  |h_1h_2;1M_{S_h}
 \rangle \otimes |q;\frac12 M_{S_q}\rangle
 \Psi_{h_1h_2}^{\Xi,\,\Omega}(r_1,r_2,r_{12})
\nonumber \\
\end{eqnarray}
where $M_J$ is the third component of the baryon total angular
momentum while $|h_1h_2;S_h,M_{S_h}\rangle$ and $|q;\frac12
M_{S_q}\rangle $ represent spin states of the $h_1h_2$ subsystem and
the light quark respectively.   For $h_1=h_2$ we need
$\Psi_{h_1h_1}^{\Xi,\,\Omega}(r_1,r_2,r_{12})
=\Psi_{h_1h_1}^{\Xi,\,\Omega}(r_2,r_1,r_{12})$ to guarantee a complete
symmetry of the wave function under the exchange of the two heavy
quarks.
\item {\it $\Xi^*_{h_1h_2},\,\Omega^*_{h_1h_2}-$type baryons: }
\begin{eqnarray}
&& \hspace*{-2.4cm}|\Xi^*_{h_1h_2},\,\Omega^*_{h_1h_2}; J=\frac32, M_J  \rangle \,=
\nonumber \\ &&\hspace*{-2.2cm} 
\sum_{M_{S_h}M_{S_q}} \left(\left. 1,\frac12,\frac32\right|M_{S_h},M_{S_q},M_J\right)\  |h_1h_2;1M_{S_h}
 \rangle \otimes |q;\frac12 M_{S_q}\rangle 
\Psi_{h_1h_2}^{\Xi^*,\,\Omega^*}(r_1,r_2,r_{12})
\nonumber \\
\end{eqnarray}
Similarly to the case before for $h_1=h_2$  we need $\Psi_{h_1h_1}
^{\Xi^*,\,
\Omega^*}(r_1,r_2,r_{12})
=\Psi_{h_1h_1}^{\Xi^*,\,\Omega^*}(r_2,r_1,r_{12})$. 
\item {\it $\Xi'_{h_1h_2},\,\Omega'_{h_1h_2}-$type baryons: }
\begin{eqnarray} \hspace*{-2.5cm}
|\Xi'_{h_1h_2},\,\Omega'_{h_1h_2}; J=\frac12, M_J  \rangle  &=& 
  |h_1h_2;00 \rangle \otimes |q;\frac12 M_{J}\rangle
\Psi_{h_1h_2}^{\Xi',\,\Omega'}(r_1,r_2,r_{12})
\end{eqnarray}
In this case $h_1\ne h_2$ and we do not need the orbital part to have a 
definite symmetry under the exchange of the two quarks. 
\end{itemize}

The spatial wave functions $\Psi(r_1,r_2,r_{12})$ in the above
expressions will be determined by the variational principle: $\delta
\langle B | H^{\rm int}| B \rangle = 0$. For simplicity, we shall
assume a Jastrow--type functional form\footnote{ A similar form lead
to very good results in the case of baryons with a single heavy
quark~\cite{Albertus:2003sx}.}:

\begin{eqnarray}
\Psi_{h_1h_2}^{B} (r_1,r_2,r_{12}) &=& N\, F^{B}(r_{12})\,
\phi_{h_1q}(r_1)\,\phi_{h_2q}(r_2)
\end{eqnarray}
where $N$ is a constant, which is determined from
normalization
\begin{eqnarray}
1&=&\int d^3r_1 \int d^3 r_2\ \left |\Psi_{h_1h_2}^{B}(r_1,r_2,r_{12})
\right |^2 
\nonumber \\
&=& 8\pi^2 \int_0^{+\infty}dr_1~r_1^2~ \int_0^{+\infty}dr_2~r_2^2
\int_{-1}^{+1} d\mu~\left |\Psi_{h_1h_2}^{B}(r_1,r_2,r_{12})
\right |^2 
\end{eqnarray}
with $\mu$ being the cosine of the angle between the vectors $\vec
{r}_1$ and $\vec {r}_2$ ($r_{12}=(~r_1^2+r_2^2-2 r_1 r_2
\mu)^{1/2}$).
 
The functions $\phi_{h_1q}$ and $\phi_{h_2q}$ will be taken as the
$S-$wave ground states $\varphi_j(r_j)$ of the single particle
Hamiltonians $H^{sp}_{j}$ of Eq.~(\ref{eq:hint}) modified at large
distances
\begin{eqnarray}
\phi_{h_jq} (r_j) &=& (1+\alpha_j\,r_j)\,\varphi_j(r_j),\ \ j=1,2
\label{eq:onebody}
\end{eqnarray}
The heavy--heavy correlation function $F^{B}$ will be given by a
linear combination of gaussians\footnote{Note that $F^{B}$ should
vanish at large distances because of the confinement potential. The
confinement potential is also responsible for the non--vanishing
values of the parameters $\alpha_j,\ j=1,2$ in
Eq.~(\protect\ref{eq:onebody}).}
\begin{eqnarray}
F^{B}(r_{12}) &=& 
\sum_{j=1}^4 a_j e^{-b_j^2(r_{12}+d_j)^2},\quad a_1=1 
\label{eq:f12}
\end{eqnarray}
The value of one of the $a_j$ parameters can be absorbed into the
normalization constant $N$, so that we fix $a_1=1$.  The values that
we get for all parameters are given in appendix~\ref{DPapp:vp}.

\subsection{Infinite heavy quark mass limit of variational wave functions}
\label{app:ihqml}

In the infinite heavy quark mass limit the wave function of the system should
look like the one for a ``meson'' composed of a light quark and a heavy
diquark. The two heavy quarks bind into a $\bar 3$ color source diquark 
that to the light degrees of
freedom appears to be pointlike~\cite{White:1991hz}. In our model the pointlike
nature of the heavy diquark  comes about
through the one-gluon exchange Coulomb potential which binds the two heavy
quarks into a distance\footnote{The relation can  only be approximate due to 
confinement and the interaction with the light
quark.}
\begin{eqnarray}
r_{h_1h_2}\propto\frac{1}{\mu_{h_1h_2}};\ 
\mu_{h_1h_2}=\frac{m_{h_1}m_{h_2}}{m_{h_1}+m_{h_2}}
\end{eqnarray} 
that tends to zero if both quark masses go to infinity.

\begin{figure}
%\vspace{-1cm}
\centering
\resizebox{12.cm}{!}{\includegraphics{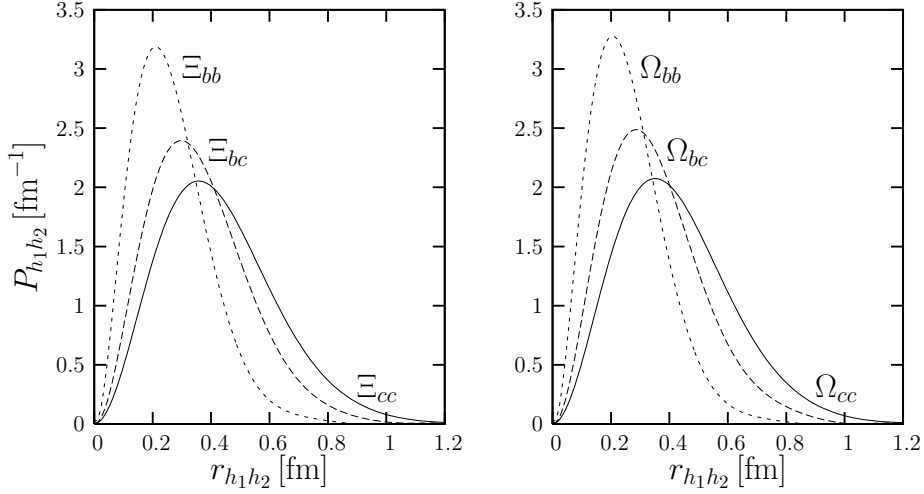}}
%\vspace{-15cm}
\caption{ $P_{h_1h_2}$ probabilities (see Eq.(\ref{eq:ph1h2})) for the
$\Xi_{cc},\,\Xi_{bc},\, \Xi_{bb}$ and 
$\Omega_{cc},\,\Omega_{bc},\, \Omega_{bb}$ baryons evaluated with the AL1
potential.  }
\label{fig:p12}
\end{figure}

For our wave functions we can define the probability distribution
$P_{h1h2}$
for the two heavy quarks
to be found at a distance $r_{h_1h_2}$
\begin{equation}
P_{h_1h_2}\,(r_{h_1h_2}) =\int d^3r_1\int d^3r_2\ \delta(r_{12}-r_{h_1h_2})
\left|\Psi^B_{h_1h_2}(r_2,r_2,r_{12})
\right|^2
\label{eq:ph1h2}
\end{equation}
In Fig.~\ref{fig:p12}, we show the $P_{h_1h_2}$ probability distributions
 for the $\Xi_{cc},\,\Xi_{bc},\, \Xi_{bb}$ and 
$\Omega_{cc},\,\Omega_{bc},\, \Omega_{bb}$ baryons evaluated 
for the AL1 potential.
We see how the maximum moves, as expected, to lower $r_{h_1h_2}$ values as the quark masses
increase.

On the other hand the relative wave function for the light quark with respect to the heavy
quark subsystem should tend to the one of a light quark relative to a pointlike
diquark. This limit is not evident in the coordinates we work as one
has  first to separate the diquark internal orbital wave function from the 
total one. 
To see that this limiting situation is in fact reached within our variational
ansatz let us introduce the new set of coordinates
\begin{eqnarray}
\vec R &=&\frac{m_{h_1}\vec x_{h_1}+m_{h_2}\vec x_{h_2}+m_{q}\vec x_{q}}
{m_{h_1}+m_{h_2}+m_{q}}\nonumber\\
\vec r_{12}&=&\vec x_{h_1}-\vec x_{h_2}\nonumber\\
\vec r_q&=&\frac{m_{h_1}\vec x_{h_1}+m_{h_2}\vec x_{h_2}}{m_{h_1}+m_{h_2}}-\vec
x_q=\frac{m_{h_1}\vec r_{1}+m_{h_2}\vec r_{2}}{m_{h_1}+m_{h_2}}
\end{eqnarray} 
in terms of which the Hamiltonian now reads
\begin{eqnarray}
H&=&
-\frac{\stackrel{\rightarrow}{\nabla}
\stackrel{}{^2}_{\hspace{-.1cm}\vec{R}}}{2 \overline M} +
H^{\rm int} \nonumber\\
 H^{\rm
int}&=&\overline M+H_{h_1h_2}+H_{qh_1h_2}
\end{eqnarray}
where
\begin{eqnarray}
\hspace*{-.6cm}
H_{h1h_2}&=&-\frac{\stackrel{\rightarrow}{\nabla}_{12}^2}{2\mu_{h_1h_2}}
+V_{h_1h_2}(\vec r_{12},\, spin)\nonumber\\
\hspace*{-.6cm} H_{qh_1h_2}&=&-\frac12\left(\frac{1}{m_{h_1}+m_{h_2}}+\frac{1}{m_q}\right)
 \stackrel{\rightarrow}{\nabla}_{q}^2
+V_{h_1q}(\vec r_q+\frac{m_{h_2}}{m_{h_1}+m_{h_2}}\vec r_{12},\,
spin)\nonumber\\
&&\hspace{4.55cm}+V_{h_2q}(\vec r_q-\frac{m_{h_1}}{m_{h_1}+m_{h_2}}\vec r_{12},\, spin)
\end{eqnarray}
with $\stackrel{\rightarrow}{\nabla}_{12}=\partial /\partial_{\vec r_{12}}$ and
$\stackrel{\rightarrow}{\nabla}_{q}=\partial /\partial_{\vec r_q}$.
Defining now
\begin{eqnarray}
\hspace*{-.9cm} H^0_{qh_1h_2}&=&-\frac12\left(\frac{1}{m_{h_1}+m_{h_2}}+\frac{1}{m_q}\right)
 \stackrel{\rightarrow}{\nabla}_{q}^2
+V_{h_1q}(\vec r_q,\,
spin)+V_{h_2q}(\vec r_q,\, spin)
\end{eqnarray}
one would have
\begin{eqnarray}
H^{\mathrm{int}}=H_{h_1h_2}+H^0_{qh_1h_2}+(H_{qh_1h_2}-H^0_{qh_1h_2})
\end{eqnarray}
$H_{h_1h_2}$ is the  Hamiltonian for the relative
 movement of the two heavy
quarks while $H^0_{qh_1h_2}$ is the Hamiltonian  for the
relative movement of the light quark with respect to a pointlike  heavy diquark
  where the two heavy quarks are  located in their center of mass. 
Both movements are coupled through the term
$(H_{qh_1h_2}-H^0_{qh_1h_2})$.
If the  heavy quark masses get arbitrarily  large  the average $r_{12}$ value
tends to zero and thus one can neglect the effect of
$H_{qh_1h_2}-H^0_{qh_1h_2}$. In that limiting situation the light and heavy quark degrees
of freedom decouple completely and the internal Hamiltonian reduces, as it should, to the sum
of the Hamiltonians $H_{h_1h_2}$, corresponding  to the relative movement of the 
two heavy quarks,  and $H^0_{qh_1h_2}$, corresponding to the  relative movement of the light quark
with respect to the pointlike heavy diquark subsystem. 

As to the wave function it should reduce in that limit to the product 
$\Phi_{h_1h_2}(r_{12})\cdot \Phi^0_{qh_1h_2}(r_q)$, being $\Phi_{h_1h_2}(r_{12}),\,\Phi^0_{qh_1h_2}(r_q)$ the ground state wave
functions for $H_{h_1h_2},$  $H^0_{qh_1h_2}$ respectively. To check that we reach
that limiting situation we have evaluated
the projection ${\cal P}$ of our variational wave functions  onto
  $\Phi_{h_1h_2}(r_{12})\cdot \Phi^0_{qh_1h_2}(r_q)$ evaluated with the actual
  heavy quark masses.
  This projection is
given by\footnote{Note $r_q$ can be expressed in terms of 
$r_1,r_2,r_{12}$ and that $d^3r_1d^3r_2=d^3r_{12}d^3r_q$}
\begin{equation}
{\cal P}=\int d^3r_1 \int d^3r_2 \left(\Psi^B_{h_1h_2}(r_1,r_2,r_{12}\right)^*
\Phi_{h_1h_2}(r_{12})\ \Phi^0_{qh_1h_2}(r_q)
\label{eq:project}
\end{equation}
and the values for $|{\cal P}|^2$ that we obtain for the $\Xi_{cc},\,\Xi_{bc},\, \Xi_{bb}$ and 
$\Omega_{cc},\,\Omega_{bc},\, \Omega_{bb}$ baryons using the AL1 potential are given in
Table~\ref{tab:project}. We see how $|{\cal P}|^2$ increases with increasing heavy
quark masses indicating that for very high heavy quark masses the total wave
function tends to the one of a light quark relative to a pointlike
diquark times the diquark internal wave function. 
\begin{table}[t]
{\begin{center}
\begin{tabular}{lccc|ccc}
&$\Xi_{cc}$&\hspace{.3cm}$\Xi_{bc}$&\hspace{.3cm}$\Xi_{bb}$
&\hspace{.3cm}$\Omega_{cc}$&\hspace{.3cm}$\Omega_{bc}$&\hspace{.3cm}$\Omega_{bb}$\\
\hline
$|{\cal P}|^2$ &0.974&0.975&0.991&0.959&0.966&0.984
\end{tabular}
\end{center}}
\caption{Absolute value square of the ${\cal P}$ projection coefficient  defined in
Eq.~(\ref{eq:project})}
\label{tab:project}
\end{table}%
 In summary, our variational wave functions respect the infinite heavy quark
 mass limit.

We note by passing that the product $\Phi_{h_1h_2}(r_{12})\cdot \Phi^0_{qh_1h_2}(r_q)$ corrected by
a correlation function  in the variable $\vec r_{12}\cdot \vec r_q$ would have
 been a good   variational orbital wave function. 
For instance,  a calculation using the AL1 potential  and  an orbital wave 
function given just by
 that product $\Phi_{h_1h_2}(r_{12})\cdot \Phi^0_{qh_1h_2}(r_q)$  gives for  the
 expectation value of $H^{\mathrm{int}}$ for the $\Xi_{cc},\,\Xi_{bc},\,\Xi_{bb}$
 baryons
 \begin{eqnarray}\hspace*{-.5cm}
 \left.\langle H^{\mathrm{int}}\rangle\right|^{AL1}_{\Xi_{cc}}=3640\,\mathrm{MeV}; \
 \left.\langle H^{\mathrm{int}}\rangle\right|^{AL1}_{\Xi_{bc}}=6943\,\mathrm{MeV};\
 \left.\langle H^{\mathrm{int}}\rangle\right|^{AL1}_{\Xi_{bb}}=10198\,\mathrm{MeV}
 \end{eqnarray}
 which are respectively 28\,MeV, 24\,MeV, and 1\,MeV larger, and
 therefore worse,
 than our best values in
 Table~\ref{tab:xiomegamasses}. The correlation function would clearly be 
 needed,
 being its role more important for the ``$cc$'' and ``$bc$'' systems.
 The improvement in going from a ``$cc$''  to a ``$bc$ system'' is not as good as
 one would naively expect, the reason being that,
 considering $\vec r_{12}$ to be a small quantity,
$H_{qh_1h_2}-H^0_{qh_1h_2}$ is first order in $\vec r_{12}$ for the ``$bc$''
system while, for symmetry reasons,  it is necessarily of second order
 for the ``$cc$'' and ``$bb$'' ones. In any case one sees how the expectation 
 values obtained with this
 simple ansatz get closer to our best values in Table~\ref{tab:xiomegamasses}
 as the heavy quark masses increase.

\section{Static Properties}\label{sec:sp}

\subsection{Masses}
\label{subsec:masses}
The mass of the baryon is simply given by the expectation value of the
intrinsic Hamiltonian. Our results (VAR) appear in
Table~\ref{tab:xiomegamasses} where we compare them with the ones
obtained in Ref.~\cite{Silvestre-Brac:1996bg} with the use of the same inter-quark
interactions but within a Faddeev approach (FAD). For that purpose we have 
eliminated from the latter
 a small three-body force contribution of the type 
$V_{123}= {\rm constant}/m_{h_1}m_{h_2}m_q$ that was also
included  in the evaluation of Ref.~\cite{Silvestre-Brac:1996bg}. We will show their full 
results in the following tables.  Whenever comparison is
possible we find an excellent agreement between the two
calculations. In some cases the variational masses are even lower than
the Faddeev ones. Besides we give predictions for states that were not
considered in the study of Ref.~\cite{Silvestre-Brac:1996bg}.
\begin{table}
\centering
\begin{tabular}{lc|ccccc}
& & AL1 & AL2 &AP1 &AP2 &BHAD  \\ \hline\tstrut

$\Xi_{cc}$\hspace{.5cm}      &VAR& 3612  & 3619  & 3629 &  3630 &  3639   \\
                &FAD~\cite{Silvestre-Brac:1996bg}& 3609  & 3616  &
                3625  &  3628 &  3633  \\ 
&&&&\\
		
$\Xi_{cc}^*$    &VAR& 3706  & 3715  & 3722  &  3729 &  3722  \\ &&&&\\
$ \Xi_{bb} $    &VAR& 10197 & 10180 & 10207 & 10179 & 10202  \\
                &FAD~\cite{Silvestre-Brac:1996bg}& 10194 & 10175 & 10204 & 10176 & 10197  \\ &&&&\\
$\Xi_{bb}^*$    &VAR& 10236 & 10219 & 10245 & 10219 & 10235 \\&&&& \\
$\Xi_{bc}$     &VAR& 6919  & 6912  & 6933  & 6917  & 6936   
\\ 
                &FAD~\cite{Silvestre-Brac:1996bg}& 6916  & 6913  & 6928  & 6907  & 6934   \\ &&&&\\
$\Xi_{bc}'$      &VAR& 6948  & 6942  & 6957  & 6944  & 6965   \\&&&&\\
$\Xi_{bc}^*$   &VAR& 6986  & 6981  & 7000  & 6987  & 6993   \\ \hline \hline

\end{tabular}\hspace{1cm}
\begin{tabular}{lc|ccccc}
 & & AL1 & AL2 &AP1 &AP2 &BHAD \\ \hline\tstrut

$\Omega_{cc}$ \hspace{.5cm}        &VAR&  3702 & 3718  & 3711 & 3710 & 3743  
\\
                     &FAD~\cite{Silvestre-Brac:1996bg}&  3711 & 3718  & 3710 & 3709 & 3741\\&&&& \\

$\Omega_{cc}^*$       &VAR&  3783 & 3802  & 3800 & 3802 & 3805 \\ &&&&\\
$\Omega_{bb}$         &VAR& 10260 & 10249 & 10259 &10226 &10274
\\
                      &FAD~\cite{Silvestre-Brac:1996bg}& 10267 & 10246 & 10258 &10224 &10271\\
		      &&&&\\

$\Omega_{bb}^*$       &VAR& 10297 & 10287 & 10301 &10269 &10302\\ &&&&\\
$\Omega_{bc}$        &VAR& 6986  & 6986  & 6990 & 6969 & 7013 
\\
                      &FAD~\cite{Silvestre-Brac:1996bg}&  7003 & 6996  & 6996 & 6971 & 7023 \\
		      &&&&\\

$\Omega_{bc}'$         &VAR& 7009  & 7010  & 7011 & 6994 & 7033 \\ &&&&\\
$\Omega_{bc}^*$      &VAR& 7046  & 7047  & 7055 & 7037 & 7057 \\ \hline %
\end{tabular}
\caption{Doubly heavy $\Xi$ and $\Omega$ baryons masses in MeV. VAR
stands for the results of our variational calculation. FAD stands for
the results obtained in Ref.~\cite{Silvestre-Brac:1996bg} using the same interquark interactions
but within a Faddeev approach.}
\label{tab:xiomegamasses}
\end{table}

In Tables~\ref{tab:ximassesdiffmod} and \ref{tab:omegamassesdiffmod}
we compare our results with other theoretical calculations\footnote{Note in
Refs.~\cite{Roncaglia:1995az,Roncaglia:1994ex} the $\Xi_{bc},\,\Xi'_{bc}$ and 
$\Omega_{bc},\,\Omega'_{bc}$ baryons are defined such that the  total spin of the light 
$q$ quark and the heavy $c$ quark are well defined, being 0 for  
$\Xi_{bc},\,\Omega_{bc}$ and 1 for $\Xi'_{bc},\,\Omega'_{bc}$. They are thus  linear
combinations of our states. The different spin functions are related by
\begin{eqnarray}
\left(|qc;0\rangle\otimes|b;\frac12\rangle\right)^{1/2}=
\frac12\left(|bc;0\rangle\otimes|q;\frac12\rangle   \right)^{1/2}
-\frac{\sqrt3}{2}\left(|bc;1\rangle\otimes|q;\frac12\rangle  
\right)^{1/2}\nonumber\\
\left(|qc;1\rangle\otimes|b;\frac12\rangle\right)^{1/2}=
-\frac{\sqrt3}{2}\left(|bc;0\rangle\otimes|q;\frac12\rangle   \right)^{1/2}
-\frac{1}{2}\left(|bc;1\rangle\otimes|q;\frac12\rangle  
\right)^{1/2}
\end{eqnarray}
In order to extract their predictions  for the $\Xi_{bc},\,\Xi'_{bc}$ and 
$\Omega_{bc},\,\Omega'_{bc}$ baryons with total spin of the two heavy
quarks well defined, 
we have assumed that the above relations,  but with  coefficients square, are also valid for
the masses. Note this may be incorrect as we are neglecting a possible interference
contribution.}.
Our
central values correspond to the results obtained with the AL1
potential, while the errors quoted take into account the variation
when using  different potentials. The same presentation is used for
the results obtained in Ref.~\cite{Silvestre-Brac:1996bg} for which we now show their full
values including the contribution of the three-body force.  All calculations give
similar results that vary within a few per cent. From the experimental
point of view the SELEX Collaboration~\cite{Mattson:2002vu} has recently
measured the value of $m_{\Xi_{cc}}$. This experimental value is 100
MeV smaller that our result. On account of what has been said in the
introduction to this chapter , one should take this experimental value with due
caution. Note also that in
Ref.~\cite{Mattson:2002vu}  the systematic
error is not given. There are also different lattice determinations for baryons with two 
equal heavy quarks~\cite{AliKhan:1999yb,Lewis:2001iz,Flynn:2003vz}. Our results agree within errors with 
the lattice data for baryons with two $c$ quarks, while they are roughly 100\,MeV below lattice
results for doubly $b$-quark baryons. The best overall agreement with  lattice
data available so far is achieved in the calculation of Ref.~\cite{Roncaglia:1995az,Roncaglia:1994ex}
where they use the Feynman-Hellmann theorem and semiempirical mass formulas
in the framework of a nonrelativistic quark model but without the use of an
explicit Hamiltonian. Within full dynamical calculations, ours and the
relativistic calculations in Refs.~\cite{Ebert:2002ig,Tong:1999qs,Ebert:1996ec} have the best
overall agreement with lattice data.

\begin{table}[h!!]
{\footnotesize
\setlength{\tabcolsep}{2pt}
\hspace*{-1.cm}
\begin{tabular}{lccccccccccccc}
& This work&
\cite{Silvestre-Brac:1996bg}&
\cite{Ebert:2002ig}&\cite{Kiselev:2001fw}&\cite{Narodetskii:2001bq}&\cite{Tong:1999qs}&\cite{Itoh:2000um}
&\cite{Vijande:2004at}&\cite{Gershtein:2000nx}&\cite{Ebert:1996ec}&\cite{Roncaglia:1995az,Roncaglia:1994ex}&\cite{Korner:1994nh}
&\cite{Mathur:2002ce}\\\hline
$\Xi_{cc}$\hspace{.5cm}  &$3612^{+17}$
&$3607^{+24}$&3620&3480&3690&3740&3646&3524&3478&3660&$3660\pm70$&3610&$3588\pm72$\\   
$\Xi^*_{cc}$  &$3706^{+23}$ & &3727&3610&&3860&3733&3548&3610&3810&$3740\pm80$&3680&\\   
$\Xi_{bb}$  &$10197^{+10}_{-17}$&$10194^{+10}_{-19}$
&10202&10090&10160&10300&&&10093&10230&$10340\pm100$\\   
$\Xi^*_{bb}$  &$10236^{+9}_{-17}$& &10237&10130&&10340&&&10133&10280&$10370\pm100$\\   
$\Xi_{bc}$  &$6919^{+17}_{-7}$ &$6915^{+17}_{-9}$ &6933&6820&6960&7010&&&6820
&6950&$6965\pm90$$^\dagger$
&&$6840\pm236$\\   
$\Xi'_{bc}$  &$6948^{+17}_{-6}$ &&6963&6850&&7070&&&6850&7000&$7065\pm90$$^\dagger$
\\   
$\Xi^*_{bc}$  &$6986^{+14}_{-5}$ &&6980&6900&&7100&&&6900&7020&$7060\pm90$ \\   
\end{tabular}\vspace{1cm}
\begin{center}
\begin{tabular}{lcccccc}
& This work&Exp.~\cite{Mattson:2002vu}
&Latt.~\cite{AliKhan:1999yb}&Latt.~\cite{Lewis:2001iz}&Latt.~\cite{Flynn:2003vz}\\\hline
$\Xi_{cc}$\hspace{.5cm}  &$3612^{+17}$
&$3519\pm1$&&$3605\pm23$&$3549\pm 95$ \\   
$\Xi^*_{cc}$  &$3706^{+23}$ &  &&$3685\pm 23$&$3641\pm97$\\   
$\Xi_{bb}$  &$10197^{+10}_{-17}$&&$10314\pm47$\\   
$\Xi^*_{bb}$  &$10236^{+9}_{-17}$& &$10333\pm 55$\\   
\end{tabular}
\end{center}
}
\caption{First panel: doubly heavy $\Xi$ masses in MeV as obtained in different
models. Our central values, and the ones of Ref.~\cite{Silvestre-Brac:1996bg}, have
been evaluated with the AL1 potential. Second panel: we compare our results with 
the experimental
value for $M_{\Xi_{cc}}$ measured by the SELEX
Collaboration~\cite{Mattson:2002vu} (Note the cautions that appear on this 
experimental mass in the introduction to this chapter), and  lattice results from 
Refs.~\cite{AliKhan:1999yb,Lewis:2001iz,Flynn:2003vz}. Entries with $^\dagger$ should be taken
with due caution (see text). }

\label{tab:ximassesdiffmod}
\end{table}

\begin{table}[h!!!]
{\setlength{\tabcolsep}{2pt}
\footnotesize
\hspace*{-.75cm}
\begin{tabular}{lcccccccccccc}
& This work &
\cite{Silvestre-Brac:1996bg}&
\cite{Ebert:2002ig}&\cite{Kiselev:2001fw}&\cite{Narodetskii:2001bq}&\cite{Tong:1999qs}&\cite{Itoh:2000um}
&\cite{Gershtein:2000nx}&\cite{Ebert:1996ec}&\cite{Roncaglia:1995az,Roncaglia:1994ex}&\cite{Korner:1994nh}
&\cite{Mathur:2002ce}
\\\hline
$\Omega_{cc}$\hspace{.5cm}  &$3702^{+41}$  &$3710^{+29}_{-2}$& 3778&3590&3860&3760&3749&
3590&3760&$3740\pm70$&3710&$3698\pm65$\\   
$\Omega^*_{cc}$ &$3783^{+22}$ &&  3872&3690&&3900&3826&3690&3890&$3820\pm80$&3760\\   
$\Omega_{bb}$  &$10260^{+14}_{-34}$ &$10267^{+4}_{-43}$&
10359&10180&10340&10340&&10180&10320&$10370\pm100$&\\   
$\Omega^*_{bb}$ &$10297^{+5}_{-28}$ &&
10389&10200&&10380&&10200&10360&$10400\pm100$&\\   
$\Omega_{bc}$   &$6986^{+27}_{-17}$
&$7003^{+20}_{-32}$&7088&6910&7130&7050&&6910&7050&$7045\pm90$$^\dagger$
&&$6954\pm225$\\   
$\Omega'_{bc}$  &$7009^{+24}_{-15}$& &7116&6930&&7110&&6930&7090&$7105\pm90$$^\dagger$
&\\   
$\Omega^*_{bc}$ &$7046^{+11}_{-9}$ &&7130&6990&&7130&&6990&7110&$7120\pm90$& \\   
\end{tabular}\vspace{1cm}
\begin{center}
\begin{tabular}{lcccc}
& This work&Latt.~\cite{AliKhan:1999yb}&Latt.~\cite{Lewis:2001iz}&Latt.~\cite{Flynn:2003vz}\\\hline
$\Omega_{cc}$\hspace{.5cm}  &$3702^{+41}$  &&$3733\pm9^{+7}_{-38}$&$3663\pm 97$\\   
$\Omega^*_{cc}$ &$3783^{+22}$ &&$3801\pm9^{+3}_{-34}$&$3734\pm 98$ \\   
$\Omega_{bb}$  &$10260^{+14}_{-34}$ &$10365\pm40^{-11}_{+12}$$^{+16}_{-0}$\\   
$\Omega^*_{bb}$ &$10297^{+5}_{-28}$ &$10383\pm39^{-8}_{+8}$$^{+12}_{-0}$\\   
\end{tabular}
\end{center}
}
\caption{Same as Table~\ref{tab:ximassesdiffmod} for doubly heavy $\Omega$
baryons.}
\label{tab:omegamassesdiffmod}
\end{table}

There are also independent determinations of mass splittings in lattice
QCD~\cite{AliKhan:1999yb,Lewis:2001iz,Flynn:2003vz}, nonrelativisitic lattice QCD~\cite{Mathur:2002ce}
and pNRQCD~\cite{Brambilla:2005yk}. In Table~\ref{tab:ms} we compare those results to
the ones obtained in the present calculation and in other models.

\begin{table}[b!]
{\footnotesize\setlength{\tabcolsep}{2pt}
\hspace*{-1cm}
\begin{tabular}{lcccccccccccc}
& This work&
\cite{Ebert:2002ig}&\cite{Kiselev:2001fw}&\cite{Tong:1999qs}&
\cite{Gershtein:2000nx}&\cite{Ebert:1996ec}
&\cite{Brambilla:2005yk} 
&\cite{Mathur:2002ce}
&Latt.~\cite{AliKhan:1999yb}&Latt.~\cite{Lewis:2001iz}&Latt.~\cite{Flynn:2003vz}
\\\hline
$M_{\Xi^*_{cc}}-M_{\Xi_{cc}}$\hspace{.5cm}&$94^{+5}_{-11}$&107&130&120&132&150&$120\pm40$&$70\pm13$&&$80\pm11$&$87\pm 19$  \\ 
$M_{\Xi^*_{bb}}-M_{\Xi_{bb}}$\hspace{.5cm}&$39^{+1}_{-6}$&35&40&40&40&50&$34\pm4$&$20\pm7$&$20\pm6$\\ 
$M_{\Xi^*_{bc}}-M_{\Xi_{B's}}$\hspace{.5cm}&$67^{+3}_{-10}$&47&80&90&80&70&&$43\pm11$\\ 
$M_{\Xi'_{bc}}-M_{\Xi_{bc}}$\hspace{.5cm}&$29^{+1}_{-5}$&30&30&60&30&50&&$9\pm7$\\ 
$M_{\Omega^*_{cc}}-M_{\Omega_{cc}}$\hspace{.5cm}&$81^{+11}_{-19}$&94&100&140&100&130&&$63\pm9$&&$68\pm7$&$67\pm 16$  \\   
$M_{\Omega^*_{bb}}-M_{\Omega_{bb}}$\hspace{.5cm}&$37^{+6}_{-9}$&30&20&40&20&40&&$19\pm5$&$20\pm5$  \\   
$M_{\Omega^*_{bc}}-M_{\Omega_{bc}}$\hspace{.5cm}&$60^{+8}_{-16}$&42&80&80&80&60&&$39\pm8$  \\   
$M_{\Omega'_{bc}}-M_{\Omega_{bc}}$\hspace{.5cm}&$23^{+2}_{-3}$&28&20&60&20&40&&$9\pm6$  \\   
\end{tabular}
}
\caption{Mass splittings in MeV for doubly heavy $\Xi$ and $\Omega$
baryons. Our central values have been obtained with the AL1 potential.}
\label{tab:ms}
\end{table}

Our central results evaluated with the AL1 potential are larger than the ones obtained in lattice QCD~\cite{AliKhan:1999yb,Lewis:2001iz,Flynn:2003vz} 
and lattice nonrelativistic QCD~\cite{Mathur:2002ce}. The agreement is better when
we use  the BHAD potential of
Ref.~\cite{Bhaduri:1981pn} for
which we  get the lowest results. Similar results are obtained by 
the relativistic calculation of
Ref.~\cite{Ebert:2002ig}, whereas for the  relativistic
calculations in Refs.~\cite{Kiselev:2001fw,Tong:1999qs,Ebert:1996ec} and the nonrelativistic
one in Ref.~\cite{Gershtein:2000nx} the
agreement  with lattice QCD and  nonrelativistic lattice QCD data   worsens. As
for the calculations in Ref.~\cite{Roncaglia:1995az,Roncaglia:1994ex} 
 we do not quote their results due
to the large theoretical errors involved.

\subsection{Charge  densities and radii}
\label{subsec:chmdr}
The baryon charge density at the point $P$ (coordinate vector $\vec{r}$ in 
the CM frame, see Fig.~\ref{fig:coor}) is given by: 
\begin{eqnarray}
\rho_e^{B} (\vec{r}\,) 
& = & \int d^3 r_1 d^3 r_2\ \Big |
\Psi_{h_1h_2}^{B}(r_1,r_2,r_{12}) \Big |^2 
\nonumber\\ && \hspace{1cm}
\left \{ e_{h_1} \delta^3
(\vec{r}-\vec{y}_{h_1}) + e_{h_2} \delta^3 (\vec{r}-\vec{y}_{h_2}) + e_{q}
\delta^3 (\vec{r}-\vec{y}_q) \right \} \nonumber \\ 
& \equiv & \rho_e^{B} (\vec{r}\,)\big|_{h_1} 
+ \rho_e^{B}(\vec{r}\,)\big|_{h_2}  +
\rho_e^{B}(\vec{r}\,)\big|_{q} \label{eq:dens}
\end{eqnarray} 
where $e_{h_1,\,h_2,\,q}$ are the quark charges in proton charge units
$e$, and from Fig.~\ref{fig:coor} we have\footnote{There exists the
obvious restriction
$m_{h_1}\vec{y}_{h_1}+m_{h_2}\vec{y}_{h_2}+m_{q}\vec{y}_q=\vec 0$.}
$\vec{y}_{h_1}=\vec{y}_q+\vec{r}_1$,
$\vec{y}_{h_2}=\vec{y}_q+\vec{r}_2$ and $\vec{y}_q= - \left
(m_{h_1}\vec{r}_1+m_{h_2}\vec{r}_2 \right)/\overline M$.  Since our
$L=0$ wave functions only depend on scalars ($r_1,r_2$ and $r_{12}$)
the charge density is spherically symmetric ($\rho_e^{B} (\vec{r}\,) =
\rho_e^{B} (|\vec{r}~|)$).

The charge form factor is defined as usual
\begin{eqnarray}
{\cal F}_e^{B} (\vec{q}~) = \int d^3 r\ e^{{\rm
i}\vec{q}\cdot\vec{r}}\rho_e^{B} (r)\label{eq:fe}
\end{eqnarray}
and it only depends on $|\vec{q}~|$. Its value at $\vec{q}=\vec 0$
 gives the baryon charge in units of the proton charge.

\begin{figure}[!t]
%\vspace{-3cm}
\centering
\resizebox{13.cm}{!}{\includegraphics{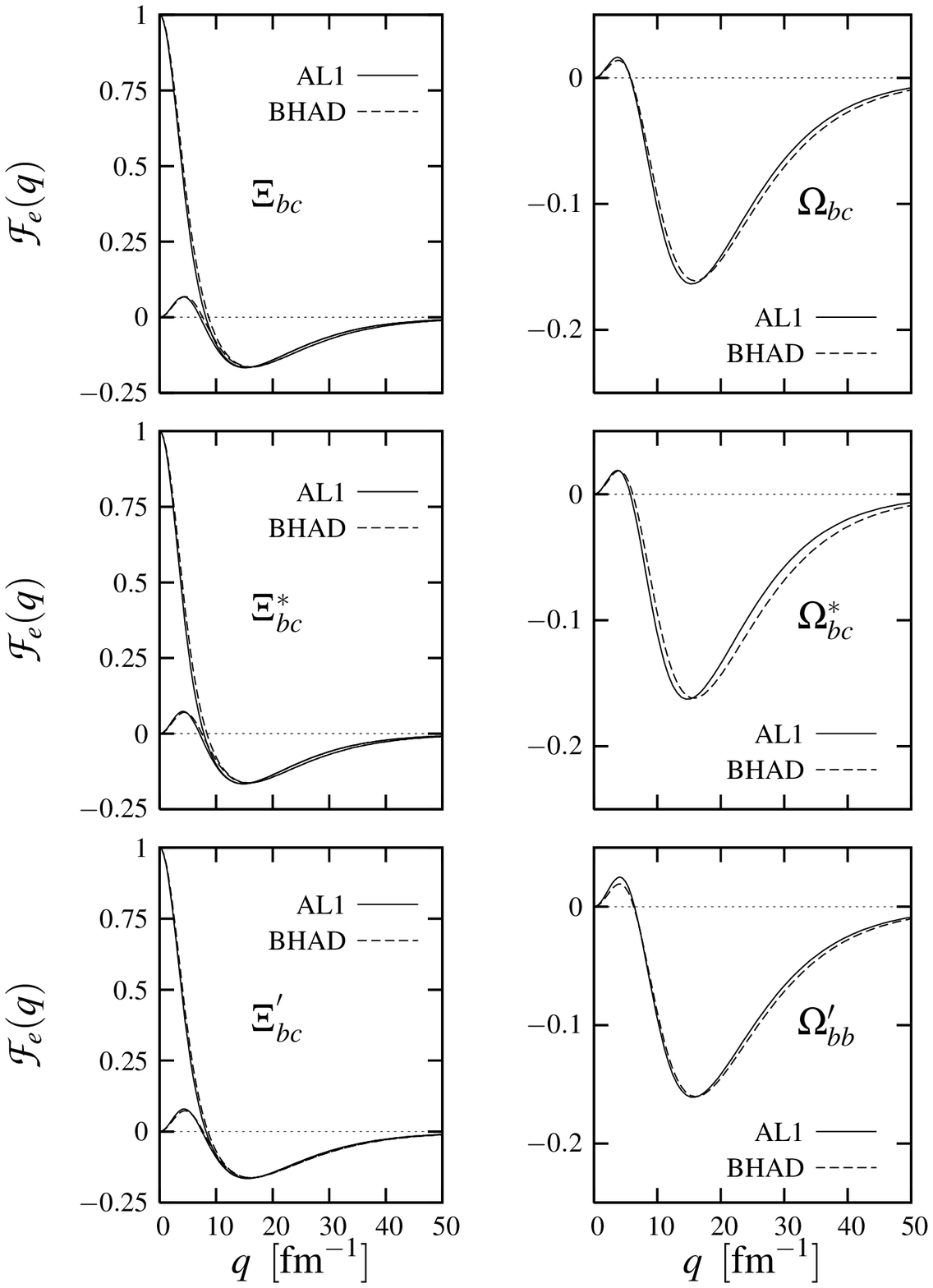}}
%\vspace{-15cm}
\caption{Charge form factor of the $\Xi_{bc},\,\Xi^*_{bc},\,\Xi'_{bc}$
and $\Omega_{bc},\,\Omega^*_{bc}\,\Omega'_{bc}$ baryons evaluated with the AL1
(solid line) and BHAD (dashed line) potentials. For the $\Xi$ baryons we show the
two possible charge states.
 }\label{fig:cffbc}
\end{figure}

The charge mean square radii are defined
\begin{equation}
\langle r^2 \rangle_e^{B} = 
\int d^3 r\ r^2 \rho_e^{B} (r) = 
4\pi \int_0^{+\infty}dr\ r^4 \rho_e^{B} (r) 
\label{eq:r2q}
\end{equation}
\begin{figure}[!t]
%\vspace{-3cm}
\centering
\resizebox{13.cm}{!}{\includegraphics{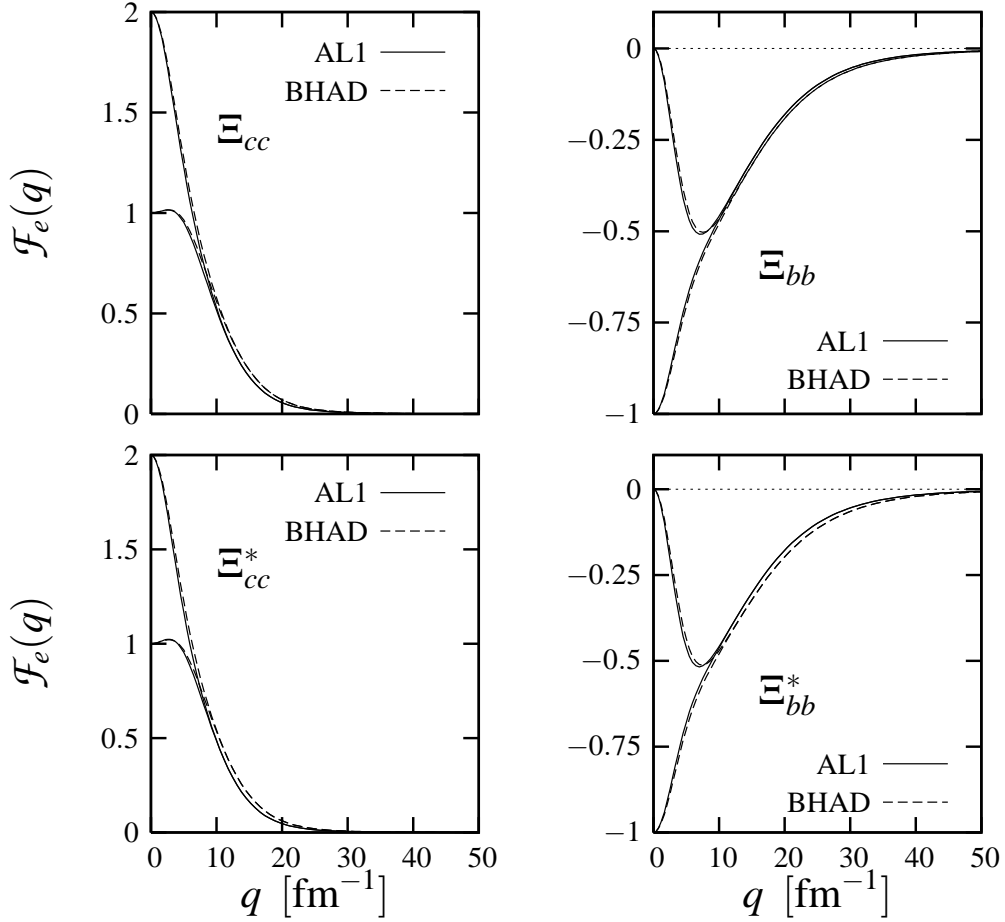}}\vspace{.5cm}\\
%\vspace{-15cm}
\caption{ Charge form factor of the $\Xi_{cc},\,\Xi_{bb}$
and $\Xi^*_{cc},\,\Xi^*_{bb}$ baryons evaluated with the AL1
(solid line) and BHAD (dashed line) potentials. We show the two
possible charge states.}
\label{fig:cffxi}
\end{figure}
\begin{figure}[!t]
\centering
\resizebox{13.cm}{!}{\includegraphics{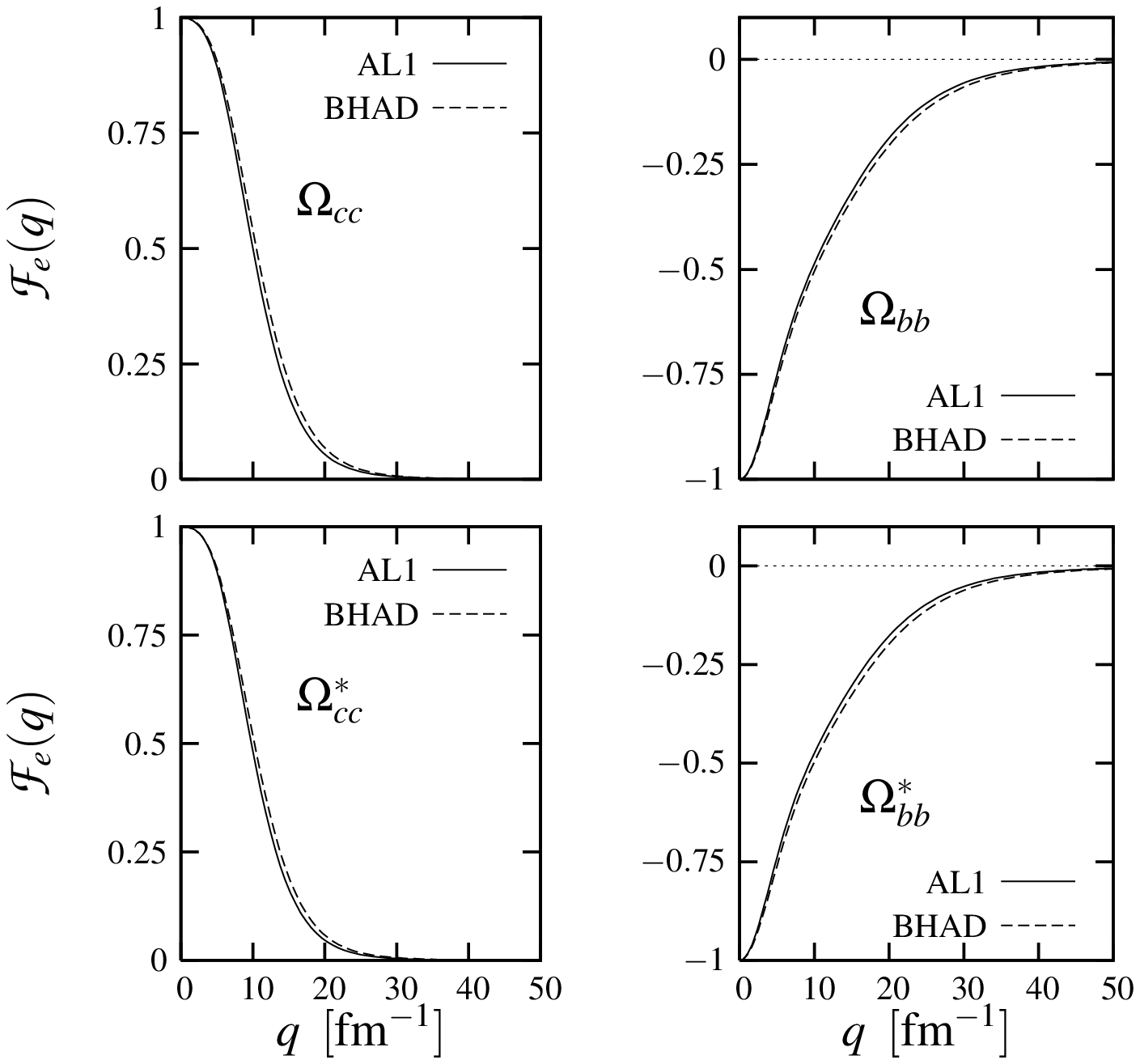}}
%\vspace{-15cm}
\caption{ Charge form factor of the 
 $\Omega_{cc},\,\Omega_{bb}$ and $\Omega^*_{cc},\,\Omega^*_{bb}$ baryons 
 evaluated with the AL1 (solid line)
and BHAD (dashed line) potentials.
}\label{fig:cffomega}
\end{figure}
In Figs.~\ref{fig:cffxi}, \ref{fig:cffomega} and \ref{fig:cffbc} we
show the charge form factors for the different doubly heavy baryons
under study including the two different charge states for the doubly
heavy $\Xi$ ones. We show the calculations with both the AL1 potential
and the BHAD potential.
 The differences between the two calculations are
minor in most cases.

In Table~\ref{tab:xiomegacr} we show the charge mean square
radii. With the exceptions of the $\Xi_{bc}^0$ and $\Omega_{bc}^0$, we
find good agreement with the results obtained in Ref.~\cite{Silvestre-Brac:1996bg}
within a Faddeev calculation. The possible presence of a $S_h=0$
contribution in the wave functions of Ref.~\cite{Silvestre-Brac:1996bg} could be the
possible explanation for this discrepancy. We also compare with the
results obtained, for a few states, in Ref.~\cite{Julia-Diaz:2004vh} with the
use a relativistic quark model in the instant form. The agreement is
bad in this case.
\begin{table}[h!!!]
{\footnotesize
\begin{tabular}{lccc}
&This work & \cite{Silvestre-Brac:1996bg}&\cite{Julia-Diaz:2004vh}\\ \hline

$\Xi_{cc}^+$\hspace{.5cm}   &$-0.030^{+0.003}_{-0.016}$ &$-0.038^{+0.004}_{-0.016}$&$0.1$ \\
$\Xi_{cc}^{++}$ & $0.298^{+0.034}_{-0.028}$  &$0.315^{+0.035}_{-0.030}$&$1.3$ \\
		 \\
$\Xi_{cc}^{*+}$    &$-0.042^{+0.007}_{-0.019}$& & \\
$\Xi_{cc}^{*++}$    &$0.341^{+0.041}_{-0.042}$&&\\ \\
$ \Xi_{bb}^0 $    &$0.221^{+0.033}_{-0.025}$& $0.242^{+0.035}_{-0.027}$&\\
$ \Xi_{bb}^- $    &$-0.133^{+0.014}_{-0.016}$& $-0.143^{+0.006}_{-0.018}$&\\ \\
$\Xi_{bb}^{*-}$    &$-0.142^{+0.018}_{-0.018}$& &\\ \\
$\Xi_{bb}^{*0}$    &$0.238^{+0.035}_{-0.031}$& &\\ \\
$\Xi_{bc}^0$     &$-0.057^{+0.006}_{-0.013}$& $-0.072^{+0.008}_{-0.017}$&\\  
$\Xi_{bc}^+$     &$0.279^{+0.026}_{-0.031}$& $0.306^{+0.035}_{-0.011}$&\\ \\
$\Xi_{bc}'^0$      &$-0.065^{+0.010}_{-0.015}$& & \\
$\Xi_{bc}'^+$      &$0.283^{+0.036}_{-0.025}$& & \\
\\
$\Xi_{bc}^{*0}$   &$-0.065^{+0.010}_{-0.018}$& &\\ 
$\Xi_{bc}^{*+}$   &$0.305^{+0.031}_{-0.039}$&& \\ 
\hline
\end{tabular} \hspace*{1cm}
\begin{tabular}{lccc}
 &This work &\cite{Silvestre-Brac:1996bg}&\cite{Julia-Diaz:2004vh} \\ \hline 

$\Omega_{cc}^+$\hspace{0.5cm} &$0.013^{+0.001}_{-0.002}$&$0.009_{-0.003}$&$0.2$\\ \\

$\Omega_{cc}^{*+}$       &$0.009^{+0.001}_{-0.002}$& & \\ \\
$\Omega_{bb}^-$         &$-0.086^{+0.008}_{-0.001}$&   $-0.090^{+0.007}_{-0.002}$&\\ \\

$\Omega_{bb}^{*-}$       &$-0.092^{+0.011}_{-0.001}$& &\\ \\
$\Omega_{bc}^0$        &$-0.016_{+0.003}$& $-0.025^{+0.002}_{-0.003}$& \\\\

$\Omega_{bc}'^0$         &$-0.019^{+0.003}_{-0.003}$&&\\ \\
$\Omega_{bc}^{*0}$      &$-0.021^{+0.004}_{-0.002}$& &\\ \hline %

\end{tabular}
}
\caption{Charge mean square radii in fm$^2$ for doubly heavy $\Xi$ and
 $\Omega$ baryons.  Our central values, and the ones of
 Ref.~\cite{Silvestre-Brac:1996bg}, have been evaluated with the AL1 potential.}
\label{tab:xiomegacr}
\end{table}

\subsection{Magnetic moments}
\label{subsec:mm}
The orbital part of the magnetic moment is defined in terms of the
velocities $\vec{v}$ of the quarks, with respect to the position of
the CM, and it reads
\begin{eqnarray}
{\mu}^{B} &=& \int  d^3 r_1 d^3 r_2 \left(\Psi_{h_1h_2}^{B}(r_1,r_2,r_{12})\right)^* \left \{
\frac{e_{h_1}}{2m_{h_1}} (\vec{y}_{h_1} \times m_{h_1}\vec{v}_{{h_1}})_z \right.\nonumber \\
&+&\left. \frac{e_{h_2}}{2m_{h_2}} (\vec{y}_{h_2} \times
m_{h_2}\vec{v}_{h_2})_z + 
\frac{e_{q}}{2m_{q}} (\vec{y}_q \times m_{q}
\vec{v}_{y_q})_z \right \}\Psi_{h_1h_2}^{B}(r_1,r_2,r_{12})
\end{eqnarray}
with
 $m_{h_1}\vec{v}_{h_1} = - i \vec{\nabla}_1$,
$m_{h_2}\vec{v}_{h_2} = - i \vec{\nabla}_2$ and $m_q\vec{v}_{q}
= i \left ( \vec{\nabla}_1 + \vec{\nabla}_2 \right)$\footnote{Note that the classical kinetic energy  has a term on 
$\vec{v}_{h_1}\cdot\vec{v}_{h_2}$ and then the operator
$m_{h_1} \vec{v}_{h_1}$ is not proportional to 
$-i\stackrel{\to}{\nabla}_{y_{h_1}}$, but it is rather given by
\bea
m_{h_1} \vec{v}_{h_1}\,=\, \frac{(\overline M-m_{h_1})}{\overline M}\cdot
(-i\stackrel{\to}{\nabla}_{y_{h_1}}) 
-\frac{m_{h_2}}{\overline M}\cdot(-i\stackrel{\to}{\nabla}_{y_{h_2}})
\,=\,(-i\stackrel{\to}{\nabla}_1)
\eea
%\bea
%m_{h_1} \vec{v}_{h_1}={(\overline M-m_{h_1})}/{\overline M}\cdot
%(-i\stackrel{\to}{\nabla}_{y_{h_1}}) 
%-{m_{h_2}}/{\overline M}\cdot(-i\stackrel{\to}{\nabla}_{y_{h_2}})
%=(-i\stackrel{\to}{\nabla}_1)
%\eea
 Similarly
 $m_{h_2} \vec{v}_{h_2}=(-i\stackrel{\to}{\nabla}_2)$.}.

 Since our orbital wave
function has $L=0$, the orbital
 magnetic moment vanishes. The magnetic moment of the baryon is then
 entirely given by the spin contribution.
\begin{eqnarray}
\langle B;\, J,\,M_J=J|\ \frac{e_{h_1}}{2m_{h_1}} (\vec{\sigma}_{h_1})_z
+ \frac{e_{h_2}}{2m_{h_2}} (\vec{\sigma}_{h_2})_z  + 
\frac{e_{q}}{2m_{q}} (\vec{\sigma}_q)_z\ | B;\,J,\, M_J=J\rangle
\end{eqnarray}
 Those matrix elements are trivially evaluated with the
results
\begin{eqnarray}
\Xi_{h1h_2},\ \Omega_{h1h_2}&\longrightarrow&
\frac{2}{3}\left( \frac{e_{h_1}}{2m_{h_1}}+\frac{e_{h_2}}{2m_{h_2}}-\frac12
\ \frac{e_{q}}{2m_{q}} \right)\nonumber\\
\Xi^*_{h1h_2},\ \Omega^*_{h1h_2}&\longrightarrow&
 \frac{e_{h_1}}{2m_{h_1}}+\frac{e_{h_2}}{2m_{h_2}}+
\frac{e_{q}}{2m_{q}} \nonumber\\
\Xi'_{h1h_2},\ \Omega'_{h1h_2}&\longrightarrow&
 \frac{e_{q}}{2m_{q}}
\end{eqnarray}

In Table~\ref{tab:xiomegamm} we give our numerical results. Our
central values, as the ones obtained in Ref.~\cite{Silvestre-Brac:1996bg} within a
Faddeev approach, have been evaluated with the use of the AL1
potential. When compared to the values obtained in Ref.~\cite{Silvestre-Brac:1996bg} we
find very good agreement with just a few exceptions
($\Xi_{bc}^0,\,\Xi_{bc}^+,\,\Omega_{bc}^0$). The discrepancy for the
latter baryons may come from a possible non negligible $S_h=0$
contribution to their wave functions in the calculation of
Ref.~\cite{Silvestre-Brac:1996bg}. In our case we have fixed $S_h=1$ which we think is
a very good approximation since in the limit of infinite heavy quark
masses the spin of the heavy quark degrees of freedom is well defined.

We also compare our results to the ones obtained in
Refs.\cite{Lichtenberg:1976fi,Faessler:2006ft,Julia-Diaz:2004vh,Oh:1991ws,Jena:1986xs,Bose:1980vy} using
different approaches. The differences between different calculations are
 in some cases large. Being
$L=0$ a good approximation the magnetic
moments are essentially determined by the spin contribution of the
quarks. With $m_b\gg m_u,\,m_d,\,m_s$, the contribution from the $b$ quarks is
negligible compared to the one of the light quark. This is also true to a lesser
extent for the $c$ quark.

\begin{table}[h!!!]
\hspace*{-1.6cm}
{\footnotesize
\begin{tabular}{lcccccccc}
& This work& \cite{Silvestre-Brac:1996bg}&\cite{Lichtenberg:1976fi}&\cite{Faessler:2006ft}&\cite{Julia-Diaz:2004vh}&\cite{Oh:1991ws}
&\cite{Jena:1986xs}&\cite{Bose:1980vy} \\ \hline

$\Xi_{cc}^+$\hspace{.5cm} &$0.785^{+0.050}_{-0.030}$ &$0.784^{+0.050}_{-0.029}$&0.806
&0.72&$0.72$&$0.89\sim0.98$&$0.778\sim0.790$&0.86\\
$\Xi_{cc}^{++}$&$-0.208^{+0.035}_{-0.086}$
&$-0.206^{+0.034}_{-0.086}$&$-0.124$&0.13&$-0.10$&$-0.47$&$-0.172\sim-0.154$&0.17\\
		 \\
$\Xi_{cc}^{*+}$ &$-0.311^{+0.052}_{-0.130}$ &$$ &$-0.186$&
&&$-1.17\sim-0.98$&&0.20\\
$\Xi_{cc}^{*++}$&$2.67^{+0.27}_{-0.15}$ &$$ &2.60 &&&$3.16\sim3.18$&&2.54\\ \\
$ \Xi_{bb}^0 $
&$-0.742^{+0.044}_{-0.091}$&$-0.742^{+0.044}_{-0.092}$&&$-0.53$&&&$-0.726\sim-0.705$&0.61\\
$ \Xi_{bb}^- $  &$0.251^{+0.045}_{-0.021}$ &$0.251^{+0.046}_{-0.021}$ &&0.18
&&&$0.226\sim0.236$&0.14\\ \\
$\Xi_{bb}^{*0}$ &$1.87^{+0.27}_{-0.13}$  &$$ && &&&&1.37\\ 
$\Xi_{bb}^{*-}$ &$-1.11^{+0.06}_{-0.14}$ &$$ &&& &&&$-0.95$\\  \\
$\Xi_{bc}^0$
&$0.518^{+0.048}_{-0.020}$&$0.058^{+0.059}_{-0.054}$&& 0.42   &&&&\\
$\Xi_{bc}^+$  &$-0.475^{+0.040}_{-0.088}$  &$-0.198^{+0.057}_{-0.056}$ &&$-0.12$
&&&& \\  \\

$\Xi_{bc}'^0$ &$-0.993^{+0.065}_{-0.137}$  &$$ &&$-0.76$&&& $-0.385\sim-0.366$ & \\
$\Xi_{bc}'^+$ &$1.99^{+0.27}_{-0.13}$  &$$ && 1.52 & &&  $1.50\sim1.54$  &\\
\\
$\Xi_{bc}^{*0}$ &$-0.712^{+0.059}_{-0.133}$  &$$ &&&&&   &$-0.39$\\ 
$\Xi_{bc}^{*+}$ &$2.27^{+0.27}_{-0.14}$ &$$ &&&&&  &2.04\\ 
\hline
\end{tabular}\vspace{1cm}\\%\hspace{0.9cm}
\hspace*{-1.6cm}
\begin{tabular}{lcccccccc}
 & This work&\cite{Silvestre-Brac:1996bg}&\cite{Lichtenberg:1976fi}&\cite{Faessler:2006ft}&\cite{Julia-Diaz:2004vh}&\cite{Oh:1991ws} 
& \cite{Jena:1986xs}&\cite{Bose:1980vy}\\ \hline 

$\Omega_{cc}^+$\hspace{.5cm}  &$0.635^{+0.012}_{-0.015}$ &$0.635^{+0.011}_{-0.015}$ &0.688&0.67
&$0.72$&$0.59\sim0.64$&$0.657\sim0.663$&0.84\\ \\

$\Omega_{cc}^{*+}$  &$0.139^{+0.009}_{-0.017}$   &$$ &0.167&
&&$-0.20\sim0.03$&&0.39\\ \\
$\Omega_{bb}^-$ &$0.101^{+0.007}_{-0.007}$ &$0.101^{+0.007}_{-0.006}$ &&0.04& &&
$0.105\sim0.108$&0.084\\ \\

$\Omega_{bb}^{*-}$  &$-0.662^{+0.022}_{-0.024}$  &$$&  & &&&&$-1.28$\\ \\
$\Omega_{bc}^0$ &$0.368^{+0.010}_{-0.011}$&$0.009^{+0.038}_{-0.029}$ &&0.45&&&&\\
		      \\

$\Omega_{bc}'^0$  &$-0.542^{+0.021}_{-0.024}$   &$$&
&$-0.61$&&& $-0.130\sim-0.125$    &\\ \\
$\Omega_{bc}^{*0}$ &$-0.261^{+0.015}_{-0.021}$  &$$& &&&&&$-0.22$\\ \hline %
%\vspace{2.32cm}

\end{tabular}
}
\caption{Magnetic moments, in nuclear magnetons ($e/2m_p$, with
  $m_p$ the proton mass), of doubly heavy $\Xi$ and $\Omega$
  baryons. Our central values, and the ones of Ref.~\cite{Silvestre-Brac:1996bg}, have
  been evaluated with the AL1 potential.}
\label{tab:xiomegamm}
\end{table}

\chapter{Doubly heavy baryons semileptonic decay}
\chaptermark{Doubly heavy baryons 2}
\label{cha:dp_sl}
\section{Introduction}

In this chapter we should use the manageable wave functions that we
got in chapter  \ref{cha:dp_masas}   to study
semileptonic decays of doubly, $J=1/2$, baryons.
All the transitions studied here involve a $b\to c$ transition at the
quark level.
 We shall evaluate
form factors, decay widths and angular asymmetry parameters. Previous
calculations of semileptonic decay widths have been done in different
relativistic quark model approaches ~\cite{Ebert:2004ck,Faessler:2001mr,Guo:1998yj},
or with the use of HQET~\cite{Sanchis-Lozano:1994vh}.

\section{Decay width and angular asymmetries}
\label{sec:sd}

 The differential decay width for a $B(1/2^+)\to B'(1/2^+)$ transition reads
\bea &&
\hspace*{-1.7cm}
{\rm d}\Gamma= 
8 |V_{cb}|^2 m_{B'} G_F^{\,2}  
 \frac{d^3p^\prime}{(2\pi)^32E^\prime_{B'}(\vec{p}') }
\frac{d^3k}{(2\pi)^32E_{\bar \nu_l}(\vec{k}) } \frac{d^3k^\prime}{(2\pi)^32E^\prime_{l}(\vec{k}')
}  \ 
\nonumber\\&&\hspace{3cm}\times
(2\pi)^4 \delta^4(p-p^\prime-k-k^\prime)
{\cal
L}^{\alpha\beta}(k,k')
{\cal H}_{\alpha\beta}(p,p')    
\eea
where $|V_{cb}|$ is the modulus of the corresponding
CKM  matrix element, $m_{B'}$ is the mass of
the final baryon, 
$p$, $p'$, $k$ and $k'$ are the four-momenta of the initial baryon,
final baryon, final anti-neutrino and final lepton respectively, and
${\cal L}$ and ${\cal H}$ are the lepton and hadron tensors.

% CAMBIADAS EPSILONS 30-01
The lepton tensor
is given as
\begin{eqnarray}
{\cal L}^{\mu\sigma}(k,k')&=& k'^\mu k^\sigma +k'^\sigma k^\mu
- g^{\mu\sigma} k\cdot k^\prime - {\rm i}
\epsilon^{\mu\sigma\alpha\beta}k'_{\alpha}k_\beta \label{eq:lep}
\end{eqnarray}

As for the hadron tensor we have
\begin{eqnarray}
\hspace*{-.6cm}
{\cal H}_{\mu\sigma}(p,p') &=& \frac12 \sum_{r,r'}  
 \left\langle B', r'\
\vec{p}^{\,\prime}\left|\,
\overline \Psi^c(0)\gamma_\mu(I-\gamma_5)\Psi^b(0)\right| B, r\ \vec{p}   \right\rangle 
\nonumber\\ &&\hspace*{2.6cm}
 \left\langle B', r'\ 
\vec{p}^{\,\prime}\left|\,\overline \Psi^c(0)\gamma_\sigma(I-\gamma_5)
\Psi^b(0) \right|  B, r\ \vec{p} \right\rangle^*
\label{eq:wmunu}
\end{eqnarray}
with $\left|B, r\ \vec p\right\rangle\, (\left|B', r'\
\vec{p}\,'\right\rangle)$ representing the initial (final) baryon with
three--mo\-men\-tum $\vec p$ ($\vec{p}\,'$) and spin third component $r$
($r'$). The baryon states are here  normalized such that $\langle r\
\vec{p}\, |\, r' \ \vec{p}^{\,\prime} \rangle = (2\pi)^3 (E(\vec
p\,)/m)\,\delta_{rr'}\, \delta^3(\vec{p}-\vec{p}^{\,\prime})$.  The
hadron matrix elements can be parametrized in terms of six form
factors as
\begin{eqnarray}
\lefteqn{\hspace*{-1.1cm}  
\left\langle B', r'\ \vec{p}^{\,\prime}\left|\,\overline \Psi^c(0)\gamma_\mu(I-\gamma_5)\Psi^b(0)
 \right| B, r\ \vec{p}
\right\rangle \,=}
\nonumber \\\nonumber &&
 =\, {\bar u}^{B'}_{r'}(\vec{p}^{\,\prime})\Big\{
\gamma_\mu\left(F_1(w)-\gamma_5 G_1(w)\right)+ v_\mu\left(F_2(w)-\gamma_5
G_2(w)\right)
\nonumber\\
&&\hspace{5.5cm}
+v^\prime_\mu\left(F_3(w)-\gamma_5 G_3(w)
\right)\Big\}u^{B}_r(\vec{p}\,) \label{eq:def_ff}
\end{eqnarray}
where $u^{B,B'}$ are dimensionless Dirac spinors, normalized as ${\bar
u} u = 1$, and $v_\mu = p_\mu/m_{B}$ ($v^\prime_\mu =
p^\prime_\mu/m_{B'}$) is the four velocity of the initial $B$ (final
$B'$) baryon. The form factors are functions of the velocity transfer
$w=v\cdot v^\prime$ or equivalently of the four momentum transfer
($q=p-p'$) square $q^2= m_{B}^2 + m_{B'}^2 - 2m_{B}m_{B'}w$. In the
decay $w$ ranges from $w=1$, corresponding to zero recoil of the final
baryon, to a maximum value given by $w=w_{\rm max}= (m_{B}^2 +
m_{B'}^2)/(2m_{B}m_{B'})$ which depends on the transition.

Neglecting lepton masses, we have for the differential decay rates
from transversely $(\Gamma_T)$ and longitudinally $(\Gamma_L)$
polarized $W$'s (the total width is
$\Gamma=\Gamma_L+\Gamma_T$)~\cite{Korner:1991ph}
\begin{eqnarray}
\frac{{\rm d}\Gamma_T}{{\rm d}w}&=&
\frac{G^2_F |V_{cb}|^2}{12\pi^3}m_{B'}^3\sqrt{w^2-1}\,q^2 
\Big\{ (w-1)|F_1(w)
|^2+(w+1)|G_1(w)|^2 \Big\} \nonumber\\
&&\nonumber\\
\frac{{\rm d}\Gamma_L}{{\rm d}w}&=&
\frac{G^2_F |V_{cb}|^2}{24\pi^3}m_{B'}^3\sqrt{w^2-1}
\Big\{(w-1)|{\cal F}^V(w)|^2 + (w+1)|{\cal
F}^A(w)|^2  \Big\} \nonumber\\
&&\nonumber\\
 {\cal F}^{V,A}(w) &=& \Big[ (m_{B}\pm m_{B'}) F_1^{V,A}(w) +
(1\pm w)\left(m_{B'} F_2^{V,A}(w)+m_{B}
F_3^{V,A}(w)\right)\Big],\nonumber\\&& \quad F_j^V \equiv F_j(w) , ~ F_j^A \equiv
G_j(w),~ j=1,2,3\nonumber\\
 \label{eq:dg}
\end{eqnarray}

One can also evaluate the polar angle distribution~\cite{Korner:1991ph}:
\begin{equation}
\frac{{\rm d}^2\Gamma}{{\rm d}w\,{\rm d}\cos\theta} = \frac38
\left(\frac{{\rm d}\Gamma_T}{{\rm d}w} +
 2\frac{{\rm d}\Gamma_L}{{\rm d}w} \right)\Big\{1+2\alpha^\prime\cos\theta +
 \alpha^{\prime\prime} \cos^2\theta \Big\}
\label{eq:asymmetry1}
\end{equation}
where $\theta$ is the angle between $\vec{k}^\prime$ and
$\vec{p}^{\,\prime}$ measured in the off--shell $W$ rest frame, and
$\alpha^\prime$ and $\alpha^{\prime\prime}$ are asymmetry parameters
given by
\begin{eqnarray}
\alpha^\prime &=&  - \frac{G^2_F
    |V_{cb}|^2}{6\pi^3} 
{m_{B'}^3}\frac{
    q^2\,(w^2-1)\,F_1(w)G_1(w)}{{\rm d}\Gamma_T/{{\rm d}w}+
    2\,{\rm d}\Gamma_L/{{\rm d}w}}  \nonumber\\%\label{eq:asy1i} \\
&&\nonumber\\
\alpha^{\prime\prime} &=& \frac{{\rm d}\Gamma_T/{{\rm d}w}  
   - 2\,{\rm d}\Gamma_L/{{\rm d}w}}
   {{\rm d}\Gamma_T/{{\rm d}w}+
    2\,{\rm d}\Gamma_L/{{\rm d}w}} %\label{eq:asy1f}
\label{eq:asymmetry2}
\end{eqnarray} 
These asymmetry parameters  are functions of the velocity
transfer $w$ and on averaging over $w$ the numerators and denominators
are integrated separately and  thus we have
\begin{eqnarray}
\langle \alpha' \rangle &=& - \frac{G^2_F |V_{cb}|^2}{6\pi^3}
\frac{m_{B'}^3}{\Gamma_T}\frac{ \int_1^{w_{\rm max}}
q^2\,(w^2-1)\,F_1(w)G_1(w) {\rm d }w}{1+2R_{L/T}} 
\nonumber \\
 \quad \langle \alpha^{\prime\prime} \rangle &=&
\frac{1-2R_{L/T}
 } {1+2R_{L/T}}
\nonumber \\
  \qquad R_{L/T} &=& \frac{\Gamma_L}{\Gamma_T} %\label{eq:rlt} 
\label{eq:averageasymmetry}
\end{eqnarray}

\section{Form factors}
\label{sec:ff}
To obtain the form factors we have to evaluate the matrix elements
\begin{equation}
\left\langle B', r'\ \vec{p}^{\,\prime}\left|\,\overline \Psi^c(0)\gamma_\mu(I-\gamma_5)\Psi^b(0)
 \right| B, r\ \vec{p}
\right\rangle 
\end{equation}
which in our model are given by
\begin{equation}
\sqrt{\frac{E_{B}(\vec p\,)}{m_{B}}}\ \sqrt{\frac{E_{B'}(\vec
p\,')\,}{m_{B'}}}\
{}_{\stackrel{}{\stackrel{}{NR}}}\left\langle B', r'\ \vec{p}^{\,\prime}\left|\,\overline \Psi^c(0)\gamma_\mu(I-\gamma_5)\Psi^b(0)
 \right| B, r\ \vec{p}
\right\rangle_{NR}
\end{equation}
where the suffix ``$NR$'' denotes our nonrelativistic states and the
factors $\sqrt{E/m}$ take into account the different normalization.
We shall work in the initial baryon rest frame so that $\vec p=\vec
0,\,\vec p\,'=-\vec q$, and take $\vec q$ in the positive $z$
direction.  Furthermore we shall use the spectator
approximation. Having all this in mind we have in momentum space
\begin{eqnarray}
&&\hspace*{-1cm}{\sqrt{\frac{E_{B'}(-\vec q\,)}{m_{B'}}}\
{}_{\stackrel{}{\stackrel{}{NR}}}\left\langle B', r'\ -\vec{q}\left|\,\overline \Psi^c(0)\gamma_\mu(I-\gamma_5)\Psi^b(0)
 \right| B, r\ \vec{0}
\right\rangle_{NR}}\,=\nonumber\\ &&
=\,2\sqrt{\frac{E_{B'}(-\vec q\,)}{m_{B'}}}\ \sum_{s_1}\sum_{s_2} \left(\frac12,\frac12, S_h
\bigg|s_1, s_2-s_1, s_2\right) \left(S_h,\frac12,\frac12
\bigg|s_2, r-s_2, r\right)\nonumber\\ && \hspace{.6cm}
\times\left(\frac12,\frac12, S'_h
\bigg|r'-r+s_1, s_2-s_1, r'-r+s_2\right) \left(S'_h,\frac12,\frac12
\bigg|r'-r+s_2, r-s_2, r'\right)\nonumber\\ && \hspace{.6cm}
\times\int d\,^3q_1\, d\,^3q_2\, 
\left(\Phi^{B'}_{c\,h_2}(\vec q_1-\frac{m_{h_2}+m_q}{\overline M\,'}\,\vec q,
\,\vec{q}_2+\frac{m_{h_2}}{\overline M\,'}\,\vec q\,)\right)^*\ 
\Phi^B_{b\,h_2}(\vec{q}_1,\vec{q}_2)
\nonumber\\ && \hspace{2.cm}
\sqrt{\frac{m_b}{E_b(\vec{q}_1)}}\sqrt{\frac{m_c}
{E_c(\vec{q}_1-\vec q\,)}}\ \bar{u}^c_{r'-r+s_1}(\vec{q}_1-\vec q\,)
\gamma_\mu(I-\gamma_5)\,u^b_{s_1}(\vec{q}_1)
\end{eqnarray}
where $\Phi^B_{b\,h_2}(\vec{q}_1,\vec{q}_2)$
($\Phi^{B'}_{c\,h_2}(\vec{q}_1,\vec{q}_2)$) is the Fourier transform
of the coordinate space wave function
$\Psi^B_{b\,h_2}(r_1,r_2,r_{12})$
($\Psi^{B'}_{c\,h_2}(r_1,r_2,r_{12})$) with $\vec q_1,\,\vec q_2$
being the conjugate momenta to the space variables $\vec r_1,\,\vec
r_2$. The factor of two comes from the fact that: i) for $bc-$baryon
decays, the charm quark resulting from the $b\to c$ transition could be
either the particle 1 or the particle 2 in the final $cc$ baryon,
while ii) for $bb-$baryon decays, there exist two equal contributions
resulting for the decay of each of the two bottom quarks of the
initial baryon.

The actual calculation is done in coordinate space where we have
\begin{eqnarray}
&&\hspace*{-1cm}{\sqrt{\frac{E_{B'}(-\vec q\,)}{m_{B'}}}\
{}_{\stackrel{}{\stackrel{}{NR}}}\left\langle B', r'\ -\vec{q}\left|\,\overline \Psi^c(0)\gamma_\mu(I-\gamma_5)\Psi^b(0)
 \right| B, r\ \vec{0}
\right\rangle_{NR}}\nonumber\\
&&
=\,2\sqrt{\frac{E_{B'}(-\vec q\,)}{m_{B'}}}\ \sum_{s_1}\sum_{s_2} \left(\frac12,\frac12, S_h
\bigg|s_1, s_2-s_1, s_2\right) \left(S_h,\frac12,\frac12
\bigg|s_2, r-s_2, r\right)\nonumber\\
&& \hspace{.6cm}
\times\left(\frac12,\frac12, S'_h
\bigg|r'-r+s_1, s_2-s_1, r'-r+s_2\right) \left(S'_h,\frac12,\frac12
\bigg|r'-r+s_2, r-s_2, r'\right)\nonumber\\
&& \hspace{.6cm}
\times\int d\,^3r_1\, d\,^3r_2\, 
\Psi^{B'}_{c\,h_2}(r_1,r_2,r_{12})
\ e^{i\frac{m_{h_2}}{\overline M\,'}\,\vec
q\cdot\vec r_2}\,
e^{-i\frac{m_{h_2}+m_q}{\overline M\,'}\,\vec
q\cdot\vec r_1}
\sqrt{\frac{m_b}{E_b(\vec{l}\ )}}\sqrt{\frac{m_c}
{E_c(\vec{l}-\vec q\,)}}
\nonumber\\ 
&& \hspace{2.9cm}
 \bar{u}^c_{r'-r+s_1}(\vec{l}-\vec q\,)
\gamma_\mu(I-\gamma_5)\,u^b_{s_1}(\vec{l}\ )\ \Psi^B_{b\,h_2}(r_1,r_2,r_{12})
\label{eq:coor}
\end{eqnarray}
where $\vec l=-i\stackrel{\rightarrow}{\nabla}_1$ represents an
internal momentum which is much smaller than the heavy quark masses
$m_b,\,m_c$. On the other hand $|\vec q\,|$ can be large\footnote{At
$q^2=0$ one has $|\vec q\,|=(m_B^2-m_{B'}^2) /2m_B$ which is $\approx
m_B/3$ for the transitions under study.}. Thus, to evaluate the above
expression we have made use of an expansion in $\vec l$, introduced in
Ref.~\cite{Albertus:2004wj}, where second order terms in $\vec l$ are
neglected, while all orders in $|\vec q\,|$ are kept. For instance
$E_c(\vec l-\vec q\,)$ is approximated by $E_c(\vec l-\vec q\,)\approx
E_c(\vec q\,)\times (1-\vec l\cdot\vec q\,/E_c^2(\vec q\,))$ with
$E_c(\vec q\,)=\sqrt{m_c^2+\vec q\,^2}$.

The three vector and three axial form factors can be extracted from the set of
equations\footnote{Remember $\vec q$ is in the $z$ direction. Notice also that,
for $\vec p=\vec 0$, $w$ is just a function of $|\vec q\,|$.}
\begin{eqnarray}
\lefteqn{\hspace*{-1cm}\left\langle B',1/2\ -\vec{q}\left|\,\overline \Psi^c(0)\gamma_1\Psi^b(0)
 \right| B, -1/2\ \vec{0}
\right\rangle}
\nonumber \\
&&=\,\sqrt{\frac{E_{B'}(-\vec q\, )+m_{B'}}{2m_{B'}}}
\,\frac{|\vec q\,|}{E_{B'}(-\vec q\, )+m_{B'}}\ 
F_1(|\vec q\,|)\nonumber\\
\lefteqn{\hspace*{-1cm}\left\langle B', 1/2\ -\vec{q}\left|\,\overline \Psi^c(0)\gamma_3\Psi^b(0)
 \right| B, 1/2\ \vec{0}
\right\rangle}
\nonumber\\
&&=\,\sqrt{\frac{E_{B'}(-\vec q\, )+m_{B'}}{2m_{B'}}}\ 
|\vec q\,|\,\left(
\frac{F_1(|\vec q\,|)}{E_{B'}(-\vec q\, )+m_{B'}}+\frac{F_3(|\vec
 q\,|)}{m_{B'}}\right)
\nonumber\\
\lefteqn{\hspace*{-1cm}\left\langle B', 1/2\ -\vec{q}\left|\,\overline \Psi^c(0)\gamma_0\Psi^b(0)
 \right| B, 1/2 \ \vec{0}
\right\rangle}
\nonumber\\
&&=\,\sqrt{\frac{E_{B'}(-\vec q\, )+m_{B'}}{2m_{B'}}}\left(
F_1(|\vec q\,|)+F_2(|\vec q\,|)+\frac{E_{B'}(-\vec q\, )}{m_{B'}}\,F_3(|\vec q\,|)\right)
\label{eq:f123}
\end{eqnarray}
for the vector form factors and
\begin{eqnarray}
\lefteqn{\hspace*{-1cm}\left\langle B',1/2\  -\vec{q}\left|\,\overline \Psi^c(0)\gamma_1\gamma_5\Psi^b(0)
 \right| B, -1/2\ \vec{0}
\right\rangle}
\nonumber \\ 
&&=\,\sqrt{\frac{E_{B'}(-\vec q\, )+m_{B'}}{2m_{B'}}}
\ (-G_1(|\vec q\,|))\nonumber\\
\lefteqn{\hspace*{-1cm}\left\langle B', 1/2\ -\vec{q}\left|\,\overline \Psi^c(0)\gamma_3\gamma_5\Psi^b(0)
 \right| B, 1/2\ \vec{0}
\right\rangle}
\nonumber\\
&&=\,\sqrt{\frac{E_{B'}(-\vec q\, )+m_{B'}}{2m_{B'}}}\  \
\left(-G_1(|\vec q\,|)+\frac{|\vec q\,|^2\,G_3(|\vec q\,|)}{m_{B'}(E_{B'}(-\vec q\,
)+m_{B'})}\right)
\nonumber\\
\lefteqn{\hspace*{-1cm}\left\langle B',1/2\  -\vec{q}\left|\,\overline \Psi^c(0)\gamma_0\gamma_5\Psi^b(0)
 \right| B,1/2\  \vec{0}
\right\rangle}
\nonumber \\&&=\,\sqrt{\frac{E_{B'}(-\vec q\, )+m_{B'}}{2m_{B'}}}\,
\frac{|\vec q\,|}{E_{B'}(-\vec q\, )+m_{B'}}\bigg(
-G_1(|\vec q\,|)+G_2(|\vec q\,|)\nonumber\\
&&\hspace{6.75cm}+\frac{E_{B'}(-\vec q\, )}{m_{B'}}\,G_3(|\vec q\,|)\bigg)
\label{eq:g123}
\end{eqnarray} 
for the axial ones. All the left hand side terms can be evaluated using Eq.(\ref{eq:coor}) with
the approximation mentioned above.

For each transition there are only two different coordinate space
integrals from which all different matrix elements can be
evaluated. Those integrals are
\begin{eqnarray}
{\cal I}^{B'B}(|\vec q\,|)&=&\int\, d\,^3r_1\, d\,^3r_2\, 
e^{i\frac{m_{h_2}}{\overline M\,'}\,\vec
q\cdot\vec r_2}\,
e^{-i\frac{m_{h_2}+m_q}{\overline M\,'}\,\vec
q\cdot\vec r_1}
\nonumber\\ && \hspace{2.2cm}
 \left[\Psi^{B'}_{c\,h_2}(r_1,r_2,r_{12})\right]^*\ \Psi^{B}_{b\,h_2}(r_1,r_2,r_{12}) \nonumber\\
{\cal K}^{B'B}(|\vec q\,|)&=&\frac{1}{|\vec q\,|^2}\int\, d\,^3r_1\, d\,^3r_2\, 
e^{i\frac{m_{h_2}}{\overline M\,'}\,\vec
q\cdot\vec r_2}\,
e^{-i\frac{m_{h_2}+m_q}{\overline M\,'}\,\vec
q\cdot\vec r_1} 
\nonumber\\ && \hspace{2.2cm}
\left[\Psi^{B'}_{c\,h_2}(r_1,r_2,r_{12})\right]^*\ 
\vec l\cdot\vec q\ \Psi^{B}_{b\,h_2}(r_1,r_2,r_{12})
\label{eq:integrales}
\end{eqnarray}
In appendix~\ref{DPapp:fg123} we relate the form factors to the
 integrals ${\cal I}^{B'B}(|\vec q\,|)$ and ${\cal K}^{B'B}(|\vec
 q\,|)$ for the different $S_h,\,S'_h$ combinations, while in
 appendix~\ref{DPapp:integrals} we give the actual expressions we use to
 evaluate those integrals.

\subsection{Current conservation}
In the limit $m_b=m_c$ and for $B'=B$ (and thus $S_h=S'_h$) vector
current conservation provides a relation among the vector $F_2$ and
$F_3$ form factors, namely
\begin{equation}
F_2(w)=F_3(w)
\label{eq:cc}
\end{equation}
On the other hand the matrix element of the zeroth component of the
vector current evaluated at $w=1$ just counts the number of heavy
quarks so that we should have
\begin{equation}
F_1(1)+F_2(1)+F_3(1)=2
\label{eq:baryonnumber}
\end{equation}
In this limiting situation  the integrals ${\cal I}^{BB}(|\vec q\,|)$
and ${\cal K}^{BB}(|\vec q\,|)$ are related by\footnote{One just has to
integrate by parts in the ${\cal K}^{BB}(|\vec q\,|)$ expression.}%
\begin{equation}
{\cal K}^{BB}(|\vec q\,|)=\frac{m_{h_2}+m_q}{2\overline M}\
{\cal I}^{BB}(|\vec q\,|)
\end{equation}
Besides one has that ${\cal I}^{BB}(0)=1$. 

Using now the relations in Eq.(\ref{eq:appf123}) in
appendix~\ref{DPapp:fg123} we see that our model satisfies the
constraint in Eq.(\ref{eq:baryonnumber}) exactly.  On the other hand
we violate current conservation .  For instance, and again using the
relations in Eq.(\ref{eq:appf123}), we obtain for $w=1$
\begin{equation}
F_2(1)=F_3(1)+2(1-\frac{m_B}{\overline M})
\end{equation}

which shows that current conservation is violated by a term
proportional to the binding energy of the baryon divided by the sum of
the masses of its constituents.  This violation disappears in the infinite
heavy quark mass limit. This deficiency is shared by the relativistic calculation
of Ref.~\cite{Ebert:2004ck} and it is avoided in two other~\cite{Guo:1998yj,Sanchis-Lozano:1994vh} by 
the neglect of binding effects\footnote{If we look at
Ref.~\cite{Ebert:2004ck} and consider for instance transitions where the initial and
 final baryons have total  heavy quark spin  equal to 0, we see that  vector
current conservation in the equal mass case would require the $h_-(\omega)$ form factor to
vanish. This is not accomplished within that model. In  Ref.~\cite{Guo:1998yj} the currents
are constructed at the diquark level and the vector part is conserved thanks to the 
neglect of binding effects with the light quark. Finally,  the calculation in
Ref.~\cite{Sanchis-Lozano:1994vh} avoids this problem by using 
 the infinite heavy quark mass
limit and thus cancelling binding effects.}. Improvements on vector current
conservation would require at minimum the introduction of two--body
currents~\cite{Buchmann:1994bt}, going thus beyond the spectator
approximation, that we have not considered in this analysis. 

Note also that, for transitions that do not conserve the spin 
of the heavy quark subsystem $S_h$
(i.e. $\Xi'^0_{bc} \to \Xi^+_{cc} l \bar\nu_l)$ we have in the 
$m_b=m_c$ limit and at zero recoil that
\begin{equation}
F_1(1)+F_2(1)+F_3(1)=0
\end{equation}
due to the orthogonality of the initial and final baryon wave--functions.

\section{Results}
\label{sec:results}
\begin{figure}[!t]
%\vspace{-3cm}
\centering
\resizebox{12cm}{!}{\includegraphics{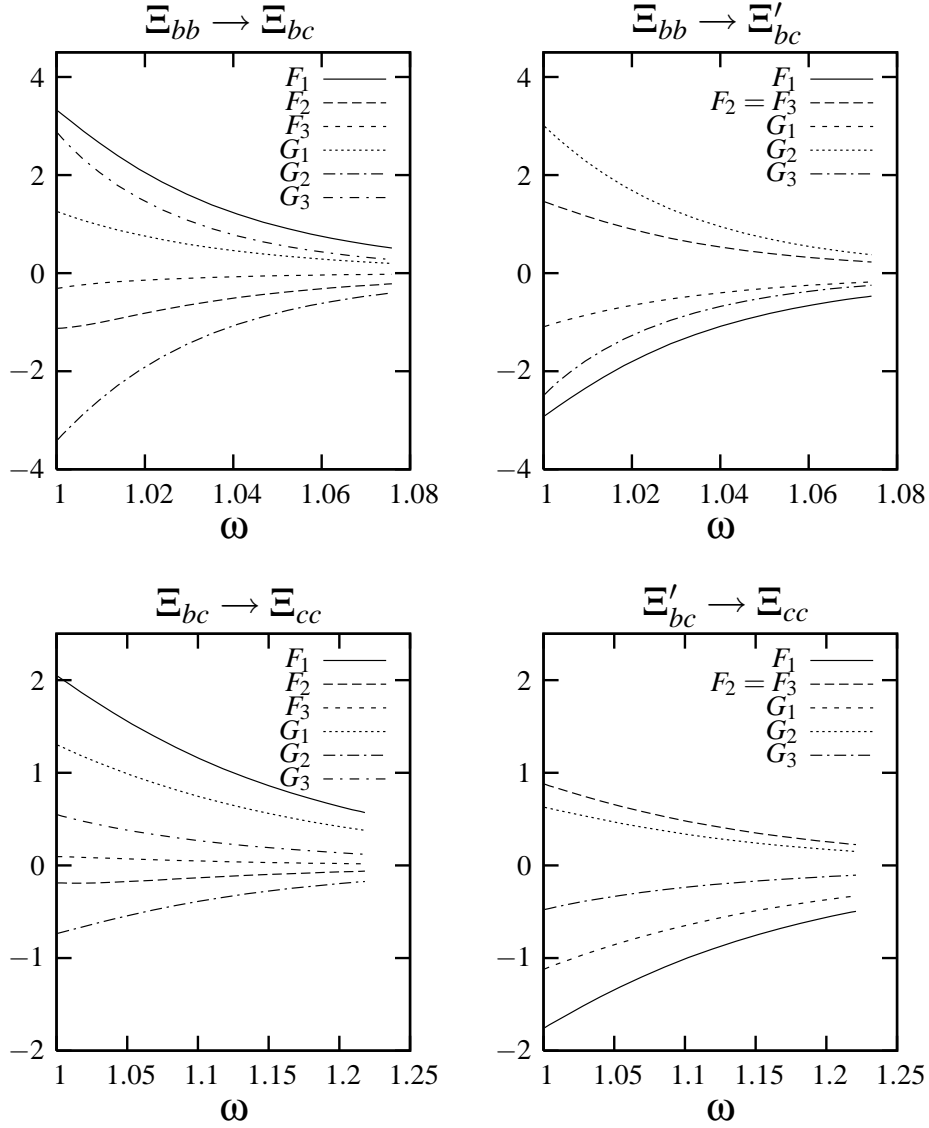}}
%\vspace{-15cm}
\caption{ Vector $F_1,\,F_2,\,F_3$ and axial $G_1,\,G_2,\,G_3$ form factors 
for doubly heavy  $\Xi(J=1/2)$ baryons
decays evaluated with the AL1 potential.}
\label{fig:ffxi}
\end{figure}
\begin{figure}[!t]
%\vspace{-1cm}
\centering
\resizebox{12.cm}{!}{\includegraphics{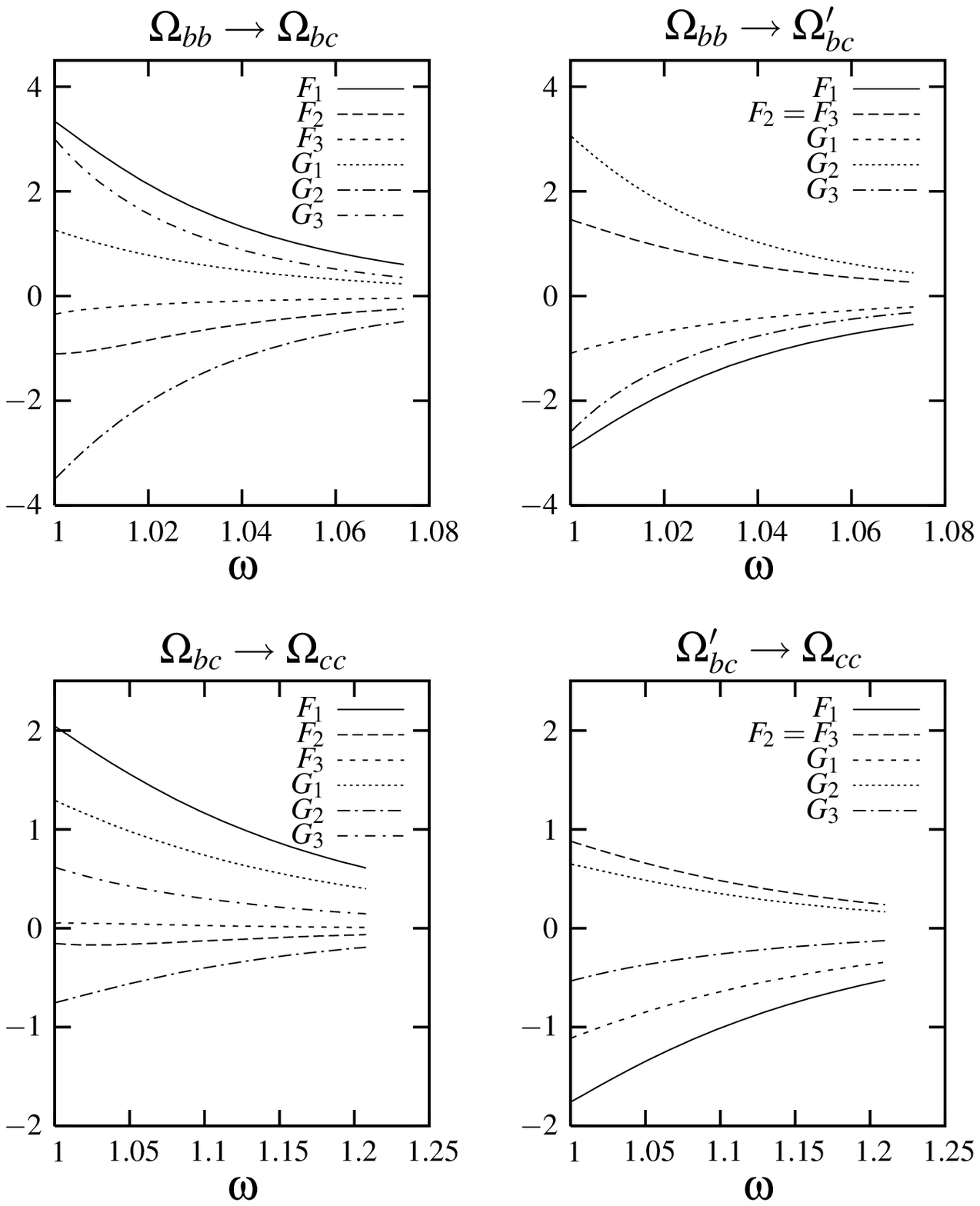}}
%\vspace{-15cm}
\caption{ Same as Fig.~\ref{fig:ffxi} for doubly heavy $\Omega(J=1/2)$ baryons
decays.
}
\label{fig:ffomega}
\end{figure}
In Figs.~\ref{fig:ffxi}, \ref{fig:ffomega} we show the form factors
for the different transitions evaluated with the AL1
potential. Variations when using a different potential are at the
level a few per cent at most. The results for doubly heavy $\Xi$
decays are almost identical to the corresponding ones for doubly heavy
$\Omega$ decays. The fact that we have two heavy quarks and that the
light one acts as a spectator makes the results almost independent of
the light quark mass.

\begin{figure}[!t]
%\vspace{-3cm}
\centering
\resizebox{13.cm}{!}{\includegraphics{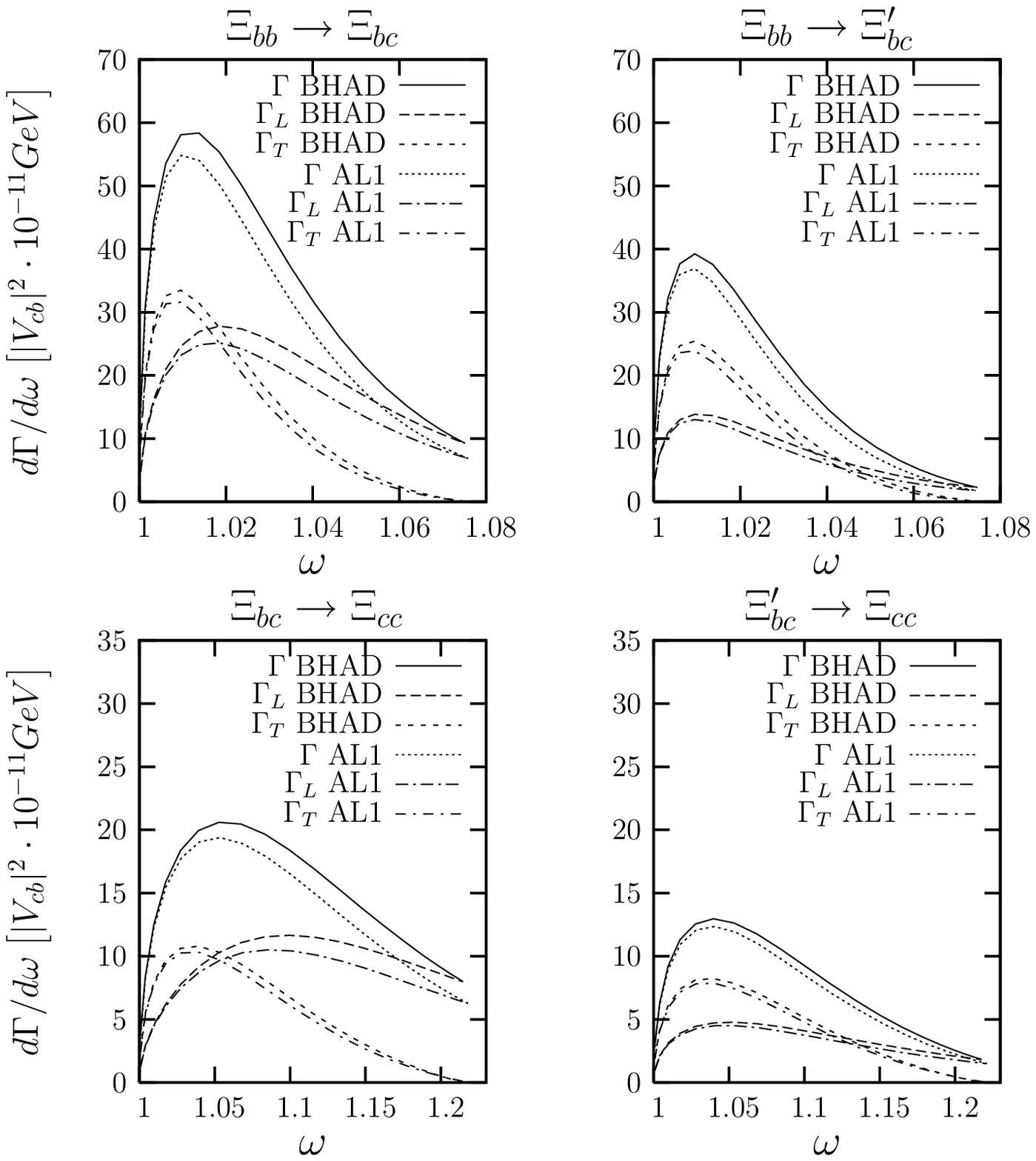}}
%\vspace{-15cm}
\caption{ $d\Gamma/dw$, $d\Gamma_L/dw$ and $d\Gamma_T/dw$ semileptonic
decay widths in units of $|V_{cb}|^2\cdot10^{-11}$\,GeV, for doubly
$\Xi(J=1/2)$ baryons decays. Solid line, long--dashed line and
short--dashed line: $d\Gamma/dw$, $d\Gamma_L/dw$ and $d\Gamma_T/dw$
evaluated with the AL1 potential; dotted line, long--dashed dotted
line and short--dashed dotted line: $d\Gamma/dw$, $d\Gamma_L/dw$ and
$d\Gamma_T/dw$ evaluated with the BHAD potential.}
\label{fig:dgdwxi}
\end{figure}

\begin{figure}[!t]
\centering
\resizebox{13.cm}{!}{\includegraphics{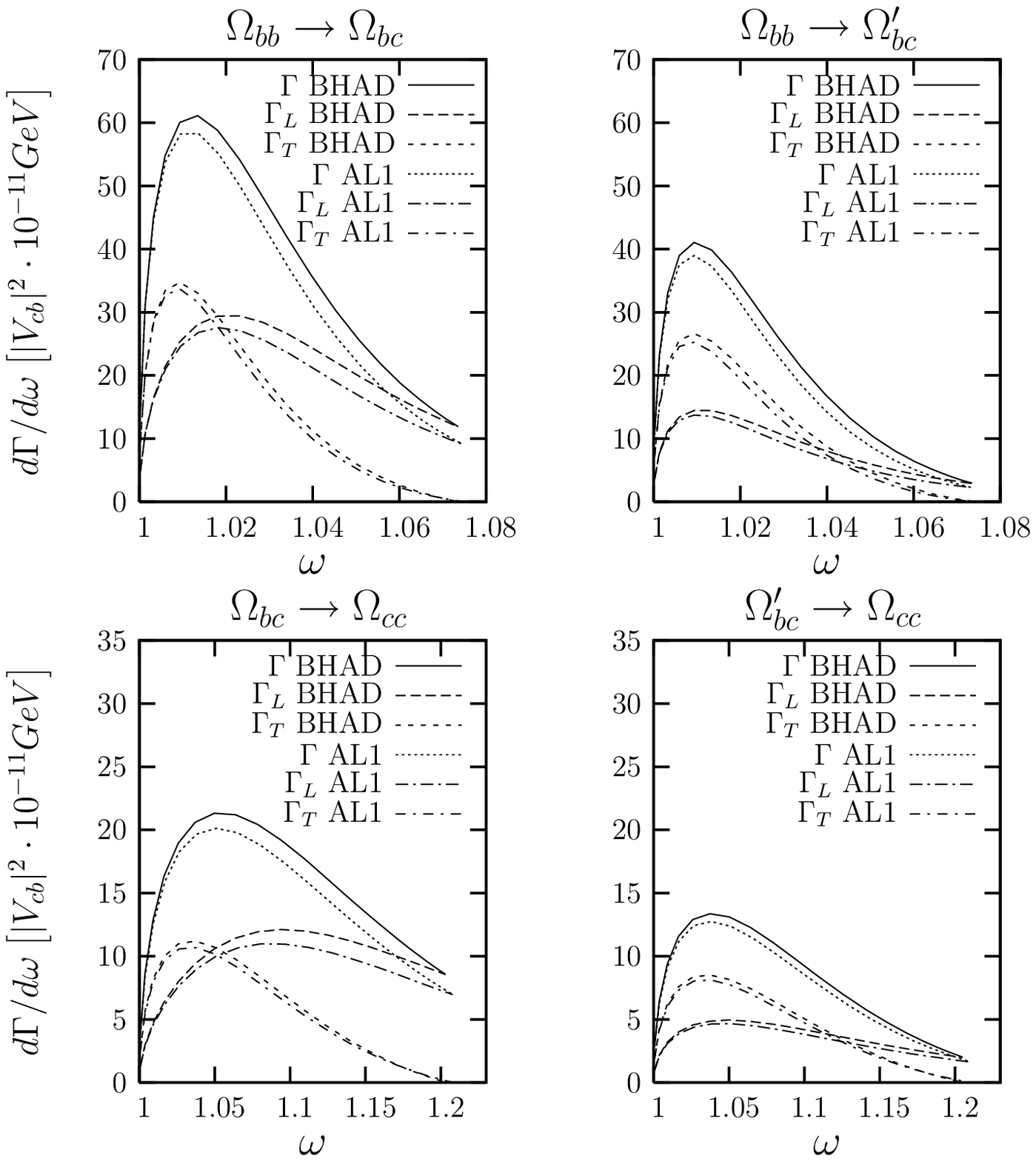}}
%\vspace{-15cm}
\caption{ Same as Fig.~\ref{fig:dgdwxi} for doubly heavy
$\Omega(J=1/2)$ baryons decays.  }
\label{fig:dgdwomega}
\end{figure}

In Figs.~\ref{fig:dgdwxi}, \ref{fig:dgdwomega} we show now our results
for the differential $d\Gamma_T/dw$, $d\Gamma_L/dw$ and $d\Gamma/dw$
decay widths evaluated with the AL1 and BHAD potentials. The differences
between the results obtained with the two inter-quark interactions
could reach 30\% for some transitions and for some regions of $w$. As
a consequence of the apparent SU(3) symmetry in the form factors we
also find that the results for doubly heavy $\Xi$ and $\Omega$ decays
are very close to each other. This apparent SU(3) symmetry goes over
to the integrated decay widths and asymmetry parameters.


In Table~\ref{tab:semi} we give our results for the semileptonic decay
width (transverse $\Gamma_T$, longitudinal $\Gamma_L$ and total
$\Gamma$) for the different processes under study. Our central values
have been evaluated using the AL1 potential while the errors show the
variations when changing the interaction. The biggest variations
appear for the BHAD potential for which one obtains results which are
larger by $7\sim12\% 
$. In Table~\ref{tab:DPgamma} we compare our
results to the ones calculated in different models. For that purpose
we need a value for $|V_{cb}|$ for which we take $|V_{cb}|=0.0413$.
Our results are in reasonable agreement with the ones in
Ref.~\cite{Ebert:2004ck} where they use a relativistic quark model
evaluated in the quark-diquark approximation, and with the
$\Gamma(\Xi_{bc}\to\Xi_{cc})$ value of Ref.~\cite{Sanchis-Lozano:1994vh} obtained
using HQET. The value for the latter width but now evaluated in the
relativistic three--quark model calculation of Ref.~\cite{Faessler:2001mr}
is much smaller than in any other calculation.  On the other hand in
Ref.~\cite{Guo:1998yj}, where they use the Bethe--Salpeter equation applied
to a quark-diquark system, they obtain much larger results for all
transitions.
\begin{table}[!t]
\centering
\begin{tabular}{lccc}
  &$\Gamma_T$ & $\Gamma_L$& $\Gamma$ \\ \hline\\
$\Xi_{bb}\to\Xi_{bc}\,l\bar\nu_l$\hspace{1cm}  & $0.97^{+0.10}_{-0.02}$ &
$1.28^{+0.19}_{-0.04} $
&$ 2.25^{+0.29}_{-0.06}$\\ \\
$\Xi_{bc}\to\Xi_{cc}\,l\bar\nu_l$  & $1.15^{+0.08}_{-0.01}$ &$ 1.86^{+0.}_{-0.02}$
&$ 3.01^{+0.30}_{-0.03} $ \\ \\
$\Xi_{bb}\to\Xi_{bc}'\,l\bar\nu_l$ & $0.73^{+0.08}_{-0.02}$ & $0.52^{+0.07}_{-0.01} $
& $1.24 ^{+0.15}_{-0.03}$ \\ \\
$\Xi_{bc}'\to\Xi_{cc}\,l\bar\nu_l$ & $0.89^{+0.05}_{-0.02}$ & $0.70^{+0.06}_{-0.01} $
 & $1.59^{+0.11}_{-0.03}$ \\ \hline
\end{tabular}\hspace{1cm}
\begin{tabular}{lccc}
  &$\Gamma_T$ & $\Gamma_L$& $\Gamma$ \\ \hline\\
$\Omega_{bb}\to\Omega_{bc}\,l\bar\nu_l$\hspace{1cm}  & $1.06^{+0.07}_{-0.01}$&
$1.45^{+0.16}_{-0.01}$ & $2.51^{+0.23}_{-0.02}$\\ \\
$\Omega_{bc}\to\Omega_{cc}\,l\bar\nu_l$  &$ 1.15^{+0.06}$ &$1.88^{+0.17}$
&$3.03^{+0.23}$\\ \\
$\Omega_{bb}\to\Omega_{bc}'\,l\bar\nu_l$  & $0.79^{+0.08}$ & $0.57^{+0.07} $
&$1.36^{+0.15}$\\ \\
$\Omega_{bc}'\to\Omega_{cc}\,l\bar\nu_l$  & $0.89^{+0.05}$ &
$0.70^{+0.05}$&$1.59^{+0.10} $ \\\hline
\end{tabular}
\caption{Semileptonic decay widths in units of
$|V_{cb}|^2\cdot10^{-11}$\,GeV. $\Gamma_T$ and $\Gamma_L$ stand for the
transverse and longitudinal contributions to the width $\Gamma$.
The central values have been obtained with the AL1 potential.
$l$ stands for a light charged lepton, $l=e,\,\mu$.}
\label{tab:semi}
\end{table}
\begin{table}[!t]
\centering
\begin{tabular}{lccccc}
  &This work &\cite{Ebert:2004ck}&\cite{Faessler:2001mr}&\cite{Guo:1998yj}&\cite{Sanchis-Lozano:1994vh}\\ \hline\\
$\Gamma(\Xi_{bb}\to\Xi_{bc}\,l\bar\nu_l)$\hspace{1cm}  &$ 3.84^{+0.49}_{-0.10}$&
3.26&&28.5&\\ \\
$\Gamma(\Xi_{bc}\to\Xi_{cc}\,l\bar\nu_l)$  &$ 5.13^{+0.51}_{-0.05} $ & 4.59&0.79&8.93&4.0\\ \\
$\Gamma(\Xi_{bb}\to\Xi_{bc}'\,l\bar\nu_l)$ &  $2.12 ^{+0.26}_{-0.05}$ &1.64&&4.28&\\ \\
$\Gamma(\Xi_{bc}'\to\Xi_{cc}\,l\bar\nu_l)$ &  $2.71^{+0.19}_{-0.05}$ & 1.76&&7.76&\\ \hline
\end{tabular}\hspace{1cm}
\begin{tabular}{lccc}
  &This work &\cite{Ebert:2004ck}&\cite{Guo:1998yj}\\ \hline\\
$\Gamma(\Omega_{bb}\to\Omega_{bc}\,l\bar\nu_l)$\hspace{1cm}  & $4.28^{+0.39}_{-0.03}$& 3.40&28.8\\ \\
$\Gamma(\Omega_{bc}\to\Omega_{cc}\,l\bar\nu_l)$ &$5.17^{+0.39}$&4.95&\\ \\
$\Gamma(\Omega_{bb}\to\Omega_{bc}'\,l\bar\nu_l)$ &$2.32^{+0.26}$& 1.66&\\ \\
$\Gamma(\Omega_{bc}'\to\Omega_{cc}\,l\bar\nu_l)$ &$2.71^{+0.17} $ &1.90&\\\hline
\end{tabular}
\caption{Semileptonic decay widths in units of $10^{-14}$\,GeV. We have
used a value $|V_{cb}|=0.0413$. $l$ stands for a light charged lepton, $l=e,\,\mu$.}
\label{tab:DPgamma}
\end{table}%

In Table~\ref{tab:asimetrias} we compile our results for the average
angular asymmetries $\alpha'$ and $\alpha''$, as well as the $R_{L/T}$
ratio, introduced in Eq.(\ref{eq:averageasymmetry}). The central
values have been obtained with the AL1 potential. Being all quantities
ratios the variation when changing the inter-quark interaction are in
most cases small.
\begin{table}[!t]
\centering
\begin{tabular}{lccc}
& $\left<\alpha'\right>$ &
  $\left<\alpha''\right>$ &$R_{L/T}$\\ \hline \\
%%%%%%%%%%%%
$\Xi_{bb}\to\Xi_{bc}\,l\bar\nu_l$\hspace{1cm}  &$-0.13^{+0.01}$&$-0.45_{-0.02}$&
$1.33^{+0.06}_{-0.01}$ \\\\

$\Xi_{bc}\to\Xi_{cc}\,l\bar\nu_l$  &$-0.12^{+0.01}$&$-0.53_{-0.01}$&
$1.62^{+0.07}$ \\\\

 $\Xi_{bb}\to\Xi_{bc}'\,l\bar\nu_l$   & $-0.19$ & $-0.17_{-0.01}$ &
$0.71^{+0.01}_{-0.01}$ \\ \\

 $\Xi_{bc}'\to\Xi_{cc}\,l\bar\nu_l$  & $-0.19$ & $-0.23_{-0.01}$ &
$0.79^{+0.02}$  \\ \hline
\end{tabular}\hspace{1cm}
%------------------------------------------------------------------
\begin{tabular}{lccc}
& $\left<\alpha'\right>$ &
  $\left<\alpha''\right>$ &$R_{L/T}$\\ \hline \\
%%%%%%%%%%%%
$\Omega_{bb}\to\Omega_{bc}\,l\bar\nu_l$ \hspace{1cm}
 &$ -0.13^{+0.01}$ & $-0.47_{-0.01}$&$1.37^{+0.06}$ \\ \\
%

%
%-----------------------------------------------------------------
 $\Omega_{bc}\to\Omega_{cc}\,l\bar\nu_l$  
 & $-0.12^{+0.01}$ &$-0.53_{-0.01}$ &$1.63^{+0.06}$
\\ \\

%-----------------------------------------------------------------
 $\Omega_{bb}\to\Omega_{bc}'\,l\bar\nu_l$ 
 & $-0.19_{-0.01}$ & $-0.18_{-0.01}$&$0.72^{+0.02}$
\\ \\
%

%------------------------------------------------------------------
 $\Omega_{bc}'\to\Omega_{cc}\,l\bar\nu_l$ 
  & $-0.19$ & $-0.23_{-0.01}$ & $0.79^{+0.02}$\\\hline

\end{tabular}
\caption{Averaged values of the asymmetry parameters $\alpha'$ and $\alpha''$
evaluated as indicated in 
Eq.(\ref{eq:averageasymmetry}). We also show the ratio $R_{L/T}=\Gamma_L/\Gamma_T$.
The central values have been obtained with the AL1 potential. 
$l$ stands for a light charged lepton, $l=e,\,\mu$.}
\label{tab:asimetrias}
\end{table}

\chapter{Strong pionic decay of heavy baryons}
\label{cha:lambdapi}

\section{Introduction}

In this chapter  we shall evaluate strong widths for the $\Sigma_c\to
\Lambda_c\, \pi$, $\Sigma_c^{*}\to \Lambda_c\, \pi$ and $\Xi_c^{*}\to
\Xi_c\, \pi$ decays.

Last decade has seen a great progress on
charmed--baryon physics and now the ground state baryons with a $c$
quark, with the exception of the $\Omega_c^*$, are well
established~\cite{Eidelman:2004wy}, and we have experimental information on the
strong one-pion decay widths for the
$\Sigma_c$~\cite{Artuso:2001us,Ammar:2000uh,Link:2001ee},
$\Sigma_c^*$~\cite{Ammar:2000uh,Athar:2004ni} and $\Xi_c^*$~\cite{Gibbons:1996yv,Avery:1995ps}.
With very little kinetic energy available in the final state these
reactions should be well described in a nonrelativistic approach. 
Although they have been analyzed before in the framework of the
constituent quark model (CQM)~\cite{Rosner:1995yu,Pirjol:1997nh},  no attempt was made
there to evaluate the full matrix elements. While there have been
dynamical calculations in other models (see references below), to our knowledge, ours is the first fully
dynamical calculation within a nonrelativistic approach.
In our calculation we will use the HQS--constrained wave functions 
evaluated in Ref.~\cite{Albertus:2003sx} using the same different interquark interactions
that have been used in this work,
 and  whose goodness have already been tested in
the study of the semileptonic $\Lambda_b\to \Lambda_c$ and $\Xi_b\to\Xi_c$ 
decays in  Ref.~\cite{Albertus:2004wj,Albertus:2004iw}. Again, the use of different quark--quark potentials
 will allow us 
 to obtain theoretical  uncertainties on the widths due to the 
 quark-quark interaction. The pion emission amplitude will
be obtained in a spectator model (one-quark pion emission) with the
use of partial conservation of axial current hypothesis (PCAC) as was
done, in the meson sector, in the  previous study of the strong
$B^*B\pi$ and $D^*D\pi$ couplings in section \ref{sect:gh*hpi}.

These reactions, and similar ones, have also been  addressed  in heavy hadron chiral 
perturbation theory (HHCPT)~\cite{Pirjol:1997nh,Yan:1992gz,Huang:1995ke,Cheng:1997rp,Chiladze:1997ev}, 
in QCDSR \cite{Grozin:1997qq,Zhu:1998vg}  
and within 
relativistic quark models like the light-front quark
model (LFQM)~\cite{Tawfiq:1998nk} and the  relativistic three-quark
model (RTQM)~\cite{Ivanov:1999bk,Ivanov:1998qe}.

\section{One-pion Strong decay width}
\label{sect:sdwpcac}
The width for the reaction $B\to B'\,\pi$ is given by
\begin{eqnarray}
\label{eq:gamma}
\Gamma&=&\frac{1}{2m_B}\int
\frac{d^{\,3}P_{B'}}{(2\pi)^3\,2E_{B'}(\vec{P}_{B'})}
\ \frac{d^{\,3}P_{\pi}}{(2\pi)^3\,2E_{\pi}(\vec{P}_{\pi})}
\ (2\pi)^4\delta^4(P_B-P_{B'}-P_{\pi})\nonumber\\
&&\hspace{3cm}\times\ \frac{1}{2 J_B+1}\sum_s\ \sum_{s'}
\left|{\cal A}_{BB'\pi}^{(s,s')}(P_{B},P_{B'})
\right|^2
\end{eqnarray}
where 
$P_B=(m_B;\vec{0}\,)$, $P_{B'}=(E_{B'}(\vec{P}_{B'}),\vec{P}_{B'})$ 
and $P_{\pi}=(E_{\pi}(\vec{P}_{\pi}),\vec{P}_{\pi})$ are the four-momenta of the particles
involved. $J_B$ is the spin of the $B$ baryon and $s$ and $s'$ are the third component of the
spin of the $B$ and $B'$ baryons in their respective center of mass systems. Finally
${\cal A}_{BB'\pi}^{(s,s')}(P_{B},P_{B'})$ is the pion emission
amplitude, that will be obtained through the non-pole part of the
matrix element of the divergence of the axial current similarly to the
case for mesons in section \ref{sect:gh*hpi}.

The quantity 
\begin{equation}
\frac{1}{2 J_B+1}\sum_s\ \sum_{s'}
\left|{\cal A}_{BB'\pi}^{(s,s')}(P_{B},P_{B'})
\right|^2
\end{equation}
 has to be a Lorentz
invariant. Denoting that invariant as $\left|{\cal M}_{BB'\pi}\right|^2$ we can perform the
integrations to give
\begin{equation}
\Gamma=\left|{\cal M_{BB'\pi}}\right|^2\ \frac{|\vec{q}\,|}{8\pi m_B^2}
\end{equation}
with $|\vec{q}\,|$ the modulus of the three-momentum of the $B'$ baryon or
the $\pi$ meson given by
\begin{equation}
|\vec{q}\,|=\frac{\lambda^{1/2}(m_B^2,m^2_{B'},m_{\pi}^2)}{2m_B}
\end{equation}

Throughout the calculation we shall use  physical masses
taken from Ref.~\cite{Eidelman:2004wy}.

\section{Description of baryon states}
\label{sect:bs}

\begin{table}
\centering
\begin{tabular}{cccccc%cc
}\hline
Baryon &~~~~$S$~~~~&~~~~$J^P$~~~~&~~~~ $I$~~~~&~~~~$S_l^\pi$~~~~& 
Quark content 
%& $M_{exp.}$ \cite{pdg02}~~~& $M_{Latt.}$~\cite{bowler}
\\
       &       &         &   &          &               %& [MeV]  & [MeV]   
\\\hline
$\Lambda_c$& 0 &$\frac12^+$& 0 &$0^+$&$udc$%& $2285 \pm 1$ & $2270 \pm 50  $
\\
$\Sigma_c$ & 0 &$\frac12^+$& 1 &$1^+$&$llc$%& $2452 \pm 1$ & $2460 \pm 80 $
\\
$\Sigma^*_c$ & 0 &$\frac32^+$& 1 &$1^+$&$llc$%& $2518 \pm 2$ &$ 2440
%\pm 70 $
\\
$\Xi_c$ & $-$1 &$\frac12^+$&$\frac12$&$0^+$&$lsc$%& $2469 \pm 3$ & $2410
%\pm 50 $
\\
$\Xi'_c$ & $-$1 &$\frac12^+$&$\frac12$&$1^+$&$lsc$%& $2576 \pm 3$ & $2570
%\pm 80$ 
\\
$\Xi^*_c$ &$-$1&$\frac32^+$&$\frac12$&$1^+$&$lsc$%& $2646 \pm 2$ & $2550
%\pm 80$
\\
$\Omega_c$ &$-$2 &$\frac12^+$& 0 &$1^+$&$ssc$%& $2698 \pm 3$ & $2680 \pm 70$
\\
$\Omega^*_c$ &$-$2 &$\frac32^+$& 0 &$1^+$&$ssc$%&     & $2660 \pm 80$
\\\hline
%$\Lambda_b$& 0 &$\frac12^+$& 0 &$0^+$&$udb$%& $ 5624 \pm 9$  & $5640
%\pm 60 $
%\\
%$\Sigma_b$ & 0 &$\frac12^+$& 1 &$1^+$&$llb$%&  & $5770 \pm 70 $
%\\
%$\Sigma^*_b$ & 0 &$\frac32^+$& 1 &$1^+$&$llb$%&  & $5780 \pm 70$
%\\
%$\Xi_b$ & $-$1 &$\frac12^+$&$\frac12$&$0^+$&$lsb$%&    & $5760 \pm 60 $
%\\
%$\Xi'_b$ & $-$1 &$\frac12^+$&$\frac12$&$1^+$&$lsb$%&    & $5900 \pm 70$
%\\
%$\Xi^*_b$ &$-$1&$\frac32^+$&$\frac12$&$1^+$&$lsb$%&  & $5900 \pm 80 $
%\\
%$\Omega_b$ &$-$2 &$\frac12^+$& 0 &$1^+$&$ssb$%&  & $5990 \pm 70 $
%\\
%$\Omega^*_b$ &$-$2 &$\frac32^+$& 0 &$1^+$&$ssb$%&     & $6000 \pm 70$
%\\\hline
\end{tabular}
\caption{Summary of the quantum numbers of  ground state baryons containing a single 
heavy $c$ quark. $I$, and
$S_l^\pi$ are the isospin, and the spin--parity of the light
degrees of freedom and $S$, $J^P$ are the strangeness and the spin--parity
of the baryon. In the last column $l$ denotes a light quark of flavor $u$ or
$d$.}
\label{tab:gscb}
\end{table}

We use a similar description for baryons as the one used in chapters
\ref{cha:dp_masas}, \ref{cha:dp_sl}, but in this case we shall work in
momentum space. The Fock space representation of a 
baryon $B$ 
with three-momentum $\vec{P}$ and spin projection $s$ 
 in the baryon center of mass is
\begin{eqnarray}
\label{lp_wf}
\lefteqn{\hspace*{0cm}\left|{B,s\,\vec{P}}\,\right\rangle_{NR}}
\nonumber\\
&&=\,\int d^{\,3}Q_1 \int d^{\,3}Q_2\ \frac{1}{\sqrt2}\sum_{\alpha_1,\alpha_2,\alpha_3}
\frac{\hat{\psi}^{(B,s)}_{\alpha_1,\alpha_2,\alpha_3}(\,\vec{Q}_1,\vec{Q}_2\,)}{(2\pi)^3\
 \sqrt{2E_{f_1}(\vec{p}_1)2E_{f_2}(\vec{p}_2)
2E_{f_3}(\vec{p}_3)}}\nonumber\\ 
&&\hspace{3cm}
\times\left|\ \alpha_1\
\vec{p}_1=\frac{m_{f_1}}{\overline{M}}\vec{P}+\vec{Q}_1\ \right\rangle
\left|\ \alpha_2\ \vec{p}_2=\frac{m_{f_2}}{\overline{M}}\vec{P}+\vec{Q}_2\ \right\rangle
\nonumber \\ &&\hspace{5cm}\times
\left|\ \alpha_3\ \vec{p}_3=\frac{m_{f_3}}{\overline{M}}\vec{P}-\vec{Q}_1
-\vec{Q}_2\ \right\rangle
 \end{eqnarray}
$\alpha_1$, $\alpha_2$
and $\alpha_3$ represent
the quantum numbers of spin (s), flavor (f) and color (c), $\alpha\equiv
(s,f,c)$, 
of the three quarks, while $(E_{f_1}(\vec{p}_1),\,\vec{p}_1)$, 
$(E_{f_2}(\vec{p}_2),\,\vec{p}_2)$ 
and $(E_{f_3}(\vec{p}_3),\,\vec{p}_3)$ are their 
respective four-momenta. $m_f$ is the mass of the quark 
 with flavor $f$, and $\overline{M}=m_{f_1}+m_{f_2}+m_{f_3}$. We choose the
 third quark to be the $c$ quark while the first two  will be the light
 ones.\\
 The normalization of the quark 
states is 
\begin{eqnarray}
\left\langle\ \alpha^{\prime}\ \vec{p}^{\ \prime}\,|\,\alpha\
\vec{p}\, \right\rangle=\delta_{\alpha^{\prime},\, \alpha}\, (2\pi)^3\,
2E(\vec{p}\,)\,\delta^3( \vec{p}^{\ \prime}-\vec{p}\,)
\end{eqnarray}
Besides, $\hat{\psi}^{\,(B,s)}_{\alpha_1,\alpha_2,\alpha_3}
(\,\vec{Q}_1,\vec{Q}_2\,)$ is the
momentum space wave
function  for the internal motion, being
$\vec{Q}_1$ and $\vec{Q}_2$  the momenta conjugate  to the positions $\vec{r}_1$ and 
$\vec{r}_2$ of the
two light quarks  with respect to the heavy one (See Fig.~\ref{lp_fig:coor}). This
wave function is
  antisymmetric under the simultaneous exchange $\alpha_1\longleftrightarrow
\alpha_2, \vec{Q}_1 \longleftrightarrow \vec{Q}_2 $, being 
    also antisymmetric under an overall exchange of
the color degrees of freedom. It is normalized such that 
\begin{equation}
\int d^{\,3}Q_1 \int d^{\,3}Q_2\ \sum_{\alpha_1,\alpha_2,\alpha_3}
\left(\hat{\psi}^{(B,s')}_{\alpha_1,\alpha_2,\alpha_3}(\,\vec{Q}_1,\vec{Q}_2\,)\right)^*
\hat{\psi}^{(B,s)}_{\alpha_1,\alpha_2,\alpha_3}(\,\vec{Q}_1,\vec{Q}_2\,)
=\delta_{s',\, s}
\end{equation}
and, thus,  the normalization of our baryon states is 
\begin{equation}
{}_{\stackrel{}{NR}}\left\langle\, {B,s'\,\vec{P}^{\,\prime}}\,|\,{B,s
\,\vec{P}}\,\right\rangle_{NR}
=\delta_{s',\,s}\,(2\pi)^3\,\delta^3(\vec{P}^{\,\prime}-\vec{P}\,)
\end{equation}
For the particular case of ground state  $\Lambda_c$, $\Sigma_c$,
$\Sigma^{*}_c$, $\Xi_c$ and $\Xi^{*}_c$ we can assume the
orbital angular momentum to be zero. We will also take advantage of HQS and
assume the light--degrees of freedom quantum numbers are well defined (See
Table~\ref{tab:gscb}). In that case we have\footnote{We only give the wave
function for the baryons involved in $\pi^+$ emission. Wave functions for
other isospin states 
of the same baryons are
easily constructed.}

{\allowdisplaybreaks
\begin{eqnarray} 
&&\hspace*{-1.7cm}\hat{\psi}^{\,(\Lambda^+_c, s)}_{\alpha_1,\,\alpha_2,\,\alpha_3}(\,\vec{Q}_1,\vec{Q}_2\,)
\,=\,\frac{1}{\sqrt{3}!}\,\varepsilon_{c_1\,c_2\,c_3}\
\ \left(\left.\frac12,\frac12,0\right|s_1,s_2,0\right)\ \delta_{s_3,\,s}\nonumber
\\*
&&\hspace{.9cm}
\times\ \delta_{f_3,\,c}\,\frac{1}{\sqrt2}
\left(
\delta_{f_1,\,u}\,\delta_{f_2,\,d}\,\widetilde{\phi}^{S_{l}=0}_{u,\,d,\,c}(\,\vec{Q}_1,\vec{Q}_2\,)\ 
-\delta_{f_1,\,d}\,\delta_{f_2,\,u}\,\widetilde{\phi}^{S_{l}=0}_{d,\,u,\,c}(\,\vec{Q}_1,\vec{Q}_2\,)\
\right)  
 \nonumber\\
&&\hspace*{-1.7cm}\hat{\psi}^{\,(\Sigma^{++}_c,s)}_{\alpha_1,\,\alpha_2,\,\alpha_3}(\,\vec{Q}_1,\vec{Q}_2\,)
\,=\,\frac{1}{\sqrt{3}!}\,\varepsilon_{c_1\,c_2\,c_3}\
\ \widetilde{\phi}^{S_{l}=1}_{u,\,u,\,c}(\,\vec{Q}_1,\vec{Q}_2\,)\ 
\delta_{f_1,\,u}\, \delta_{f_2,\,u}\, \delta_{f_3,\,c}\nonumber\\*
&&\hspace{.9cm} \times\ \sum_m \left(\left.\frac12,\frac12,1\right|s_1,s_2,m\right)\ 
\left(\left.1,\frac12,\frac12\right|m,s_3,s\right)\nonumber\\
&&\hspace*{-1.7cm}\hat{\psi}^{\,(\Sigma^{*\,++}_c,s)}_{\alpha_1,\,\alpha_2,\,\alpha_3}(\,\vec{Q}_1,\vec{Q}_2\,)
\,=\,\frac{1}{\sqrt{3}!}\,\varepsilon_{c_1\,c_2\,c_3}\
\ \widetilde{\phi}^{S_{l}=1}_{u,\,u,\,c}(\,\vec{Q}_1,\vec{Q}_2\,)\ 
\delta_{f_1,\,u}\,\delta_{f_2,\,u}\, \delta_{f_3,\,c}\nonumber\\*
&&\hspace{.9cm} \times\ \sum_m 
\left(\left.\frac12,\frac12,1\right|s_1,s_2,m\right)\ 
\left(\left.1,\frac12,\frac32\right|m,s_3,s\right)\nonumber\\
&&\hspace*{-1.7cm}\hat{\psi}^{\,(\Xi^0_c,s)}_{\alpha_1,\,\alpha_2,\,\alpha_3}(\,\vec{Q}_1,\vec{Q}_2\,)
\,=\,\frac{1}{\sqrt{3}!}\,\varepsilon_{c_1\,c_2\,c_3}\
\ 
\left(\left.\frac12,\frac12,0\right|s_1,s_2,0\right)\ \delta_{s_3,\,c}\nonumber\\*
&&\hspace{.9cm} \times\ \delta_{f_3,\,c}\,\frac{1}{\sqrt2}\, \left(
\delta_{f_1,\,d}\,\delta_{f_2,\,s}\ \widetilde{\phi}^{S_{l}=0}_{d,s,c}(\,\vec{Q}_1, \vec{Q}_2\,)
-\delta_{f_1,\,s}\,\delta_{f_2,\,d}\ \widetilde{\phi}^{S_{l}=0}_{s,d,c}(\,\vec{Q}_1, \vec{Q}_2\,)
\right)\nonumber\\
&&\hspace*{-1.7cm}\hat{\psi}^{\,(\Xi^{*\,+}_c,s)}_{\alpha_1,\,\alpha_2,\,\alpha_3}(\,\vec{Q}_1,\vec{Q}_2\,)
\,=\,\frac{1}{\sqrt{3}!}\,\varepsilon_{c_1\,c_2\,c_3}\
\ \sum_m \left(\left.\frac12,\frac12,1\right|s_1,s_2,m\right)\,\left(\left.1,\frac12,\frac32\right|m,s_3,s\right)
\nonumber\\
&&\hspace{.9cm} \times\ \delta_{f_3,\,c}\,\frac{1}{\sqrt2}\,\left(
\delta_{f_1,\,u}\,\delta_{f_2,\,s}\ \widetilde{\phi}^{S_{l}=1}_{u,s,c}(\,\vec{Q}_1, \vec{Q}_2\,)
+\delta_{f_1,\,s}\,\delta_{f_2,\,u}\ \widetilde{\phi}^{S_{l}=1}_{s,u,c}(\,\vec{Q}_1, \vec{Q}_2\,)
\right)\nonumber\\
\end{eqnarray}
} %Del allowdisplaybreacks
\\
Here $\varepsilon_{c_1 c_2 c_3}$ is the fully antisymmetric tensor on color
indices being $\varepsilon_{c_1 c_2 c_3}/\sqrt{3!}$  the antisymmetric color wave
function and 
the
$\widetilde{\phi}^{S_l}_{f_1,\,f_2,\,f_3}(\,\vec{Q}_1, \vec{Q}_2\,)$, with
$S_l$ the total spin of the light degrees
of freedom, are the  Fourier transform of the corresponding  coordinate space 
wave functions ($\phi_{f_1,f_2,f_3}^{S_l}(\vec{r}_1,\vec{r}_2)$, 
see below). Their dependence on momenta is through  $|\vec{Q}_1|$, $|\vec{Q}_2|$ 
and $\vec{Q}_1\cdot\vec{Q}_2$ alone, and they are symmetric under the
simultaneous exchange $f_1\longleftrightarrow
f_2, \vec{Q}_1 \longleftrightarrow \vec{Q}_2 $. Their normalization
is given by
\begin{equation}
\int d^{\,3}Q_1 \int d^{\,3}Q_2\ 
\left|\widetilde{\phi}^{S_l}_{f_1,f_2,f_3}(\,\vec{Q}_1,\vec{Q}_2\,)\right|^2
=1
\end{equation}
\section{Coordinate space wave functions}
Wave functions for baryons with a heavy quark are quite similar to the
ones presented in chapter \ref{cha:dp_masas}, but there are some
differences. In this section we briefly explain how they are constructed in
Ref. \cite{Albertus:2003sx}. 

\subsection{Intrinsic Hamiltonian}
\label{sect:ih}
\begin{figure}[!t]
\centering
\resizebox{10cm}{!}{\includegraphics{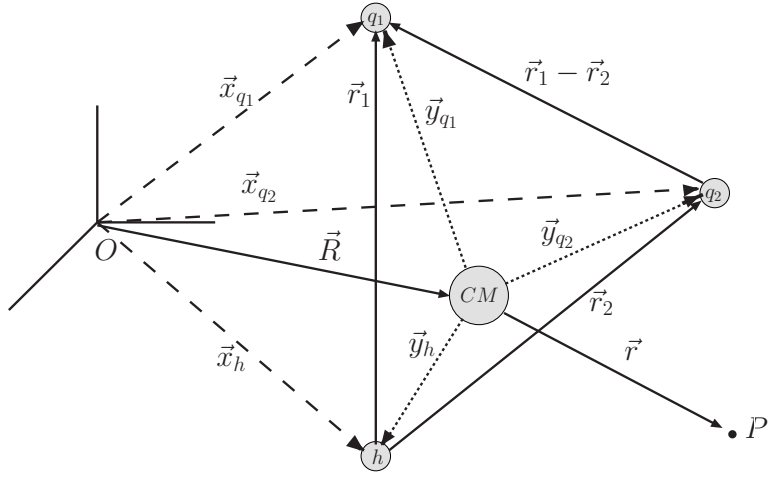}}
\caption{ Definition of the coordinates used.}\label{lp_fig:coor}
\end{figure}

Working in the heavy quark frame $(\vec{R},\vec{r}_1,\vec{r}_2)$ (See
Fig.~\ref{lp_fig:coor}), where $\vec{R}$, $\vec{r}_1$ and $\vec{r}_2$ are
the center of mass position in the laboratory frame and the relative
positions of the light $q_1$ and $q_2$ quarks with respect to the
heavy one $h$, the center of mass motion can be separated from the
intrinsic Hamiltonian, 
\begin{eqnarray}
H&=&-\frac{\vec{\nabla}^2_{\vec{R}}}{2\overline{M}}+ H_{int}\nonumber\\
H_{int}&=&\overline{M}+\sum_{j=1,2}H_j^{sp}+V_{q_1,q_2}(\vec{r}_1-\vec{r}_2,spin)
-\frac{\vec{\nabla}_1\cdot\vec{\nabla}_2}{m_h}\nonumber\\
H_j^{sp}&=&-\frac{\vec{\nabla}^2_j}{2\mu_j}+V_{h,q_j}(\vec{r}_j,spin), \hspace{1cm}
j=1,2
\end{eqnarray}
with $\overline{M}=m_{q_1}+m_{q_2}+m_h$, $\mu_{j}=\left(1/m_{q_j}+1/m_h\right)^{-1}$ and
$\vec{\nabla}_{j}=\partial/\partial_{\vec{r}_j}$, $j=1,2$.\\
Apart from $\overline{M}$, $H_{int}$ consists of the  sum of two single particle Hamiltonians ($H_j^{sp}$), describing
the dynamics of the light quarks in the mean field created by the heavy quark,
plus the light-light interaction term that includes the Hughes-Eckart term
$(\vec{\nabla}_1\cdot\vec{\nabla}_2)$.

For the inter-quark interaction $V$, the  potentials presented in chapter
\ref{cha:potential} were used.  

\subsection{Variational coordinate--space wave function and its Fourier
transform}
\label{sect:vwf}
To solve the three-body problem the following variational ansazt was used in
Ref.~\cite{Albertus:2003sx} for the spatial part of the wave function
\begin{equation}
\phi^{S_l}_{q_1,q_2,h}(\vec{r}_1,\vec{r}_2)= N \phi_{hq_1}(r_1)
 \phi_{hq_2}(r_2) F^{S_l}(r_{12})
 \end{equation}
 with $r_j=|\vec{r}_j|$, $r_{12}=|\vec{r}_1-\vec{r}_2|$.
 $N$ is a normalization constant, $\phi_{hq_j}$ is the s-wave ground state
 solution (\,$\psi_{hq_j}$\,) of the single particle Hamiltonian (\,$H_j^{sp}$\,) 
 corrected at large distances in the form
 \begin{equation}
 \phi_{hq_j}(r_j)=(1+\alpha_j\, r_j)\,\psi_{hq_j}(r_j), \hspace{1cm} j=1,2
 \end{equation}
 and finally $F^{S_l}_{q_1,q_2}(r_{12})$ is a Jastrow correlation function in the
 relative distance of the two light quarks and was constructed as a
 sum of gaussians
 \begin{eqnarray}
 F^{S_l}(r_{12})=\sum_{j=1}^4\,
 a_j\,e^{-b^2_j\,(r_{12}+d_j)^2} , \hspace{1cm} a_1=1
 \end{eqnarray}

 The variational parameters for the different states obtained with the different
  potentials can be
 found in the appendix of Ref.~\cite{Albertus:2003sx}. The $\psi_{hq_j}$ 
 functions can be
 easily obtained through a Numerov algorithm.

In momentum space we will have
\begin{eqnarray}
\widetilde{\phi}^{S_l}_{q_1,q_2,h}(\vec{Q}_1,\vec{Q}_2)
&=&\frac{1}{(2\pi)^3}\int d^{\,3}r_1\int d^{\,3}r_2\
e^{i\,\vec{Q}_1\cdot\vec{r}_1}\ e^{i\,\vec{Q}_2\cdot\vec{r}_2}
\ N \phi_{hq_1}(r_1)
 \phi_{hq_2}(r_2) F^{S_l}(r_{12})\nonumber\\
 &=&\frac{1}{(2\pi)^{3/2}}\int d^{\,3}P\
N\, \widetilde{\phi}_{hq_1}(|\vec{P}|)\
 \widetilde{\phi}_{hq_2}(|\vec{Q}_1+\vec{Q}_2-\vec{P}|)\
 \widetilde{F}^{S_l}(|\vec{Q}_1-\vec{P}|) 
\nonumber\\
\end{eqnarray}
being $\widetilde{\phi}_{hq_1}$, $\widetilde{\phi}_{hq_2}$ and
$\widetilde{F}^{S_l}$ the Fourier transforms of 
${\phi}_{hq_1}$, ${\phi}_{hq_2}$ and ${F}^{S_l}$ respectively. They
depend solely on the modulus of the momentum. Clearly
\begin{eqnarray}
\phi^{S_l}_{q_1,q_2,h}(R\vec{Q}_1,R\vec{Q}_2)
=\phi^{S_l}_{q_1,q_2,h}(\vec{Q}_1,\vec{Q}_2)
\end{eqnarray}
with $R$ a rotation. For the purpose of evaluation we can take
$\vec{Q}_1$ in the $z$ direction and $\vec{Q}_2$ in the $OXZ$ plane.

\section{Results}  %la cambio entera por la del paper
\subsection{$\Sigma_c\to\Lambda_c \,\pi$ decay}
\label{sect:sig}
Let us do the $\Sigma_c^{++}\to\Lambda_c^+ \,\pi^+$ case. 
The matrix element of the divergence $q^\mu J_{A\,\mu}^{d\, u}(0)$ of  the axial 
 current determines the $\pi^+$ emission  amplitude as
\begin{eqnarray}
\label{eq:amsiglam}
{\cal A}^{(s,s')}_{\Sigma_c^{++}\Lambda_c^+\pi^+}(P,P')&=&\frac{-i}{f_\pi}\ 
\left\langle \Lambda_c^+,\ s'\,\vec{P}'\,\left|\,q^\mu\, J_{A\,\mu}^{d\, u}(0)
\,\right|
\,\Sigma_c^{++},\ s\,\vec{P}\,\right\rangle_{non-pole}\nonumber\\
&=&ig_{\Sigma_c^{++}\Lambda_c^+\pi^+}
\ \overline{u}_{\Lambda_c^+\ s'}(\vec{P}'\,)\
\gamma_5\
u_{\Sigma_c^{++}\ s}(\vec{P})
\end{eqnarray}
where the coupling constant $g_{\Sigma_c^{++}\Lambda_c^+\pi^+}$,
 in analogy to the pion coupling to nucleons and nucleon resonances,
has been chosen to be dimensionless, and $u_{\Sigma_c^{++}\ s}(\vec{P})$,
  $u_{\Lambda_c^{+}\ s'}(\vec{P}')$ are Dirac spinors
 normalized to twice the energy ($u^\dagger u\,=\,2E$).  
 The width is given by
\begin{equation}
\label{eq:gsiglam}
\Gamma(\Sigma_c^{++}\to\Lambda_c^+\pi^+)=\frac{|\vec{q}\,|}{8\pi M_{\Sigma_c^{++}}^2 }\
g_{\Sigma_c^{++}\Lambda_c^+\pi^+}^2\left(\ (
M_{\Sigma_c^{++}}-M_{\Lambda_c^+})^2-m^2_{\pi}\right)
\end{equation}
with $|\vec{q}\,|$ the modulus of the final baryon or pion three-momentum. From Eq.~(\ref{eq:amsiglam}), taking
$\vec{P}=\vec{0}$,  $\vec{P}'=-|\vec{q}\,|\,\vec{k}$ in the $z$ direction,
$s=s'=1/2$, and taking into account the different normalization of our
nonrelativistic states, we have 
\begin{eqnarray}
\hspace*{-.9cm}g_{\Sigma_c^{++}\Lambda_c^+\pi^+}
&=&\frac{-1}{f_\pi}\ 
\frac{\sqrt{E_{\Lambda_c^+}(\vec{q}\,)+M_{\Lambda_c^+}}
\sqrt{2M_{\Sigma_c^{++}} 2E_{\Lambda_c^+}(\vec{q}\,) }}{|\vec{q}\,|
\sqrt{2M_{\Sigma_c^{++}} }}
\nonumber\\ &&
\hspace*{.7cm}
\times 
\left(
(M_{\Sigma_c^{++}}- E_{\Lambda_c^+}(\vec{q}\,) )\ A_{\Sigma_c^{++}\Lambda_c^+
,\ 0}^{1/2,1/2}
+|\vec{q}\,|\ A_{\Sigma_c^{++}\Lambda_c^+
,\ 3}^{1/2,1/2}
\right)
\end{eqnarray}
with
\begin{eqnarray}
A_{\Sigma_c^{++}\Lambda_c^+
,\ \mu}^{1/2,1/2}={}_{\stackrel{}{NR}}\left\langle \Lambda_c^+,\
1/2\ -|\vec{q}|\,\vec{k}\ \left|\  J_{A\,\mu}^{d\, u}(0)\ \right|
\,\Sigma_c^{++},\ 1/2\ \vec{0}\,\right\rangle_{NR,\ non-pole}
\end{eqnarray}
The $A_{\Sigma_c^{++}\Lambda_c^+
,\ \mu}^{1/2,1/2}$  are easily evaluated using one-body current operators and their 
expressions  can be found in  appendix  \ref{Lambdapiapp}.

\begin{table}[t]
{\footnotesize \setlength{\tabcolsep}{2pt}
\begin{center}
\hspace*{-.7cm}
\begin{tabular}{l|c c c c}
\hline
 & \hspace{.25cm}$g_{\Sigma_c^{++}\Lambda_c^+\pi^+}$\hspace{.25cm} &
 \hspace{.25cm}$\Gamma(\Sigma_c^{++}\to\Lambda_c^+\pi^+)$ \hspace{.25cm} &
 \hspace{.25cm} $\Gamma(\Sigma_c^{+}\to\Lambda_c^+\pi^0)$ \hspace{.25cm} &
 \hspace{.25cm}$\Gamma(\Sigma_c^{0}\to\Lambda_c^+\pi^-)$ \hspace{.25cm} 
 \\
 &&[MeV]&[MeV]&[MeV]\\
\hline
This work 
%avg. 
& $21.73\pm0.32\pm0.08$&$2.41\pm0.07\pm0.02$ &$2.79\pm0.08\pm0.02$ 
&$2.37\pm0.07\pm0.02$\\
\hline
Experiment &  & $2.3\pm0.2\pm 0.3$~\cite{Artuso:2001us} & $< 4.6$ (CL=90\%)~\cite{Ammar:2000uh} & 
$2.5\pm0.2\pm0.3$~\cite{Artuso:2001us} \\
 &  &$2.05^{+0.41}_{-0.38}\pm0.38$~\cite{Link:2001ee} & &
 $1.55^{+0.41}_{-0.37}\pm0.38$~\cite{Link:2001ee}\\
 \hline
Theory & &  & &\\
CQM
& & $1.31\pm 0.04$~\cite{Rosner:1995yu} &$1.31\pm 0.04$~\cite{Rosner:1995yu}& 
$1.31\pm 0.04$~\cite{Rosner:1995yu}\\
 & &$2.025^{+1.134}_{-0.987}$~\cite{Pirjol:1997nh} & &
 $1.939^{+1.114}_{-0.954}$~\cite{Pirjol:1997nh} \\
HHCPT &22, 29.3~\cite{Yan:1992gz} & 2.47, 4.38~\cite{Yan:1992gz}& 2.85,
5.06~\cite{Yan:1992gz}&2.45, 4.35~\cite{Yan:1992gz}\\
 & & 2.5~\cite{Huang:1995ke}&  3.2~\cite{Huang:1995ke}& 2.4~\cite{Huang:1995ke}\\
 & & &  &$1.94\pm0.57$~\cite{Cheng:1997rp}\\
LFQM & & 1.64 ~\cite{Tawfiq:1998nk}&1.70 ~\cite{Tawfiq:1998nk} & 1.57 ~\cite{Tawfiq:1998nk}\\
RTQM & &$2.85\pm0.19$~\cite{Ivanov:1999bk,Ivanov:1998qe} &$3.63\pm0.27$~\cite{Ivanov:1999bk,Ivanov:1998qe}
  & $2.65\pm0.19$~\cite{Ivanov:1999bk,Ivanov:1998qe} \\
\hline 
\end{tabular}
\end{center}
}
\caption{ Coupling constant  $g_{\Sigma_c^{++}\Lambda_c^+\pi^+}$ and 
total widths $\Gamma(\Sigma_c^{++}\to\Lambda_c^+\pi^+)$,
$\Gamma(\Sigma_c^{+}\to\Lambda_c^+\pi^0)$ and
$\Gamma(\Sigma_c^{0}\to\Lambda_c^+\pi^-)$
(See text for details).
  Experimental data and different theoretical
calculations are also shown.}
\label{tab:siglam}
\end{table}

In Table~\ref{tab:siglam} we present the results for 
$g_{\Sigma_c^{++}\Lambda_c^+\pi^+}$ and the widths 
$\Gamma(\Sigma_c^{++}\to\Lambda_c^+\pi^+)$,
$\Gamma(\Sigma_c^{+}\to\Lambda_c^+\pi^0)$ and 
$\Gamma(\Sigma_c^{0}\to\Lambda_c^+\pi^-)$. 
To get the  values for 
$\Gamma(\Sigma_c^{+}\to\Lambda_c^+\pi^0)$ and
$\Gamma(\Sigma_c^{0}\to\Lambda_c^+\pi^-)$ we use 
$g_{\Sigma_c^{++}\Lambda_c^+\pi^+}$ and make the appropriate mass changes in the rest
of factors in Eq.~(\ref{eq:gsiglam}).

 Our results show two types of errors. The second one
 results from the Monte Carlo evaluation of the integrals needed to obtain the
 $g_{\Sigma_c^{++}\Lambda_c^+\pi^+}$ coupling constant. The first one
 is the usual  theoretical uncertainty coming from the use of
 different potentials. 
The results are in very good agreement with the experimental data by the CLEO
Collaboration in Refs.~\cite{Artuso:2001us,Ammar:2000uh}. The value for
 $\Gamma(\Sigma_c^{++}\to\Lambda_c^+\pi^+)$ also agrees with the experimental
 data by the FOCUS Collaboration in Ref.~\cite{Link:2001ee}. The agreement with FOCUS
 data is not  good for the $\Gamma(\Sigma_c^{0}\to\Lambda_c^+\pi^-)$ case,
 although our result is still within experimental errors. 
 Our results show variations as large as $\approx$17\%      between different charge configurations. This is due to the little
 kinetic energy available in the final state that makes the widths very
 sensitive to the precise masses of the hadrons involved.
In this respect there is a new precise determination of the
 $\Lambda_c^+$ mass by the {\sl BABAR} collaboration
 $M_{\Lambda_c}=2286.46\pm0.14$\ MeV/c$^2$~\cite{Aubert:2005gt}, which is roughly 1.5
 MeV/c$^2$ above the value quoted by the Particle Data Group in 
 Ref.~\cite{Eidelman:2004wy}. With this new value our calculated widths would get 
 reduced by $9\%
 $. This reduction comes from  phase space factors while the coupling 
 $g_{\Sigma_c^{++}\Lambda_c^+\pi^+}$ changes only at the level of
 0.1\%.

As for the
 other theoretical determinations, the CQM calculation in Ref.~\cite{Rosner:1995yu} uses exact $s \longleftrightarrow c$
symmetry to relate the $\Sigma_c\to \Lambda_c\pi$ decay to the non--charmed 
 $\Sigma^*\to \Lambda\pi$ analogue decay.  Their results are smaller than the
 experimental data obtained by the CLEO Collaboration. In
 Ref.~\cite{Pirjol:1997nh} a unique coupling constant is fixed in order to reproduce
 all experimental information on $\Sigma_c^*\to\Lambda_c\pi$ widths. That
 coupling is latter used to predict the
 $\Sigma_c\to\Lambda_c\pi$ widths. This coupling suffers from large
 uncertainties and thus the theoretical errors on the predicted widths
 are also very large. In the HHCPT calculation of Ref.~\cite{Yan:1992gz}  a
 simple CQM argument is used in order to obtain the unknown coupling constant in
 the HHCPT Lagrangian. Furthermore the 
 authors allow for a renormalization of the axial coupling $g_A^{ud}$ for light 
 quarks. The
 largest of the two values quoted corresponds to the case where that coupling
 is unrenormalized and then is given by $g_A^{ud}=1$. The smaller number quoted
 corresponds to the use of a renormalized value of $g_A^{ud}=0.75$. The case $g_A^{ud}=1$ is the one that compares with our
 calculation. Their results for the widths almost double ours and are not in
 agreement with experiment. Their simple determination of the coupling constant
 can not be correct. The values obtained in the HHCPT calculation of
 Ref.~\cite{Huang:1995ke} are closer to our results and experimental data. There the 
 authors determine
 the needed coupling constants from the analysis of analogue decays involving
 non--charmed baryons. In Ref.~\cite{Cheng:1997rp}, also within the HHCPT approach,
 and similarly to Ref.~\cite{Pirjol:1997nh},
 the authors fix the unknown coupling in the Lagrangian using the experimental
 data for the $\Sigma_c^*\to\Lambda_c\pi$ decays. From there they predict the
  $\Gamma(\Sigma_c^0\to\Lambda_c^+\pi^-)$ obtaining a value very close to the
 one in Ref.~\cite{Pirjol:1997nh}, and that  suffers also  from large uncertainties.
 The two relativistic quark model calculations of
 Refs.~\cite{Tawfiq:1998nk,Ivanov:1999bk,Ivanov:1998qe} give results that differ by almost a factor
 of two. Our results are closer to the ones obtained within the RTQM
of Ref.~\cite{Ivanov:1999bk,Ivanov:1998qe}.
\subsection{$\Sigma_c^{*}\to \Lambda_c\pi$  decay}
\label{sect:sig*}
Let us  analyze the case with a $\pi^+$ in the final state,
$\Sigma_c^{*\,++}\to \Lambda_c^+\pi^+$.
Similarly to the $\Sigma_c$ decay  before we now have
\begin{eqnarray}
\label{eq:amsig*lam}
{\cal A}^{(s,s')}_{\Sigma_c^{*\,++}\Lambda_c^+\pi^+}(P,P')&=&\frac{-i}{f_\pi}
\left\langle \Lambda_c^+,\ s'\,\vec{P}'\,\left|\,q^\mu\, J_{A\,\mu}^{d\, u}(0)
\,\right|
\,\Sigma_c^{*\,++},\ s\,\vec{P}\,\right\rangle_{non-pole}\nonumber\\
&=&i\frac{g_{\Sigma_c^{*\,++}\Lambda_c^+\pi^+}}{2M_{\Lambda_c^+}}
\ q_\nu\ \overline{u}_{\Lambda_c^+\ s'}(\vec{P}'\,)\
u_{\Sigma_c^{*\,++}\ s}^\nu(\vec{P})
\end{eqnarray}
where we have introduced the dimensionless coupling constant
$g_{\Sigma_c^{*\,++}\Lambda_c^+\pi^+}$
and $u_{\Sigma_c^{*\,++}\ s}^\nu(\vec{P})$ is a Rarita-Schwinger
spinor normalized to twice the energy $(u^{\nu\dagger}u_\nu=-2E)$.
The width is given by
\begin{equation}
\label{eq:gammasig*lam}
\Gamma(\Sigma_c^{*\,++}\to\Lambda_c^+\pi^+)=\frac{|\vec{q}\,|^3}{24\pi M_{\Sigma_c^{*\,++}}^2 }\
\frac{g_{\Sigma_c^{*\,++}\Lambda_c^+\pi^+}^2}{4M_{\Lambda_c^{+}}^2} \left((
M_{\Sigma_c^{*\,++}}+M_{\Lambda_c^+})^2-m^2_{\pi}\right)
\end{equation}
\begin{table}[t]
{\footnotesize \setlength{\tabcolsep}{3pt}
\begin{center}
\hspace*{-.50cm}
\begin{tabular}{l|c c c c}
\hline
 & \hspace{.25cm}$g_{\Sigma_c^{*\, ++}\Lambda_c^+\pi^+}$\hspace{.25cm}
 & \hspace{.25cm}$\Gamma(\Sigma_c^{*\,++}\to\Lambda_c^+\pi^+)$ \hspace{.25cm}
 & \hspace{.25cm}$\Gamma(\Sigma_c^{*\,+}\to\Lambda_c^+\pi^0)$ \hspace{.25cm}
 & \hspace{.25cm}$\Gamma(\Sigma_c^{*\,0}\to\Lambda_c^+\pi^-)$ \hspace{.25cm}\\
 &&[MeV]&[MeV]&[MeV]\\
\hline
This work 
%avg. 
&$36.20\pm0.75\pm0.13$&$17.52\pm0.74\pm0.12$&$17.31\pm0.73\pm0.12$ &
$16.90\pm0.71\pm0.12$\\
\hline
Experiment &  & $14.1^{+1.6}_{-1.5}\pm 1.4$~\cite{Athar:2004ni}& $< 17$ (CL=90\%)~\cite{Ammar:2000uh} & $16.6^{+1.9}_{-1.7}\pm1.4$~\cite{Athar:2004ni}\\
\hline
Theory &&&&\\
QCDSR &$13.8\div 24.2$~\cite{Grozin:1997qq}&&&\\
&$32.5\pm2.1\pm6.9$~\cite{Zhu:1998vg} & & &\\
CQM& & 20~\cite{Rosner:1995yu}&20~\cite{Rosner:1995yu} &20~\cite{Rosner:1995yu}\\
HHCPT   & & 25~\cite{Huang:1995ke}&25~\cite{Huang:1995ke} &25~\cite{Huang:1995ke}\\
LFQM   & & 12.84~\cite{Tawfiq:1998nk} & & 12.40~\cite{Tawfiq:1998nk}\\
RTQM& &$21.99\pm0.87$~\cite{Ivanov:1999bk,Ivanov:1998qe} &
  & $21.21\pm0.81$~\cite{Ivanov:1999bk,Ivanov:1998qe} \\
\hline
\end{tabular}
\end{center}
}
\caption{ Coupling constant $g_{\Sigma_c^{*\,++}\Lambda_c^+\pi^+}$ and 
total widths $\Gamma(\Sigma_c^{*\,++}\to\Lambda_c^+\pi^+)$,
$\Gamma(\Sigma_c^{*\,+}\to\Lambda_c^+\pi^0)$ and
$\Gamma(\Sigma_c^{*\,0}\to\Lambda_c^+\pi^-)$. %obtained 
%with different potentials.
 Experimental data and different theoretical
calculations are also shown. Note that in order to compare with our definition
of $g_{\Sigma_c^{*\,++}\Lambda_c^+\pi^+}$ we have multiplied the coupling
constants evaluated in Refs.~\cite{Grozin:1997qq,Zhu:1998vg} by $2M_{\Lambda_c^+}/f_\pi$.
}
\label{tab:sig*lam}
\end{table}
Taking again 
$\vec{P}=\vec{0}$, $\vec{P}'=-|\vec{q}\,|\,\vec{k}$ in the $z$ direction,
 and $s=s'=1/2$ we obtain from Eq.(\ref{eq:amsig*lam})
 \begin{eqnarray}
\label{eq:gsig*lam}
\hspace{-0.9cm}g_{\Sigma_c^{*\,++}\Lambda_c^+\pi^+}
&=&\frac{\sqrt3}{f_\pi\sqrt2}\ \frac{2M_{\Lambda_c^+}
\sqrt{2M_{\Sigma_c^{*\,++}} 2E_{\Lambda_c^+}(\vec{q}\,) }}
{|\vec{q}\,|\sqrt{2M_{\Sigma_c^{*\,++}} \left(
E_{\Lambda_c^+}(\vec{q}\,)+M_{\Lambda_c^+}\right)}}
\nonumber\\&& \hspace*{.4cm}
\times
\left(
(M_{\Sigma_c^{*\,++}}- E_{\Lambda_c^+}(\vec{q}\,) )\ A_{\Sigma_c^{*\,++}
\Lambda_c^+
,\ 0}^{1/2,1/2}
+|\vec{q}\,|\ A_{\Sigma_c^{*\,++}\Lambda_c^+
,\ 3}^{1/2,1/2}
\right)
\end{eqnarray}
with
\begin{eqnarray}
A_{\Sigma_c^{*\,++}\Lambda_c^+
,\ \mu}^{1/2,1/2}={}_{\stackrel{}{NR}}\left\langle \Lambda_c^+,\
1/2\ -|\vec{q}\,|\,\vec{k}\,\left|\  J_{A\,\mu}^{d\, u}(0)\ \right|
\,\Sigma_c^{*\,++},\ 1/2\ \vec{0}\,\right\rangle_{NR,\ non-pole}
\end{eqnarray}
The expressions for $A_{\Sigma_c^{*\,++}\Lambda_c^+
,\ \mu}^{1/2,1/2}$ ($\mu=0,3$) can be found in  appendix  \ref{Lambdapiapp}.\\
Results for $g_{\Sigma_c^{*\,++}\Lambda_c^+\pi^+}$ and the total widths 
$\Gamma(\Sigma_c^{*\,++}\to\Lambda_c^+\pi^+)$, 
$\Gamma(\Sigma_c^{*\,+}\to\Lambda_c^+\pi^0)$ and
$\Gamma(\Sigma_c^{*\,-}\to\Lambda_c^+\pi^-)$
appear in Table~\ref{tab:sig*lam}. Our value for the latter two are obtained with the use
of
$g_{\Sigma_c^{*\,++}\Lambda_c^+\pi^+}$ and with the appropriate mass
changes in the rest of  factors in Eq.~(\ref{eq:gammasig*lam}).
Our central value for $\Gamma(\Sigma_c^{*\,++}\to\Lambda_c^+\pi^+)$ is above 
the central value of the
latest experimental determination by the CLEO Collaboration in 
Ref.~\cite{Athar:2004ni}. For some of the potentials used, AP1 and AP2, the results  obtained are within experimental errors. The central value for
$\Gamma(\Sigma_c^{*\,+}\to\Lambda_c^+\pi^0)$ is
slightly above the upper experimental bound determined also by the 
CLEO Collaboration in Ref.~\cite{Ammar:2000uh}, but again, we obtain results 
which are below the experimental bound using the AP1 and AP2 potentials.
  As for  $\Gamma(\Sigma_c^{*\,-}\to\Lambda_c^+\pi^-)$
we get a nice agreement with experiment. Our results for the different charge
configurations differ by 4\%
 at most. With the new value for $M_{\Lambda_c^+}$ given by the {\sl BABAR} 
Collaboration in Ref.~\cite{Aubert:2005gt} they would get reduced by 3\%. 
 Our results are
globally in better agreement with experiment than the ones obtained by other
 theoretical
calculations~\footnote{We did not show those cases were data on
$\Sigma_c^*\to\Lambda_c\pi$ widths were used to fit parameters of the models.}
with perhaps the exception of the QCDSR calculation of Ref.~\cite{Zhu:1998vg}.
\subsection{$\Xi_c^{*}\to \Xi_c\pi$ decay}
\label{sect:xi*}
Once more we analyze  the case with a $\pi^+$ in the final state, 
$\Xi_c^{*\,+}\to \Xi_c^0\pi^+$.
 What we obtain is
\begin{eqnarray}
\label{eq:amcas}
{\cal A}^{(s,s')}_{\Xi_c^{*\,+}\Xi_c^0\pi^+}(P,P')&=&\frac{-i}{f_\pi}
\left\langle \Xi_c^0,\ s'\,\vec{P}'\,|\,q^\mu\, J_{A\,\mu}^{d\, u}(0)\,|
\,\Xi_c^{*\,+},\ s\,\vec{P}\,\right\rangle_{non-pole}\nonumber\\
&=&i\frac{g_{\Xi_c^{*\,+}\Xi_c^+\pi^+}}{M_{\Xi_c^+}+M_{\Xi_c^0}}
\ q_\nu\ \overline{u}_{\Xi_c^0\ s'}(\vec{P}'\,)\
u_{\Xi_c^{*\,+}\ s}^\nu(\vec{P})
\end{eqnarray}
where again we have introduced a dimensionless coupling 
 $g_{\Xi_c^{*\,+}\Xi_c^0\pi^+}$.
The width is given as
\begin{equation}
\label{eq:gammacas*cas}
\Gamma(\Xi_c^{*\,+}\to\Xi_c^0\pi^+)=\frac{|\vec{q}\,|^3}{24\pi M_{\Xi_c^{*\,+}}^2 }\
\frac{g_{\Xi_c^{*\,+}\Xi_c^0\pi^+}^2}{(M_{\Xi_c^{+}}+M_{\Xi_c^{0}})^2} \left((
M_{\Xi_c^{*\,+}}+M_{\Xi_c^0})^2-m^2_{\pi}\right)
\end{equation}
\begin{table}[t]
{\tiny
\begin{center}
\hspace*{-1.cm} \setlength{\tabcolsep}{2pt}
\begin{tabular}{l|c c c c c}
\hline
 &\hspace{.01cm} $\ g_{\Xi_c^{*\,+}\Xi_c^0\pi^+}$\hspace{.01cm} 
 &\hspace{.01cm} $\Gamma(\Xi_c^{*\,+}\to\Xi_c^0\pi^+)$ \hspace{.01cm} 
 &\hspace{.01cm} $\Gamma(\Xi_c^{*\,+}\to\Xi_c^+\pi^0)$ \hspace{.01cm} 
 &\hspace{.01cm} $\Gamma(\Xi_c^{*\,0}\to\Xi_c^+\pi^-)$ \hspace{.01cm}
 &\hspace{.01cm} $\Gamma(\Xi_c^{*\,0}\to\Xi_c^0\pi^0)$ \hspace{.01cm} \\
& &[MeV] & [MeV]&[MeV]&[MeV]\\ \hline
 & &&&&\\
This work  
%avg. 
&$-28.83\pm0.50\pm0.10$ &$1.84\pm0.06\pm0.01$&
$1.34\pm0.04\pm0.01$&$2.07\pm0.07\pm0.01$&$0.956\pm0.030\pm0.007$\\
\hline
Theory  &&&&\\
LFQM  & &$1.12$~\cite{Tawfiq:1998nk} & $0.69$~\cite{Tawfiq:1998nk} &
  $1.16$~\cite{Tawfiq:1998nk}& $0.72$~\cite{Tawfiq:1998nk}\\
RTQM& &$1.78\pm0.33$~\cite{Ivanov:1999bk,Ivanov:1998qe} &$1.26\pm0.17$~\cite{Ivanov:1999bk,Ivanov:1998qe}&$2.11\pm0.29$~\cite{Ivanov:1999bk,Ivanov:1998qe}
  & $1.01\pm0.15$~\cite{Ivanov:1999bk,Ivanov:1998qe} \\
\hline
&&&&&\\
\multicolumn{1}{l|}{}&&\multicolumn{2}{c}
{$\Gamma(\Xi_c^{*\,+}\to\Xi_c^0\pi^++\Xi_c^+\pi^0)$}
&\multicolumn{2}{c}{$\Gamma(\Xi_c^{*\,0}\to\Xi_c^+\pi^-+\Xi_c^0\pi^0)$}\\
 &&\multicolumn{2}{c}{[MeV]}
&\multicolumn{2}{c}{[MeV]}\\
\hline
\multicolumn{1}{l|}{This work}&&\multicolumn{2}{c}{$3.18\pm0.10\pm0.01$}&
\multicolumn{2}{c}{$3.03\pm0.10\pm0.01$}\\
\hline
\multicolumn{1}{l|}{Experiment}& & \multicolumn{2}{c}
{$<3.1$ (CL=90\%)~\cite{Gibbons:1996yv}}&\multicolumn{2}{c}{$<5.5$ 
(CL=90\%)~\cite{Avery:1995ps}}\\
\hline
Theory  &&&&\\
CQM& & \multicolumn{2}{c}{$<2.3\pm0.1$~\cite{Rosner:1995yu}}  &
\multicolumn{2}{c}{$<2.3\pm0.1$~\cite{Rosner:1995yu}}\\
 & & \multicolumn{2}{c}{1.191\,--\, 3.971~\cite{Pirjol:1997nh}} & 
 \multicolumn{2}{c}{1.230\, --\, 4.074~\cite{Pirjol:1997nh}}\\
 HHCPT && \multicolumn{2}{c}{$2.44\pm0.85$~\cite{Cheng:1997rp}} &
  \multicolumn{2}{c}{$2.51\pm0.88$~\cite{Cheng:1997rp}}\\
LFQM & & \multicolumn{2}{c}{1.81~\cite{Tawfiq:1998nk}} &
 \multicolumn{2}{c}{1.88~\cite{Tawfiq:1998nk}}\\
RTQM & &\multicolumn{2}{c}{$3.04\pm0.50$~\cite{Ivanov:1999bk,Ivanov:1998qe}} &
 \multicolumn{2}{c}{$3.12\pm0.33 $~\cite{Ivanov:1999bk,Ivanov:1998qe}}\\
 \hline
\end{tabular}
\end{center}
}
\caption{ Values for  the coupling $g_{\Xi_c^{*\,+}\Xi_c^0\pi^+}$ and 
decay widths $\Gamma(\Xi_c^{*\,+}\to\Xi_c^0\pi^+)$,
$\Gamma(\Xi_c^{*\,+}\to\Xi_c^+\pi^0)$,
$\Gamma(\Xi_c^{*\,0}\to\Xi_c^+\pi^-)$ and
$\Gamma(\Xi_c^{*\,0}\to\Xi_c^0\pi^0)$.
%obtained with different potentials. 
Experimental upper bounds for the total $\Xi_c^{*\,+}$ and 
$\Xi_c^{*\,0}$ widths, and different theoretical
calculations are also shown.}
\label{tab:cas*cas}
\end{table}
Taking now
$\vec{P}=\vec{0}$,  $\vec{P}'=-|\vec{q}\,|\,\vec{k}$ in the $z$ direction,
and $s=s'=1/2$, $g_{\Xi_c^{*\,+}\Xi_c^0\pi^+}$ is evaluated from Eq.(\ref{eq:amcas}) to be
\begin{eqnarray}
\label{eq:gcas*cas}
\hspace{-0.5cm}g_{\Xi_c^{*\,+}\Xi_c^0\pi^+}
&=&\frac{\sqrt3}{f_\pi\sqrt2}\ \frac{(M_{\Xi_c^+}+M_{\Xi_c^0})
\sqrt{2M_{\Xi_c^{*\,+}} 2E_{\Xi_c^0}(\vec{q}\,) }}
{|\vec{q}\,|\sqrt{2M_{\Xi_c^{*\,+}} \left(
E_{\Xi_c^0}(\vec{q}\,)+M_{\Xi_c^0}\right)}}
\nonumber\\&& \hspace{.5cm}
\times
\left(
(M_{\Xi_c^{*\,+}}- E_{\Xi_c^0}(\vec{q}\,) )\ A_{\Xi_c^{*\,+}
\Xi_c^0
,\ 0}^{1/2,1/2}
+|\vec{q}\,|\ A_{\Xi_c^{*\,+}\Xi_c^0
,\ 3}^{1/2,1/2}
\right)
\end{eqnarray}
with
\begin{eqnarray}
A_{\Xi_c^{*\,+}\Xi_c^0
,\ \mu}^{1/2,1/2}={}_{\stackrel{}{NR}}\left\langle \Xi_c^0,\
1/2\ -|\vec{q}\,|\,\vec{k}\,|\  J_{A\,\mu}^{d\, u}(0)\ |
\,\Xi_c^{*\,+},\ 1/2\ \vec{0}\,\right\rangle_{NR,\ non-pole}
\end{eqnarray}
which expressions can be found in  appendix  \ref{Lambdapiapp}.\\
 Results for the coupling $g_{\Xi_c^{*\,+}\Xi_c^0\pi^+}$, 
the widths $\Gamma(\Xi_c^{*\,+}\to\Xi_c^0\pi^+)$,
$\Gamma(\Xi_c^{*\,+}\to\Xi_c^+\pi^0)$,
$\Gamma(\Xi_c^{*\,0}\to\Xi_c^+\pi^-)$ and
$\Gamma(\Xi_c^{*\,0}\to\Xi_c^0\pi^0)$,  and the total
  widths  
$\Gamma(\Xi_c^{*\,+}\to\Xi_c^0\pi^++\Xi_c^+\pi^0)$  and
$\Gamma(\Xi_c^{*\,0}\to\Xi_c^0\pi^0+\Xi_c^+\pi^-)$ 
appear in Table~\ref{tab:cas*cas}. Our values for 
$\Gamma(\Xi_c^{*\,+}\to\Xi_c^+\pi^0)$,
$\Gamma(\Xi_c^{*\,0}\to\Xi_c^+\pi^-)$ and
$\Gamma(\Xi_c^{*\,0}\to\Xi_c^0\pi^0)$
are obtained with the use of $g_{\Xi_c^{*\,+}\Xi_c^0\pi^+}
/(M_{\Xi_c^+}+M_{\Xi_c^0})$, and with the appropriate mass changes
in the rest of  factors in Eq.~(\ref{eq:gammacas*cas}).
For $\Gamma(\Xi_c^{*\,+}\to\Xi_c^+\pi^0)$ and
$\Gamma(\Xi_c^{*\,0}\to\Xi_c^0\pi^0)$ an extra $1/2$ isospin factor 
should be included.
Our results for $\Gamma(\Xi_c^{*\,+}\to\Xi_c^0\pi^++\Xi_c^+\pi^0)$ are slightly
above the experimental bound obtained by the CLEO Collaboration~\cite{Gibbons:1996yv}.
As for the $\Sigma_c^*$ decay  case above, the AP1 and AP2 potentials gives results closer 
to experiment. For $\Gamma(\Xi_c^{*\,0}\to\Xi_c^+\pi^-+\Xi_c^0\pi^0)$ our result
is well below the CLEO Collaboration experimental bound in Ref.~\cite{Avery:1995ps}.
Isospin breaking due to mass effects is clearly seen when comparing the
predictions for $\Gamma(\Xi_c^{*\,+}\to\Xi_c^0\pi^-)$ and
$\Gamma(\Xi_c^{*\,+}\to\Xi_c^+\pi^0)$. One finds a factor 1.4 difference when
 a factor of two would be 
expected from isospin symmetry. Again, the fact that there is little phase space
available makes the results very sensitive to the actual mass values.
Compared to other theoretical calculations our results agree nicely with the
ones obtained within the RTQM of Refs.~\cite{Ivanov:1999bk,Ivanov:1998qe}, while they are larger
than most other  determinations.

\chapter{Conclusions}

This work has been devoted to the study of different properties of
hadrons with one and two heavy quarks $c$ and/or $b$.  All
calculations have been done in the framework of a nonrelativistic
constituent quark model.  In order to check the sensitivity of our
results to the inter--quark interaction we have used five different
quark--quark potentials. The spread in results gives an estimation of
theoretical uncertainties and so has been quoted. 
  Most observables studied change only at the level of a few
per cent when changing the interaction potential.
 Another source of theoretical uncertainty is the use of
nonrelativistic kinematics in the evaluation of the orbital wave
functions and the construction of our nonrelativistic states, although we think that a good part of
these relativistic effects are contained in an effective way in the
parameters of the quark-quark potentials, which were adjusted in the
original works to reproduce the experimental spectra of mesons. Mesons
wave functions have been evaluated using 
 a Numerov algorithm, while for baryons the three body problem has
been solved using a variational approach with  ansatz wave functions that
include simplifications that arise from HQS and HQSS. To compute
hadron decays we have relied on the  impulse approximation, which
induces small axial and vector current conservation violations due to binding
effects. These binding effects, though they affect  some of the weak form factors
analyzed, have little influence on the decay widths.

\bigskip
$\bullet$ In chapter \ref{cha:BD}
we have evaluated  leptonic decay constants
of $D$, $D^*$, $B$ and $B^*$ mesons, and form factors and decay widths
for the semileptonic $B\to D\,l\,\overline\nu$ and $B\to
D^*\,l\,\overline\nu$.

Our analysis of leptonic decay constants shows that in a
nonrelativistic calculation the equality $f_VM_V= f_PM_P$ is satisfied
within 2\%. This equality is expected in HQS in the limit where the
heavy quark masses go to infinity. The nonrelativistic result suggests
that the HQS infinite mass limit sets in already at the $m_c$ scale,
something that is not supported by lattice data for $D$ mesons where
one finds deviations as large as 20\% from the above equality.

One also finds problems in the semileptonic $B\to D$ and $B\to D^*$
decays. We have seen how the $h_-(w)$ form factor of the $B\to D$
decay and the $h_{A_2}(w)$ form factor of the $B\to D^*$ decay are not
reliably calculated in the nonrelativistic quark model where one finds
large deviations from the HQET relations of Eq.(\ref{hiw}). We have
tried to remedy this failure by evaluating the Isgur-Wise function
from a form factor whose calculation in the quark model we trust:
$h_+(w)$ for the $B\to D$ decay, and $h_{A_1}(w)$ for the $B\to D^*$
decay. Those Isgur-Wise function were later used together with HQET
constraints in Eq.(\ref{hiw}) to recalculate all other form
factors. The two Isgur-Wise functions thus determined show an overall
reasonable agreement with lattice data but in both cases the slope
seems to be too small. This deficiency multiply its effects when one
goes to larger $w$ values and as a consequence the quantities
$F_D(w)\,|V_{cb}|$ and $F_{D^*}(w)\,|V_{cb}|$ go above experimental
data for $w$ values larger than 1.2, and for any reasonable value of
$|V_{cb}|$. A failure of some kind is expected in a nonrelativistic
calculation as $w$ increases. For $w=1.2$ the three momenta of the
final meson amounts to 66\% of its mass and relativistic corrections
in the wave function could start to be important. On the other hand we
believe our results are sound at zero recoil ($w=1$). That enables us
to obtain $|V_{cb}|=0.040\pm0.006$ from the $B\to D$ decay. This
result is in perfect agreement with the value $|V_{cb}|=0.040\pm0.005$
previously obtained using the same model in the analysis of the
$\Lambda_b\to \Lambda_c$ semileptonic decay in Ref
\cite{Albertus:2004wj}, and in good agreement with a recent
determination $|V_{cb}|=0.0414\pm0.0012\pm0.0021\pm0.0018$ by the
DELPHI Collaboration \cite{Abdallah:2004rz}, and with the average
value $|V_{cb}|=0.0416\pm0.0006$ favored by the PDG \cite{Yao:2006px}.
 The experimental situation concerning the $B\to D^*$ reaction is not
so clear as different experiments give values for
$F_{D^*}(w)\,|V_{cb}|$ which are hardly compatible. From DELPHI
Collaboration data \cite{Abdallah:2004rz} we would get
$|V_{cb}|=0.040\pm0.003$ in agreement with the above result.

Finally we have made use of PCAC to evaluate the strong coupling
constants $g_{B^*B\pi}$ and $g_{D^*D\pi}$. Our results are larger than
experimental data or the results provided by lattice and QCDSR
calculations. In this case
the final meson is nearly at rest and one would  expect a nonrelativistic 
calculation to perform better. The main difference with the other observables
analyzed is that here the two active quarks are the light ones. 
 On the other hand the value for the 
 ratio $R=\frac{g_{B^*B\pi}(0)\ f_{B^*}\,\sqrt{m_D}}
{g_{D^*D\pi}(0)\ f_{D^*}\,\sqrt{m_B}} $ in
 Eq.~(\ref{R}) agrees with the  HQS prediction and with the one evaluated using
 a combination of lattice and experimental data, being also close to a QCDSR
 determination.

Results of this chapter were published in Ref.\cite{Albertus:2005vd}

%---------------------------------------------------
%---------------------------------------------------
%---------------------------------------------------
%---------------------------------------------------

% BC

\bigskip

$\bullet$ In chapter \ref{cha:Bc} we have made a comprehensive and
exhaustive study of exclusive semileptonic and nonleptonic two--meson
decays of the $B_c$ meson. We have left out semileptonic processes
involving a $b\to u$ transition at the quark level to avoid known
deficiencies both at high and low $q^2$ transfers (see section
\ref{Bc_intro}). For similar reasons we have only considered
two--meson nonleptonic decay channels that include a $c\bar c$ or $B$
meson in the final state.  Our model respects HQSS constraints in the
infinite heavy quark mass limit but hints at sizable corrections away
from that limit for some form factors. Unfortunately such corrections
have not been worked out in perturbative QCD as they have been for
heavy--light mesons~\cite{Neubert:1992tg}.

Our results for the observables analyzed are in a general good
agreement (whenever comparison is possible) with the results obtained
within the quasi-potential approach to the relativistic quark model of
Ebert {\it et al.}~\cite{Ebert:2003cn,Ebert:2003wc}.

The branching ratios for the leptonic $B_c^-\to c\bar c$ and
 $B_c^-\to\overline{B}$ decays are also in reasonable agreement with
the relativistic constituent quark model results of Ivanov {\it et
 al.}~\cite{Ivanov:2006ni,Ivanov:2005fd}.

For the nonleptonic $B_c^-\to \eta_c\,M_F^-$ and $B_c^-\to
J/\Psi\,M_F^-$ two--meson decay channels with
$M_F^-=\pi^-,\,\rho^-,\,K^-,\,K^{*-}$, we find also reasonable
agreement (better for the $J/\Psi$ channel) with the Bethe--Salpeter
calculation by El-Hady {\it et al.}~\cite{AbdEl-Hady:1999xh} and the
light front calculation by Anisimov {\it et
al.}~\cite{Anisimov:1998xv}, while our results are a factor of two
smaller than the ones by Ivanov {\it et al.}~\cite{Ivanov:2006ni}, and
Chang {\it et al.}  ~\cite{Chang:1992pt,Chang:2001pm,Chang:2001vq},
the latter obtained within the nonrelativistic approach to the
Bethe--Salpeter equation. For the two--meson decay channels with
$\chi_{c0},\,\chi_{c1},\,h_c,\,\chi_{c2}$ or $\Psi(3836)$ as the final
$c\bar c$ meson and $M_F=\pi^-,\,\rho^-,\,K^-,\,K^{*-}$ our results
are generally a factor of two smaller than the ones of Ivanov {\it et
al.}~\cite{Ivanov:2006ni}, whereas for some channels
($\chi_{c0}\,\pi^-$, $\chi_{c0}\,\rho^-$, $h_c\,\pi^-$, $h_c\,\rho^-$,
$\chi_{c2}\,\pi^-$ and $\chi_{c2}\,\rho^-$) we find very good
agreement with the results by Chang {\it et
al.}~\cite{Chang:1992pt,Chang:2001pm,Chang:2001vq}. The disagreement
with Ivanov {\it et al.}  extends to the two-meson decay channels with
 final $c\bar c$ and $D$ mesons. There we find good agreement with
the results by El-Hady {\it et al.}~\cite{AbdEl-Hady:1999xh} and the
ones obtained within the sum rules of QCD and nonrelativistic QCD by
Kiselev~\cite{Kiselev:2002vz}.

As for the two--meson nonleptonic decay channels $B_c^-\to
\overline{B}\, M_F$ with $M_F=\pi^-,\,\rho^-,\,K^-,\,K^{*-}$ we are in
a reasonable good agreement with the results by Anisimov {\it et
al.}~\cite{Anisimov:1998xv}.  For the case of a final
$\overline{B}^{0}_s$ or $\overline{B}^{*0}_s$ the agreement with the
results by Ivanov {\it et al.}~\cite{Ivanov:2006ni} is very good. For
the $B_c^-\to B^-\ M_F$ case with $M_F=\pi^0,\,\rho^0,\,K^0,\,K^{*0}$,
and apart from the results by Ebert {\it et al.}~\cite{Ebert:2003wc},
we find no global agreement with other calculations, neither do they
agree with each other.

From the above comparison one sees that there are different models
producing sometimes very different results for the same
observables. Accurate experimental data is needed to  shed light into this
issue.

Results in this chapter were published in Ref. \cite{Hernandez:2006gt}.

%---------------------------------------------------------------------
%---------------------------------------------------------------------
%---------------------------------------------------------------------
%---------------------------------------------------------------------

% AQUI VAN LAS PROPIEDADES ESTATICAS

\bigskip

$\bullet$  In chapter \ref{cha:dp_masas}
we have computed wave functions, spectra and other static properties 
of doubly heavy baryons. In order to
build our wave functions we have made use of the constraints imposed
by the infinite heavy quark mass limit. In this limit the spin--spin
interactions vanish and the total spin of the two heavy quarks is well
defined. With this approximation we have used a simple variational
approach, with Jastrow type orbital wave functions, to solve the
involved three-body problem. We have also checked that our  wave
functions have the correct behavior in the infinite heavy quark mass limit.

Among the static properties, our results for the masses are in very
good agreement with previous results obtained with the same
inter-quark interactions but within a more complicated Faddeev
approach~\cite{Silvestre-Brac:1996bg}. In some cases we even get
lower, and thus better, masses. We have calculated charge densities
(charge form factors) finding that the corresponding mean square radii
are again in good agreement with the Faddeev calculation of
Ref.~\cite{Silvestre-Brac:1996bg}.  We have also evaluated magnetic
moments. Being the total orbital angular momentum of the baryon $L=0$,
the magnetic moments come from the spin contributions alone.  With the
exception of $\Xi_{bc}^0$, $\Xi_{bc}^+$ and $\Omega_{bc}^0$ we agree
perfectly with the Faddeev calculation in
Ref.~\cite{Silvestre-Brac:1996bg}. For the magnetic moments of
$\Xi_{bc}^0$, $\Xi_{bc}^+$ and $\Omega_{bc}^0$ the discrepancies
between the two calculation are very large. The origin might be
attributed to the possible presence of a non--negligible $S_h=0$
component in the wave functions of Ref.~\cite{Silvestre-Brac:1996bg}.
In our case we have $S_h=1$ which we think is a good approximation
based on the infinite heavy quark mass limit. This assertion seems to
be corroborated by the results obtained in the relativistic
calculation of Ref.~\cite{Faessler:2006ft}, at least for the
$\Xi_{bc}^0$ and $\Omega_{bc}^0$ cases.

Results in this chapter were published in \cite{Albertus:2006ya}.

% AQUI VAN LAS DESINTEGRACIONES 
\bigskip $\bullet$ In chapter
\ref{cha:dp_sl}, and using the wave functions obtained in chapter
\ref{cha:dp_masas}, we have evaluated form factors, decay widths and
angular asymmetry parameters for semileptonic decays involving the
ground state of doubly heavy $\Xi$ and $\Omega$ baryons.  We have a
small vector current violation by an amount given by the binding
energy over the mass of the baryon that disappears in the infinite
heavy quark mass limit.

Our results for the total decay widths are in reasonable agreement
with the ones obtained in Ref.~\cite{Ebert:2004ck} using a
relativistic quark model in the quark--diquark approximation, while
they are much smaller than the ones obtained in Ref.~\cite{Guo:1998yj}
by means of the Bethe--Salpeter equation applied to a quark-diquark
system.

For the weak form factors the results exhibit an apparent SU(3)
symmetry when going from $\Xi$ to $\Omega$ baryons. This is due to the
fact that we have two heavy quarks and the light one acts as a
spectator in the weak transition. This apparent symmetry appears also
in the decay widths and asymmetry parameters. On the other hand SU(3)
violating effects are clearly visible in some static quantities like
charge form factors and radii and magnetic moments, that depend
strongly on the light quark charge and/or mass.

Results in this chapter were published in \cite{Albertus:2006ya}.

%--------------------------------------------------------------
%--------------------------------------------------------------
%--------------------------------------------------------------

\bigskip

$\bullet$ Finally in chapter \ref{cha:lambdapi} we have evaluated the
widths for the strong decays $\Sigma_c^*\to\Lambda_c\pi$,
$\Sigma_c\to\Lambda_c\pi$ and $\Xi_c^*\to\Xi_c\pi$. We use HQS
constrained wave functions taken from Ref \cite{Albertus:2003sx} and
that were obtained solving the nonrelativistic three--body problem in
a similar way as the one used in chapter \ref{cha:dp_masas}.

The pion emission amplitude has been obtained with
the use of PCAC from the analysis of weak current matrix elements.
Our results are rather stable against the quark--quark interaction,
with variations in the decay widths at the level of 6--8\%. We find an
overall good agreement with experiment for the three reactions. This
agreement is, in most cases, better than the one obtained by other
models.  

Results in this chapter were published in Ref. \cite{Albertus:2005zy}.

\bigskip

For all single and double heavy hadrons studied in this thesis, we
have shown that NRQM calculations respect the HQS and HQSS constraints
obtained in the infinite heavy quark mass limit. On the other hand the
deviations from that limit for the actual heavy quark masses are not
always in agreement with HQET predictions, and we had to include
corrections at some instances to improve the NRQM results.  On the
other hand we find NRQM results for many observables agree with
experiment, lattice results and quark model calculations where
relativity is explicitly included.

\appendix
%\chapter{Particle states with definite helicity}
%\input{helicity.inp}
% Los del BD (los meto a lo burro) 
\input{epsilons_app.inp}
{\small
\chapter{Matrix elements for the leptonic decay of pseudo scalars and vector
mesons}
\chaptermark{Matrix elements for leptonic decays}
\label{app:lep}
The  matrix element needed for the evaluation of the pseudoscalar decay constant
is given by
\begin{eqnarray}
\nonumber  %\hspace{-1.7cm}
\left\langle 0 \left|\, J_{A\, 0}^{f_1\,f_2}(0)
\,\right|\, P, \vec{0}\,
\right\rangle_{NR}%&&=
%\\
&&= \sqrt3\int\,d^3p\sum_{s_1, s_2}
\hat{\phi}^{(P)}_{(s_1,f_1),\,(s_2,f_2)}(\,\vec{p}\,)
\frac{(-1)^{\frac{1}{2}-s_2}}{(2\pi)^{\frac{3}{2}}
\sqrt{2E_{f_1}(\vec{p}\,)2E_{f_2}(\vec{p}\,)}}\nonumber\\
&&\hspace{3cm}\bar{v}_{s_2,f_2}(\,\vec{p}\,)\, 
\gamma_0\gamma_5\,u_{s_1,f_1}(-\vec{p}\,)
\nonumber\\
&&=\sqrt3\int\,d^3p\sum_{s_1, s_2}
\hat{\phi}^{(P)}_{(s_1,f_1),\,(s_2,f_2)}(\,\vec{p}\,)
\frac{(-1)^{\frac{1}{2}-s_2}}{(2\pi)^{\frac{3}{2}}
\sqrt{2E_{f_1}(\vec{p}\,)2E_{f_2}(\vec{p}\,)}}\nonumber\\
&&\hspace{3cm}\bar{u}_{-s_2,f_2}(\,\vec{p}\,)\, 
\gamma_0\,u_{s_1,f_1}(-\vec{p}\,)\nonumber\\
&&=i\,\frac{\sqrt3}{\pi}\int_0^\infty d|\vec{p}\,| \
\hat{\Phi}^{(P)}_{f_1,\,f_2}(|\vec{p}\,|)\,|\vec{p}\,|^2
\sqrt{\frac{(E_{f_1}(\vec{p}\,)+m_{f_1})(E_{f_2}(\vec{p}\,)+m_{f_2})}
{4E_{f_1}(\vec{p}\,)E_{f_2}(\vec{p}\,)}}\nonumber\\
&&\hspace{3cm}\left(
1-\frac{|\vec{p}\,|^2}{(E_{f_1}(\vec{p}\,)+m_{f_1})(E_{f_2}(\vec{p}\,)+m_{f_2})}\right)
\end{eqnarray}
where we have used the fact that $v_{s,f}(\,\vec{p}\,)=
\gamma_5\,u_{-s,f}(\,\vec{p}\,)$.\\
\newpage
Similarly for the vector meson case
\begin{eqnarray}
\left\langle 0 \left|\, J_{V\, 3}^{f_1\,f_2}(0)
\,\right|\, V,\, 0\, \vec{0}\,
\right\rangle_{NR}
&&\hspace{-.6cm}= \sqrt3\int\,d^3p\sum_{s_1, s_2}
\hat{\phi}^{(V,\,0)}_{(s_1,f_1),\,(s_2,f_2)}(\,\vec{p}\,)
\frac{(-1)^{\frac{1}{2}-s_2}}{(2\pi)^{\frac{3}{2}}
\sqrt{2E_{f_1}(\vec{p}\,)2E_{f_2}(\vec{p}\,)}}\nonumber\\
&&\hspace{-.6cm}\hspace{2cm}\bar{v}_{s_2,f_2}(\,\vec{p}\,)\, 
\gamma_3\,u_{s_1,f_1}(-\vec{p}\,)
\nonumber\\
&&\hspace{-.6cm}=\sqrt3\int\,d^3p\sum_{s_1, s_2}
\hat{\phi}^{(V,0)}_{(s_1,f_1),\,(s_2,f_2)}(\,\vec{p}\,)
\frac{(-1)^{\frac{1}{2}-s_2}}{(2\pi)^{\frac{3}{2}}
\sqrt{2E_{f_1}(\vec{p}\,)2E_{f_2}(\vec{p}\,)}}\nonumber\\
&&\hspace{-.6cm}\hspace{2cm}\bar{u}_{-s_2,f_2}(\,\vec{p}\,)\, 
\gamma_3\gamma_5\,u_{s_1,f_1}(-\vec{p}\,)\nonumber\\
&&\hspace{-.6cm}=\frac{-\sqrt3}{\pi}\int_0^\infty d|\vec{p}\,| \
\hat{\Phi}^{(V)}_{f_1,\,f_2}(|\vec{p}\,|)\,|\vec{p}\,|^2
\sqrt{\frac{(E_{f_1}(\vec{p}\,)+m_{f_1})(E_{f_2}(\vec{p}\,)+m_{f_2})}
{4E_{f_1}(\vec{p}\,)E_{f_2}(\vec{p}\,)}}\nonumber\\
&&\hspace{-.6cm}\hspace{2cm}\left(
1+\frac{|\vec{p}\,|^2}{3(E_{f_1}(\vec{p}\,)+m_{f_1})(E_{f_2}(\vec{p}\,)+m_{f_2})}\right)
\end{eqnarray}

%-------------------------------------------------------------------
%-------------------------------------------------------------------
%-------------------------------------------------------------------

\chapter{Expression for the $V^\mu(|\vec{q}\,|)$ matrix element}
\chaptermark{Expression for  $V^\mu(|\vec{\lowercase{q}}\,|)$ }
\label{app:v}

In this appendix we give the expressions 
for the $V^\mu(|\vec{q}\,|)$ matrix elements introduced in chapter
\ref{cha:BD} for the  
 $B\to D$ decay. They are just a particular
case of the ones presented in appendix \ref{Bcapp:va}

\begin{eqnarray}
&&\hspace*{-1.8cm}V^\mu(|\vec{q}\,|)\,=\,\sqrt{2m_B2E_D(-\vec{q})}\
_{\stackrel{}{NR}}
\left\langle\, D,\,
-|\vec{q}\,|\,{\vec{k}}\,\bigg|\, (J^{c\,b}_V)^\mu(0)\,
\bigg| \, B,\,\vec{0}\right\rangle_{NR}\nonumber\\
&&\hspace*{-1.5cm}=\,\sqrt{2m_B2E_D(-\vec{q})}\ \int\,d^3p\sum_{s_1, s_2}
\left(\hat{\phi}^{(D)}_{(s_1,c),\,(s_2,f_2)}(\,\frac{m_{f_2}}{m_c+m_{f_2}}\,
|\vec{q}\,|{\vec{k}}+
\vec{p}\,)\right)^*\ 
%\nonumber\\
%&&\hspace{3.75cm}
\sum_{s_1^\prime}\hat{\phi}^{(B)}_{(s_1^\prime,b),\,(s_2,f_2)}(\,\vec{p}\,)
\hspace*{-2cm}
\nonumber\\
&&\hspace{1.75cm}\times\frac{1}{\sqrt{2E_c(|\vec{q}\,|{\vec{k}}+\vec{p}\,)2E_b(\vec{p}\,)}}%\nonumber\\
%&&\hspace{3cm}
\,\bar{u}_{s_1,c}(-|\vec{q}\,|\,
{\vec{k}}-\vec{p}\,)\, 
\gamma^\mu\,u_{s_1^\prime,b}(-\vec{p}\,)
%V^3=\sqrt{2M_B2E_D(|\vec{q}\,|)}\ {}_{\stackrel{}{NR}}\left\langle\, D,\,-|\vec{q}\,|\,\vec{k}\,\left|\, (J^{c\,b}_V)^3(0)\,
%\right| \, B,\,\vec{0}\right\rangle_{NR}
\end{eqnarray}
where $f_2$ represents  a light $u$ or $d$ quark. 

Going a little further we have
\begin{eqnarray}
V^0(|\vec{q}\,|)&=&\sqrt{2m_B2E_D(-\vec{q})}\ \int\,d^3p\ \frac{1}{4\pi}
\left(\hat{\phi}^{(D)}_{c,\,f_2}(|\,\frac{m_{f_2}}{m_c+m_{f_2}}\,
|\vec{q}\,|{\vec{k}}+
\vec{p}\,|)\right)^*\,
 \hat{\phi}^{(B)}_{b,\,f_2}(|\,\vec{p}\,|)\nonumber\\
&&\hspace{1cm}\times\sqrt{\frac{\bigg(E_{c}(|\vec{q}\,|{\vec{k}}+\vec{p}\,)
+m_{c}\bigg)\ 
\bigg(E_{b}(\vec{p}\,)+m_{b}\bigg)}{4E_{c}(|\vec{q}\,|{\vec{k}}+\vec{p}\,)
\ E_{b}(\vec{p}\,)}}
\nonumber\\
&&\hspace{2cm}\times\left(
1+\frac{|\vec{p}\,|^2 +p_z\,|\vec{q}\,|}{\bigg(E_{c}(|\vec{q}\,|
{\vec{k}}+\vec{p}\,)+m_{c}\bigg)\ \bigg(E_{b}(\vec{p}\,)+m_{b}\bigg)}\right) 
\end{eqnarray}
In the case of equal masses $m_b=m_c$ and for $|\vec{q}\,|=0\ (w=1)$ we will
obtain
\begin{eqnarray}
\left.V^0(|\vec{q}\,|)\right| _{|\vec{q}\,|=0}&=&2m_B
\end{eqnarray}
Similarly
\begin{eqnarray}
V^3(|\vec{q}\,|)&=&-\sqrt{2m_B2E_D(-\vec{q})}\ \int\,d^3p\ \frac{1}{4\pi}
\left(\hat{\phi}^{(D)}_{c,\,f_2}(|\,\frac{m_{f_2}}{m_c+m_{f_2}}\,
|\vec{q}\,|{\vec{k}}+
\vec{p}\,|)\right)^*\,
 \hat{\phi}^{(B)}_{b,\,f_2}(|\,\vec{p}\,|)\nonumber\\
&&\hspace{1cm}\times\sqrt{\frac{\bigg(E_{c}(|\vec{q}\,|{\vec{k}}+\vec{p}\,)
+m_{c}\bigg)\ 
\bigg(E_{b}(\vec{p}\,)+m_{b}\bigg)}{4E_{c}(|\vec{q}\,|{\vec{k}}+\vec{p}\,)\ E_{b}(\vec{p}\,)}}
\nonumber\\ 
&& \hspace{2cm} \times
\left(\frac{p_z}{E_{b}(\vec{p}\,)+m_{b}}+
\frac{p_z +|\vec{q}\,|}{E_{c}(|\vec{q}\,|
{\vec{k}}+\vec{p}\,)+m_{c}}
\right) 
\end{eqnarray}

%-------------------------------------------------------------------------------
%-------------------------------------------------------------------------------
%-------------------------------------------------------------------------------
%-------------------------------------------------------------------------------

\chapter{Expressions for the $V^{(*)}_{\lambda,\,\mu}$ and 
$A^{(*)}_{\lambda,\,\mu}$ matrix elements}
\chaptermark{Expressions for  $V^{(*)}_{\lambda,\,\mu}$ and 
$A^{(*)}_{\lambda,\,\mu}$}
\label{app:va}

In this appendix 
we present expressions for the $V^{(*)}_{\lambda,\,\mu}$ and 
$A^{(*)}_{\lambda,\,\mu}$ matrix elements
introduced in chapter \ref{cha:BD} for the semileptonic
 $B\to D^*$ decay. They are just a particular
case of the ones presented in appendix \ref{Bcapp:va}

\begin{eqnarray}
%\label{va123}
&&\hspace*{-.8cm }V^{(*)}_{\lambda,\,\mu}(|\vec{q}\,|)\,=\,\sqrt{2m_B2E_{D^*}(-\vec{q})}\ {}_{\stackrel{}{NR}}
\left\langle\, D^*,\,\lambda\ 
-|\vec{q}\,|\,{\vec{k}}\,\bigg|\, (J^{c\,b}_V)_\mu(0)\,
\bigg| \, B,\,\vec{0}\right\rangle_{NR}\nonumber\\
&&\hspace*{-.5cm}
=\,\sqrt{2m_B2E_{D^*}(-\vec{q})}\ \int\,d^3p\sum_{s_1, s_2}
\left(\hat{\phi}^{(D^*,\,\lambda)}_{(s_1,c),\,(s_2,f_2)}(\,
\frac{m_{f_2}}{m_c+m_{f_2}}\,
|\vec{q}\,|{\vec{k}}+
\vec{p}\,)\right)^*
%\nonumber\\
%&&\hspace{3.9cm}
\ \sum_{s_1^\prime}\hat{\phi}^{(B)}_{(s_1^\prime,b),\,(s_2,f_2)}(\,\vec{p}\,)
\hspace*{-3cm}\nonumber\\
&&\hspace*{-.5cm}\hspace{3.9cm}\,\frac{1}{\sqrt{2E_c(|\vec{q}\,|\,
{\vec{k}}+\vec{p}\,)\,2E_b(\vec{p}\,)}}%\nonumber\\
%&&\hspace{3cm}
\,\bar{u}_{s_1,c}(-|\vec{q}\,|\,
{\vec{k}}-\vec{p}\,)\, 
\gamma_\mu\,u_{s_1^\prime,b}(-\vec{p}\,)\nonumber\\
&&\hspace*{-.8cm} A^{(*)}_{\lambda,\,\mu}(|\vec{q}\,|)\,=\,\sqrt{2m_B2E_{D^*}(-\vec{q})}\ {}_{\stackrel{}{NR}}
\left\langle\, D^*,\,\lambda\ 
-|\vec{q}\,|\,{\vec{k}}\,\bigg|\, (J^{c\,b}_A)_\mu(0)\,
\bigg| \, B,\,\vec{0}\right\rangle_{NR}\nonumber\\
&&\hspace{-.5cm}
=\,\sqrt{2m_B2E_{D^*}(-\vec{q})}\ \int\,d^3p\sum_{s_1, s_2}
\left(\hat{\phi}^{(D^*,\,\lambda)}_{(s_1,c),\,(s_2,f_2)}(\,
\frac{m_{f_2}}{m_c+m_{f_2}}\,
|\vec{q}\,|{\vec{k}}+
\vec{p}\,)\right)^*
%\nonumber\\
%&&
\ \sum_{s_1^\prime}\hat{\phi}^{(B)}_{(s_1^\prime,b),\,(s_2,f_2)}(\,\vec{p}\,)
\nonumber\\
&&\hspace*{-.5cm}\hspace{3.9cm}\,\frac{1}{\sqrt{2E_c(|\vec{q}\,|\,
{\vec{k}}+\vec{p}\,)\,2E_b(\vec{p}\,)}}%\nonumber\\
%&&\hspace{3cm}
\,\bar{u}_{s_1,c}(-|\vec{q}\,|\,
{\vec{k}}-\vec{p}\,)\, 
\gamma_\mu\gamma_5\,u_{s_1^\prime,b}(-\vec{p}\,)\nonumber\\
%V^3=\sqrt{2M_B2E_D(|\vec{q}\,|)}\ {}_{\stackrel{}{NR}}\left\langle\, D,\,-|\vec{q}\,|\,\vec{k}\,\left|\, (J^{c\,b}_V)^3(0)\,
%\right| \, B,\,\vec{0}\right\rangle_{NR}
\end{eqnarray}
So that
\begin{eqnarray}
V^{(*)}_{-1,\,2}(|\vec{q}\,|)
&=&-\frac{1}{\sqrt2}\sqrt{2m_B2E_{D^*}(-\vec{q})}\ \int\,d^3p\ \frac{1}{4\pi}
\left(\hat{\phi}^{(D^*)}_{c,\,f_2}(|\,
\frac{m_{f_2}}{m_c+m_{f_2}}\,
|\vec{q}\,|{\vec{k}}+
\vec{p}\,|\,)\right)^*
\hat{\phi}^{(B)}_{b,\,f_2}(|\,\vec{p}\,|)\nonumber\\
&&\hspace{1cm}\times\sqrt{\frac{\bigg(E_{c}(|\vec{q}\,|{\vec{k}}+\vec{p}\,)
+m_{c}\bigg)\ 
\bigg(E_{b}(\vec{p}\,)+m_{b}\bigg)}{4E_{c}(|\vec{q}\,|{\vec{k}}+\vec{p}\,)\ E_{b}(\vec{p}\,)}}
\nonumber\\ &&\hspace{2cm}\times
\left(
\frac{p_z}{E_{b}(\vec{p}\,)+m_{b}}-
\frac{p_z+|\vec{q}\,|}{E_{c}(|\vec{q}\,|
{\vec{k}}+\vec{p}\,)+m_{c}}\right)\nonumber\\
\end{eqnarray}%
\begin{eqnarray}
A^{(*)}_{-1,\,1}(|\vec{q}\,|)
&=&-\frac{i}{\sqrt2}\sqrt{2m_B2E_{D^*}(-\vec{q})}\ \int\,d^3p\ \frac{1}{4\pi}
\left(\hat{\phi}^{(D^*)}_{c,\,f_2}(|\,
\frac{m_{f_2}}{m_c+m_{f_2}}\,
|\vec{q}\,|{\vec{k}}+
\vec{p}\,|\,)\right)^*
\hat{\phi}^{(B)}_{b,\,f_2}(|\,\vec{p}\,|)\nonumber\\
&&\hspace{1cm}\times \sqrt{\frac{\bigg(E_{c}(|\vec{q}\,|{\vec{k}}+\vec{p}\,)
+m_{c}\bigg)\ 
\bigg(E_{b}(\vec{p}\,)+m_{b}\bigg)}{4E_{c}(|\vec{q}\,|{\vec{k}}+\vec{p}\,)\ E_{b}(\vec{p}\,)}}
\nonumber\\  &&\hspace{2cm}\times
\left(1+
\frac{2p_x^2-|\,\vec{p}\,|^2-p_z\,|\vec{q}\,|}{\bigg(E_{c}(|\vec{q}\,|
{\vec{k}}+\vec{p}\,)+m_{c}\bigg)\,\bigg(E_{b}(\vec{p}\,)+m_{b}\bigg)}
\right)\nonumber\\ 
\end{eqnarray}%
%
%\vspace{2cm}
\begin{eqnarray}
A^{(*)}_{0,\,0}(|\vec{q}\,|)
&=&-i\sqrt{2m_B2E_{D^*}(-\vec{q})}\ \int\,d^3p\ \frac{1}{4\pi}
\left(\hat{\phi}^{(D^*)}_{c,\,f_2}(|\,
\frac{m_{f_2}}{m_c+m_{f_2}}\,
|\vec{q}\,|{\vec{k}}+
\vec{p}\,|\,)\right)^*
\hat{\phi}^{(B)}_{b,\,f_2}(|\,\vec{p}\,|)\nonumber\\
&&\hspace{1cm}\times \sqrt{\frac{\bigg(E_{c}(|\vec{q}\,|{\vec{k}}+\vec{p}\,)
+m_{c}\bigg)\ 
\bigg(E_{b}(\vec{p}\,)+m_{b}\bigg)}{4E_{c}(|\vec{q}\,|{\vec{k}}+\vec{p}\,)\
 E_{b}(\vec{p}\,)}}
\nonumber \\  &&\hspace{2cm}\times 
\left(\frac{p_z}{E_{b}(\vec{p}\,)+m_{b}}+
\frac{p_z+|\vec{q}\,|}{E_{c}(|\vec{q}\,|
{\vec{k}}+\vec{p}\,)+m_{c}}\right)\nonumber\\
\end{eqnarray}
\begin{eqnarray}
A^{(*)}_{0,\,3}(|\vec{q}\,|)
&=&-i\sqrt{2m_B2E_{D^*}(-\vec{q})}\ \int\,d^3p\ \frac{1}{4\pi}
\left(\hat{\phi}^{(D^*)}_{c,\,f_2}(|\,
\frac{m_{f_2}}{m_c+m_{f_2}}\,
|\vec{q}\,|{\vec{k}}+
\vec{p}\,|\,)\right)^*
\hat{\phi}^{(B)}_{b,\,f_2}(|\,\vec{p}\,|)\nonumber\\
&&\hspace{1cm}\times \sqrt{\frac{\bigg(E_{c}(|\vec{q}\,|{\vec{k}}+\vec{p}\,)
+m_{c}\bigg)\ 
\bigg(E_{b}(\vec{p}\,)+m_{b}\bigg)}{4E_{c}(|\vec{q}\,|{\vec{k}}+\vec{p}\,)\ E_{b}(\vec{p}\,)}}
\nonumber\\  &&\hspace{2cm}\times
\left(1+
\frac{2p_z^2-|\,\vec{p}\,|^2+p_z\,|\vec{q}\,|}{\bigg(E_{c}(|\vec{q}\,|
{\vec{k}}+\vec{p}\,)+m_{c}\bigg)\,\bigg(E_{b}(\vec{p}\,)+m_{b}\bigg)}
\right)
\end{eqnarray}
}  %HECHO      
%             APENDICES APENDICES APENDICES APENDICES
%
{\small

%
% APENDICE b -- APENDICE b --  APENDICE b -- APENDICE b -- APENDICE b 
%
%
\chapter{Expressions for the $V^\mu(|\vec{q}\,|),\,V^\mu_{(\lambda)}(|\vec{q}\,|)
,\,V^\mu_{T(\lambda)}(|\vec{q}\,|)$ and 
$A^\mu(|\vec{q}\,|)$,$\,A^\mu_{(\lambda)}(|\vec{q}\,|)$,
$\,A^\mu_{T(\lambda)}(|\vec{q}\,|)$ 
matrix elements}
\chaptermark{$V^\mu(|\vec{\lowercase{q}}\,|),\,V^\mu_{(\lambda)}(|\vec{\lowercase{q}}\,|)
,\,V^\mu_{T(\lambda)}(|\vec{\lowercase{q}}\,|) 
A^\mu(|\vec{\lowercase{q}}\,|),\,A^\mu_{(\lambda)}(|\vec{\lowercase{q}}\,|)
,\,A^\mu_{T(\lambda)}(|\vec{\lowercase{q}}\,|)$ 
}

\label{Bcapp:va}

\section{Transitions involving quarks}

Here we give general expressions valid for transitions between a pseudoscalar
meson $M_I$ at rest with quark content $q_{f_1}\overline{q}_{f_2}$ and a final 
$M_F$ meson with total angular momentum and parity $J^\pi=0^-,0^+,1^-,1^+,2^-,2^+$,
three-momentum $-|\vec{q}\,|\vec{k}$ and  quark content
$q_{f'_1}\overline{q}_{f_2}$. In the  transition it is the quark that
changes flavor. The phases of the wave functions are the ones chosen in
Eqs.~(\ref{eq:0pm},\ref{eq:1pm},\ref{eq:2pm}).
We generally have
\begin{eqnarray}
&&\hspace*{-.7cm}{\cal V}^\mu(|\vec{q}\,|)-{\cal A}^\mu(|\vec{q}\,|)\,=\,\sqrt{2m_{I}2E_{F}(-\vec{q\,})}\
{}_{\stackrel{}{\stackrel{}{NR}}}
\left\langle\, M_F(J^\pi),\,\lambda\,
-|\vec{q}\,|\,\vec{k}\,\bigg|\, J^{f'_1\,f_1\, \mu}(0)\,
\bigg| \, M_I(0^-),\,\vec{0}\right\rangle_{NR}\nonumber\\
&&\hspace*{-.7cm}=\,\sqrt{2m_{I}2E_{F}(-\vec{q\,})}\ \int\,d^3p\sum_{s'_1}\sum_{s_1,s_2}
\left(\hat{\phi}^{(M_F(J^\pi),\lambda)}_{(s'_1,f'_1),\,(s_2,f_2)}(\vec{p}\,)\right)^*\ 
\hat{\phi}^{(M_I(0^-))}_{(s_1,f_1),\,(s_2,f_2)}(\vec{p}
-\frac{m_{f_2}}{m_{f'_1}+m_{f_2}}\,|\vec{q}\,|\vec{k})
\nonumber\\
&&\hspace*{-.7cm}\hspace{1cm}\times\frac{1}
{\sqrt{2E_{f'_1}\,2E_{f_1}}}\ \bar{u}_{s'_1,f'_1}(-\frac{m_{f'_1}}{m_{f'_1}+m_{f_2}}
|\vec{q}\,|\vec{k}-\vec{p}\,)\, 
\gamma^\mu(I-\gamma_5)\,u_{s_1,f_1}(\frac{m_{f_2}}{m_{f'_1}+m_{f_2}}
|\vec{q}\,|\vec{k}-\vec{p}\,)\nonumber\\
\end{eqnarray}
where ${\cal V}^\mu ({\cal A}^\mu)$ represent any of the $V^\mu\, (A^\mu)$,
$V^\mu_{(\lambda)}\, (A^\mu_{(\lambda)})$ or
$V^\mu_{T(\lambda)}\, (A^\mu_{T(\lambda)})$,
and where $E_{f'_1}$ and $E_{f_1}$ are shorthand notations for
$E_{f'_1}(-\frac{m_{f'_1}}{m_{f'_1}+m_{f_2}}
|\vec{q}\,|\vec{k}-\vec{p}\,)$ and $E_{f_1}(\frac{m_{f_2}}{m_{f'_1}+m_{f_2}}
|\vec{q}\,|\vec{k}-\vec{p}\,)$ respectively. Defining also
$\widehat{E}_{f'_1}=E_{f'_1}+m_{f'_1}$ and $\widehat{E}_{f_1}=E_{f_1}+m_{f_1}$
 we arrive at the following final expressions:
\newpage
\begin{itemize}
\item {\large Case $J^\pi=0^-$}
\end{itemize}
\begin{eqnarray}
&&\hspace*{-.7cm}V^0(|\vec{q}\,|)\,=\,\sqrt{2m_I2E_F(-\vec{q}\,)}\ \int\,d^3p\ \frac{1}{4\pi}
\left(\hat{\phi}^{(M_F(0^-))}_{f'_1,\,f_2}(|\vec{p}\,|)\right)^*\,
\hat{\phi}^{(M_I(0^-))}_{f_1,\,f_2}\left(\bigg|\,\vec{p}-\frac{m_{f_2}}{m_{f'_1}+m_{f_2}}
|\vec{q}\,|
\vec{k} \bigg|\right)\hspace*{-2cm}
\nonumber\\
&&\hspace{3cm}\times\sqrt{\frac{\widehat{E}_{f'_1}\widehat{E}_{f1}}{4E_{f'_1}
E_{f_1}}}
%\nonumber\\
\left(
1+\frac{(-\frac{m_{f'_1}}{m_{f'_1}+m_{f_2}}\,|\vec{q}\,|\vec{k}-\vec{p}\,)
\cdot(\frac{m_{f_2}}{m_{f'_1}+m_{f_2}}\,|\vec{q}\,|\vec{k}-\vec{p}\,)}{\widehat{E}_{f'_1}\widehat{E}_{f_1}}
\right) \nonumber\\
&&\hspace*{-.7cm}V^3(|\vec{q}\,|)\,=\,\sqrt{2m_I2E_F(-\vec{q}\,)}\ \int\,d^3p\ \frac{1}{4\pi}
\left(\hat{\phi}^{(M_F(0^-))}_{f'_1,\,f_2}(|\vec{p}\,|)\right)^*\,
 \hat{\phi}^{(M_I
 (0^-))}_{f_1,\,f_2}\left(\bigg|\,\vec{p}-\frac{m_{f_2}}{m_{f'_1}+m_{f_2}}
|\vec{q}\,|
\vec{k} \bigg|\right) \hspace*{-2cm}
 \nonumber\\
&&\hspace{3cm}\times\sqrt{\frac{\widehat{E}_{f'_1}\widehat{E}_{f1}}{4E_{f'_1}
E_{f_1}}}%\nonumber\\
\left(\frac{\frac{m_{f_2}}{m_{f'_1}+m_{f_2}}\,|\vec{q}\,|-p_z}{\widehat{E}_{f_1}}+
\frac{-\frac{m_{f'_1}}{m_{f'_1}+m_{f_2}}\,|\vec{q}\,|-p_z}{\widehat{E}_{f'_1}}
\right) \nonumber\\
\end{eqnarray}
\begin{itemize}
\item {\large Case $J^\pi= 0^+$}
\end{itemize}
\begin{eqnarray}
&&\hspace*{-.9cm}
A^0(|\vec{q}\,|)\,=\,\sqrt{2m_I2E_F(-\vec{q}\,)}\ \int\,d^3p\ \frac{1}{4\pi
|\vec{p}\,|}
\left(\hat{\phi}^{(M_F(0^+))}_{f'_1,\,f_2}(|\vec{p}\,|)\right)^*\,
\hat{\phi}^{(M_I
(0^-))}_{f_1,\,f_2}\left(\bigg|\,\vec{p}-\frac{m_{f_2}}{m_{f'_1}+m_{f_2}}
|\vec{q}\,|
\vec{k} \bigg|\right)\nonumber\\
&&\hspace{1cm}\times\sqrt{\frac{\widehat{E}_{f'_1}\widehat{E}_{f1}}{4E_{f'_1}
E_{f_1}}}
%\nonumber\\
\left(\frac{\vec{p}\cdot(\frac{m_{f_2}}{m_{f'_1}+m_{f_2}}\,|\vec{q}\,|\vec{k}-\vec{p}\,)}{\widehat{E}_{f_1}}
+\frac{\vec{p}\cdot(-\frac{m_{f'_1}}{m_{f'_1}+m_{f_2}}\,|\vec{q}\,|\vec{k}-\vec{p}\,)}
{\widehat{E}_{f'_1}}
\right) \nonumber\\
&&\hspace*{-.9cm}
A^3(|\vec{q}\,|)\,=\,\sqrt{2m_I2E_F(-\vec{q}\,)}\ \int\,d^3p\
\frac{1}{4\pi|\vec{p}\,|}
\left(\hat{\phi}^{(M_F (0^+))}_{f'_1,\,f_2}(|\vec{p}\,|)\right)^*\,
\hat{\phi}^{(M_I(0^-))}_{f_1,\,f_2}\left(\bigg|\,\vec{p}-\frac{m_{f_2}}{m_{f'_1}+m_{f_2}}
|\vec{q}\,|
\vec{k} \bigg|\right)
 \nonumber\\
&&\hspace{1cm}\times\sqrt{\frac{\widehat{E}_{f'_1}\widehat{E}_{f1}}{4E_{f'_1}
E_{f_1}}}\bigg\{\
p_z\bigg(  
1-\frac{(-\frac{m_{f'_1}}{m_{f'_1}+m_{f_2}}\,|\vec{q}\,|\vec{k}-\vec{p}\,)
\cdot(\frac{m_{f_2}}{m_{f'_1}+m_{f_2}}\,|\vec{q}\,|\vec{k}-\vec{p}\,)}{\widehat{E}_{f'_1}\widehat{E}_{f_1}}
\bigg)\nonumber\\ 
&&\hspace{3.cm}+\frac{1}{\widehat{E}_{f'_1}\widehat{E}_{f1}}
\bigg[\hspace{.35cm} (-\frac{m_{f'_1}}{m_{f'_1}+m_{f_2}}|\vec{q}\,|-p_z)\ \ \
\vec{p}\cdot\bigg(\hspace{.4cm}\frac{m_{f_2}}{m_{f'_1}+m_{f_2}}|\vec{q}\,|\vec{k}-\vec{p}
\bigg)\nonumber\\
&&\hspace{4.75cm}+(\hspace{.4cm} \frac{m_{f_2}}{m_{f'_1}+m_{f_2}}|\vec{q}\,|-p_z
)\ \ \
\vec{p}\cdot\bigg(-\frac{m_{f'_1}}{m_{f'_1}+m_{f_2}}|\vec{q}\,|\vec{k}-\vec{p}
\bigg)
\bigg]\bigg\}\nonumber\\
\end{eqnarray}

\begin{itemize}
\item {\large Case $J^\pi=1^-$}
\end{itemize}

\begin{eqnarray}
\lefteqn{\hspace*{-1.5cm}V^{(1^-)\,1}_{\lambda=-1}(|\vec{q}\,|)\,=}
\nonumber\\&&\hspace*{-1.3cm}
=\,\frac{-i}{\sqrt2}
\sqrt{2m_I2E_F(-\vec{q}\,)}\ \int\,d^3p\ \frac{1}{4\pi}
\left(\hat{\phi}^{(M_F(1^-))}_{f'_1,\,f_2}(|\vec{p}\,|)\right)^*\,
\hat{\phi}^{(M_I(0^-))}_{f_1,\,f_2}\left(\bigg|\,\vec{p}-\frac{m_{f_2}}{m_{f'_1}+m_{f_2}}
|\vec{q}\,|
\vec{k} \bigg|\right)\nonumber \hspace*{-2cm}\\
&&\hspace{1.8cm}\times\sqrt{\frac{\widehat{E}_{f'_1}\widehat{E}_{f1}}{4E_{f'_1}
E_{f_1}}}
%\nonumber\\
\left(-\frac{\frac{m_{f_2}}{m_{f'_1}+m_{f_2}}\,|\vec{q}\,|-p_z}{\widehat{E}_{f_1}}+
\frac{-\frac{m_{f'_1}}{m_{f'_1}+m_{f_2}}\,|\vec{q}\,|-p_z}{\widehat{E}_{f'_1}}
\right)\nonumber\\
\end{eqnarray}
and similarly

{\allowdisplaybreaks
\begin{eqnarray}
\lefteqn{\hspace*{-0.8cm}A^{(1^-)\,0}_{\lambda=0}(|\vec{q}\,|)\,=}
\nonumber \\*
&&\hspace*{-1.3cm}=\,
i\
\sqrt{2m_I2E_F(-\vec{q}\,)}\ \int\,d^3p\ \frac{1}{4\pi}
\left(\hat{\phi}^{(M_F(1^-))}_{f'_1,\,f_2}(|\vec{p}\,|)\right)^*\,
\hat{\phi}^{(M_I(0^-))}_{f_1,\,f_2}\left(\bigg|\,\vec{p}-\frac{m_{f_2}}{m_{f'_1}+m_{f_2}}
|\vec{q}\,|
\vec{k} \bigg|\right)\nonumber\\*
&&\hspace{3cm}\times\sqrt{\frac{\widehat{E}_{f'_1}\widehat{E}_{f1}}{4E_{f'_1}
E_{f_1}}}
%\nonumber\\
\left(\frac{\frac{m_{f_2}}{m_{f'_1}+m_{f_2}}\,|\vec{q}\,|-p_z}{\widehat{E}_{f_1}}+
\frac{-\frac{m_{f'_1}}{m_{f'_1}+m_{f_2}}\,|\vec{q}\,|-p_z}{\widehat{E}_{f'_1}}
\right)\nonumber\\
\lefteqn{\hspace*{-.8cm}A^{(1^-)\,1}_{\lambda=-1}(|\vec{q}\,|)}
\nonumber\\* &&\hspace*{-1.3cm}=\,
\frac{i}{\sqrt2}
\sqrt{2m_I2E_F(-\vec{q}\,)}\ \int\,d^3p\ \frac{1}{4\pi}
\left(\hat{\phi}^{(M_F(1^-))}_{f'_1,\,f_2}(|\vec{p}\,|)\right)^*\,
\hat{\phi}^{(M_I(0^-))}_{f_1,\,f_2}\left(\bigg|\,\vec{p}-\frac{m_{f_2}}{m_{f'_1}+m_{f_2}}
|\vec{q}\,|
\vec{k} \bigg|\right)\nonumber\\*
&&\hspace{3cm}\times\sqrt{\frac{\widehat{E}_{f'_1}\widehat{E}_{f1}}{4E_{f'_1}
E_{f_1}}}
\left(1+\frac{2p_x^2-(-\frac{m_{f'_1}}{m_{f'_1}+m_{f_2}}\,|\vec{q}\,|\vec{k}-\vec{p}\,)
\cdot(\frac{m_{f_2}}{m_{f'_1}+m_{f_2}}\,|\vec{q}\,|\vec{k}-\vec{p}\,)}
{\widehat{E}_{f'_1}\widehat{E}_{f_1}}
\right) \hspace*{-4cm}
\nonumber\\
\lefteqn{\hspace*{-.8cm}A^{(1^-)\,3}_{\lambda=0}(|\vec{q}\,|)\,=}
\nonumber \\* &&\hspace*{-1.3cm}=\,
i
\sqrt{2m_I2E_F(-\vec{q}\,)}\ \int\,d^3p\ \frac{1}{4\pi}
\left(\hat{\phi}^{(M_F(1^-))}_{f'_1,\,f_2}(|\vec{p}\,|)\right)^*\,
\hat{\phi}^{(M_I(0^-))}_{f_1,\,f_2}\left(\bigg|\,\vec{p}-\frac{m_{f_2}}{m_{f'_1}+m_{f_2}}
|\vec{q}\,|
\vec{k} \bigg|\right)\nonumber\\*
&&\hspace{3cm}\times\sqrt{\frac{\widehat{E}_{f'_1}\widehat{E}_{f1}}{4E_{f'_1}
E_{f_1}}}
\ \Bigg(
1+\frac{2(-\frac{m_{f'_1}}{m_{f'_1}+m_{f_2}}\,|\vec{q}\,|-p_z\,)
\cdot(\frac{m_{f_2}}{m_{f'_1}+m_{f_2}}\,|\vec{q}\,|-p_z\,)}
{\widehat{E}_{f'_1}\widehat{E}_{f_1}}\nonumber\\*
&&\hspace{5.45cm}-\frac{(-\frac{m_{f'_1}}{m_{f'_1}+m_{f_2}}\,|\vec{q}\,|\vec{k}-\vec{p}\,)
\cdot(\frac{m_{f_2}}{m_{f'_1}+m_{f_2}}\,|\vec{q}\,|\vec k-\vec{p}\,)}
{\widehat{E}_{f'_1}\widehat{E}_{f_1}} 
\Bigg) \hspace*{-3cm}
\nonumber\\* \nonumber\\*
\end{eqnarray}
} %Del allowdisplaybreaks

\begin{itemize}
\item {\large Case $J^\pi=1^+$}
\end{itemize}
{\allowdisplaybreaks
\begin{eqnarray}
\lefteqn{\hspace*{-.8cm}V^{(1^+,S_{q\bar{q}}=0)\,0}_{\lambda=0}(|\vec{q}\,|)\,=}
\nonumber\\* &&\hspace*{-1.3cm}=\,
i\sqrt3
\sqrt{2m_I2E_F(-\vec{q}\,)}\ \int\,d^3p\ \frac{1}{4\pi|\vec p\,|}
\left(\hat{\phi}^{(M_F(1^+,S_{q\bar{q}}=0))}_{f'_1,\,f_2}(|\vec{p}\,|)\right)^*\,
\hat{\phi}^{(M_I(0^-))}_{f_1,\,f_2}\left(\bigg|\,\vec{p}-\frac{m_{f_2}}{m_{f'_1}+m_{f_2}}
|\vec{q}\,|
\vec{k} \bigg|\right)\nonumber\\*
&&\hspace{3cm}\times\sqrt{\frac{\widehat{E}_{f'_1}\widehat{E}_{f1}}{4E_{f'_1}
E_{f_1}}}\,p_z\,
\left(
1+\frac{(-\frac{m_{f'_1}}{m_{f'_1}+m_{f_2}}\,|\vec{q}\,|\vec{k}-\vec{p}\,)
\cdot(\frac{m_{f_2}}{m_{f'_1}+m_{f_2}}\,|\vec{q}\,|\vec{k}-\vec{p}\,)}{\widehat{E}_{f'_1}\widehat{E}_{f_1}}
\right)
\nonumber\\
\lefteqn{\hspace*{-0.8cm}V^{(1^+,S_{q\bar{q}}=1)\,0}_{\lambda=0}(|\vec{q}\,|)\,=}
\nonumber\\* &&\hspace*{-1.3cm}=\,
-i\sqrt{\frac{3}{2}}
\sqrt{2m_I2E_F(-\vec{q}\,)}\ \int\,d^3p\ \frac{1}{4\pi|\vec p\,|}
\left(\hat{\phi}^{(M_F(1^+,S_{q\bar{q}}=1))}_{f'_1,\,f_2}(|\vec{p}\,|)\right)^*\,
\hat{\phi}^{(M_I(0^-))}_{f_1,\,f_2}\left(\bigg|\,\vec{p}-\frac{m_{f_2}}{m_{f'_1}+m_{f_2}}
|\vec{q}\,|
\vec{k} \bigg|\right)\nonumber\\*
&&\hspace{3cm}\times\sqrt{\frac{\widehat{E}_{f'_1}\widehat{E}_{f1}}{4E_{f'_1}
E_{f_1}}}
\ \frac{|\vec{q}\,| (p_z^2-\vec{p}^{\,2})}{\widehat{E}_{f'_1}\widehat{E}_{f1}}\nonumber\\
\lefteqn{\hspace*{-.8cm}V^{(1^+,S_{q\bar{q}}=0)\,1}_{\lambda=-1}(|\vec{q}\,|)\,=}
\nonumber\\* &&\hspace*{-1.3cm}
-i\sqrt{\frac{3}{2}}
\sqrt{2m_I2E_F(-\vec{q}\,)}\ \int\,d^3p\ \frac{1}{4\pi|\vec{p}\,|}
\left(\hat{\phi}^{(M_F(1^+,S_{q\bar{q}}=0))}_{f'_1,\,f_2}(|\vec{p}\,|)\right)^*\,
\hat{\phi}^{(M_I(0^-))}_{f_1,\,f_2}\left(\bigg|\,\vec{p}-\frac{m_{f_2}}{m_{f'_1}+m_{f_2}}
|\vec{q}\,|
\vec{k} \bigg|\right)\nonumber\\*
&&\hspace{3cm}\times\sqrt{\frac{\widehat{E}_{f'_1}\widehat{E}_{f1}}{4E_{f'_1}
E_{f_1}}}
%\nonumber\\
\ p_x^2\,\left(\frac{1}{\widehat{E}_{f_1}}+\frac{1}{\widehat{E}_{f'_1}}
\right)\nonumber\\%
\lefteqn{\hspace*{-.8cm}V^{(1^+,S_{q\bar{q}}=1)\,1}_{\lambda=-1}(|\vec{q}\,|)}
\nonumber\\* &&\hspace*{-1.3cm}=\,
i\frac{\sqrt{3}}{2}
\sqrt{2m_I2E_F(-\vec{q}\,)}\ \int\,d^3p\ \frac{1}{4\pi|\vec{p}\,|}
\left(\hat{\phi}^{(M_F(1^+,S_{q\bar{q}}=1))}_{f'_1,\,f_2}(|\vec{p}\,|)\right)^*\,
\hat{\phi}^{(M_I(0^-))}_{f_1,\,f_2}\left(\bigg|\,\vec{p}-\frac{m_{f_2}}
{m_{f'_1}+m_{f_2}}
|\vec{q}\,|
\vec{k} \bigg|\right)\nonumber\\*
&&\hspace{3cm}\times\sqrt{\frac{\widehat{E}_{f'_1}\widehat{E}_{f1}}{4E_{f'_1}
E_{f_1}}}
%\nonumber\\
\ \left(\frac{p_y^2+p_z^2+p_z|\vec{q}\,|
\frac{m_{f'_1}}{m_{f'_1}+m_{f_2}}
}{\widehat{E}_{f'_1}}-\frac{p_y^2+p_z^2-p_z|\vec{q}\,|
\frac{m_{f_2}}{m_{f'_1}+m_{f_2}}}{\widehat{E}_{f_1}}
\right)\nonumber\\
\lefteqn{\hspace*{-.8cm}V^{(1^+,S_{q\bar{q}}=0)\,3}_{\lambda=0}(|\vec{q}\,|)\,=}
\nonumber \\* && \hspace*{-1.3cm}=\,
i\sqrt3
\sqrt{2m_I2E_F(-\vec{q}\,)}\ \int\,d^3p\ \frac{1}{4\pi|\vec{p}\,|}
\left(\hat{\phi}^{(M_F(1^+,S_{q\bar{q}}=0))}_{f'_1,\,f_2}(|\vec{p}\,|)\right)^*\,
\hat{\phi}^{(M_I(0^-))}_{f_1,\,f_2}\left(\bigg|\,\vec{p}-\frac{m_{f_2}}{m_{f'_1}+m_{f_2}}
|\vec{q}\,|
\vec{k} \bigg|\right)\nonumber\\*
&&\hspace{3cm}\times\sqrt{\frac{\widehat{E}_{f'_1}\widehat{E}_{f1}}{4E_{f'_1}
E_{f_1}}}\
%\nonumber\\
p_z\,\Bigg(\frac{\frac{m_{f_2}}{m_{f'_1}+m_{f_2}}\,|\vec{q}\,|-p_z}{\widehat{E}_{f_1}}+
\frac{-\frac{m_{f'_1}}{m_{f'_1}+m_{f_2}}\,|\vec{q}\,|-p_z}{\widehat{E}_{f'_1}}
\Bigg)\nonumber\\%
\lefteqn{\hspace*{-.8cm}V^{(1^+,S_{q\bar{q}}=1)\,3}_{\lambda=0}(|\vec{q}\,|)\,=}
\nonumber \\* &&\hspace*{-1.3cm}=\,
-i\sqrt{\frac{3}{2}}
\sqrt{2m_I2E_F(-\vec{q}\,)}\ \int\,d^3p\ \frac{1}{4\pi|\vec{p}\,|}
\left(\hat{\phi}^{(M_F(1^+,S_{q\bar{q}}=1))}_{f'_1,\,f_2}(|\vec{p}\,|)\right)^*\,
\hat{\phi}^{(M_I(0^-))}_{f_1,\,f_2}\left(\bigg|\,\vec{p}-\frac{m_{f_2}}{m_{f'_1}+m_{f_2}}
|\vec{q}\,|
\vec{k} \bigg|\right)\nonumber\\*
&&\hspace{3cm}\times\sqrt{\frac{\widehat{E}_{f'_1}\widehat{E}_{f1}}{4E_{f'_1}
E_{f_1}}}
%\nonumber\\
\ (p_x^2+p_y^2)\,\left(\frac{1}{\widehat{E}_{f_1}}-\frac{1}{\widehat{E}_{f'_1}}
\right)\nonumber\\*\nonumber\\*
\end{eqnarray}
}%DEL allowdisplaybreaks
and similarly
{\allowdisplaybreaks
\begin{eqnarray}
\lefteqn{\hspace*{-0.8cm}A^{(1^+,S_{q\bar{q}}=0)\,1}_{\lambda=-1}(|\vec{q}\,|)\,=}
\nonumber\\* &&\hspace*{-1.3cm}=\,
-i\sqrt{\frac{3}{2}}
\sqrt{2m_I2E_F(-\vec{q}\,)}\ \int\,d^3p\ \frac{1}{4\pi|\vec{p}\,|}
\left(\hat{\phi}^{(M_F(1^+,S_{q\bar{q}}=0))}_{f'_1,\,f_2}(|\vec{p}\,|)\right)^*\,
\hat{\phi}^{(M_I(0^-))}_{f_1,\,f_2}\left(\bigg|\,\vec{p}-\frac{m_{f_2}}{m_{f'_1}+m_{f_2}}
|\vec{q}\,|
\vec{k} \bigg|\right)\nonumber\\*
&&\hspace{3cm}\times\sqrt{\frac{\widehat{E}_{f'_1}\widehat{E}_{f1}}{4E_{f'_1}
E_{f_1}}}
%\nonumber\\
\ \frac{p_y^2|\vec{q}\,|}{\widehat{E}_{f_1}\widehat{E}_{f'_1}}
\nonumber\\%
\lefteqn{\hspace*{-.8cm}A^{(1^+,S_{q\bar{q}}=1)1}_{\lambda=-1}(|\vec{q}\,|)\,=}
\nonumber\\* &&\hspace*{-1.3cm}=\,
i\frac{\sqrt3}{2}\sqrt{2m_I2E_F(-\vec{q}\,)}\ \int\,d^3p\
\frac{1}{4\pi|\vec{p}\,|}
\left(\hat{\phi}^{(M_F (1^+,S_{q\bar{q}}=1))}_{f'_1,\,f_2}(|\vec{p}\,|)
\right)^*\,
\hat{\phi}^{(M_I(0^-))}_{f_1,\,f_2}\left(\bigg|\,\vec{p}-\frac{m_{f_2}}{m_{f'_1}+m_{f_2}}
|\vec{q}\,|
\vec{k} \bigg|\right)
 \nonumber\\*
&&\hspace{3cm}\times\sqrt{\frac{\widehat{E}_{f'_1}\widehat{E}_{f1}}{4E_{f'_1}
E_{f_1}}}\bigg\{\
p_z\bigg(  
1-\frac{(-\frac{m_{f'_1}}{m_{f'_1}+m_{f_2}}\,|\vec{q}\,|\vec{k}-\vec{p}\,)
\cdot(\frac{m_{f_2}}{m_{f'_1}+m_{f_2}}\,|\vec{q}\,|\vec{k}-\vec{p}\,)}{\widehat{E}_{f'_1}\widehat{E}_{f_1}}
\bigg)\nonumber\\* 
&&\hspace{6.5cm}+\frac{m_{f_2}-m_{f'_1}}{m_{f'_1}+m_{f_2}}
\frac{p_x^2|\vec{q}\,|}{\widehat{E}_{f'_1}\widehat{E}_{f1}}
\bigg\}\nonumber\\*\nonumber\\*
\end{eqnarray}
} %Del allow...

\begin{itemize}
\item {\large Case $J^\pi= 2^-$}
\end{itemize}
{\allowdisplaybreaks
\begin{eqnarray}
\lefteqn{\hspace*{-.8cm}V^{(2^-)\,0}_{T\lambda=0}(|\vec{q}\,|)\,=}
\nonumber\\* &&\hspace*{-1,3cm} =\,
\sqrt{\frac{15}{2}}
\sqrt{2m_I2E_F(-\vec{q}\,)}\ \int\,d^3p\ \frac{1}{4\pi|\vec{p}\,|^2}
\left(\hat{\phi}^{(M_F(2^-))}_{f'_1,\,f_2}(|\vec{p}\,|)\right)^*\,
\hat{\phi}^{(M_I(0^-))}_{f_1,\,f_2}\left(\bigg|\,\vec{p}-\frac{m_{f_2}}{m_{f'_1}+m_{f_2}}
|\vec{q}\,|
\vec{k} \bigg|\right)\nonumber\\*
&&\hspace{3cm}\times\sqrt{\frac{\widehat{E}_{f'_1}\widehat{E}_{f1}}{4E_{f'_1}
E_{f_1}}}\
\frac{p_z\left(p_x^2+p_y^2\right)|\vec{q}\,|}
{\widehat{E}_{f'_1}\widehat{E}_{f1}}\nonumber\\
\lefteqn{\hspace*{-.8cm}V^{(2^-)\,1}_{T\lambda=+1}(|\vec{q}\,|)\,=}
\nonumber \\* &&\hspace*{-1.3cm} =\,
i\,\frac{\sqrt5}{2}
\sqrt{2m_I2E_F(-\vec{q}\,)}\ \int\,d^3p\ \frac{1}{4\pi|\vec{p}\,|^2}
\left(\hat{\phi}^{(M_F(2^-))}_{f'_1,\,f_2}(|\vec{p}\,|)\right)^*\,
\hat{\phi}^{(M_I(0^-))}_{f_1,\,f_2}\left(\bigg|\,\vec{p}-\frac{m_{f_2}}{m_{f'_1}+m_{f_2}}
|\vec{q}\,|
\vec{k} \bigg|\right)\nonumber\\*
&&\hspace{3cm}\times\sqrt{\frac{\widehat{E}_{f'_1}\widehat{E}_{f1}}{4E_{f'_1}
E_{f_1}}}
%\nonumber\\
\Bigg\{\left(p_z^2-p_x^2\right)
\left(\ \frac{-p_z-\frac{m_{f'_1}}{m_{f'_1}+m_{f_2}}|\vec{q}\,|}
{\widehat{E}_{f'_1}}
-\frac{-p_z+\frac{m_{f_2}}{m_{f'_1}+m_{f_2}}|\vec{q}\,|}
{\widehat{E}_{f_1}}
\right)\nonumber\\*
&&\hspace{5cm}-p_zp_y^2\left(\frac{1}{\widehat{E}_{f'_1}}-\frac{1}{\widehat{E}_{f_1}}
\right)\Bigg\}\nonumber\\
\lefteqn{\hspace*{-.8cm}V^{(2^-)\,3}_{T\lambda=0}(|\vec{q}\,|)\,=}
\nonumber\\*&&\hspace*{-1.3cm}=\,
i\,\sqrt{\frac{15}{2}}
\sqrt{2m_I2E_F(-\vec{q}\,)}\ \int\,d^3p\ \frac{1}{4\pi|\vec{p}\,|^2}
\left(\hat{\phi}^{(M_F(2^-))}_{f'_1,\,f_2}(|\vec{p}\,|)\right)^*\,
\hat{\phi}^{(M_I(0^-))}_{f_1,\,f_2}\left(\bigg|\,\vec{p}-\frac{m_{f_2}}{m_{f'_1}+m_{f_2}}
|\vec{q}\,|
\vec{k} \bigg|\right)\nonumber\\*
&&\hspace{3cm}\times\sqrt{\frac{\widehat{E}_{f'_1}\widehat{E}_{f1}}{4E_{f'_1}
E_{f_1}}}\ p_z\left(p_x^2+p_y^2\right)
\left(
\frac{1}
{\widehat{E}_{f'_1}}-\frac{1}
{\widehat{E}_{f_1}}
\right)\nonumber\\*\nonumber\\*
\end{eqnarray}
}%del allow
and similarly
\begin{eqnarray}
\lefteqn{\hspace*{-.8cm}A^{(2^-)\,1}_{T\lambda=+1}(|\vec{q}\,|)\,=}
\nonumber\\* &&\hspace*{-1.3cm}=\,
i\,\frac{\sqrt5}{2}
\sqrt{2m_I2E_F(-\vec{q}\,)}\ \int\,d^3p\ \frac{1}{4\pi|\vec{p}\,|^2}
\left(\hat{\phi}^{(M_F(2^-))}_{f'_1,\,f_2}(|\vec{p}\,|)\right)^*\,
\hat{\phi}^{(M_I(0^-))}_{f_1,\,f_2}\left(\bigg|\,\vec{p}-\frac{m_{f_2}}{m_{f'_1}+m_{f_2}}
|\vec{q}\,|
\vec{k} \bigg|\right)\nonumber\\
&&\hspace{2cm}\times\sqrt{\frac{\widehat{E}_{f'_1}\widehat{E}_{f1}}{4E_{f'_1}
E_{f_1}}}
%\nonumber\\
\Bigg\{\left(p_z^2-p_y^2\right)
\left(1-
\frac{(-\frac{m_{f'_1}}{m_{f'_1}+m_{f_2}}\,|\vec{q}\,|\vec{k}-\vec{p}\,)
\cdot(\frac{m_{f_2}}{m_{f'_1}+m_{f_2}}\,|\vec{q}\,|\vec k-\vec{p}\,)}{\widehat{E}_{f'_1}\widehat{E}_{f1}}
\right)
\nonumber\\
&&\hspace{4cm}-p_zp_x^2|\vec{q}\,|\,\frac{m_{f'_1}-m_{f_2}}{m_{f'_1}+m_{f_2}}
\,\frac{1}
{\widehat{E}_{f'_1}\widehat{E}_{f_1}}\ 
\Bigg\}\nonumber\\*\nonumber\\*
\end{eqnarray}

\begin{itemize}
\item  {\large  Case $J^\pi= 2^+$}
\end{itemize}
\begin{eqnarray}
\lefteqn{V^{(2^+)\,1}_{T\lambda=+1}(|\vec{q}\,|)\,=}
\nonumber\\&&\hspace{-1.3cm}=\,
i\,\frac{\sqrt3}{2}
\sqrt{2m_I2E_F(-\vec{q}\,)}\ \int\,d^3p\ \frac{1}{4\pi|\vec{p}\,|}
\left(\hat{\phi}^{(M_F(2^+))}_{f'_1,\,f_2}(|\vec{p}\,|)\right)^*\,
\hat{\phi}^{(M_I(0^-))}_{f_1,\,f_2}\left(\bigg|\,\vec{p}-\frac{m_{f_2}}{m_{f'_1}+m_{f_2}}
|\vec{q}\,|
\vec{k} \bigg|\right)\nonumber\\
&&\hspace{3cm}\times\sqrt{\frac{\widehat{E}_{f'_1}\widehat{E}_{f1}}{4E_{f'_1}
E_{f_1}}}
%\nonumber\\
\left(\frac{p_y^2-p_z^2-p_z|\vec{q}\,|\frac{m_{f'_1}}{m_{f'_1}+m_{f_2}}}
{\widehat{E}_{f'_1}}
-\frac{p_y^2-p_z^2+p_z|\vec{q}\,|\frac{m_{f_2}}{m_{f'_1}+m_{f_2}}}
{\widehat{E}_{f_1}}
\right)\nonumber\\*\nonumber\\*
\end{eqnarray}
and similarly
{\allowdisplaybreaks
\begin{eqnarray}
\lefteqn{\hspace*{-.8cm}A^{(2^+)\,0}_{T\lambda=0}(|\vec{q}\,|)\,=}
\nonumber\\*&&\hspace*{-1.3cm}=\,
\frac{-i}{\sqrt2}
\sqrt{2m_I2E_F(-\vec{q}\,)}\ \int\,d^3p\ \frac{1}{4\pi|\vec{p}\,|}
\left(\hat{\phi}^{(M_F(2^+))}_{f'_1,\,f_2}(|\vec{p}\,|)\right)^*\,
\hat{\phi}^{(M_I(0^-))}_{f_1,\,f_2}\left(\bigg|\,\vec{p}-\frac{m_{f_2}}{m_{f'_1}+m_{f_2}}
|\vec{q}\,|
\vec{k} \bigg|\right)\nonumber\\*
&&\hspace{.5cm}\times\sqrt{\frac{\widehat{E}_{f'_1}\widehat{E}_{f1}}{4E_{f'_1}
E_{f_1}}}
%\nonumber\\
\left(\frac{p_x^2+p_y^2-2p_z^2-2p_z|\vec{q}\,|\frac{m_{f'_1}}{m_{f'_1}+m_{f_2}}}
{\widehat{E}_{f'_1}}
+\frac{p_x^2+p_y^2-2p_z^2+2p_z|\vec{q}\,|\frac{m_{f_2}}{m_{f'_1}+m_{f_2}}}
{\widehat{E}_{f_1}}
\right)\nonumber\\
\lefteqn{\hspace*{-.8cm}A^{(2^+)\,1}_{T\lambda=+1}(|\vec{q}\,|)\,=}
\nonumber\\*&&\hspace*{-1.3cm}=\,
i\,\frac{\sqrt3}{2}
\sqrt{2m_I2E_F(-\vec{q}\,)}\ \int\,d^3p\ \frac{1}{4\pi|\vec{p}\,|}
\left(\hat{\phi}^{(M_F(2^+))}_{f'_1,\,f_2}(|\vec{p}\,|)\right)^*\,
\hat{\phi}^{(M_I(0^-))}_{f_1,\,f_2}\left(\bigg|\,\vec{p}-\frac{m_{f_2}}{m_{f'_1}+m_{f_2}}
|\vec{q}\,|
\vec{k} \bigg|\right)\nonumber\\*
&&\hspace{3cm}\times\sqrt{\frac{\widehat{E}_{f'_1}\widehat{E}_{f1}}{4E_{f'_1}
E_{f_1}}}\Bigg\{\ \,p_z\,\left(1-
\frac{(-\frac{m_{f'_1}}{m_{f'_1}+m_{f_2}}\,|\vec{q}\,|\vec{k}-\vec{p}\,)
\cdot(\frac{m_{f_2}}{m_{f'_1}+m_{f_2}}\,|\vec{q}\,|\vec k-\vec{p}\,)}{\widehat{E}_{f'_1}\widehat{E}_{f1}}
\right)\nonumber\\
&&\hspace{5.cm}
+\frac{4p_zp_x^2-p_x^2|\vec{q}\,|
\frac{m_{f_2}-m_{f'_1}}{m_{f'_1}+m_{f_2}}}
{\widehat{E}_{f'_1}\widehat{E}_{f1}}\Bigg\}\nonumber\\
\lefteqn{\hspace*{-.8cm} A^{(2^+)\,3}_{T\lambda=0}(|\vec{q}\,|)\,=}
\nonumber\\*&&\hspace*{-1.3cm}=\,
-i\,\sqrt2
\sqrt{2m_I2E_F(-\vec{q}\,)}\ \int\,d^3p\ \frac{1}{4\pi|\vec{p}\,|}
\left(\hat{\phi}^{(M_F(2^+))}_{f'_1,\,f_2}(|\vec{p}\,|)\right)^*\,
\hat{\phi}^{(M_I(0^-))}_{f_1,\,f_2}\left(\bigg|\,\vec{p}-\frac{m_{f_2}}{m_{f'_1}+m_{f_2}}
|\vec{q}\,|
\vec{k} \bigg|\right)\nonumber\\*
&&\hspace{2cm}\sqrt{\frac{\widehat{E}_{f'_1}\widehat{E}_{f1}}{4E_{f'_1}
E_{f_1}}}\Bigg\{\ \,p_z\,\left(1-
\frac{(-\frac{m_{f'_1}}{m_{f'_1}+m_{f_2}}\,|\vec{q}\,|\vec{k}-\vec{p}\,)
\cdot(\frac{m_{f_2}}{m_{f'_1}+m_{f_2}}\,|\vec{q}\,|\vec k-\vec{p}\,)}{\widehat{E}_{f'_1}\widehat{E}_{f1}}
\right)\nonumber\\*
&&\hspace{4.cm}
+\frac{1}
{\widehat{E}_{f'_1}\widehat{E}_{f1}}
\Bigg[
 \ \,2p_z(-\frac{m_{f'_1}}{m_{f'_1}+m_{f_2}}\,|\vec{q}\,|-p_z)
\cdot (\frac{m_{f_2}}{m_{f'_1}+m_{f_2}}\,|\vec{q}\,|-p_z)\nonumber\\*
&&\hspace{5.75cm}+\left(p_x^2+p_y^2\right)\bigg(
-p_z+\frac{m_{f_2}-m_{f'_1}}{2(m_{f'_1}+m_{f_2})}|\vec{q}\,|
\bigg)
\Bigg]
\ \Bigg\}\nonumber\\*\nonumber\\*
\end{eqnarray}
}%Del Allow...
%\end{itemize}

%-------------------------------------------------------------------------------
%-------------------------------------------------------------------------------

\section{Transitions involving antiquarks} 
\label{Bcapp:sign}
When it is the antiquark that suffers the decay the expressions are modified
as described below for a general transition between
a pseudoscalar
meson $M_I$ at rest with quark content $q_{f_1}\overline{q}_{f_2}$ and a final 
$M_F$ meson with total angular momentum and parity $J^\pi=0^-,0^+,1^-,1^+,2^-,2^+$,
three-momentum $-|\vec{q}\,|\vec{k}$ and  quark content
$q_{f_1}\overline{q}_{f'_2}$. 
We have
\begin{eqnarray}
&& \hspace*{-.7cm}
{\cal V}^\mu(|\vec{q}\,|)-{\cal A}^\mu(|\vec{q}\,|)\,=\,\sqrt{2m_{I}2E_{F}(-\vec{q\,})}\
{}_{\stackrel{}{\stackrel{}{NR}}}
\left\langle\, M_F(J^\pi),\,\lambda\,
-|\vec{q}\,|\,\vec{k}\,\bigg|\, J^{f_2\,f'_2\, \mu}(0)\,
\bigg| \, M_I(0^-),\,\vec{0}\right\rangle_{NR}
\hspace*{-1cm}
\nonumber\\
&&\hspace*{-.7cm}=\,-\sqrt{2m_{I}2E_{F}(-\vec{q\,})}\ \int\,d^3p\sum_{s'_2}\sum_{s_1,s_2}
\left(\hat{\phi}^{(M_F(J^\pi),\lambda)}_{(s_1,f_1),\,(s'_2,f'_2)}(\vec{p}\,)\right)^*\ 
\hat{\phi}^{(M_I(0^-))}_{(s_1,f_1),\,(s_2,f_2)}(\vec{p}
+\frac{m_{f_1}}{m_{f_1}+m_{f'_2}}\,|\vec{q}\,|\vec{k})
\hspace*{-1cm}
\nonumber\\
&&\hspace*{-.7cm}\hspace{1cm}\frac{(-1)^{s_2-s'_2}}
{\sqrt{2E_{f'_2}\,2E_{f_2}}}\ \bar{v}_{s_2,f_2}(\frac{m_{f_1}}{m_{f_1}+m_{f'_2}}
|\vec{q}\,|\vec{k}+\vec{p}\,)\, 
\gamma^\mu(I-\gamma_5)\,v_{s'_2,f'_2}(-\frac{m_{f'_2}}{m_{f_1}+m_{f'_2}}
|\vec{q}\,|\vec{k}+\vec{p}\,)\nonumber\\
\end{eqnarray}
where ${\cal V}^\mu ({\cal A}^\mu)$ represent any of the $V^\mu\, (A^\mu)$,
$V^\mu_{(\lambda)}\, (A^\mu_{(\lambda)})$ or
$V^\mu_{T(\lambda)}\, (A^\mu_{T(\lambda)})$,
and $E_{f'_2},\,E_{f_2}$ are shorthand notation for
$E_{f'_2}(-\frac{m_{f'_2}}{m_{f_1}+m_{f'_2}}
|\vec{q}\,|\vec{k}+\vec{p}\,)$, $E_{f_2}(\frac{m_{f_1}}{m_{f_1}+m_{f'_2}}
|\vec{q}\,|\vec{k}+\vec{p}\,)$.
We can use now that
\begin{eqnarray}
v_{s,f}(\vec{p}\,)=(-1)^{(1/2)-s}\, {\cal C}\, \bar{u}^T_{s,f}(\vec{p}\,)
\hspace{.5cm};\hspace{.5cm}
\bar{v}_{s,f}(\vec{p}\,)=-(-1)^{(1/2)-s}\,  {u}^T_{s,f}(\vec{p}\,)\, {\cal C}^\dagger
\end{eqnarray}
where ${\cal C}$ is a matrix given in the Fermi--Dirac representation that we
use by
\begin{eqnarray}
{\cal C}=i\gamma^2\gamma^0
\end{eqnarray}
and that satisfies
\begin{eqnarray}
{\cal C}=-{\cal C}^{-1}=-{\cal C}^\dagger=-{\cal C}^T\hspace{.5cm};\hspace{.5cm}
{\cal C}\gamma_\mu^T{\cal C}^\dagger=-\gamma_\mu
\end{eqnarray}
Using the above information and making the change of variable
$\vec p\to -\vec p$\ \ we can rewrite
\begin{eqnarray}
&&\hspace{-.7cm}{\cal V}^\mu(|\vec{q}\,|)-{\cal A}^\mu(|\vec{q}\,|)\,=\,
\sqrt{2m_{I}2E_{F}(-\vec{q\,})}\
\nonumber\\ 
&&
\hspace{-.3cm} \int\,d^3p\sum_{s'_2}\sum_{s_1,s_2}
\left(\hat{\phi}^{(M_F(J^\pi),\,\lambda)}_{(s_1,f_1),\,(s'_2,f'_2)}(-\vec{p}\,)
\right)^*\ 
\hat{\phi}^{(M_I(0^-))}_{(s_1,f_1),\,(s_2,f_2)}\bigg(-\bigg(\vec{p}
-\frac{m_{f_1}}{m_{f_1}+m_{f'_2}}\,|\vec{q}\,|\vec{k}\bigg)\bigg)
\nonumber\\
&&\hspace{-.7cm}\hspace{1cm}\frac{1}
{\sqrt{2E_{f'_2}\,2E_{f_2}}}\ \bar{u}_{s'_2,f'_2}(-\frac{m_{f'_2}}{m_{f_1}+m_{f'_2}}
|\vec{q}\,|\vec{k}-\vec{p}\,)\, 
\gamma^\mu(-I-\gamma_5)\,u_{s_2,f_2}(\frac{m_{f_1}}{m_{f_1}+m_{f'_2}}
|\vec{q}\,|\vec{k}-\vec{p}\,)\nonumber\\
\end{eqnarray}%
By comparison with the corresponding expressions involving  quarks  we find 
that, apart from the changes in the masses involved, there is an
 extra minus sign for the vector part, and, due to Clebsch--Gordan
re-arrangements and the fact that $Y_{lm}(-\vec{p}\,)=(-1)^l\,Y_{lm}(\vec{p}\,)$,
a global sign given by $(-1)^{l_I+s_I-l_F-s_F}$ where $l_I,\,s_I$\,($l_F,\,s_F$) are 
the orbital and spin angular momenta of the initial (final) meson.

In any case this implies a change of sign in the relative phase between 
 vector and  axial contributions, which in its term produces 
a sign change in the tensor helicity
components combination $H_P$ due to the fact that 
${\cal H}_{+1\,+1}$ goes into ${\cal H}_{-1\,-1}$ and
vice versa. All other tensor helicity
components combinations defined in Eq.(\ref{eq:combinaciones}) keep their signs.\\

A simple way of anticipating the above result is the following: the current for 
$\bar q_{f_2}$ decay into $\bar q_{f'_2}$ is
\begin{eqnarray}
\overline{\Psi}_{f_2}(0)\gamma^\mu (I-\gamma_5)\Psi_{f'_2}(0)
\end{eqnarray} 
But for antiquarks the fields that play the similar role as the $\Psi$ fields
play for quarks are the charge conjugate ones $ \Psi^{\cal C}$. In terms of the
latter the above current is written as
}
{\small
\begin{eqnarray}
\overline{\Psi}_{f_2}(0)\gamma^\mu (I-\gamma_5)\Psi_{f'_2}(0)=
\overline{\Psi}_{f'_2}^{\cal C}(0)\gamma^\mu (-I-\gamma_5)\Psi_{f_2}^{\cal C}(0)
\end{eqnarray} 
Now this is similar to the current for  quark decay but with an extra minus
sign in the vector part. Whatever other changes might come from reorderings
in the wave functions we will have an extra relative sign between the vector and
axial part.

\chapter{Expressions for the helicity components of the hadron tensor}
\chaptermark{Helicity components}
\label{Bcapp:hcht}
In this appendix we give the expressions for the non--zero helicity components 
${\cal H}_{rs}$ of the
hadron tensor, as defined in Eq.(\ref{eq:hcht}), corresponding to a $B_c^-\to
c\bar c$ transition. The different cases correspond
to the ones discussed in the main text.
{\allowdisplaybreaks
\begin{itemize}
\item {\large Case $0^-\to 0^-,0^+$}
\begin{eqnarray}
&& \hspace*{-1.7cm}
{\cal H}_{t\,t}({P}_{B_c},{P}_{c\bar{c}})\,=\,
\left(\frac{m^2_{B_c}-m^2_{c\bar c}}{\sqrt{q^2}}\,F_+(q^2)+\sqrt{q^2}\,
F_-(q^2)
\right)^2\nonumber
\nonumber \\&& \hspace*{-1.7cm}
{\cal H}_{t\,0}({P}_{B_c},{P}_{c\bar{c}})\,=\,
{\cal H}_{0\,t}({P}_{B_c},{P}_{c\bar{c}})=
\lambda^{1/2}(q^2,m^2_{B_c},m^2_{c\bar c})\left(
\frac{m^2_{B_c}-m^2_{c\bar c}}{{q^2}}\,F_+^2(q^2)+
F_+(q^2)F_-(q^2)\right)\nonumber\\
&& \hspace*{-1.7cm}
{\cal H}_{0\,0}({P}_{B_c},{P}_{c\bar{c}})\,=\,
\frac{\lambda(q^2,m^2_{B_c},m^2_{c\bar c})}{q^2}\, F^2_+(q^2)
\end{eqnarray}
\item {\large Case $0^-\to 1^-,1^+$}.
\begin{eqnarray}
&&\hspace*{-1.7cm}
{\cal H}_{t\,t}({P}_{B_c},{P}_{c\bar{c}})\,=\,
\frac{\lambda(q^2,m^2_{B_c},m^2_{c\bar c})}{4m^2_{c\bar c}{q^2}}\,
\left((m_{B_c}-m_{c\bar c})\left(A_0(q^2)-A_+(q^2)\right)
-\frac{q^2}{m_{B_c}+m_{c\bar c}}A_-(q^2)\right)^2\nonumber\\
&&\hspace*{-1.7cm}
{\cal H}_{t\,0}({P}_{B_c},{P}_{c\bar{c}})\,=\,
{\cal H}_{0\,t}({P}_{B_c},{P}_{c\bar{c}})\nonumber\\*
&&\hspace*{-.7cm}=\,\frac{\lambda^{1/2}(q^2,m^2_{B_c},m^2_{c\bar c})}{2m_{c\bar c}\sqrt{q^2}}\,
\left[(m_{B_c}-m_{c\bar c})\left(A_0(q^2)-A_+(q^2)\right)
-\frac{q^2}{m_{B_c}+m_{c\bar c}}A_-(q^2)\right]\nonumber\\*
&&\hspace*{-.0cm}\times\ \left[(m_{B_c}-m_{c\bar c})\frac{m^2_{B_c}-q^2-m^2_{c\bar c}}
{2m_{c\bar c}\sqrt{q^2}}
\, A_0(q^2)-\frac{\lambda(q^2,m^2_{B_c},m^2_{c\bar c})}
{2m_{c\bar c}\,\sqrt{q^2}}\,\frac{A_+(q^2)}
{m_{B_c}+m_{c\bar c}}\right]\nonumber\\
&&\hspace*{-1.7cm}{\cal H}_{+1\,+1}({P}_{B_c},{P}_{c\bar{c}})\,=\,
\left(\frac{\lambda^{1/2}(q^2,m^2_{B_c},m^2_{c\bar c})}{m_{B_c}+m_{c\bar c}}\,
V(q^2)+(m_{B_c}-m_{c\bar c})\,A_0(q^2)\right) ^2\nonumber\\
&&\hspace*{-1.7cm}
{\cal H}_{-1\,-1}({P}_{B_c},{P}_{c\bar{c}})\,=\,
\left(-\frac{\lambda^{1/2}(q^2,m^2_{B_c},m^2_{c\bar c})}{m_{B_c}+m_{c\bar c}}\,
V(q^2)+(m_{B_c}-m_{c\bar c})\,A_0(q^2)\right) ^2\nonumber\\
&&\hspace*{-1.7cm}
{\cal H}_{0\,0}({P}_{B_c},{P}_{c\bar{c}})\,=\,
\left( (m_{B_c}-m_{c\bar c})\frac{m^2_{B_c}-q^2-m^2_{c\bar c}}{2m_{c\bar c}\sqrt{q^2}}\,
A_0(q^2)-\frac{\lambda(q^2,m^2_{B_c},m^2_{c\bar c})}{2m_{c\bar
c}\sqrt{q^2}}\,\frac{A_+(q^2)}{m_{B_c}+m_{c\bar c}}
\right)^2 \nonumber \\*
\end{eqnarray}
\end{itemize}
}% Del allow...

For a $B_c^-\to B$ transition (with $B$ representing any of the 
$B=\overline B^0,\,\overline B^{*0},\,B^-,\,B^{*-}$), where it is the 
$\bar c$ antiquark that decays, we have
to change the mass of the final meson in the expressions above and 
take into account the changes in the form factors that derive from the
discussions  in appendix~\ref{Bcapp:sign}.
{\allowdisplaybreaks
\begin{itemize}
\item {\large Case $0^-\to 2^-,2^+$}.
\begin{eqnarray} 
&&\hspace*{-1.7cm}
{\cal H}_{t\,t}({P}_{B_c},{P}_{c\bar{c}})\,=\,
\frac{\lambda^{2}(q^2,m^2_{B_c},m^2_{c\bar c})}{24\,m^4_{c\bar c}\,q^2}\,
\left(\, T_1(q^2)+(m^2_{B_c}-m^2_{c\bar c})\,T_2(q^2)+q^2\,T_3(q^2)\,
\right)^2\nonumber\\
&&\hspace*{-1.7cm}
{\cal H}_{t\,0}({P}_{B_c},{P}_{c\bar{c}})\,=\,
{\cal H}_{0\,t}({P}_{B_c},{P}_{c\bar{c}})\nonumber\\
&&\hspace*{-1.7cm}=\,\frac{\lambda^{3/2}(q^2,m^2_{B_c},m^2_{c\bar c})}{24\,m^4_{c\bar c}\,q^2}\,
\left(\, T_1(q^2)+(m^2_{B_c}-m^2_{c\bar c})\,T_2(q^2)+q^2\,T_3(q^2)\
\right)\nonumber\\
&&\hspace*{-1.7cm}\hspace{2.55cm}\times\,\left(\,(m^2_{B_c}-q^2-m^2_{c\bar c})\,T_1(q^2)+\lambda(q^2,m^2_{B_c},m^2_{c\bar c})\,T_2(q^2)\,
\right)
\nonumber\\
&&\hspace*{-1.7cm}
{\cal H}_{+1\,+1}({P}_{B_c},{P}_{c\bar{c}})\,=\,
\frac{\lambda(q^2,m^2_{B_c},m^2_{c\bar c})}{8\,m^2_{c\bar c}}\,
\left(\,T_1(q^2)-\lambda^{1/2}(q^2,m^2_{B_c},m^2_{c\bar c})\,T_4(q^2)\,
\right)^2\nonumber\\
&&\hspace*{-1.7cm}
{\cal H}_{-1\,-1}({P}_{B_c},{P}_{c\bar{c}})\,=\,
\frac{\lambda(q^2,m^2_{B_c},m^2_{c\bar c})}{8\,m^2_{c\bar c}}\,
\left(\,T_1(q^2)+\lambda^{1/2}(q^2,m^2_{B_c},m^2_{c\bar c})\,T_4(q^2)\,
\right)^2\nonumber\\
&&\hspace*{-1.7cm}
{\cal H}_{0\,0}({P}_{B_c},{P}_{c\bar{c}})\,=\,
\frac{\lambda(q^2,m^2_{B_c},m^2_{c\bar c})}{24\,m^4_{c\bar c}\,q^2}\,
\left(\,(m^2_{B_c}-q^2-m^2_{c\bar c})\,T_1(q^2)+\lambda(q^2,m^2_{B_c},m^2_{c\bar c})\,T_2(q^2)\,
\right)^2\nonumber\\
\end{eqnarray}
\end{itemize}
}% del allow
}  %HECHO
{\small
%
% APendices para los capitulos de Doble peado
%

\chapter{Variational wave function parameters for doubly heavy baryons}
\chaptermark{Variational parameters}
\label{DPapp:vp}
%All VAR results presented in the paper have been obtained with
%heavy-heavy correlation functions, $F^{B}$, constructed from a
%linear combination of four gaussians (see eq.~(\ref{eq:fbig12})). In
%Table~\ref{tab:ng} and for the $\Lambda_c$ baryon with an AL1
%inter-quark interaction, we present the dependence of the mass and the
%value of the wave function at the origin on the number of gaussians
%used to build up the correlation function.
%
%
%\begin{table}
%\begin{tabular}{c|c|c|c}
%Number of gaussians & $\Lambda_c$ Mass [MeV]  & $\eta^{\Lambda_c}_{ll}(0)$
%[fm$^{-3}$] &  $\eta^{\Lambda_c}_{cl}(0)$ [fm$^{-3}$] \\\hline\tstrut
%1 & 2302.1  & 16.4 & 5.2 \\
%2 & 2295.0  & 17.0 & 5.7 \\
%3 & 2294.8  & 18.6 & 6.2 \\
%4 & 2294.6  & 18.6 & 6.5 \\\hline\tstrut
%FAD & 2296  & 14.8 & 7.4 \\\hline
%\end{tabular}
%\caption{$\Lambda_c$ baryon mass and wave functions at the origin (see
%Eq.~(\ref{eq:eta0})) as a function of the number of gaussians included
%in the light--light correlation function. Results have been obtained
%with the AL1 inter--quark interaction  and
%for comparison, the Faddeev results of Ref.\protect\cite{Si96}
%(denoted by FAD) have been also compiled.  }\label{tab:ng}
%\end{table}
%
In Tables~\ref{tab:xiparam} (doubly heavy $\Xi$ baryons) and
\ref{tab:omegaparam} (doubly heavy $\Omega$ baryons) we give the  variational parameters of the 
orbital wave functions
evaluated with the different quark--quark
interactions analyzed in this work.

\begin{table}[h]
{\scriptsize
\begin{tabular}{ll|rrrrrrrrrrrrr}
%             &&\multicolumn{12}{c}{Charm} \\\tstrut
             &         &$b_1$ &$d_1$ &~$a_2$   &$b_2$ &  $d_2$&~$a_3$
  &$b_3$ &$d_3$ &~$a_4$   &$b_4$ &$d_4$
  &~$\alpha_{h_1}$&~$\alpha_{h_2}$\\\hline\tstrut
%--
  $\Xi_{cc}$ &AL1     & 1.78 & 0.37 & 0.38 & 1.04& 0.28 &  0.57 &1.24
  &0.66 & 1.36 & 1.56 & 0.08 & 0.44 & 0.44 \\ 
             &AL2     & 2.48 & 0.09 & 1.25 & 1.50 & 0.01 & 0.34 & 1.55
             & 1.46 & 2.20 & 2.21 &1.82 &0.45 &0.45 \\ 
%--
               & AP1 & 2.14 & 0.14 & 0.83 & 1.45 & -0.01 & 0.67 & 1.95
               & 1.77 & 2.69 & 2.98 & 1.34 & 0.40 & 0.40\\
%-----------
               & AP2 & 1.86 & 0.11 & 0.48 & 1.25 & 0.14 & 0.62 & 1.62
               & 1.17 & 1.92 & 2.20 & 1.26 & 0.41 & 0.41 \\
%-------------
               &BD& 1.71 & 0.18 & 0.37 & 1.23 & 0.16 & 0.85 & 1.59 &
               0.96 & 1.46 & 2.07 & 1.22 &0.44 & 0.44\\

\hline\tstrut 
%-------------
  $\Xi_{cc}^*$ & AL1 & 1.26 & 0.40 & -0.03 & 0.82 & -0.30 & 0.55 &0.89
  &0.54 &0.76 &1.66 &0.15 &0.83 &0.83 \\
%--
               & AL2 & 1.18 & 0.53 & -0.02 & 0.65 & -0.36 & 0.58 &0.85
               & 0.58 &1.29 & 1.71 & 0.12 & 0.83 & 0.83 \\
%---------------

               & AP1 & 1.93 & 0.17 & 0.66 & 1.37 & 0.00 & 0.67 & 2.29
               & 1.56 & 1.66 & 2.99 & 1.28 & 0.76 & 0.76\\
%------------------
               & AP2 & 1.18 & 0.46 & -0.02 & 0.64 & -0.38 & 0.43 &
               0.79 & 0.65 & 1.59 & 1.73 & 0.12 & 0.75 & 0.75\\
%-----------------
	       &BD& 1.23 & 0.31 & -0.01 & 0.72 & -0.22 & 0.53 & 0.97 &
	       0.63 & 1.22 & 1.79 & 0.17 &0.73 & 0.73 \\

 \hline \tstrut 
%---------------
 $\Xi_{bb}$ & AL1 & 2.37 & 0.23 & 1.26 & 1.31 & 0.85 & 1.47 & 
1.93 & 1.04 & 1.17 &2.09 &2.13 & 0.39 & 0.39 \\ 
%--
            & AL2 & 2.35 & 1.81 & 1.54 & 1.57 & 1.41 & 2.03 & 2.11 &
            0.27 & 0.69 & 1.85 & 2.20 &0.39 &0.39 \\
%--
            &AP1 & 2.66 & 1.52 & 1.57 & 1.48 & 1.42 & 1.99 & 1.94 &
            0.37 & 0.85 & 1.77 & 2.31 & 0.35 & 0.35 \\
%-------
	    &AP2 & 2.22 & 1.93 &1.63 & 1.53 & 1.59 & 2.01 & 2.23 &
	    0.19 & 0.69 & 1.82 & 2.08 & 0.36 & 0.36 \\
%-----------------------
	       &BD & 2.74 & 1.35 & 1.42 & 1.42 &1.20 & 2.04 & 1.90 &
	       0.43 & 0.85 & 1.80 & 2.54 & 0.40 & 0.40\\

\hline\tstrut
%-----------------
  $\Xi_{bb}^*$ &AL1 &3.17 & 0.15 & 1.04 & 1.89 & 0.24 & 0.14 & 1.47 &
  1.79 & 2.25 & 2.84 & 1.47 & 0.54 &0.54 \\ 
%--
               & AL2 & 2.82 & 0.11 & 0.56 & 1.56 & 0.39 & 0.57 & 2.14
               & 1.47 & 1.47 & 2.80 & 1.89 & 0.52 & 0.52 \\
%------------

               &AP1 & 2.69 & 0.18 & 0.88 & 1.73 & 0.33 & 0.75 & 2.33
               & 1.80 & 1.90 & 3.14 & 1.76 & 0.48 & 0.48 \\
%--------------------
               &AP2 & 2.78 & 0.08 & 0.41 & 1.60 & 0.33 & 0.71 & 2.05 &
               1.45 & 1.56 & 2.87 & 1.91 & 0.49 & 0.49 \\

%--------------------------

	       &BD & 3.40 & 0.13 & 0.19 & 2.74 & -0.15 & -0.07 & 2.13
	       & 2.04 & 2.64 & 3.85 & 3.02 &0.51 & 0.51\\

\hline\tstrut
%-----------------
 $\Xi_{bc}'$ & AL1 & 2.24 & 0.21 & 0.39 & 1.73 & 0.01 & -0.01 & 2.18 &
 -0.78 & 1.62 & 0.77 & 1.84 & 0.01 &1.51 \\ 
%--
            & AL2 & 2.40 & 0.12 & 0.28 & 1.67 & -0.06 & -0.01 & 2.15 &
            -0.82 & 1.29 & 0.87 & 1.49 & 0.01 & 1.53 \\
%--
             & AP1 &4.81 & -0.01& 2.78 & 2.03 & 0.05 & -0.15 & 3.45 &
             0.52 & 2.31 & 0.91 & 1.09 & 0.01 & 1.37\\
%--
             & AP2 & 2.85 & 0.05 & 0.98 & 1.91 & -0.03 & -0.15 & 2.66
             & -0.19 & 0.68 & 0.99 & 1.55 & 0.02 & 1.34 \\  
%--
             & BD  & 2.60 & 0.14 & 0.58 & 1.62 & 0.05 & -0.04 & 2.47 &
             -0.66 & 1.65 & 0.94 & 1.28 & -0.03 & 1.54 \\

\hline\tstrut
%------------------
 $\Xi_{bc}$ &AL1 & 2.82 & 0.15 & 0.46 & 2.37 & -0.15 & -0.09 & 1.06
 & -0.37 & 1.31 &0.70 & 1.35 &-0.02 &1.09 \\ 
%--
             & AL2 & 2.97 & 0.08 & 0.57 &2.14 &-0.11 & -0.05 & 1.15 &
             -0.40 & 0.92 & 0.72 & 1.39 & -0.06 & 1.20 \\
%-------------
             & AP1 & 2.91 & 0.10 & 0.56 & 2.16 & -0.10 & -0.08 & 1.13
             & -0.50 & 1.21 & 0.72 & 1.25 & -0.06 & 1.05\\
%-------------
             & AP2 & 4.75 & -0.02 & 1.98  & 2.74 & 0.06 & 0.38 &
             1.47 & -0.16 & 3.03 & 2.13 & -0.04 & -0.07 &1.09 \\
%---------
             & BD & 2.64 & 0.21 & 0.48 & 2.38 & -0.08 & -0.12 & 1.02 &
             -0.57 & 1.25 & 0.56 & 1.74 &-0.07 & 1.20\\

\hline \tstrut
%------------------
 $\Xi_{bc}^* $ & AL1 & 1.85 & 0.21 & -0.19 & 2.20 & 0.10 & -0.02 &
 1.26 & -0.66 &1.27 & 0.70 &1.71 &0.07 &1.72 \\ 
%--

                &AL2 & 1.93 & 0.13 & -0.14 &2.25 & -0.09 & -0.05 &1.10 &
             -0.56 & 0.92 & 0.62 & 1.67 & 0.06 &1.74 \\
%--------
                &AP1 & 1.93 & 0.17 & 0.06 & 2.54 & 0.17 & -0.10 & 0.97
                & -0.57 & 1.07 & 0.55 & 1.76 & 0.06 &1.56\\
%-----------------
                & AP2 & 2.16 & 0.07 & -0.12 & 1.91 & -0.08 & -0.09 &
                1.23 & -0.78 & 0.84 & 0.66 & 1.14 & 0.07 & 1.48\\
%---
                & BD & 2.79 & 0.12 & 0.55 & 2.14 & -0.05 & -0.03 &
                1.65 & -0.83 & 2.08 & 0.79 & 1.39 & -0.01 & 1.70\\

\hline \tstrut

\end{tabular}
}
\caption{Variational parameters  of the three body orbital wave function for 
doubly heavy $\Xi$ baryons.  $a's$ are dimensionless, $d's$ have
dimensions of fm and $b's$ and $\alpha's$ have dimensions of fm$^{-1}$.
Whenever the two heavy quark are different
the suffix $h_1$ stands for a $b$ quark and $h_2$ stands for a $c$ quark.}
\label{tab:xiparam}
\end{table}
%
%
%\vspace{2cm}
\begin{table}[h]
{\scriptsize
\begin{tabular}{ll|rrrrrrrrrrrrr}
%             &&\multicolumn{12}{c}{Charm} \\\tstrut
             &         &$b_1$ &$d_1$ &~$a_2$   &$b_2$ &  $d_2$&~$a_3$
  &$b_3$ &$d_3$ &~$a_4$   &$b_4$ &$d_4$
  &~$\alpha_{h_1}$&~$\alpha_{h_2}$\\\hline\tstrut
%---------------------------------------------------------------------
  $\Omega_{cc}$ & AL1 & 1.60 & 0.29 & 0.30 & 1.29 & 1.20 & 1.70 & 1.27
  & 0.25 & 0.94 & 0.85 & 1.33 & 0.53 & 0.53 \\ 
%--
                & AL2 & 1.21 & 0.17 & 0.25 & 1.35 & 1.22 & 2.04 & 1.66
                & 0.22 & 0.69 & 0.98 & 1.63 & 0.48 & 0.48 \\
%--
                &AP1 &  2.30 & 0.11 & 0.48 & 1.70 & 1.26 & 1.72 & 1.36
                & 0.12 & 0.08 & 0.56 & 2.55 &0.47 &0.47\\
%--
                &AP2 & 1.63 & -0.24 & 0.62 & 1.75 &2.01 & 2.86 & 2.16
                & 0.10 & 0.16 & 1.35 & 2.56 &0.49 & 0.49 \\
%--
                & BD & 1.34& 0.09 & 0.37 & 1.46 & 1.30 & 1.91 & 1.69 &
                0.28 & 0.73 & 1.17 & 1.55 &0.47 &0.47\\

\hline\tstrut 
%------------------------------------------------
  $\Omega_{cc}^*$ & AL1 & 1.29 & 0.16 &-0.01 & 0.90 &  -0.49 & 0.96 &
  0.99 & 0.75 & 0.67 &2.05 & 0.13 & 0.95 & 0.95 \\
%--
                  & AL2 & 1.41 & 0.10 & 0.01 & 0.72 & -0.47 & 0.99
                  &1.00 & 0.77 & 0.71 & 2.00 & 0.15 & 0.89 & 0.89 \\
%-----------
                  &AP1 & 1.81 & 0.10 & 0.15 & 1.69 & -0.49 & 1.35 &
                  1.06 & 0.86 & -0.08 & 2.66 & 0.41 & 0.94 &0.94 \\
%------------------
                  &AP2 & 2.32 & 0.04 & 0.22 & 1.97 & -0.29 & 1.33 &
                  1.07 & 0.47 & -0.65 & 2.31 & 1.70 & 0.92 & 0.92\\
%-------
                  &BD & 1.66 & 0.15 & 0.16 & 1.41 & -0.27 & 1.14 &
                  1.26 & 0.80 & 0.30 & 2.21 & 0.71 & 0.73 & 0.73\\

\hline\tstrut
%----------------------------------------------------------------
 $\Omega_{bb}$ & AL1 & 1.28 &0.91 & -0.07 & 0.71 & 1.05 &1.86 & 2.32
 & 0.26& 1.38 & 0.99 &1.27 & 0.51 & 0.51 \\
%--
               & AL2 &  1.20 & 0.82 & -0.04 & 0.65 & 1.28 & 1.78 &
               2.56 & 0.15 & 1.22 & 1.03 & 1.41 &0.45 & 0.45 \\
%--------
               &AP1 & 1.18 & 0.90 & -0.09 & 0.72 & 1.24 & 1.70 & 2.48
               & 0.20 & 1.43 & 1.12 & 1.12 & 0.44 & 0.44\\
%-----------
               &AP2 & 2.00 & 0.03 & 0.22 & 0.58 & 1.70 & 2.81 & 3.22 &
               0.06 & -0.63 & 0.42 & 3.32 & 0.47 & 0.47 \\
%----
               &BD& 1.47 & 0.58 & -0.09 & 0.79 & 1.16 & 1.68 & 2.48 &
               0.25 & 1.25 & 1.00 & 1.52 & 0.46 & 0.46\\
%%%

\hline\tstrut
%---------------------------------------------
 $\Omega_{bb}^*$& AL1 & 1.14 & 1.01 & 0.01 & 1.03 & 0.44 & 3.05 & 2.03
 & 0.36 & 1.20 & 2.48 & 1.37 & 0.68 & 0.68 \\
%--
                & AL2 &  0.97 & 1.30 & 0.08 & 0.98 & 0.75 & 2.79 &
                2.36 & 0.19 & 1.21 & 2.40 & 1.38 & 0.63 & 0.63 \\
%--
                & AP1 & 1.05 & 1.16 & 0.11 & 1.29 & 0.56 & 3.25 & 2.16
                & 0.29 & 0.86 & 2.54 & 1.31 & 0.63 & 0.63 \\
%-------
                & AP2 & 0.94 & 1.30 & 0.33 & 1.11 & 0.76 & 3.39 & 2.55
                & 0.11 & 0.79 & 2.45 & 1.75 &0.65 &0.65 \\
%--------
                & BD & 1.52 & 1.12 & 0.19 & 1.15 & 0.74 & 2.91 & 2.05
                & 0.37 & 1.29 & 2.44 & 1.41 & 0.57 & 0.57\\

\hline\tstrut
%----------------------------------------------
 $\Omega_{bc}' $ &AL1 & 2.20 &0.32 & 1.00 & 1.92 & 0.05 & -0.09 & 2.89
 & -0.24 &1.79 &0.75 & 1.87 &-0.14 &2.42 \\ 
%--
                &AL2 & 2.21 & 0.22 & 0.76 & 2.03 & -0.07 & -0.16 &
                2.94 & -0.30 & 1.56 & 0.78 & 1.92 & -0.18 &2.34 \\
%-------
                &AP1 & 5.34 & 0.04 & 2.24 & 2.83 & 0.01 & -1.34 & 4.37
                & 0.29 & 0.64 & 2.14 & -0.28 & -0.17 & 2.35\\
%----------    
                &AP2 & 3.33 & 0.01 & 1.71 & 2.14 & -0.04 & -0.36 & 3.00
                & -0.20 & 1.23 & 0.71 & 1.88 & -0.17 & 2.34\\ 
%---------------
                & BD & 3.06 & 0.14 & 0.83 & 1.94 & 0.18 & 0.44 &2.16 &
                -0.11 & 1.67 & 0.75 & 1.78 & -0.24 & 2.26 \\

\hline\tstrut
%--------------------------------------------------------------
 $\Omega_{bc}$ &AL1 &2.15 &0.14 & -0.04 & 1.22 &-0.32 &-0.02 & 2.33
 &-1.04 &1.10 &0.84 &0.98 &-0.20 &1.84 \\ 
%--
                &AL2 & 2.03 & 0.15 & 0.03 & 1.39 &-0.36 & -0.01 & 2.37
                & -0.81 & 1.22 & 0.75 & 1.78 & -0.25 & 1.84 \\
%--
                &AP1 & 1.89 & 0.18 & -0.02& 1.49 & -0.40 & 0.04 & 2.10
                & -0.61 & 1.01 & 0.81 & 1.90 & -0.24 & 1.78\\
% -- 
		&AP2 & 2.34 & 0.02 & 0.51 & 4.50 & -0.01 & 0.53 & 1.82
		& -0.08 & 3.95 & 2.69 & 5.47 &-0.23 & 1.78 \\
%------
                & BD & 2.17 & 0.18 & 0.15 & 1.50 & -0.06 & -0.01 &
                2.46 & -0.89 & 1.41 & 0.75 & 1.97 & -0.27 & 1.85 \\

\hline\tstrut
%-----------------------------------------------------------------
 $\Omega_{bc}^*$ &AL1 &2.59 &0.10 & 1.01 & 1.90 &-0.01 & -0.27 &2.88
 &-0.13 &0.87 &0.51 &2.88 &-0.06 &2.54  \\ 
%--
                  &AL2 & 2.69 & 0.11 & 1.15 & 1.65 & 0.10 & 0.08 &
                  2.80 &-0.07 & 1.28 & 0.64 & 2.90 & -0.13 & 2.51 \\
%-------
                  & AP1 &  2.59 & 0.14 & 0.96 & 1.88 & -0.01 & -0.08 &
                  2.82 & -0.23 & 1.15 & 0.52 & 3.09 & -0.10 & 2.59\\
%-----
                  &AP2 & 3.80 & -0.01 & 1.75 & 1.81 & 0.03 & 0.69 &
                  2.31 & 0.03 & 2.48 & 0.56 & 2.94 & -0.11 &2.60 \\
%%_---------
		  & BD & 2.67 & 0.12 & 1.25 & 1.88 & 0.06 & -0.21 &
		  2.97 & -0.11 & 1.15 & 0.52 & 3.16 & -0.20 & 2.35 \\

\hline\tstrut

\end{tabular}
}
\caption{Same as Table~\protect\ref{tab:xiparam} for doubly heavy $\Omega$ baryons.}
\label{tab:omegaparam}
\end{table}
\chapter{Expressions for the $F_1,F_2,F_3$ and $G_1,G_2,G_3$ form
  factors}
\chaptermark{$F_1,F_2,F_3$ and $G_1,G_2,G_3$}

In this appendix we study the vector  $F_1,\,F_2,\,F_3$
and axial 
$G_1,\,G_2,\,G_3$
form factors, defined in chapter \ref{cha:dp_sl} to study
semileptonic decays of doubly heavy baryons.

\section[Form factors in terms of the ${\cal I}^{B'B}(|\vec q\,|)$ 
and ${\cal K}^{B'B}(|\vec q\,|)$
integrals]{Form factors in terms of the ${\cal I}^{B'B}(|\vec q\,|)$ 
and ${\cal K}^{B'B}(|\vec q\,|)$
integrals\sectionmark{Form factors in terms of ${\cal
    I}^{B'B}(|\vec{\lowercase{q}}\,|)$,${\cal
    K}^{B'B}(|\vec{\lowercase{q}}\,|)$,  }}
\sectionmark{Form factors in terms of ${\cal
    I}^{B'B}(|\vec{\lowercase{q}}\,|)$,${\cal
    K}^{B'B}(|\vec{\lowercase{q}}\,|)$,  }
\label{DPapp:fg123}

In this section  we relate the vector  $F_1,\,F_2,\,F_3$
and axial 
$G_1,\,G_2,\,G_3$
form factors
 that we evaluate in the center of mass of the decaying baryon, to the integrals ${\cal I}^{B'B}(|\vec q\,|),
\,{\cal K}^{B'B}(|\vec q\,|)$ defined in Eq.(\ref{eq:integrales}).  To simplify the expressions it is convenient
to introduce
\begin{eqnarray}
\widehat F_j(|\vec q\,|)&=&\sqrt{\frac{E_{B'}(-\vec q\, )+m_{B'}}{2E_{B'}(-\vec q\, )}}
\,\sqrt{\frac{2E_{c}(\vec q\, )}{E_{c}(\vec q\, )+m_{c}}}\ F_j(|\vec q\,|)\ \ ,\ j=1,2,3\nonumber\\
\widehat G_j(|\vec q\,|)&=&\sqrt{\frac{E_{B'}(-\vec q\, )+m_{B'}}{2E_{B'}(-\vec q\, )}}
\,\sqrt{\frac{2E_{c}(\vec q\, )}{E_{c}(\vec q\, )+m_{c}}}\ G_j(|\vec q\,|)\ \ ,\ j=1,2,3\nonumber\\
\end{eqnarray}

\begin{itemize}
\item Cases $S_h=1,\,S'_h=0$ or $S_h=0,\,S'_h=1$:
%\end{itemize}
\begin{eqnarray}
&&
\frac{\widehat F_1(|\vec q\,|)}{E_{B'}(-\vec q\, )+m_{B'}}\ 
\,=\,-\frac{2}{\sqrt3}\,
\left(\frac{{\cal I}^{B'B}(|\vec q\,|)}{E_{c}(\vec q\, )+m_c}
-\frac{{\cal K}^{B'B}(|\vec q\,|)}{2}
\left(\frac{m_c}{E^2_{c}(\vec q\, )}-\frac{1}{m_b}
\right)
\right)
\nonumber\\
&&
\frac{\widehat F_1(|\vec q\,|)}{E_{B'}(-\vec q\,
  )+m_{B'}}+\frac{\widehat F_3(|\vec q\,|)}{m_{B'}}\,=\,0
\nonumber\\
&&
\widehat F_1(|\vec q\,|)+\widehat F_2(|\vec q\,|)+\frac{E_{B'}(-\vec
  q\, )}{m_{B'}}\,\widehat F_3(|\vec q\,|)\,=\,0
\nonumber\\
\label{eq:appf123s10}
\end{eqnarray} 
where $E_c(\vec q\,)=\sqrt{m_c^2+\vec q^{\,2}}$. From the above expressions
we have that $F_2=F_3$.
{\allowdisplaybreaks
\begin{eqnarray}
\lefteqn{\hspace*{-2.2cm}
-\widehat G_1(|\vec q\,|)\, 
=\,\frac{2}{\sqrt3}\,
\left(\,{\cal I}^{B'B}(|\vec q\,|)+\frac{|\vec q\,|^2\,{\cal K}^{B'B}(|\vec q\,|)}{2\left(E_{c}(\vec q\, )+m_c\right)}
\left(\frac{m_c}{E^2_{c}(\vec q\, )}+\frac{1}{m_b}
\right)
\right)}\nonumber\\
\lefteqn{\hspace*{-2.2cm}
\left(-\widehat G_1(|\vec q\,|)+\frac{|\vec q\,|^2\,\widehat G_3(|\vec q\,|)}{m_{B'}(E_{B'}(-\vec q\,
)+m_{B'})}\right)\,=}
\nonumber \\* && \hspace*{-0.0cm} 
=\,\frac{2}{\sqrt3}\,
\left(\,{\cal I}^{B'B}(|\vec q\,|)+\frac{|\vec q\,|^2\,{\cal K}^{B'B}(|\vec q\,|)}{2\left(E_{c}(\vec q\, )+m_c\right)}
\left(\frac{m_c}{E^2_{c}(\vec q\, )}-\frac{1}{m_b}
\right)
\right)\nonumber\\
\lefteqn{\hspace*{-2.2cm}
\frac{-\widehat G_1(|\vec q\,|)+\widehat G_2(|\vec q\,|)+\frac{E_{B'}(-\vec q\, )}{m_{B'}}\,
\widehat G_3(|\vec q\,|)}{E_{B'}(-\vec q\, )+m_{B'}}\,=}
\nonumber \\* && \hspace*{-0.0cm} 
=\,\frac{2}{\sqrt3}\,
\left(\,\frac{{\cal I}^{B'B}(|\vec q\,|)}{E_{c}(\vec q\, )+m_c}
-\frac{{\cal K}^{B'B}(|\vec q\,|)}{2}
\left(\frac{m_c}{E^2_{c}(\vec q\, )}+\frac{1}{m_b}
\right)
\right)\nonumber\\*\nonumber\\* 
%\label{eq:appg123}
\end{eqnarray} 
}% Del allow...

\item Case $S_h=1,\,S'_h=1$

{\allowdisplaybreaks
\begin{eqnarray}
\lefteqn{\hspace*{-2.2cm}
\frac{\widehat F_1(|\vec q\,|)}{E_{B'}(-\vec q\, )+m_{B'}}\,=\,\frac{4}{3}\,
\left(\frac{{\cal I}^{B'B}(|\vec q\,|)}{E_{c}(\vec q\, )+m_c}
-\frac{{\cal K}^{B'B}(|\vec q\,|)}{2}
\left(\frac{m_c}{E^2_{c}(\vec q\, )}-\frac{1}{m_b}
\right)
\right)}
\nonumber\\
\lefteqn{\hspace*{-2.2cm}
\frac{\widehat F_1(|\vec q\,|)}{E_{B'}(-\vec q\, )+m_{B'}}+
\frac{\widehat F_3(|\vec q\,|)}{m_{B'}}\,=}
\nonumber\\*&&
=\,
2\,
\left(\frac{{\cal I}^{B'B}(|\vec q\,|)}{E_{c}(\vec q\, )+m_c}
-\frac{{\cal K}^{B'B}(|\vec q\,|)}{2}
\left(\frac{m_c}{E^2_{c}(\vec q\, )}+\frac{1}{m_b}
\right)
\right)
\nonumber\\
\lefteqn{\hspace*{-2.2cm}
\widehat F_1(|\vec q\,|)+\widehat F_2(|\vec q\,|)+\frac{E_{B'}(-\vec
  q\, )}{m_{B'}}\,\widehat F_3(|\vec q\,|)\,=}
\nonumber\\* && =\,
2\,
\left(\,{\cal I}^{B'B}(|\vec q\,|)
+\frac{|\vec q\,|^2\,{\cal K}^{B'B}(|\vec q\,|)}
{2\left(E_{c}(\vec q\, )+m_c\right)}
\left(\frac{m_c}{E^2_{c}(\vec q\, )}-\frac{1}{m_b}
\right)
\right)
\nonumber\\
\label{eq:appf123}
\end{eqnarray} 
}%Del Allow...

{\allowdisplaybreaks
\begin{eqnarray}
\lefteqn{\hspace*{-2.2cm}
\widehat G_1(|\vec q\,|)\,=\,\frac{4}{3}\,
\left(\,{\cal I}^{B'B}(|\vec q\,|)+\frac{|\vec q\,|^2\,{\cal K}^{B'B}(|\vec q\,|)}
{2\left(E_{c}(\vec q\, )+m_c\right)}
\left(\frac{m_c}{E^2_{c}(\vec q\, )}+\frac{1}{m_b}
\right)
\right)}\nonumber\\
\lefteqn{\hspace*{-2.2cm}
\left(-\widehat G_1(|\vec q\,|)+\frac{|\vec q\,|^2\,\widehat G_3(|\vec q\,|)}{m_{B'}(E_{B'}(-\vec q\,
)+m_{B'})}\right)\,=}
\nonumber\\* &&
=\,-\frac{4}{3}\,
\left(\,{\cal I}^{B'B}(|\vec q\,|)+\frac{|\vec q\,|^2\,{\cal K}^{B'B}(|\vec q\,|)}
{2\left(E_{c}(\vec q\, )+m_c\right)}
\left(\frac{m_c}{E^2_{c}(\vec q\, )}-\frac{1}{m_b}
\right)
\right)\nonumber\\
\lefteqn{\hspace*{-2.2cm}
\frac{-\widehat G_1(|\vec q\,|)+\widehat G_2(|\vec q\,|)+\frac{E_{B'}(-\vec q\, )}{m_{B'}}\,
\widehat G_3(|\vec q\,|)}{E_{B'}(-\vec q\, )+m_{B'}}\,=}
\nonumber\\*&&
=\,-\frac{4}{3}\,
\left(\,\frac{{\cal I}^{B'B}(|\vec q\,|)}{E_{c}(\vec q\, )+m_c}
-\frac{{\cal K}^{B'B}(|\vec q\,|)}{2}
\left(\frac{m_c}{E^2_{c}(\vec q\, )}+\frac{1}{m_b}
\right)
\right)\nonumber\\ 
%\label{eq:appg123}
\end{eqnarray} 
} % del allow

\end{itemize}

\section{${\cal I}^{B'B}(|\vec q\,|)$ and ${\cal K}^{B'B}(|\vec q\,|)$
  integrals} 
\chaptermark{${\cal I}^{B'B}(|\vec{\lowercase{q}}\,|)$ and ${\cal K}^{B'B}(|\vec{\lowercase{q}}\,|)$
  integrals}
\label{DPapp:integrals}
To evaluate  ${\cal I}^{B'B}(|\vec q\,|)$ and ${\cal K}^{B'B}(|\vec q\,|)$
we use  a partial wave expansion of the orbital wave functions:
\begin{eqnarray}
\Psi^{B}_{b\,h_2}(r_1,r_2,r_{12})=\sum_{l=0}^\infty f_l^{B}(r_1,r_2) P_l
(\mu)\nonumber\\
\Psi^{B'}_{c\,h_2}(r_1,r_2,r_{12})=\sum_{l=0}^\infty f_l^{B'}(r_1,r_2) P_l (\mu)
\end{eqnarray}
where $\mu$ is the cosine of
the angle made by $\vec r_1$ and $\vec r_2$ and $P_{l}(\mu)$ is a  Legendre 
polynomial of rank $l$.
The radial functions $f_l^{B}(r_1,r_2),\,f_l^{B'}(r_1,r_2),\,$ are evaluated as
\begin{eqnarray}
f_l^{B}(r_1,r_2)&=&\frac{2l+1}{2}\int_{-1}^{+1}d\mu\
P_l(\mu)\, \Psi^{B}_{b\,h_2}(r_1,r_2,r_{12})\nonumber\\
f_{l}^{B'}(r_1,r_2)&=&\frac{2l+1}{2}\int_{-1}^{+1}d\mu\
P_{l}(\mu)\, \Psi^{B'}_{c\,h_2}(r_1,r_2,r_{12})
\end{eqnarray}
In terms of those we have
\begin{eqnarray}
&&\hspace*{-.7cm}
{\cal I}^{B'B}(|\vec q\,|)\,=
\nonumber\\ && \hspace*{-.7cm}
=\,
16\pi^2\sum_l\sum_{l'}\sum_{l''}
\left(\left.l,l',l''\right|0,0,0\right)^2
\int_0^\infty d\,r_1\, r_1^2\ j_{l''}(\,\frac{m_{h_2}+m_q}{\overline M'} 
|\vec q\,|r_1)
\int_0^\infty d\,r_2\, r_2^2\ j_{l''}(\,\frac{m_{h_2}}{\overline M'} 
|\vec q\,|r_2)\nonumber\\
&&\hspace{5cm}\times f_{l'}^{B'}(r_1,r_2)\,f_{l}^{B}(r_1,r_2)
\end{eqnarray}
with  $j's$ being spherical Bessel functions.

And
\begin{eqnarray} &&\hspace*{-.7cm}
{\cal K}^{B'B}(|\vec q\,|)
\nonumber\\&&\hspace*{-.7cm}
=\,-\frac{16\pi^2}{\sqrt3\,|\vec q\,|}
\sum_l\sum_{l'}\sum_{l''}\sum_{l'''}\sum_L
(-1)^{(l{''}+l{'''}+1)/2}\sqrt{(2L+1)(2l''+1)(2l'''+1)}\nonumber\\
\hspace*{-.7cm}&&
\times\left(\left.l,l',l''\right|0,0,0\right)\,\left(\left.l'',l''',1\right|0,0,0\right)
\,\left(\left.l',L,l'''\right|0,0,0\right)\,W(l''l'1L;ll''')\nonumber\\
\hspace*{-.7cm}&& 
\times\int_0^\infty d\,r_1\, r_1^2\ j_{l'''}(\,\frac{m_{h_2}+m_q}{\overline M'} 
|\vec q\,|r_1)
\int_0^\infty d\,r_2\, r_2^2\ j_{l''}(\,\frac{m_{h_2}}{\overline M'} 
|\vec q\,|r_2)\
 f_{l'}^{B'}(r_1,r_2)\,\Omega_L\,f_{l}^{B}(r_1,r_2)\hspace{1cm}
\end{eqnarray}
with $W(l''l'1L;l,l''')$ being a Racah coefficient and $\Omega_L$ the
differential operator\footnote{Note that the Racah and Clebsch-Gordan
coefficients restrict $L$ to the two possible values  $L=l\pm1$.}
\begin{eqnarray}
\Omega_{L=l+1}&=&-\sqrt{\frac{l+1}{2l+1}}\left(
\frac{\partial}{\partial r_1}-\frac{l}{r_1}\right)\nonumber\\
\Omega_{L=l-1}&=&\ \sqrt{\frac{l}{2l+1}}\left(
\frac{\partial}{\partial r_1}+\frac{l+1}{r_1}\right)
\end{eqnarray}

For the actual evaluation we restrict the $l,\,l'$ values to
$l,\,l'=0,\cdots,6$.

}
{\small
\chapter{Expressions for the $A^{1/2,1/2}_{BB',\, \mu}$  matrix elements}
\label{Lambdapiapp}
The values for the $A^{1/2,1/2}_{BB',\, \mu}$  are evaluated using one-body current
operators and their expressions are given by
\begin{eqnarray}
%\label{va123}
\lefteqn{%\hspace*{-.7cm}
A^{1/2,1/2}_{\Sigma_c^{++}\Lambda_c^+,\, \mu}\,=}
\nonumber \\&& \hspace*{-.5cm}
=\,
\frac{\sqrt2}{\sqrt3}
\int d^{\,3}Q_1\ d^{\,3}Q_2\ 
\phi^{S_l=1}_{u,u,c}(\vec{Q}_1,\vec{Q}_2)
\left(
\phi^{S_l=0}_{d,u,c}(\vec{Q}_1-\frac{m_u+m_c}{\overline{M}_{\Lambda_c^+}}\, |
\vec{q}\,|\,\vec{k},\ \vec{Q}_2+\frac{m_u}{\overline{M}_{\Lambda_c^+}}\, |
\vec{q}\,|\,\vec{k} )
\right)^*\nonumber\\
&&\times\ \sum_{s_1}\,
\left(\left.\frac12,\frac12,1\right|s_1,-s_1,0\right)\,\left(\left.\frac12,\frac12,0\right|s_1,-s_1,0\right)
\nonumber\\
&&\hspace{1.2cm}\times\ \frac{1}{\sqrt{2E_d(\vec{Q}_1-|
\vec{q}\,|\,\vec{k} )2E_u(\vec{Q}_1)}}\ 
\overline{u}_{d\ s_1}(\vec{Q}_1-|
\vec{q}\,|\,\vec{k} )\ \gamma_\mu\gamma_5\ u_{u\, s_1}(\vec{Q}_1)
\end{eqnarray}
With $\overline{M}_{\Lambda_c^+}=m_u+m_d+m_c$. For $\mu=0,3$ we get the final 
expressions

\begin{eqnarray}
&&\hspace*{-.6cm}
A^{1/2,1/2}_{\Sigma_c^{++}\Lambda_c^+,\, 0}\,=
\nonumber\\ && \hspace*{-.5cm}
=\,
\frac{\sqrt2}{\sqrt3}
\int d^{\,3}Q_1\ d^{\,3}Q_2\ 
\phi^{S_l=1}_{u,u,c}(\vec{Q}_1,\vec{Q}_2)
\left(
\phi^{S_l=0}_{d,u,c}(\vec{Q}_1-\frac{m_u+m_c}{\overline{M}_{\Lambda_c^+}}\, |
\vec{q}\,|\,\vec{k},\ \vec{Q}_2+\frac{m_u}{\overline{M}_{\Lambda_c^+}}\, |
\vec{q}\,|\,\vec{k} )
\right)^*\nonumber\\
&&\hspace{2cm}\times\ \sqrt{\frac{\left(E_d(\vec{Q}_1-|
\vec{q}\,|\,\vec{k} )+m_d\right)\left(E_u(\vec{Q}_1)+m_u\right)}{{2E_d(\vec{Q}_1-|
\vec{q}\,|\,\vec{k} )2E_u(\vec{Q}_1)}}}\ 
\nonumber \\&& \hspace*{3cm}
\times
\left(
\frac{Q_1^z}{E_u(\vec{Q}_1)+m_u}+
\frac{Q_1^z-|\vec{q}\,|}{E_d(\vec{Q}_1-|
\vec{q}\,|\,\vec{k} )+m_d}
\right)
\end{eqnarray}

\begin{eqnarray}
&&\hspace*{-.6cm}
A^{1/2,1/2}_{\Sigma_c^{++}\Lambda_c^+,\, 3}\,=
\nonumber\\ && \hspace*{-.5cm}
=\,
-\frac{\sqrt2}{\sqrt3}
\int d^{\,3}Q_1\ d^{\,3}Q_2\ 
\phi^{S_l=1}_{u,u,c}(\vec{Q}_1,\vec{Q}_2)
\left(
\phi^{S_l=0}_{d,u,c}(\vec{Q}_1-\frac{m_u+m_c}{\overline{M}_{\Lambda_c^+}}\, |
\vec{q}\,|\,\vec{k},\ \vec{Q}_2+\frac{m_u}{\overline{M}_{\Lambda_c^+}}\, |
\vec{q}\,|\,\vec{k} )
\right)^*\nonumber\\
&&\times\ \sqrt{\frac{\left(E_d(\vec{Q}_1-|
\vec{q}\,|\,\vec{k} )+m_d\right)\left(E_u(\vec{Q}_1)+m_u\right)}{{2E_d(\vec{Q}_1-|
\vec{q}\,|\,\vec{k} )2E_u(\vec{Q}_1)}}}\nonumber\\ 
&&\times\ \left(1+\frac{1}{\left(E_d(\vec{Q}_1-|
\vec{q}\,|\,\vec{k} )+m_d\right)\left(E_u(\vec{Q}_1)+m_u\right)}\left(
2(Q_1^z)^2-|\vec{Q}_1|^2-Q_1^z|
\vec{q}\,|\right)
\right)
\nonumber\\
\end{eqnarray}
%
%
%
%\vspace{1cm}\\
For $A^{1/2,1/2}_{\Sigma_c^{*\,++}\Lambda_c^+,\, \mu}$ we just have
\begin{equation}
A^{1/2,1/2}_{\Sigma_c^{*\,++}\Lambda_c^+,\, \mu}
=-\sqrt2\ A^{1/2,1/2}_{\Sigma_c^{++}\Lambda_c^+,\, \mu}
\end{equation}
Finally
\begin{eqnarray}
&&\hspace*{-.6cm}
A^{1/2,1/2}_{\Xi_c^{*\, +}\Xi_c^0,\, \mu}\,=
\nonumber\\ && \hspace*{-.5cm}
=\,
\frac{\sqrt2}{\sqrt3}
\int d^{\,3}Q_1\ d^{\,3}Q_2\ 
\phi^{S_l=1}_{u,s,c}(\vec{Q}_1,\vec{Q}_2)
\left(
\phi^{S_l=0}_{d,s,c}(\vec{Q}_1-\frac{m_s+m_c}{\overline{M}_{\Xi_c^0}}\, |
\vec{q}\,|\,\vec{k},\ \vec{Q}_2+\frac{m_s}{\overline{M}_{\Xi_c^0}}\, |
\vec{q}\,|\,\vec{k} )
\right)^*\nonumber\\
&&\times\ \sum_{s_1}\,
\left(\left.\frac12,\frac12,1\right|s_1,-s_1,0\right)\,\left(\left.\frac12,\frac12,0\right|s_1,-s_1,0\right)
\nonumber\\
&&\hspace{1.2cm}\times\ \frac{1}{\sqrt{2E_d(\vec{Q}_1-|
\vec{q}\,|\,\vec{k} )2E_u(\vec{Q}_1)}}\ 
\overline{u}_{d\ s_1}(\vec{Q}_1-|
\vec{q}\,|\,\vec{k} )\ \gamma_\mu\gamma_5\ u_{u\, s_1}(\vec{Q}_1)
\end{eqnarray}
with $\overline{M}_{\Xi_c^0}=m_d+m_s+m_c$. For $\mu=0,3$ we get the final
expressions
\begin{eqnarray}
&&\hspace*{-.6cm}
A^{1/2,1/2}_{\Xi_c^{*\, +}\Xi_c^0,\, 0} \,=
\nonumber\\ && \hspace*{-.5cm}
=\,\frac{\sqrt2}{\sqrt3}
\int d^{\,3}Q_1\ d^{\,3}Q_2\ 
\phi^{S_l=1}_{u,s,c}(\vec{Q}_1,\vec{Q}_2)
\left(
\phi^{S_l=0}_{d,s,c}(\vec{Q}_1-\frac{m_s+m_c}{\overline{M}_{\Xi_c^0}}\, |
\vec{q}\,|\,\vec{k},\ \vec{Q}_2+\frac{m_s}{\overline{M}_{\Xi_c^0}}\, |
\vec{q}\,|\,\vec{k} )
\right)^*\nonumber\\
&&\hspace{2cm}\times\ \sqrt{\frac{\left(E_d(\vec{Q}_1-|
\vec{q}\,|\,\vec{k} )+m_d\right)\left(E_u(\vec{Q}_1)+m_u\right)}{{2E_d(\vec{Q}_1-|
\vec{q}\,|\,\vec{k} )2E_u(\vec{Q}_1)}}}\ 
\nonumber\\ &&\hspace{3cm}\times
\left(
\frac{Q_1^z}{E_u(\vec{Q}_1)+m_u}+
\frac{Q_1^z-|\vec{q}\,|}{E_d(\vec{Q}_1-|
\vec{q}\,|\,\vec{k} )+m_d}
\right)
\end{eqnarray}
\begin{eqnarray}
&&\hspace*{-.6cm}
A^{1/2,1/2}_{\Xi_c^{*\,+}\Xi_c^0,\, 3}\,=
\nonumber\\ && \hspace*{-.5cm}
=\,-\frac{\sqrt2}{\sqrt3}
\int d^{\,3}Q_1\ d^{\,3}Q_2\ 
\phi^{S_l=1}_{u,s,c}(\vec{Q}_1,\vec{Q}_2)
\left(
\phi^{S_l=0}_{d,s,c}(\vec{Q}_1-\frac{m_s+m_c}{\overline{M}_{\Xi_c^0}}\, |
\vec{q}\,|\,\vec{k},\ \vec{Q}_2+\frac{m_s}{\overline{M}_{\Xi_c^0}}\, |
\vec{q}\,|\,\vec{k} )
\right)^*\nonumber\\
&&\times\ \sqrt{\frac{\left(E_d(\vec{Q}_1-|
\vec{q}\,|\,\vec{k} )+m_d\right)\left(E_u(\vec{Q}_1)+m_u\right)}{{2E_d(\vec{Q}_1-|
\vec{q}\,|\,\vec{k} )2E_u(\vec{Q}_1)}}}\nonumber\\ 
&&\times\ \left(1+\frac{1}{\left(E_d(\vec{Q}_1-|
\vec{q}\,|\,\vec{k} )+m_d\right)\left(E_u(\vec{Q}_1)+m_u\right)}\left(
2(Q_1^z)^2-|\vec{Q}_1|^2-Q_1^z|
\vec{q}\,|\right)
\right)
\nonumber\\ \nonumber\\
\end{eqnarray}
}
%
% No lo pongo, aunque molar mola
%\input{camposapp.inp}

%\listoftables
%\listoffigures

%\chapter
%[\hspace{79pt}Reducción no relativista]
%{Reducción no relativista}
%\label{rnr}
%\input{norel.inp}
%--------------------------------

%-----------------------------------------------------------------------------------------------
%LA L que uso en los indices
%\mathcal{L}
%\nocite{*}
%\cite{FF1}
%\bibliographystyle{alpha} % Esta es con iniciales
%La otra posibilidad seria
%\bibliographystyle{plain} %Esta es con numeros
%\bibliographystyle{mieeetr}
\bibliographystyle{JHEP} % ESTA ES LA QUE ME CONVENCE
\bibliography{tesis}  

\providecommand{\href}[2]{#2}\begingroup\raggedright\begin{thebibliography}{10%
0}

\bibitem{Bali:2000gf}
G.~S. Bali, {\it \uppercase{QCD} forces and heavy quark bound states},  {\em
  Phys. Rept.} {\bf 343} (2001) 1--136,
  [\href{http://xxx.lanl.gov/abs/hep-ph/0001312}{{\tt hep-ph/0001312}}].

\bibitem{Politzer:1973fx}
H.~D. Politzer, {\it Reliable perturbative results for strong interactions?},
  {\em Phys. Rev. Lett.} {\bf 30} (1973) 1346--1349.

\bibitem{Gross:1973id}
D.~J. Gross and F.~Wilczek, {\it Ultraviolet behavior of non-abelian gauge
  theories},  {\em Phys. Rev. Lett.} {\bf 30} (1973) 1343--1346.

\bibitem{Gross:1973ju}
D.~J. Gross and F.~Wilczek, {\it Asymptotically free gauge theories. 1},  {\em
  Phys. Rev.} {\bf D8} (1973) 3633--3652.

\bibitem{Fritzsch:1972jv}
H.~Fritzsch and M.~Gell-Mann, {\it Current algebra: Quarks and what else?},
  {\em Proc. XVI Int. Conf. on High Energy Physics} {\bf Vol. 2} (1972) 135.

\bibitem{Fritzsch:1973pi}
H.~Fritzsch, M.~Gell-Mann, and H.~Leutwyler, {\it Advantages of the color octet
  gluon picture},  {\em Phys. Lett.} {\bf B47} (1973) 365--368.

\bibitem{Glashow:1970gm}
S.~L. Glashow, J.~Iliopoulos, and L.~Maiani, {\it Weak interactions with
  lepton-hadron symmetry},  {\em Phys. Rev.} {\bf D2} (1970) 1285--1292.

\bibitem{DeRujula:1975ge}
A.~De~Rujula, H.~Georgi, and S.~L. Glashow, {\it Hadron masses in a gauge
  theory},  {\em Phys. Rev.} {\bf D12} (1975) 147--162.

\bibitem{Aubert:1974js}
{\bf E598} Collaboration, J.~J. Aubert {\em et~al.}, {\it Experimental
  observation of a heavy particle \uppercase{J}},  {\em Phys. Rev. Lett.} {\bf
  33} (1974) 1404--1406.

\bibitem{Augustin:1974xw}
{\bf SLAC-SP-017} Collaboration, J.~E. Augustin {\em et~al.}, {\it Discovery of
  a narrow resonance in $e^+$ $e^-$ annihilation},  {\em Phys. Rev. Lett.} {\bf
  33} (1974) 1406--1408.

\bibitem{Goldhaber:1976xn}
G.~Goldhaber {\em et~al.}, {\it Observation in $e^+$ $e^-$ annihilation of a
  narrow state at $1865$ $\uppercase{M}e\uppercase{V}/c^2$ decaying to
  $\uppercase{K} \pi$ and $\uppercase{K} \pi \pi \pi$},  {\em Phys. Rev. Lett.}
  {\bf 37} (1976) 255--259.

\bibitem{Peruzzi:1976sv}
I.~Peruzzi {\em et~al.}, {\it Observation of a narrow charged state at $1876$
  $\uppercase{M}e\uppercase{V}/c^2$ decaying to an exotic combination of
  $\uppercase{K} \pi\ pi$},  {\em Phys. Rev. Lett.} {\bf 37} (1976) 569--571.

\bibitem{Kobayashi:1973fv}
M.~Kobayashi and T.~Maskawa, {\it Cp violation in the renormalizable theory of
  weak interaction},  {\em Prog. Theor. Phys.} {\bf 49} (1973) 652--657.

\bibitem{herb:1977ek}
S.~W. Herb {\em et~al.}, {\it Observation of a dimuon resonance at $9.5$
  $\uppercase{G}e\uppercase{V}$ in $400$ $\uppercase{G}e\uppercase{V}$
  proton-nucleus collisions},  {\em Phys. Rev. Lett.} {\bf 39} (1977) 252--255.

\bibitem{Abe:1994st}
{\bf CDF} Collaboration, F.~Abe {\em et~al.}, {\it Evidence for top quark
  production in $\bar{p}p$ collisions at $\sqrt{s} = 1.8$
  $\uppercase{T}e\uppercase{V}$},  {\em Phys. Rev.} {\bf D50} (1994)
  2966--3026.

\bibitem{Close:1979bt}
F.~E. Close, {\it An introduction to quarks and partons}, . Academic
  Press/london 1979, 481p.

\bibitem{Isgur:1977ef}
N.~Isgur and G.~Karl, {\it Hyperfine interactions in negative parity baryons},
  {\em Phys. Lett.} {\bf B72} (1977) 109.

\bibitem{Isgur:1978xi}
N.~Isgur and G.~Karl, {\it Symmetry breaking in baryons},  {\em Phys. Lett.}
  {\bf B74} (1978) 353.

\bibitem{Isgur:1978xj}
N.~Isgur and G.~Karl, {\it P wave baryons in the quark model},  {\em Phys.
  Rev.} {\bf D18} (1978) 4187.

\bibitem{Isgur:1978wd}
N.~Isgur and G.~Karl, {\it Positive parity excited baryons in a quark model
  with hyperfine interactions},  {\em Phys. Rev.} {\bf D19} (1979) 2653.

\bibitem{Isgur:1978xb}
N.~Isgur, G.~Karl, and R.~Koniuk, {\it Violations of \uppercase{SU}(6)
  selection rules from quark hyperfine interactions},  {\em Phys. Rev. Lett.}
  {\bf 41} (1978) 1269.

\bibitem{Isgur:1979be}
N.~Isgur and G.~Karl, {\it Ground state baryons in a quark model with hyperfine
  interactions},  {\em Phys. Rev.} {\bf D20} (1979) 1191--1194.

\bibitem{Lucha:1991vn}
W.~Lucha, F.~F. Schoberl, and D.~Gromes, {\it Bound states of quarks},  {\em
  Phys. Rept.} {\bf 200} (1991) 127--240.

\bibitem{Oka:1981ri}
M.~Oka and K.~Yazaki, {\it Short range part of baryon baryon interaction in a
  quark model. 1. formulation},  {\em Prog. Theor. Phys.} {\bf 66} (1981)
  556--571.

\bibitem{Oka:1981rj}
M.~Oka and K.~Yazaki, {\it Short range part of baryon baryon interaction in a
  quark model. 2. numerical results for \uppercase{S}-wave},  {\em Prog. Theor.
  Phys.} {\bf 66} (1981) 572--587.

\bibitem{Faessler:1983yd}
A.~Faessler, F.~Fernandez, G.~Lubeck, and K.~Shimizu, {\it The nucleon nucleon
  interaction and the role of the (42) orbital six quark symmetry},  {\em Nucl.
  Phys.} {\bf A402} (1983) 555--568.

\bibitem{Manohar:1983md}
A.~Manohar and H.~Georgi, {\it Chiral quarks and the nonrelativistic quark
  model},  {\em Nucl. Phys.} {\bf B234} (1984) 189.

\bibitem{Shimizu:1985mg}
K.~Shimizu, {\it One pion exchange potential based on a quark model},  {\em
  Phys. Lett.} {\bf B148} (1984) 418--422.

\bibitem{Maltman:1985mc}
K.~Maltman, {\it One pion exchange effects in few nucleon systems},  {\em Nucl.
  Phys.} {\bf A446} (1985) 623.

\bibitem{Fernandez:1986zn}
F.~Fernandez and E.~Oset, {\it A model of the double spin flip n n amplitude
  based on a one pion exchange potential with quark exchange},  {\em Nucl.
  Phys.} {\bf A455} (1986) 720--736.

\bibitem{Obukhovsky:1990tx}
I.~T. Obukhovsky and A.~M. Kusainov, {\it The nucleon nucleon scattering and
  the baryon spectrum in the quark cluster model with two scales of
  interaction},  {\em Phys. Lett.} {\bf B238} (1990) 142--148.

\bibitem{Fernandez:1993hx}
F.~Fernandez, A.~Valcarce, U.~Straub, and A.~Faessler, {\it The nucleon-nucleon
  interaction in terms of quark degrees of freedom},  {\em J. Phys.} {\bf G19}
  (1993) 2013--2026.

\bibitem{Valcarce:1994nr}
A.~Valcarce, F.~Fernandez, A.~Buchmann, and A.~Faessler, {\it Can one
  simultaneously describe the deuteron properties and the nucleon-nucleon phase
  shifts in the quark cluster model?},  {\em Phys. Rev.} {\bf C50} (1994)
  2246--2249.

\bibitem{Glozman:1995fu}
L.~Y. Glozman and D.~O. Riska, {\it The spectrum of the nucleons and the
  strange hyperons and chiral dynamics},  {\em Phys. Rept.} {\bf 268} (1996)
  263--303, [\href{http://xxx.lanl.gov/abs/hep-ph/9505422}{{\tt
  hep-ph/9505422}}].

\bibitem{Isgur:1999jv}
N.~Isgur, {\it Critique of a pion exchange model for interquark forces},  {\em
  Phys. Rev.} {\bf D62} (2000) 054026,
  [\href{http://xxx.lanl.gov/abs/nucl-th/9908028}{{\tt nucl-th/9908028}}].

\bibitem{Buchmann:1995ek}
A.~J. Buchmann, G.~Wagner, K.~Tsushima, A.~Faessler, and L.~Y. Glozman, {\it
  The d'-dibaryon in the nonrelativistic quark model},  {\em PiN Newslett.}
  {\bf 10} (1995) 60--66, [\href{http://xxx.lanl.gov/abs/nucl-th/9508011}{{\tt
  nucl-th/9508011}}].

\bibitem{Wagner:1995id}
G.~Wagner, L.~Y. Glozman, A.~Buchmann, and A.~Faessler, {\it Constituent quark
  model calculation for a possible \uppercase{J}(p) = 0-, \uppercase{T} = 0
  dibaryon},  {\em Prog. Part. Nucl. Phys.} {\bf 34} (1995) 133--135.

\bibitem{Mota:2001ee}
R.~D. Mota, A.~Valcarce, F.~Fernandez, D.~R. Entem, and H.~Garcilazo, {\it
  Nonlocal calculation for nonstrange dibaryons and tribaryons},  {\em Phys.
  Rev.} {\bf C65} (2002) 034006,
  [\href{http://xxx.lanl.gov/abs/nucl-th/0112059}{{\tt nucl-th/0112059}}].

\bibitem{Fernandez-Carames:2006ra}
T.~Fernandez-Carames, A.~Valcarce, H.~Garcilazo, and P.~Gonzalez, {\it Strange
  tribaryons},  {\em Phys. Rev.} {\bf C73} (2006) 034004,
  [\href{http://xxx.lanl.gov/abs/hep-ph/0601252}{{\tt hep-ph/0601252}}].

\bibitem{Vijande:2003ki}
J.~Vijande, F.~Fernandez, A.~Valcarce, and B.~Silvestre-Brac, {\it Tetraquarks
  in a chiral constituent quark model},  {\em Eur. Phys. J.} {\bf A19} (2004)
  383, [\href{http://xxx.lanl.gov/abs/hep-ph/0310007}{{\tt hep-ph/0310007}}].

\bibitem{Vijande:2006jf}
J.~Vijande, A.~Valcarce, and K.~Tsushima, {\it Dynamical study of
  $\uppercase{Q} \uppercase{Q} - \overline{u} \overline{d}$ mesons},  {\em
  Phys. Rev.} {\bf D74} (2006) 054018,
  [\href{http://xxx.lanl.gov/abs/hep-ph/0608316}{{\tt hep-ph/0608316}}].

\bibitem{Barnea:1}
N.~Barnea, J.~Vijande, and A.~Valcarce, {\it Four-quark spectroscopy within the
  hyperspherical formalism},  {\em Phys. Rev.} {\bf D73} (2006) 054004,
  [\href{http://xxx.lanl.gov/abs/hep-ph/0604010}{{\tt hep-ph/0604010}}].

\bibitem{Valcarce:2005em}
A.~Valcarce, H.~Garcilazo, F.~Fernandez, and P.~Gonzalez, {\it Quark-model
  study of few-baryon systems},  {\em Rept. Prog. Phys.} {\bf 68} (2005)
  965--1042, [\href{http://xxx.lanl.gov/abs/hep-ph/0502173}{{\tt
  hep-ph/0502173}}].

\bibitem{Nag:1987cv}
R.~Nag, S.~Sanyal, and S.~N. Mukherjee, {\it Electromagnetic structure of the
  proton and baryon spectrum in the nonrelativistic quark model},  {\em Phys.
  Rev.} {\bf D36} (1987) 2788--2799.

\bibitem{Giannini:1990pc}
M.~M. Giannini, {\it Electromagnetic excitations in the constituent quark
  model},  {\em Rept. Prog. Phys.} {\bf 54} (1990) 453--530.

\bibitem{Buchmann:1991cy}
A.~Buchmann, E.~Hernandez, and K.~Yazaki, {\it Gluon and pion exchange currents
  in the nucleon},  {\em Phys. Lett.} {\bf B269} (1991) 35--42.

\bibitem{Buchmann:1994bt}
A.~Buchmann, E.~Hernandez, and K.~Yazaki, {\it Gluon, pion and confinement
  exchange currents in the nucleon},  {\em Nucl. Phys.} {\bf A569} (1994)
  661--688.

\bibitem{Buchmann:1996bd}
A.~J. Buchmann, E.~Hernandez, and A.~Faessler, {\it Electromagnetic properties
  of the $\uppercase{\Delta}(1232)$},  {\em Phys. Rev.} {\bf C55} (1997)
  448--463, [\href{http://xxx.lanl.gov/abs/nucl-th/9610040}{{\tt
  nucl-th/9610040}}].

\bibitem{Buchmann:1998yj}
A.~J. Buchmann, U.~Meyer, A.~Faessler, and E.~Hernandez, {\it N $\to$
  $\uppercase{\Delta}(1232)$ \uppercase{E}2 transition and
  \uppercase{S}iegert's theorem},  {\em Phys. Rev.} {\bf C58} (1998)
  2478--2488.

\bibitem{Meyer:2001js}
U.~Meyer, E.~Hernandez, and A.~J. Buchmann, {\it Exchange currents in nucleon
  electroexcitation},  {\em Phys. Rev.} {\bf C64} (2001) 035203.

\bibitem{Julia-Diaz:2004eg}
B.~Julia-Diaz and D.~O. Riska, {\it D-state configurations in the
  electromagnetic form factors of the nucleon and the
  $\uppercase{\Delta}(1232)$ resonance},  {\em Nucl. Phys.} {\bf A757} (2005)
  441--455, [\href{http://xxx.lanl.gov/abs/nucl-th/0411012}{{\tt
  nucl-th/0411012}}].

\bibitem{Beyer:1985iq}
M.~Beyer and S.~K. Singh, {\it The nucleon axial vector form-factor in an
  improved constituent quark model},  {\em Phys. Lett.} {\bf B160} (1985)
  26--31.

\bibitem{Liu:1995bu}
J.~Liu, N.~C. Mukhopadhyay, and L.-s. Zhang, {\it Nucleon to delta weak
  excitation amplitudes in the nonrelativistic quark model},  {\em Phys. Rev.}
  {\bf C52} (1995) 1630--1647,
  [\href{http://xxx.lanl.gov/abs/hep-ph/9506389}{{\tt hep-ph/9506389}}].

\bibitem{Barquilla-Cano:2002cm}
D.~Barquilla-Cano, A.~J. Buchmann, and E.~Hernandez, {\it Partial conservation
  of axial current and axial exchange currents in the nucleon},  {\em Nucl.
  Phys.} {\bf A714} (2003) 611--631,
  [\href{http://xxx.lanl.gov/abs/nucl-th/0204067}{{\tt nucl-th/0204067}}].

\bibitem{Julia-Diaz:2004qr}
B.~Julia-Diaz, D.~O. Riska, and F.~Coester, {\it Axial transition form factors
  and pion decay of baryon resonances},  {\em Phys. Rev.} {\bf C70} (2004)
  045204, [\href{http://xxx.lanl.gov/abs/nucl-th/0406015}{{\tt
  nucl-th/0406015}}].

\bibitem{Barquilla-Cano:2006ws}
D.~Barquilla-Cano, A.~J. Buchmann, and E.~Hernandez, {\it Axial exchange
  currents and nucleon spin},  {\em Eur. Phys. J.} {\bf A27} (2006) 365--372,
  [\href{http://xxx.lanl.gov/abs/hep-ph/0611248}{{\tt hep-ph/0611248}}].

\bibitem{Nussinov:1986hw}
S.~Nussinov and W.~Wetzel, {\it Comparison of exclusive decay rates for b $\to$
  u and b $\to$ c transitions},  {\em Phys. Rev.} {\bf D36} (1987) 130.

\bibitem{Shifman:1986sm}
M.~A. Shifman and M.~B. Voloshin, {\it On annihilation of mesons built from
  heavy and light quark and anti-\uppercase{B}0 $\longleftrightarrow$
  \uppercase{B}0 oscillations},  {\em Sov. J. Nucl. Phys.} {\bf 45} (1987) 292.

\bibitem{Politzer:1988wp}
H.~D. Politzer and M.~B. Wise, {\it Leading logarithms of heavy quark masses in
  processes with light and heavy quarks},  {\em Phys. Lett.} {\bf B206} (1988)
  681.

\bibitem{Politzer:1988bs}
H.~D. Politzer and M.~B. Wise, {\it Effective field theory approach to
  processes involving both light and heavy fields},  {\em Phys. Lett.} {\bf
  B208} (1988) 504.

\bibitem{Isgur:1989vq}
N.~Isgur and M.~B. Wise, {\it Weak decays of heavy mesons in the static quark
  approximation},  {\em Phys. Lett.} {\bf B232} (1989) 113.

\bibitem{Isgur:1989ed}
N.~Isgur and M.~B. Wise, {\it Weak transition form-factors between heavy
  mesons},  {\em Phys. Lett.} {\bf B237} (1990) 527.

\bibitem{Georgi:1990um}
H.~Georgi, {\it An effective field theory for heavy quarks at low energies},
  {\em Phys. Lett.} {\bf B240} (1990) 447--450.

\bibitem{Neubert:1993mb}
M.~Neubert, {\it Heavy quark symmetry},  {\em Phys. Rept.} {\bf 245} (1994)
  259--396, [\href{http://xxx.lanl.gov/abs/hep-ph/9306320}{{\tt
  hep-ph/9306320}}].

\bibitem{Korner:1994nh}
J.~G. Korner, M.~Kramer, and D.~Pirjol, {\it Heavy baryons},  {\em Prog. Part.
  Nucl. Phys.} {\bf 33} (1994) 787--868,
  [\href{http://xxx.lanl.gov/abs/hep-ph/9406359}{{\tt hep-ph/9406359}}].

\bibitem{Booth:1993zb}
{\bf UKQCD} Collaboration, S.~P. Booth {\em et~al.}, {\it The
  \uppercase{I}sgur-\uppercase{W}ise function from the lattice},  {\em Phys.
  Rev. Lett.} {\bf 72} (1994) 462--465,
  [\href{http://xxx.lanl.gov/abs/hep-lat/9308019}{{\tt hep-lat/9308019}}].

\bibitem{Burford:1995fc}
{\bf UKQCD} Collaboration, D.~R. Burford {\em et~al.}, {\it Form-factors for
  $\uppercase{B} \to \pi l\bar{\nu}_l$ and $\uppercase{B} \to
  \uppercase{K}^*\gamma$ decays on the lattice},  {\em Nucl. Phys.} {\bf B447}
  (1995) 425--440, [\href{http://xxx.lanl.gov/abs/hep-lat/9503002}{{\tt
  hep-lat/9503002}}].

\bibitem{Bowler:1994zr}
{\bf UKQCD} Collaboration, K.~C. Bowler {\em et~al.}, {\it An 'improved'
  lattice study of semileptonic decays of \uppercase{D} mesons},  {\em Phys.
  Rev.} {\bf D51} (1995) 4905--4923,
  [\href{http://xxx.lanl.gov/abs/hep-lat/9410012}{{\tt hep-lat/9410012}}].

\bibitem{Flynn:1995dc}
{\bf UKQCD} Collaboration, J.~M. Flynn {\em et~al.}, {\it Lattice study of the
  decay $\bar{\uppercase{b}}^0 \to \rho^+ l^- \bar{\nu}_l $: Model independent
  determination of $|\uppercase{V}_{ub}|$},  {\em Nucl. Phys.} {\bf B461}
  (1996) 327--349, [\href{http://xxx.lanl.gov/abs/hep-ph/9506398}{{\tt
  hep-ph/9506398}}].

\bibitem{DelDebbio:1997kr}
{\bf UKQCD} Collaboration, L.~Del~Debbio, J.~M. Flynn, L.~Lellouch, and
  J.~Nieves, {\it Lattice-constrained parametrizations of form factors for
  semileptonic and rare radiative \uppercase{B} decays},  {\em Phys. Lett.}
  {\bf B416} (1998) 392--401,
  [\href{http://xxx.lanl.gov/abs/hep-lat/9708008}{{\tt hep-lat/9708008}}].

\bibitem{Thacker:1990bm}
B.~A. Thacker and G.~P. Lepage, {\it Heavy quark bound states in lattice
  $\uppercase{QCD}$},  {\em Phys. Rev.} {\bf D43} (1991) 196--208.

\bibitem{Jenkins:1992nb}
E.~Jenkins, M.~E. Luke, A.~V. Manohar, and M.~J. Savage, {\it Semileptonic
  $\uppercase{B}_c$ decay and heavy quark spin symmetry},  {\em Nucl. Phys.}
  {\bf B390} (1993) 463--473,
  [\href{http://xxx.lanl.gov/abs/hep-ph/9204238}{{\tt hep-ph/9204238}}].

\bibitem{Bhaduri:1981pn}
R.~K. Bhaduri, L.~E. Cohler, and Y.~Nogami, {\it A unified potential for mesons
  and baryons},  {\em Nuovo Cim.} {\bf A65} (1981) 376--390.

\bibitem{Semay:1994ht}
C.~Semay and B.~Silvestre-Brac, {\it Diquonia and potential models},  {\em Z.
  Phys.} {\bf C61} (1994) 271--275.

\bibitem{Silvestre-Brac:1996bg}
B.~Silvestre-Brac, {\it Spectrum and static properties of heavy baryons},  {\em
  Few Body Syst.} {\bf 20} (1996) 1--25.

\bibitem{Gutbrod:1983yi}
F.~Gutbrod and I.~Montvay, {\it Scaling of the quark - anti-quark potential and
  improved actions in \uppercase{SU}(2) lattice gauge theory},  {\em Phys.
  Lett.} {\bf B136} (1984) 411.

\bibitem{FabreDeLaRipelle:1988zr}
M.~Fabre De La~Ripelle, {\it A confining potential for quarks},  {\em Phys.
  Lett.} {\bf B205} (1988) 97--102.

\bibitem{Ono:1982ft}
S.~Ono and F.~Schoberl, {\it A simultaneous and systematic study of meson and
  baryon spectra in the quark model},  {\em Phys. Lett.} {\bf B118} (1982) 419.

\bibitem{Narodetsky:1992zy}
I.~M. Narodetsky, R.~Ceuleneer, and C.~Semay, {\it Hyperfine interaction in the
  nonrelativistic quark model and the convergence of harmonic oscillator
  variational method},  {\em J. Phys.} {\bf G18} (1992) 1901--1909.

\bibitem{Albertus:2004iw}
C.~Albertus, E.~Hernandez, and J.~Nieves, {\it Study of the semileptonic decay
  $\uppercase{\Lambda}_b^0 \to \uppercase{\Lambda}_c^+ l^- \bar\nu_l$},  {\em
  Nucl. Phys. Proc. Suppl.} {\bf 142} (2005) 27--30,
  [\href{http://xxx.lanl.gov/abs/hep-ph/0408065}{{\tt hep-ph/0408065}}].

\bibitem{Isgur:1989qw}
N.~Isgur and M.~B. Wise, {\it Influence of the $\uppercase{B}^*$ resonance on
  $\bar{\uppercase{b}}\to \pi e \bar{\nu}_e$},  {\em Phys. Rev.} {\bf D41}
  (1990) 151.

\bibitem{Isgur:1988gb}
N.~Isgur, D.~Scora, B.~Grinstein, and M.~B. Wise, {\it Semileptonic
  \uppercase{B} and \uppercase{D} decays in the quark model},  {\em Phys. Rev.}
  {\bf D39} (1989) 799.

\bibitem{Scora:1995ty}
D.~Scora and N.~Isgur, {\it Semileptonic meson decays in the quark model: An
  update},  {\em Phys. Rev.} {\bf D52} (1995) 2783--2812,
  [\href{http://xxx.lanl.gov/abs/hep-ph/9503486}{{\tt hep-ph/9503486}}].

\bibitem{Capstick:1989ra}
S.~Capstick and S.~Godfrey, {\it Pseudoscalar decay constants in the
  relativized quark model and measuring the \uppercase{CKM} matrix elements},
  {\em Phys. Rev.} {\bf D41} (1990) 2856.

\bibitem{Barik:1993aw}
N.~Barik and P.~C. Dash, {\it Weak leptonic decay of light and heavy
  pseudoscalar mesons in an independent quark model},  {\em Phys. Rev.} {\bf
  D47} (1993) 2788--2795.

\bibitem{Hwang:1995uy}
D.~S. Hwang and G.-H. Kim, {\it Ratios of \uppercase{B} and \uppercase{D} meson
  decay constants in relativistic quark model},  {\em Phys. Rev.} {\bf D53}
  (1996) 3659--3663, [\href{http://xxx.lanl.gov/abs/hep-ph/9507340}{{\tt
  hep-ph/9507340}}].

\bibitem{Hwang:1995vt}
D.~S. Hwang, C.~S. Kim, and W.~Namgung, {\it Decay constants and semileptonic
  decays of heavy mesons in relativistic quark model},  {\em Phys. Rev.} {\bf
  D53} (1996) 4951--4956, [\href{http://xxx.lanl.gov/abs/hep-ph/9506476}{{\tt
  hep-ph/9506476}}].

\bibitem{Micu:1996hj}
L.~Micu, {\it The decay constants of pseudoscalar mesons in a relativistic
  quark model},  {\em Phys. Rev.} {\bf D55} (1997) 4151--4156,
  [\href{http://xxx.lanl.gov/abs/hep-ph/9608385}{{\tt hep-ph/9608385}}].

\bibitem{AbdEl-Hady:1997rj}
A.~Abd El-Hady, A.~Datta, and J.~P. Vary, {\it Decay constants, semi-leptonic
  and non-leptonic \uppercase{B} decays in a
  \uppercase{B}ethe-\uppercase{S}alpeter model},  {\em Phys. Rev.} {\bf D58}
  (1998) 014007, [\href{http://xxx.lanl.gov/abs/hep-ph/9711338}{{\tt
  hep-ph/9711338}}].

\bibitem{Morenas:1997rx}
V.~Morenas, A.~Le~Yaouanc, L.~Oliver, O.~Pene, and J.~C. Raynal, {\it Decay
  constants in the heavy quark limit in models a la \uppercase{B}akamjian and
  \uppercase{T}homas},  {\em Phys. Rev.} {\bf D58} (1998) 114019,
  [\href{http://xxx.lanl.gov/abs/hep-ph/9710298}{{\tt hep-ph/9710298}}].

\bibitem{Melikhov:2000yu}
D.~Melikhov and B.~Stech, {\it Weak form factors for heavy meson decays: An
  update},  {\em Phys. Rev.} {\bf D62} (2000) 014006,
  [\href{http://xxx.lanl.gov/abs/hep-ph/0001113}{{\tt hep-ph/0001113}}].

\bibitem{Wang:2004xs}
Z.-G. Wang, W.-M. Yang, and S.-L. Wan, {\it Decay constants of the pseudoscalar
  mesons in the framework of the coupled
  \uppercase{S}chwinger--\uppercase{D}yson equation and
  \uppercase{B}ethe--\uppercase{S}alpeter equation},  {\em Nucl. Phys.} {\bf
  A744} (2004) 156--167, [\href{http://xxx.lanl.gov/abs/hep-ph/0403259}{{\tt
  hep-ph/0403259}}].

\bibitem{DeVito:2004zs}
M.~De~Vito and P.~Santorelli, {\it $\uppercase{B} \to \uppercase{D}^{(*)}$
  transitions in a quark model},  {\em Eur. Phys. J.} {\bf C40} (2005)
  193--197, [\href{http://xxx.lanl.gov/abs/hep-ph/0412388}{{\tt
  hep-ph/0412388}}].

\bibitem{Jaus:1989au}
W.~Jaus, {\it Semileptonic decays of \uppercase{B} and \uppercase{D} mesons in
  the light front formalism},  {\em Phys. Rev.} {\bf D41} (1990) 3394.

\bibitem{Jaus:1991cy}
W.~Jaus, {\it Relativistic constituent quark model of electroweak properties of
  light mesons},  {\em Phys. Rev.} {\bf D44} (1991) 2851--2859.

\bibitem{Jaus:1996np}
W.~Jaus, {\it Semileptonic, radiative, and pionic decays of $\uppercase{B}$,
  $\uppercase{B}^*$ and $\uppercase{D}$, $\uppercase{D}^*$ mesons},  {\em Phys.
  Rev.} {\bf D53} (1996) 1349--1365.

\bibitem{Faustov:1995xc}
R.~N. Faustov and V.~O. Galkin, {\it Heavy quark 1/m(\uppercase{Q}) expansion
  of meson weak decay form factors in the relativistic quark model},  {\em Z.
  Phys.} {\bf C66} (1995) 119--127.

\bibitem{Barik:1996xf}
N.~Barik and P.~C. Dash, {\it Exclusive semileptonic decay of \uppercase{D} and
  \uppercase{B} mesons in the independent quark model},  {\em Phys. Rev.} {\bf
  D53} (1996) 1366--1377.

\bibitem{Ishida:1997ft}
M.~Ishida, S.~Ishida, and M.~Oda, {\it Spectra of exclusive semi-leptonic
  decays of \uppercase{B} meson in the covariant oscillator quark model},  {\em
  Prog. Theor. Phys.} {\bf 98} (1997) 159--168,
  [\href{http://xxx.lanl.gov/abs/hep-ph/9705346}{{\tt hep-ph/9705346}}].

\bibitem{Ebert:1997mg}
D.~Ebert, R.~N. Faustov, and V.~O. Galkin, {\it Exclusive nonleptonic decays of
  \uppercase{B} mesons},  {\em Phys. Rev.} {\bf D56} (1997) 312--320,
  [\href{http://xxx.lanl.gov/abs/hep-ph/9701218}{{\tt hep-ph/9701218}}].

\bibitem{Ebert:2002qa}
D.~Ebert, R.~N. Faustov, and V.~O. Galkin, {\it Decay constants of heavy-light
  mesons in the relativistic quark model},  {\em Mod. Phys. Lett.} {\bf A17}
  (2002) 803--808, [\href{http://xxx.lanl.gov/abs/hep-ph/0204167}{{\tt
  hep-ph/0204167}}].

\bibitem{Ivanov:1999tv}
M.~A. Ivanov, P.~Santorelli, and N.~Tancredi, {\it The semileptonic form
  factors of \uppercase{B} and \uppercase{D} mesons in the quark confinement
  model},  {\em Eur. Phys. J.} {\bf A9} (2000) 109--114,
  [\href{http://xxx.lanl.gov/abs/hep-ph/9905209}{{\tt hep-ph/9905209}}].

\bibitem{Close:1993yk}
F.~E. Close and A.~Wambach, {\it Quark model form-factors for heavy quark
  effective theory},  {\em Nucl. Phys.} {\bf B412} (1994) 169--180,
  [\href{http://xxx.lanl.gov/abs/hep-ph/9307260}{{\tt hep-ph/9307260}}].

\bibitem{Kiselev:1994ay}
V.~V. Kiselev, {\it Semileptonic $ \uppercase{B} \to \uppercase{D}^* l\nu $:
  The slope of \uppercase{I}sgur--\uppercase{W}ise function and
  $|\uppercase{V}_{bc}|$ value in potential quark model},  {\em Mod. Phys.
  Lett.} {\bf A10} (1995) 1049--1056,
  [\href{http://xxx.lanl.gov/abs/hep-ph/9409348}{{\tt hep-ph/9409348}}].

\bibitem{LeYaouanc:1995wv}
A.~Le~Yaouanc, L.~Oliver, O.~Pene, and J.~C. Raynal, {\it Covariant quark model
  of form factors in the heavy mass limit},  {\em Phys. Lett.} {\bf B365}
  (1996) 319--326, [\href{http://xxx.lanl.gov/abs/hep-ph/9507342}{{\tt
  hep-ph/9507342}}].

\bibitem{Morenas:1996bn}
V.~Morenas, A.~Le~Yaouanc, L.~Oliver, O.~Pene, and J.~C. Raynal, {\it Slope of
  the \uppercase{I}sgur--\uppercase{W}ise function in the heavy mass limit of
  quark models a la \uppercase{B}akamjian--\uppercase{T}homas},  {\em Phys.
  Lett.} {\bf B408} (1997) 357--366,
  [\href{http://xxx.lanl.gov/abs/hep-ph/9705324}{{\tt hep-ph/9705324}}].

\bibitem{Morenas:1997nk}
V.~Morenas, A.~Le~Yaouanc, L.~Oliver, O.~Pene, and J.~C. Raynal, {\it
  Quantitative predictions for $\uppercase{B}$ semileptonic decays into $
  \uppercase{D}$, $\uppercase{D}^*$ and the orbitally excited
  $\uppercase{D}^{**}$ in quark models a la
  \uppercase{B}akamjian--\uppercase{T}homas},  {\em Phys. Rev.} {\bf D56}
  (1997) 5668--5680, [\href{http://xxx.lanl.gov/abs/hep-ph/9706265}{{\tt
  hep-ph/9706265}}].

\bibitem{Deandrea:1998uz}
A.~Deandrea, N.~Di~Bartolomeo, R.~Gatto, G.~Nardulli, and A.~D. Polosa, {\it A
  constituent quark-meson model for heavy-meson processes},  {\em Phys. Rev.}
  {\bf D58} (1998) 034004, [\href{http://xxx.lanl.gov/abs/hep-ph/9802308}{{\tt
  hep-ph/9802308}}].

\bibitem{Choi:1999nu}
H.-M. Choi and C.-R. Ji, {\it Light-front quark model analysis of exclusive
  \mbox{0- $\to$ 0-} semileptonic heavy meson decays},  {\em Phys. Lett.} {\bf
  B460} (1999) 461--466, [\href{http://xxx.lanl.gov/abs/hep-ph/9903496}{{\tt
  hep-ph/9903496}}].

\bibitem{Krutov:2000kt}
A.~F. Krutov, O.~I. Shro, and V.~E. Troitsky, {\it
  \uppercase{I}sgur--\uppercase{W}ise function in a relativistic model of
  constituent quarks},  {\em Phys. Lett.} {\bf B502} (2001) 140--146,
  [\href{http://xxx.lanl.gov/abs/hep-ph/0011071}{{\tt hep-ph/0011071}}].

\bibitem{Miller:1988tz}
G.~A. Miller and P.~Singer, {\it Radiative and pionic decays of the
  $\uppercase{D}^*$ mesons and the magnetic moment of the charmed quark},  {\em
  Phys. Rev.} {\bf D37} (1988) 2564.

\bibitem{O'Donnell:1994ey}
P.~J. O'Donnell and Q.~P. Xu, {\it Strong and radiative $\uppercase{D}^*$
  decays},  {\em Phys. Lett.} {\bf B336} (1994) 113--118,
  [\href{http://xxx.lanl.gov/abs/hep-ph/9406300}{{\tt hep-ph/9406300}}].

\bibitem{Colangelo:1994jc}
P.~Colangelo, F.~De~Fazio, and G.~Nardulli, {\it $\uppercase{D}^*$ radiative
  decays and strong coupling of heavy mesons with soft pions in a
  \uppercase{QCD} relativistic potential model},  {\em Phys. Lett.} {\bf B334}
  (1994) 175--179, [\href{http://xxx.lanl.gov/abs/hep-ph/9406320}{{\tt
  hep-ph/9406320}}].

\bibitem{Becirevic:1999fr}
D.~Becirevic and A.~L. Yaouanc, {\it $\hat{g}$ coupling
  ($g_{\uppercase{b}^*\uppercase{b}\pi}, g_{\uppercase{d}^*\uppercase{d}\pi}$):
  A quark model with \uppercase{D}irac equation},  {\em JHEP} {\bf 03} (1999)
  021, [\href{http://xxx.lanl.gov/abs/hep-ph/9901431}{{\tt hep-ph/9901431}}].

\bibitem{Eidelman:2004wy}
{\bf Particle Data Group} Collaboration, S.~Eidelman {\em et~al.}, {\it Review
  of particle physics},  {\em Phys. Lett.} {\bf B592} (2004) 1.

\bibitem{Burdman:1994ip}
G.~Burdman, T.~Goldman, and D.~Wyler, {\it Radiative leptonic decays of heavy
  mesons},  {\em Phys. Rev.} {\bf D51} (1995) 111--117,
  [\href{http://xxx.lanl.gov/abs/hep-ph/9405425}{{\tt hep-ph/9405425}}].

\bibitem{Lellouch:1994dy}
L.~Lellouch, {\it Weak decays of \uppercase{B} mesons and lattice
  \uppercase{QCD}},  {\em Acta Phys. Polon.} {\bf B25} (1994) 1679--1730,
  [\href{http://xxx.lanl.gov/abs/hep-ph/9412284}{{\tt hep-ph/9412284}}].

\bibitem{Bowler:2000xw}
{\bf UKQCD} Collaboration, K.~C. Bowler {\em et~al.}, {\it Decay constants of
  \uppercase{B} and \uppercase{D} mesons from non-perturbatively improved
  lattice \uppercase{QCD}},  {\em Nucl. Phys.} {\bf B619} (2001) 507--537,
  [\href{http://xxx.lanl.gov/abs/hep-lat/0007020}{{\tt hep-lat/0007020}}].

\bibitem{Narison:2001pu}
S.~Narison, {\it c, b quark masses and $f_{\uppercase{d}_s}$,
  $f_{\uppercase{b}_s}$ decay constants from pseudoscalar sum rules in full
  \uppercase{QCD} to order $\alpha_s^2$},  {\em Phys. Lett.} {\bf B520} (2001)
  115--123, [\href{http://xxx.lanl.gov/abs/hep-ph/0108242}{{\tt
  hep-ph/0108242}}].

\bibitem{Narison:book}
S.~Narison, {\em QCD as a Theory of Hadrons: From Partons to Confinement}.
\newblock Cambridge Monographs on Particle Physics, Nuclear Physics and
  Cosmology, 2002.

\bibitem{Aubin:2005ar}
C.~Aubin {\em et~al.}, {\it Charmed meson decay constants in three-flavor
  lattice qcd},  {\em Phys. Rev. Lett.} {\bf 95} (2005) 122002,
  [\href{http://xxx.lanl.gov/abs/hep-lat/0506030}{{\tt hep-lat/0506030}}].

\bibitem{Wingate:2003gm}
M.~Wingate, C.~T.~H. Davies, A.~Gray, G.~P. Lepage, and J.~Shigemitsu, {\it The
  $\uppercase{B}_s$ and $\uppercase{D}_s$ decay constants in three--flavor
  lattice \uppercase{QCD}},  {\em Phys. Rev. Lett.} {\bf 92} (2004) 162001,
  [\href{http://xxx.lanl.gov/abs/hep-ph/0311130}{{\tt hep-ph/0311130}}].

\bibitem{Artuso:2005ym}
{\bf CLEO} Collaboration, M.~Artuso {\em et~al.}, {\it Improved measurement of
  $\uppercase{B}(\uppercase{D}^+ \to \mu^+ \nu)$ and the pseudoscalar decay
  constant $f_{\uppercase{d}^+}$},  {\em Phys. Rev. Lett.} {\bf 95} (2005)
  251801, [\href{http://xxx.lanl.gov/abs/hep-ex/0508057}{{\tt
  hep-ex/0508057}}].

\bibitem{Chadha:1997zh}
{\bf CLEO} Collaboration, M.~Chadha {\em et~al.}, {\it Improved measurement of
  the pseudoscalar decay constant $f_{\uppercase{d}_s}$},  {\em Phys. Rev.}
  {\bf D58} (1998) 032002, [\href{http://xxx.lanl.gov/abs/hep-ex/9712014}{{\tt
  hep-ex/9712014}}].

\bibitem{Heister:2002fp}
{\bf ALEPH} Collaboration, A.~Heister {\em et~al.}, {\it Leptonic decays of the
  $\uppercase{D}_s$ meson},  {\em Phys. Lett.} {\bf B528} (2002) 1--18,
  [\href{http://xxx.lanl.gov/abs/hep-ex/0201024}{{\tt hep-ex/0201024}}].

\bibitem{Abbiendi:2001nb}
{\bf OPAL} Collaboration, G.~Abbiendi {\em et~al.}, {\it Measurement of the
  branching ratio for $\uppercase{D}_s^- \to \tau^- \bar{\nu}_\tau$},  {\em
  Phys. Lett.} {\bf B516} (2001) 236--248,
  [\href{http://xxx.lanl.gov/abs/hep-ex/0103012}{{\tt hep-ex/0103012}}].

\bibitem{Alexandrov:2000ns}
{\bf BEATRICE} Collaboration, Y.~Alexandrov {\em et~al.}, {\it Measurement of
  the $\uppercase{D}_s\to\mu \nu_\mu$ branching fraction and of the
  $\uppercase{D}_s$ decay constant},  {\em Phys. Lett.} {\bf B478} (2000)
  31--38.

\bibitem{Ablikim:2004ry}
{\bf BES} Collaboration, M.~Ablikim {\em et~al.}, {\it Direct measurement of
  the pseudoscalar decay constant $f_{\uppercase{d}^+}$},  {\em Phys. Lett.}
  {\bf B610} (2005) 183--191,
  [\href{http://xxx.lanl.gov/abs/hep-ex/0410050}{{\tt hep-ex/0410050}}].

\bibitem{Kodama:1996xq}
{\bf Fermilab E653} Collaboration, K.~Kodama {\em et~al.}, {\it Measurement of
  $\uppercase{B}(\uppercase{D}_s^+ \to \mu^+ \nu_\mu)/
  \uppercase{B}(\uppercase{D}_s^+ \to \phi \mu^+ \nu_\mu)$ and determination of
  the decay constant $f_{\uppercase{d}_s}$},  {\em Phys. Lett.} {\bf B382}
  (1996) 299--304, [\href{http://xxx.lanl.gov/abs/hep-ex/9606017}{{\tt
  hep-ex/9606017}}].

\bibitem{Ikado:2006un}
K.~Ikado {\em et~al.}, {\it Evidence of the purely leptonic decay
  $\uppercase{B}^- \to\tau^-\,\overline{\nu}_\tau$},  {\em Phys. Rev. Lett.}
  {\bf 97} (2006) 251802, [\href{http://xxx.lanl.gov/abs/hep-ex/0604018}{{\tt
  hep-ex/0604018}}].

\bibitem{Bernard:2000ki}
C.~W. Bernard, {\it Heavy quark physics on the lattice},  {\em Nucl. Phys.
  Proc. Suppl.} {\bf 94} (2001) 159--176,
  [\href{http://xxx.lanl.gov/abs/hep-lat/0011064}{{\tt hep-lat/0011064}}].

\bibitem{Hashimoto:2004hn}
S.~Hashimoto, {\it Recent results from lattice calculations},  {\em Int. J.
  Mod. Phys.} {\bf A20} (2005) 5133--5144,
  [\href{http://xxx.lanl.gov/abs/hep-ph/0411126}{{\tt hep-ph/0411126}}].

\bibitem{Burdman:1993es}
G.~Burdman, Z.~Ligeti, M.~Neubert, and Y.~Nir, {\it The decay $\uppercase{B}
  \to \pi l \nu$ in heavy quark effective theory},  {\em Phys. Rev.} {\bf D49}
  (1994) 2331--2345, [\href{http://xxx.lanl.gov/abs/hep-ph/9309272}{{\tt
  hep-ph/9309272}}].

\bibitem{Neubert:1992hb}
M.~Neubert, {\it Subleading \uppercase{I}sgur--\uppercase{W}ise form-factors
  from \uppercase{QCD} sum rules},  {\em Phys. Rev.} {\bf D46} (1992)
  3914--3928.

\bibitem{Neubert:1992tg}
M.~Neubert, {\it Short distance expansion of heavy quark currents},  {\em Phys.
  Rev.} {\bf D46} (1992) 2212--2227.

\bibitem{Luke:1990eg}
M.~E. Luke, {\it Effects of subleading operators in the heavy quark effective
  theory},  {\em Phys. Lett.} {\bf B252} (1990) 447--455.

\bibitem{Bowler:2002zh}
{\bf UKQCD} Collaboration, K.~C. Bowler, G.~Douglas, R.~D. Kenway, G.~N.
  Lacagnina, and C.~M. Maynard, {\it Semi-leptonic decays of heavy mesons and
  the \uppercase{I}sgur--\uppercase{W}ise function in quenched lattice
  \uppercase{QCD}},  {\em Nucl. Phys.} {\bf B637} (2002) 293--310,
  [\href{http://xxx.lanl.gov/abs/hep-lat/0202029}{{\tt hep-lat/0202029}}].

\bibitem{Neubert:1991td}
M.~Neubert, {\it Model independent extraction of $\uppercase{V}_{cb}$ from
  semileptonic decays},  {\em Phys. Lett.} {\bf B264} (1991) 455--461.

\bibitem{Bartelt:1998dq}
{\bf CLEO} Collaboration, J.~E. Bartelt {\em et~al.}, {\it Measurement of the
  $\uppercase{B} \to \uppercase{D} l \nu$ branching fractions and form factor},
   {\em Phys. Rev. Lett.} {\bf 82} (1999) 3746,
  [\href{http://xxx.lanl.gov/abs/hep-ex/9811042}{{\tt hep-ex/9811042}}].

\bibitem{Boyd:1997kz}
C.~G. Boyd, B.~Grinstein, and R.~F. Lebed, {\it Precision corrections to
  dispersive bounds on form factors},  {\em Phys. Rev.} {\bf D56} (1997)
  6895--6911, [\href{http://xxx.lanl.gov/abs/hep-ph/9705252}{{\tt
  hep-ph/9705252}}].

\bibitem{Abe:2001yf}
{\bf Belle} Collaboration, K.~Abe {\em et~al.}, {\it Measurement of
  $\uppercase{B}(\bar{\uppercase{b}}^0 \to \uppercase{D}^+ l^-\bar{\nu})$ and
  determination of $|\uppercase{V}_{cb}|$},  {\em Phys. Lett.} {\bf B526}
  (2002) 258--268, [\href{http://xxx.lanl.gov/abs/hep-ex/0111082}{{\tt
  hep-ex/0111082}}].

\bibitem{Caprini:1995wq}
I.~Caprini and M.~Neubert, {\it Improved bounds for the slope and curvature of
  $\bar{\uppercase{b}}\to \uppercase{D}^* l\bar\nu$ form factors},  {\em Phys.
  Lett.} {\bf B380} (1996) 376--384,
  [\href{http://xxx.lanl.gov/abs/hep-ph/9603414}{{\tt hep-ph/9603414}}].

\bibitem{Hashimoto:1998ia}
S.~Hashimoto {\em et~al.}, {\it $\uppercase{B} \to \uppercase{D} l \nu$ form
  factors and the determination of $|\uppercase{V}_{cb}|$},  {\em Nucl. Phys.
  Proc. Suppl.} {\bf 73} (1999) 399--401,
  [\href{http://xxx.lanl.gov/abs/hep-lat/9810056}{{\tt hep-lat/9810056}}].

\bibitem{Albertus:2004wj}
C.~Albertus, E.~Hernandez, and J.~Nieves, {\it Nonrelativistic constituent
  quark model and \uppercase{HQET} combined study of semileptonic decays of
  $\uppercase{\Lambda}_b$ and $\uppercase{\Xi}_b$ baryons},  {\em Phys. Rev.}
  {\bf D71} (2005) 014012, [\href{http://xxx.lanl.gov/abs/nucl-th/0412006}{{\tt
  nucl-th/0412006}}].

\bibitem{Adam:2002uw}
{\bf CLEO} Collaboration, N.~E. Adam {\em et~al.}, {\it Determination of the
  $\bar{\uppercase{b}} \to \uppercase{D}^* l \bar\nu$ decay width and
  $|\uppercase{V}_{cb}|$.},  {\em Phys. Rev.} {\bf D67} (2003) 032001,
  [\href{http://xxx.lanl.gov/abs/hep-ex/0210040}{{\tt hep-ex/0210040}}].

\bibitem{Aubert:2004nz}
{\bf BABAR} Collaboration, B.~Aubert {\em et~al.}, {\it Measurement of
  $\uppercase{B} \to \uppercase{D}^*$ form factors in the semileptonic decay
  $\bar{\uppercase{b}}^0 \to \uppercase{D}^{*+} l^-\bar\nu$},
  \href{http://xxx.lanl.gov/abs/hep-ex/0409047}{{\tt hep-ex/0409047}}.

\bibitem{Caprini:1997mu}
I.~Caprini, L.~Lellouch, and M.~Neubert, {\it Dispersive bounds on the shape of
  $\bar{\uppercase{b}} \to\uppercase{D}^* l \bar\nu$ form factors},  {\em Nucl.
  Phys.} {\bf B530} (1998) 153--181,
  [\href{http://xxx.lanl.gov/abs/hep-ph/9712417}{{\tt hep-ph/9712417}}].

\bibitem{Grinstein:2001yg}
B.~Grinstein and Z.~Ligeti, {\it Heavy quark symmetry in $\uppercase{B} \to
  \uppercase{D}^* l \bar\nu$ spectra},  {\em Phys. Lett.} {\bf B526} (2002)
  345--354, [\href{http://xxx.lanl.gov/abs/hep-ph/0111392}{{\tt
  hep-ph/0111392}}].

\bibitem{Richman:1995wm}
J.~D. Richman and P.~R. Burchat, {\it Leptonic and semileptonic decays of charm
  and bottom hadrons},  {\em Rev. Mod. Phys.} {\bf 67} (1995) 893--976,
  [\href{http://xxx.lanl.gov/abs/hep-ph/9508250}{{\tt hep-ph/9508250}}].

\bibitem{Abe:2001cs}
{\bf BELLE} Collaboration, K.~Abe {\em et~al.}, {\it Determination of
  $|\uppercase{V}_{cb}|$ using the semileptonic decay $ \bar{\uppercase{b}}^0
  \to \uppercase{D}^{*+} e^- \bar\nu$},  {\em Phys. Lett.} {\bf B526} (2002)
  247--257, [\href{http://xxx.lanl.gov/abs/hep-ex/0111060}{{\tt
  hep-ex/0111060}}].

\bibitem{Abdallah:2004rz}
{\bf DELPHI} Collaboration, J.~Abdallah {\em et~al.}, {\it Measurement of
  $|\uppercase{V}_{cb}|$ using the semileptonic decay $\bar{\uppercase{b}}_d^0
  \to \uppercase{D}^{*+} l^- \bar\nu_l$},  {\em Eur. Phys. J.} {\bf C33} (2004)
  213--232, [\href{http://xxx.lanl.gov/abs/hep-ex/0401023}{{\tt
  hep-ex/0401023}}].

\bibitem{Aubert:2004bw}
{\bf BABAR} Collaboration, B.~Aubert {\em et~al.}, {\it Measurement of the
  $\bar{\uppercase{b}}^0 \to \uppercase{D}^{*+} l^- \bar{\nu}_l$ decay rate and
  $|\uppercase{V}_{cb}|$},  {\em Phys. Rev.} {\bf D71} (2005) 051502,
  [\href{http://xxx.lanl.gov/abs/hep-ex/0408027}{{\tt hep-ex/0408027}}].

\bibitem{Hashimoto:2001nb}
S.~Hashimoto, A.~S. Kronfeld, P.~B. Mackenzie, S.~M. Ryan, and J.~N. Simone,
  {\it Lattice calculation of the zero recoil form factor of
  $\bar{\uppercase{b}} \to \uppercase{D}^* l \bar\nu$: Toward a model
  independent determination of $|\uppercase{V}_{cb}|$},  {\em Phys. Rev.} {\bf
  D66} (2002) 014503, [\href{http://xxx.lanl.gov/abs/hep-ph/0110253}{{\tt
  hep-ph/0110253}}].

\bibitem{Anastassov:2001cw}
{\bf CLEO} Collaboration, A.~Anastassov {\em et~al.}, {\it First measurement of
  $\uppercase{\Gamma}(\uppercase{D}^{*+})$ and precision measurement of
  $m(\uppercase{D}^{*+}) - m(\uppercase{D}^0)$},  {\em Phys. Rev.} {\bf D65}
  (2002) 032003, [\href{http://xxx.lanl.gov/abs/hep-ex/0108043}{{\tt
  hep-ex/0108043}}].

\bibitem{Abada:2002xe}
A.~Abada {\em et~al.}, {\it First lattice \uppercase{QCD} estimate of the
  $g_{\uppercase{d}^* \uppercase{d} \pi}$ coupling},  {\em Phys. Rev.} {\bf
  D66} (2002) 074504, [\href{http://xxx.lanl.gov/abs/hep-ph/0206237}{{\tt
  hep-ph/0206237}}].

\bibitem{Abada:2003un}
A.~Abada {\em et~al.}, {\it Lattice measurement of the couplings $\hat
  g_\infty$ and $g_{\uppercase{b}^*\uppercase{b}\pi}$},  {\em JHEP} {\bf 02}
  (2004) 016, [\href{http://xxx.lanl.gov/abs/hep-lat/0310050}{{\tt
  hep-lat/0310050}}].

\bibitem{Navarra:2001ju}
F.~S. Navarra, M.~Nielsen, and M.~E. Bracco, {\it $\uppercase{D}^*$
  $\uppercase{D} \pi$ form factor revisited},  {\em Phys. Rev.} {\bf D65}
  (2002) 037502, [\href{http://xxx.lanl.gov/abs/hep-ph/0109188}{{\tt
  hep-ph/0109188}}].

\bibitem{Belyaev:1994zk}
V.~M. Belyaev, V.~M. Braun, A.~Khodjamirian, and R.~Ruckl, {\it
  $\uppercase{D}^*\,\uppercase{D}\, \pi$ and $\uppercase{B}^* \,\uppercase{B}\,
  \pi$ couplings in \uppercase{QCD}},  {\em Phys. Rev.} {\bf D51} (1995)
  6177--6195, [\href{http://xxx.lanl.gov/abs/hep-ph/9410280}{{\tt
  hep-ph/9410280}}].

\bibitem{Abe:1998fb}
{\bf CDF} Collaboration, F.~Abe {\em et~al.}, {\it Observation of
  $\uppercase{B}_c$ mesons in $p\bar{p}$ collisions at $\sqrt{s} = 1.8$
  \uppercase{T}e\uppercase{V}},  {\em Phys. Rev.} {\bf D58} (1998) 112004,
  [\href{http://xxx.lanl.gov/abs/hep-ex/9804014}{{\tt hep-ex/9804014}}].

\bibitem{Abe:1998wi}
{\bf CDF} Collaboration, F.~Abe {\em et~al.}, {\it Observation of the
  $\uppercase{B}_c$ meson in $p\bar{p}$ collisions at $\sqrt{s} = 1.8$
  \uppercase{T}e\uppercase{V}},  {\em Phys. Rev. Lett.} {\bf 81} (1998)
  2432--2437, [\href{http://xxx.lanl.gov/abs/hep-ex/9805034}{{\tt
  hep-ex/9805034}}].

\bibitem{Albertus:2005ud}
C.~Albertus, J.~M. Flynn, E.~Hernandez, J.~Nieves, and J.~M. Verde-Velasco,
  {\it Semileptonic $\uppercase{B} \to \pi$ decays from an \uppercase{O}mnes
  improved nonrelativistic constituent quark model},  {\em Phys. Rev.} {\bf
  D72} (2005) 033002, [\href{http://xxx.lanl.gov/abs/hep-ph/0506048}{{\tt
  hep-ph/0506048}}].

\bibitem{Ivanov:2006ni}
M.~A. Ivanov, J.~G. Korner, and P.~Santorelli, {\it Exclusive semileptonic and
  nonleptonic decays of the $\uppercase{B}_c$ meson},  {\em Phys. Rev.} {\bf
  D73} (2006) 054024, [\href{http://xxx.lanl.gov/abs/hep-ph/0602050}{{\tt
  hep-ph/0602050}}].

\bibitem{Ivanov:2005fd}
M.~A. Ivanov, J.~G. Korner, and P.~Santorelli, {\it Semileptonic decays of
  $\uppercase{B}_c$ mesons into charmonium states in a relativistic quark
  model},  {\em Phys. Rev.} {\bf D71} (2005) 094006,
  [\href{http://xxx.lanl.gov/abs/hep-ph/0501051}{{\tt hep-ph/0501051}}].
  \protect{Erratum Phys. Rev. \bf{D75}, 019901(\lowercase{e})}.

\bibitem{Ivanov:2000aj}
M.~A. Ivanov, J.~G. Korner, and P.~Santorelli, {\it The semileptonic decays of
  the $\uppercase{B}_c$ meson},  {\em Phys. Rev.} {\bf D63} (2001) 074010,
  [\href{http://xxx.lanl.gov/abs/hep-ph/0007169}{{\tt hep-ph/0007169}}].

\bibitem{Ebert:2003cn}
D.~Ebert, R.~N. Faustov, and V.~O. Galkin, {\it Weak decays of the
  $\uppercase{B}_c$ meson to charmonium and $\uppercase{D}$ mesons in the
  relativistic quark model},  {\em Phys. Rev.} {\bf D68} (2003) 094020,
  [\href{http://xxx.lanl.gov/abs/hep-ph/0306306}{{\tt hep-ph/0306306}}].

\bibitem{Ebert:2003wc}
D.~Ebert, R.~N. Faustov, and V.~O. Galkin, {\it Weak decays of the
  $\uppercase{B}_c$ meson to $\uppercase{B}_s$ and $\uppercase{B}$ mesons in
  the relativistic quark model},  {\em Eur. Phys. J.} {\bf C32} (2003) 29--43,
  [\href{http://xxx.lanl.gov/abs/hep-ph/0308149}{{\tt hep-ph/0308149}}].

\bibitem{Chang:1992pt}
C.-H. Chang and Y.-Q. Chen, {\it The decays of $\uppercase{B}_c$ meson},  {\em
  Phys. Rev.} {\bf D49} (1994) 3399--3411.

\bibitem{Chang:2001pm}
C.-H. Chang, Y.-Q. Chen, G.-L. Wang, and H.-S. Zong, {\it Decays of the meson
  $\uppercase{B}_c$ to a \uppercase{P}-wave charmonium state $\chi_c$ or
  $h_c$},  {\em Phys. Rev.} {\bf D65} (2002) 014017,
  [\href{http://xxx.lanl.gov/abs/hep-ph/0103036}{{\tt hep-ph/0103036}}].

\bibitem{Chang:2001vq}
C.-H. Chang, Y.-Q. Chen, G.-L. Wang, and H.-S. Zong, {\it Semileptonic decays
  of $\uppercase{B}_c$ meson to a \uppercase{P}-wave charmonium state $\chi_c$
  or $h_c$},  {\em Commun. Theor. Phys.} {\bf 35} (2001) 395--398,
  [\href{http://xxx.lanl.gov/abs/hep-ph/0102150}{{\tt hep-ph/0102150}}].

\bibitem{AbdEl-Hady:1999xh}
A.~Abd El-Hady, J.~H. Munoz, and J.~P. Vary, {\it Semileptonic and non-leptonic
  $\uppercase{B}_c$ decays},  {\em Phys. Rev.} {\bf D62} (2000) 014019,
  [\href{http://xxx.lanl.gov/abs/hep-ph/9909406}{{\tt hep-ph/9909406}}].

\bibitem{Liu:1997hr}
J.-F. Liu and K.-T. Chao, {\it $\uppercase{B}_c$ meson weak decays and
  \uppercase{CP} violation},  {\em Phys. Rev.} {\bf D56} (1997) 4133--4145.

\bibitem{Kiselev:1999sc}
V.~V. Kiselev, A.~K. Likhoded, and A.~I. Onishchenko, {\it Semileptonic
  $\uppercase{B}_c$ meson decays in sum rules of \uppercase{QCD} and
  \uppercase{NRQCD}},  {\em Nucl. Phys.} {\bf B569} (2000) 473--504,
  [\href{http://xxx.lanl.gov/abs/hep-ph/9905359}{{\tt hep-ph/9905359}}].

\bibitem{Kiselev:2000pp}
V.~V. Kiselev, A.~E. Kovalsky, and A.~K. Likhoded, {\it $\uppercase{B}_c$
  decays and lifetime in \uppercase{QCD} sum rules},  {\em Nucl. Phys.} {\bf
  B585} (2000) 353--382, [\href{http://xxx.lanl.gov/abs/hep-ph/0002127}{{\tt
  hep-ph/0002127}}].

\bibitem{Kiselev:2002vz}
V.~V. Kiselev, {\it Exclusive decays and lifetime of $\uppercase{B}_c$ meson in
  \uppercase{QCD} sum rules},
  \href{http://xxx.lanl.gov/abs/hep-ph/0211021}{{\tt hep-ph/0211021}}.

\bibitem{Kiselev:2001zb}
V.~V. Kiselev, O.~N. Pakhomova, and V.~A. Saleev, {\it Two-particle decays of
  $\uppercase{B}_c$ meson into charmonium states},  {\em J. Phys.} {\bf G28}
  (2002) 595--606, [\href{http://xxx.lanl.gov/abs/hep-ph/0110180}{{\tt
  hep-ph/0110180}}].

\bibitem{Colangelo:1999zn}
P.~Colangelo and F.~De~Fazio, {\it Using heavy quark spin symmetry in
  semileptonic $\uppercase{B}_c$ decays},  {\em Phys. Rev.} {\bf D61} (2000)
  034012, [\href{http://xxx.lanl.gov/abs/hep-ph/9909423}{{\tt
  hep-ph/9909423}}].

\bibitem{Anisimov:1998xv}
A.~Y. Anisimov, P.~Y. Kulikov, I.~M. Narodetsky, and K.~A. Ter-Martirosian,
  {\it Exclusive and inclusive decays of the $\uppercase{B}_c$ meson in the
  light-front isgw model},  {\em Phys. Atom. Nucl.} {\bf 62} (1999) 1739--1753,
  [\href{http://xxx.lanl.gov/abs/hep-ph/9809249}{{\tt hep-ph/9809249}}].

\bibitem{Nobes:2000pm}
M.~A. Nobes and R.~M. Woloshyn, {\it Decays of the $\uppercase{B}_c$ meson in a
  relativistic quark-meson model},  {\em J. Phys.} {\bf G26} (2000) 1079--1094,
  [\href{http://xxx.lanl.gov/abs/hep-ph/0005056}{{\tt hep-ph/0005056}}].

\bibitem{Sanchis-Lozano:1994vh}
M.~A. Sanchis-Lozano, {\it Weak decays of doubly heavy hadrons},  {\em Nucl.
  Phys.} {\bf B440} (1995) 251--278,
  [\href{http://xxx.lanl.gov/abs/hep-ph/9502359}{{\tt hep-ph/9502359}}].

\bibitem{LopezCastro:2002ud}
G.~Lopez~Castro, H.~B. Mayorga, and J.~H. Munoz, {\it Non-leptonic decays of
  the $\uppercase{B}_c$ into tensor mesons},  {\em J. Phys.} {\bf G28} (2002)
  2241--2248, [\href{http://xxx.lanl.gov/abs/hep-ph/0205273}{{\tt
  hep-ph/0205273}}].

\bibitem{Lu:1995ug}
G.-G. Lu, Y.-D. Yang, and H.-B. Li, {\it Semileptonic $\uppercase{B}_c$ decay
  within heavy quark spin symmetry},  {\em Phys. Lett.} {\bf B341} (1995)
  391--396.

\bibitem{Abulencia:2005us}
{\bf CDF} Collaboration, D.~Abulencia {\em et~al.}, {\it Evidence for the
  exclusive decay $\uppercase{B}_c^\pm \to \uppercase{J}/\psi \pi^\pm$ and
  measurement of the mass of the $\uppercase{B}_c$ meson},  {\em Phys. Rev.
  Lett.} {\bf 96} (2006) 082002,
  [\href{http://xxx.lanl.gov/abs/hep-ex/0505076}{{\tt hep-ex/0505076}}].

\bibitem{Abulencia:2006zu}
{\bf CDF} Collaboration, A.~Abulencia {\em et~al.}, {\it Measurement of the
  $\uppercase{B}_c^+$ meson lifetime using $\uppercase{B}_c^+ \to
  \uppercase{J}/\psi e^+ \nu_e$},  {\em Phys. Rev. Lett.} {\bf 97} (2006)
  012002, [\href{http://xxx.lanl.gov/abs/hep-ex/0603027}{{\tt
  hep-ex/0603027}}].

\bibitem{Edwards:2000bb}
{\bf CLEO} Collaboration, K.~W. Edwards {\em et~al.}, {\it Study of
  \uppercase{B} decays to charmonium states $\uppercase{B} \to \eta_c\,
  \uppercase{K}$ and $\uppercase{B}\to\chi_{c0}\,\uppercase{K}$},  {\em Phys.
  Rev. Lett.} {\bf 86} (2001) 30--34,
  [\href{http://xxx.lanl.gov/abs/hep-ex/0007012}{{\tt hep-ex/0007012}}].

\bibitem{Hwang:1997ie}
D.~S. Hwang and G.-H. Kim, {\it Decay constant ratios
  $f_{\eta_c}/f_{\uppercase{j}/\psi}$ and
  $f_{\eta_b}/f_{\uppercase{\Upsilon}}$},  {\em Z. Phys.} {\bf C76} (1997)
  107--110, [\href{http://xxx.lanl.gov/abs/hep-ph/9703364}{{\tt
  hep-ph/9703364}}].

\bibitem{Caso:1998tx}
{\bf Particle Data Group} Collaboration, C.~Caso {\em et~al.}, {\it Review of
  particle physics},  {\em Eur. Phys. J.} {\bf C3} (1998) 1--794.

\bibitem{Becirevic:1998ua}
D.~Becirevic {\em et~al.}, {\it Non-perturbatively improved heavy-light mesons:
  Masses and decay constants},  {\em Phys. Rev.} {\bf D60} (1999) 074501,
  [\href{http://xxx.lanl.gov/abs/hep-lat/9811003}{{\tt hep-lat/9811003}}].

\bibitem{Yao:2006px}
{\bf Particle Data Group} Collaboration, W.~M. Yao {\em et~al.}, {\it Review of
  particle physics},  {\em J. Phys.} {\bf G33} (2006) 1--1232.

\bibitem{Korner:1989qb}
J.~G. Korner and G.~A. Schuler, {\it Exclusive semileptonic heavy meson decays
  including lepton mass effects},  {\em Z. Phys.} {\bf C46} (1990) 93.

\bibitem{Jenkins:PC}
E.~Jenkins and A.~Manohar Private comunication.

\bibitem{Lichtenberg:1976fi}
D.~B. Lichtenberg, {\it Magnetic moments of charmed baryons in the quark
  model},  {\em Phys. Rev.} {\bf D15} (1977) 345.

\bibitem{Savage:1990di}
M.~J. Savage and M.~B. Wise, {\it Spectrum of baryons with two heavy quarks},
  {\em Phys. Lett.} {\bf B248} (1990) 177--180.

\bibitem{White:1991hz}
M.~J. White and M.~J. Savage, {\it Semileptonic decay of baryons with two heavy
  quarks},  {\em Phys. Lett.} {\bf B271} (1991) 410--414.

\bibitem{Brambilla:2005yk}
N.~Brambilla, A.~Vairo, and T.~Rosch, {\it Effective field theory lagrangians
  for baryons with two and three heavy quarks},  {\em Phys. Rev.} {\bf D72}
  (2005) 034021, [\href{http://xxx.lanl.gov/abs/hep-ph/0506065}{{\tt
  hep-ph/0506065}}].

\bibitem{Mattson:2002vu}
{\bf SELEX} Collaboration, M.~Mattson {\em et~al.}, {\it First observation of
  the doubly charmed baryon $\uppercase{\Xi}_{cc}^+$},  {\em Phys. Rev. Lett.}
  {\bf 89} (2002) 112001, [\href{http://xxx.lanl.gov/abs/hep-ex/0208014}{{\tt
  hep-ex/0208014}}].

\bibitem{Kiselev:2002an}
V.~V. Kiselev and A.~K. Likhoded, {\it Comment on 'first observation of doubly
  charmed baryon $\xi_{cc}^+$'},
  \href{http://xxx.lanl.gov/abs/hep-ph/0208231}{{\tt hep-ph/0208231}}.

\bibitem{Ratti:2003ez}
S.~P. Ratti, {\it New results on c-baryons and a search for cc-baryons in
  \uppercase{FOCUS}},  {\em Nucl. Phys. Proc. Suppl.} {\bf 115} (2003) 33--36.
  \protect{See also www-focus.fnal.gov/xicc/xicc\_focus.html}.

\bibitem{Aubert:2006qw}
{\bf BABAR} Collaboration, B.~Aubert {\em et~al.}, {\it Search for doubly
  charmed baryons $\uppercase{\Xi}_{cc}^+$ and $\uppercase{\Xi}_{cc}^{++}$ in
  \uppercase{BABAR}},  {\em Phys. Rev.} {\bf D74} (2006) 011103,
  [\href{http://xxx.lanl.gov/abs/hep-ex/0605075}{{\tt hep-ex/0605075}}].

\bibitem{Lesiak:2006sk}
{\bf Belle} Collaboration, T.~Lesiak, {\it Charmed baryon spectroscopy with
  \uppercase{B}elle},  \href{http://xxx.lanl.gov/abs/hep-ex/0605047}{{\tt
  hep-ex/0605047}}.

\bibitem{Albertus:2003sx}
C.~Albertus, J.~E. Amaro, E.~Hernandez, and J.~Nieves, {\it Charmed and bottom
  baryons: A variational approach based on heavy quark symmetry},  {\em Nucl.
  Phys.} {\bf A740} (2004) 333--361,
  [\href{http://xxx.lanl.gov/abs/nucl-th/0311100}{{\tt nucl-th/0311100}}].

\bibitem{Ebert:2002ig}
D.~Ebert, R.~N. Faustov, V.~O. Galkin, and A.~P. Martynenko, {\it Mass spectra
  of doubly heavy baryons in the relativistic quark model},  {\em Phys. Rev.}
  {\bf D66} (2002) 014008, [\href{http://xxx.lanl.gov/abs/hep-ph/0201217}{{\tt
  hep-ph/0201217}}].

\bibitem{Kiselev:2001fw}
V.~V. Kiselev and A.~K. Likhoded, {\it Baryons with two heavy quarks},  {\em
  Phys. Usp.} {\bf 45} (2002) 455--506,
  [\href{http://xxx.lanl.gov/abs/hep-ph/0103169}{{\tt hep-ph/0103169}}].

\bibitem{Narodetskii:2001bq}
I.~M. Narodetskii and M.~A. Trusov, {\it The heavy baryons in the
  nonperturbative string approach},  {\em Phys. Atom. Nucl.} {\bf 65} (2002)
  917--924, [\href{http://xxx.lanl.gov/abs/hep-ph/0104019}{{\tt
  hep-ph/0104019}}].

\bibitem{Tong:1999qs}
S.-P. Tong {\em et~al.}, {\it Spectra of baryons containing two heavy quarks in
  potential model},  {\em Phys. Rev.} {\bf D62} (2000) 054024,
  [\href{http://xxx.lanl.gov/abs/hep-ph/9910259}{{\tt hep-ph/9910259}}].

\bibitem{Itoh:2000um}
C.~Itoh, T.~Minamikawa, K.~Miura, and T.~Watanabe, {\it Doubly charmed baryon
  masses and quark wave functions in baryons},  {\em Phys. Rev.} {\bf D61}
  (2000) 057502.

\bibitem{Vijande:2004at}
J.~Vijande, H.~Garcilazo, A.~Valcarce, and F.~Fernandez, {\it Spectroscopy of
  doubly charmed baryons},  {\em Phys. Rev.} {\bf D70} (2004) 054022,
  [\href{http://xxx.lanl.gov/abs/hep-ph/0408274}{{\tt hep-ph/0408274}}].

\bibitem{Gershtein:2000nx}
S.~S. Gershtein, V.~V. Kiselev, A.~K. Likhoded, and A.~I. Onishchenko, {\it
  Spectroscopy of doubly heavy baryons},  {\em Phys. Rev.} {\bf D62} (2000)
  054021.

\bibitem{Ebert:1996ec}
D.~Ebert, R.~N. Faustov, V.~O. Galkin, A.~P. Martynenko, and V.~A. Saleev, {\it
  Heavy baryons in the relativistic quark model},  {\em Z. Phys.} {\bf C76}
  (1997) 111--115, [\href{http://xxx.lanl.gov/abs/hep-ph/9607314}{{\tt
  hep-ph/9607314}}].

\bibitem{Roncaglia:1995az}
R.~Roncaglia, D.~B. Lichtenberg, and E.~Predazzi, {\it Predicting the masses of
  baryons containing one or two heavy quarks},  {\em Phys. Rev.} {\bf D52}
  (1995) 1722--1725, [\href{http://xxx.lanl.gov/abs/hep-ph/9502251}{{\tt
  hep-ph/9502251}}].

\bibitem{Roncaglia:1994ex}
R.~Roncaglia, A.~Dzierba, D.~B. Lichtenberg, and E.~Predazzi, {\it Predicting
  the masses of heavy hadrons without an explicit hamiltonian},  {\em Phys.
  Rev.} {\bf D51} (1995) 1248--1257,
  [\href{http://xxx.lanl.gov/abs/hep-ph/9405392}{{\tt hep-ph/9405392}}].

\bibitem{Mathur:2002ce}
N.~Mathur, R.~Lewis, and R.~M. Woloshyn, {\it Charmed and bottom baryons from
  lattice nrqcd},  {\em Phys. Rev.} {\bf D66} (2002) 014502,
  [\href{http://xxx.lanl.gov/abs/hep-ph/0203253}{{\tt hep-ph/0203253}}].

\bibitem{AliKhan:1999yb}
A.~Ali~Khan {\em et~al.}, {\it Heavy-light mesons and baryons with b quarks},
  {\em Phys. Rev.} {\bf D62} (2000) 054505,
  [\href{http://xxx.lanl.gov/abs/hep-lat/9912034}{{\tt hep-lat/9912034}}].

\bibitem{Lewis:2001iz}
R.~Lewis, N.~Mathur, and R.~M. Woloshyn, {\it Charmed baryons in lattice
  \uppercase{QCD}},  {\em Phys. Rev.} {\bf D64} (2001) 094509,
  [\href{http://xxx.lanl.gov/abs/hep-ph/0107037}{{\tt hep-ph/0107037}}].

\bibitem{Flynn:2003vz}
{\bf UKQCD} Collaboration, J.~M. Flynn, F.~Mescia, and A.~S.~B. Tariq, {\it
  Spectroscopy of doubly-charmed baryons in lattice \uppercase{QCD}},  {\em
  JHEP} {\bf 07} (2003) 066,
  [\href{http://xxx.lanl.gov/abs/hep-lat/0307025}{{\tt hep-lat/0307025}}].

\bibitem{Faessler:2006ft}
A.~Faessler {\em et~al.}, {\it Magnetic moments of heavy baryons in the
  relativistic three-quark model},  {\em Phys. Rev.} {\bf D73} (2006) 094013,
  [\href{http://xxx.lanl.gov/abs/hep-ph/0602193}{{\tt hep-ph/0602193}}].

\bibitem{Julia-Diaz:2004vh}
B.~Julia-Diaz and D.~O. Riska, {\it Baryon magnetic moments in relativistic
  quark models},  {\em Nucl. Phys.} {\bf A739} (2004) 69--88,
  [\href{http://xxx.lanl.gov/abs/hep-ph/0401096}{{\tt hep-ph/0401096}}].

\bibitem{Oh:1991ws}
Y.-s. Oh, D.-P. Min, M.~Rho, and N.~N. Scoccola, {\it Massive quark baryons as
  skyrmions: Magnetic moments},  {\em Nucl. Phys.} {\bf A534} (1991) 493--512.

\bibitem{Jena:1986xs}
S.~N. Jena and D.~P. Rath, {\it Magnetic moments of light, charmed and b
  flavored baryons in a relativistic logarithmic potential},  {\em Phys. Rev.}
  {\bf D34} (1986) 196--200.

\bibitem{Bose:1980vy}
S.~K. Bose and L.~P. Singh, {\it Magnetic moments of charmed and b flavored
  hadrons in mit bag model},  {\em Phys. Rev.} {\bf D22} (1980) 773.

\bibitem{Ebert:2004ck}
D.~Ebert, R.~N. Faustov, V.~O. Galkin, and A.~P. Martynenko, {\it Semileptonic
  decays of doubly heavy baryons in the relativistic quark model},  {\em Phys.
  Rev.} {\bf D70} (2004) 014018,
  [\href{http://xxx.lanl.gov/abs/hep-ph/0404280}{{\tt hep-ph/0404280}}].

\bibitem{Faessler:2001mr}
A.~Faessler, T.~Gutsche, M.~A. Ivanov, J.~G. Korner, and V.~E. Lyubovitskij,
  {\it Semileptonic decays of double heavy baryons},  {\em Phys. Lett.} {\bf
  B518} (2001) 55--62, [\href{http://xxx.lanl.gov/abs/hep-ph/0107205}{{\tt
  hep-ph/0107205}}].

\bibitem{Guo:1998yj}
X.-H. Guo, H.-Y. Jin, and X.-Q. Li, {\it Weak semileptonic decays of heavy
  baryons containing two heavy quarks},  {\em Phys. Rev.} {\bf D58} (1998)
  114007, [\href{http://xxx.lanl.gov/abs/hep-ph/9805301}{{\tt
  hep-ph/9805301}}].

\bibitem{Korner:1991ph}
J.~G. Korner and M.~Kramer, {\it Polarization effects in exclusive semileptonic
  $\uppercase{\Lambda}_b$ and $\uppercase{\Lambda}_c$ charm and bottom baryon
  decays},  {\em Phys. Lett.} {\bf B275} (1992) 495--505.

\bibitem{Artuso:2001us}
{\bf CLEO} Collaboration, M.~Artuso {\em et~al.}, {\it Measurement of the
  masses and widths of the $\uppercase{\Sigma}_c^{++}$ and
  $\uppercase{\Sigma}_c^0$ charmed baryons},  {\em Phys. Rev.} {\bf D65} (2002)
  071101, [\href{http://xxx.lanl.gov/abs/hep-ex/0110071}{{\tt
  hep-ex/0110071}}].

\bibitem{Ammar:2000uh}
{\bf CLEO} Collaboration, R.~Ammar {\em et~al.}, {\it First observation of the
  $\uppercase{\Sigma}_c^{*+}$ baryon and a new measurement of the
  $\uppercase{\Sigma}_c^+$ mass},  {\em Phys. Rev. Lett.} {\bf 86} (2001)
  1167--1170, [\href{http://xxx.lanl.gov/abs/hep-ex/0007041}{{\tt
  hep-ex/0007041}}].

\bibitem{Link:2001ee}
{\bf FOCUS} Collaboration, J.~M. Link {\em et~al.}, {\it Measurement of natural
  widths of $\uppercase{\Sigma}_c^0$ and $\uppercase{\Sigma}_c^{++}$ baryons},
  {\em Phys. Lett.} {\bf B525} (2002) 205--210,
  [\href{http://xxx.lanl.gov/abs/hep-ex/0111027}{{\tt hep-ex/0111027}}].

\bibitem{Athar:2004ni}
{\bf CLEO} Collaboration, S.~B. Athar {\em et~al.}, {\it A new measurement of
  the masses and widths of the $\uppercase{\Sigma}_c^{*+}$ and
  $\uppercase{\Sigma}_c^{*0}$ charmed baryons},  {\em Phys. Rev.} {\bf D71}
  (2005) 051101, [\href{http://xxx.lanl.gov/abs/hep-ex/0410088}{{\tt
  hep-ex/0410088}}].

\bibitem{Gibbons:1996yv}
{\bf CLEO} Collaboration, L.~Gibbons {\em et~al.}, {\it Observation of an
  excited charmed baryon decaying into $\uppercase{\Xi}_c^0 \pi^+$},  {\em
  Phys. Rev. Lett.} {\bf 77} (1996) 810--813.

\bibitem{Avery:1995ps}
{\bf CLEO} Collaboration, P.~Avery {\em et~al.}, {\it Observation of a narrow
  state decaying into $\uppercase{\Xi}_c^+ \pi^-$},  {\em Phys. Rev. Lett.}
  {\bf 75} (1995) 4364--4368,
  [\href{http://xxx.lanl.gov/abs/hep-ex/9508010}{{\tt hep-ex/9508010}}].

\bibitem{Rosner:1995yu}
J.~L. Rosner, {\it Charmed baryons with \uppercase{J} = 3/2},  {\em Phys. Rev.}
  {\bf D52} (1995) 6461--6465,
  [\href{http://xxx.lanl.gov/abs/hep-ph/9508252}{{\tt hep-ph/9508252}}].

\bibitem{Pirjol:1997nh}
D.~Pirjol and T.-M. Yan, {\it Predictions for s-wave and p-wave heavy baryons
  from sum rules and constituent quark model: Strong interactions},  {\em Phys.
  Rev.} {\bf D56} (1997) 5483--5510,
  [\href{http://xxx.lanl.gov/abs/hep-ph/9701291}{{\tt hep-ph/9701291}}].

\bibitem{Yan:1992gz}
T.-M. Yan {\em et~al.}, {\it Heavy quark symmetry and chiral dynamics},  {\em
  Phys. Rev.} {\bf D46} (1992) 1148--1164.

\bibitem{Huang:1995ke}
M.-Q. Huang, Y.-B. Dai, and C.-S. Huang, {\it Decays of excited charmed lambda
  type and sigma type baryons in heavy hadron chiral perturbation theory},
  {\em Phys. Rev.} {\bf D52} (1995) 3986--3992.

\bibitem{Cheng:1997rp}
H.-Y. Cheng, {\it Remarks on the strong coupling constants in heavy hadron
  chiral lagrangians},  {\em Phys. Lett.} {\bf B399} (1997) 281--286,
  [\href{http://xxx.lanl.gov/abs/hep-ph/9701234}{{\tt hep-ph/9701234}}].

\bibitem{Chiladze:1997ev}
G.~Chiladze and A.~F. Falk, {\it Phenomenology of new baryons with charm and
  strangeness},  {\em Phys. Rev.} {\bf D56} (1997) 6738--6741,
  [\href{http://xxx.lanl.gov/abs/hep-ph/9707507}{{\tt hep-ph/9707507}}].

\bibitem{Grozin:1997qq}
A.~G. Grozin and O.~I. Yakovlev, {\it Couplings of heavy hadrons with soft
  pions from \uppercase{QCD} sum rules},  {\em Eur. Phys. J.} {\bf C2} (1998)
  721--727, [\href{http://xxx.lanl.gov/abs/hep-ph/9706421}{{\tt
  hep-ph/9706421}}].

\bibitem{Zhu:1998vg}
S.-L. Zhu and Y.-B. Dai, {\it Couplings of pions with heavy baryons from
  light-cone \uppercase{QCD} sum rules in the leading order of
  \uppercase{HQET}},  {\em Phys. Lett.} {\bf B429} (1998) 72--78,
  [\href{http://xxx.lanl.gov/abs/hep-ph/9802226}{{\tt hep-ph/9802226}}].

\bibitem{Tawfiq:1998nk}
S.~Tawfiq, P.~J. O'Donnell, and J.~G. Korner, {\it Charmed baryon strong
  coupling constants in a light-front quark model},  {\em Phys. Rev.} {\bf D58}
  (1998) 054010, [\href{http://xxx.lanl.gov/abs/hep-ph/9803246}{{\tt
  hep-ph/9803246}}].

\bibitem{Ivanov:1999bk}
M.~A. Ivanov, J.~G. Korner, V.~E. Lyubovitskij, and A.~G. Rusetsky, {\it Strong
  and radiative decays of heavy flavored baryons},  {\em Phys. Rev.} {\bf D60}
  (1999) 094002, [\href{http://xxx.lanl.gov/abs/hep-ph/9904421}{{\tt
  hep-ph/9904421}}].

\bibitem{Ivanov:1998qe}
M.~A. Ivanov, J.~G. Korner, V.~E. Lyubovitskij, and A.~G. Rusetsky, {\it
  One-pion charm baryon transitions in a relativistic three--quark model},
  {\em Phys. Lett.} {\bf B442} (1998) 435--442,
  [\href{http://xxx.lanl.gov/abs/hep-ph/9807519}{{\tt hep-ph/9807519}}].

\bibitem{Aubert:2005gt}
{\bf BABAR} Collaboration, B.~Aubert {\em et~al.}, {\it A precision measurement
  of the $\uppercase{\Lambda}_c^+$ baryon mass},  {\em Phys. Rev.} {\bf D72}
  (2005) 052006, [\href{http://xxx.lanl.gov/abs/hep-ex/0507009}{{\tt
  hep-ex/0507009}}].

\bibitem{Albertus:2005vd}
C.~Albertus, E.~Hernandez, J.~Nieves, and J.~M. Verde-Velasco, {\it Study of
  the leptonic decays of pseudoscalar \uppercase{B}, \uppercase{D} and vector
  $\uppercase{B}^*$, $\uppercase{D}^*$ mesons and of the semileptonic
  $\uppercase{B} \to \uppercase{D}$ and $\uppercase{B}\to \uppercase{D}^*$
  decays},  {\em Phys. Rev.} {\bf D71} (2005) 113006,
  [\href{http://xxx.lanl.gov/abs/hep-ph/0502219}{{\tt hep-ph/0502219}}].

\bibitem{Hernandez:2006gt}
E.~Hernandez, J.~Nieves, and J.~M. Verde-Velasco, {\it Study of exclusive
  semileptonic and non-leptonic decays of $\uppercase{B}_c^-$ in a
  nonrelativistic quark model},  {\em Phys. Rev.} {\bf D74} (2006) 074008,
  [\href{http://xxx.lanl.gov/abs/hep-ph/0607150}{{\tt hep-ph/0607150}}].

\bibitem{Albertus:2006ya}
C.~Albertus, E.~Hernandez, J.~Nieves, and J.~M. Verde-Velasco, {\it Static
  properties and semileptonic decays of doubly heavy baryons in a
  nonrelativistic quark model},  {\em Eur. Phys. J.} {\bf A} (2007)
  [\href{http://xxx.lanl.gov/abs/hep-ph/0610030}{{\tt hep-ph/0610030}}]. In
  press.

\bibitem{Albertus:2005zy}
C.~Albertus, E.~Hernandez, J.~Nieves, and J.~M. Verde-Velasco, {\it Study of
  the strong $\uppercase{\Sigma}_c \to \uppercase{\Lambda}_c\pi$,
  $\uppercase{\Sigma}_c^*\to \uppercase{\Lambda}_c \pi$ and
  $\uppercase{\Xi}_c^* \to \uppercase{\Xi}_c \pi$ decays in a nonrelativistic
  quark model},  {\em Phys. Rev.} {\bf D72} (2005) 094022,
  [\href{http://xxx.lanl.gov/abs/hep-ph/0507256}{{\tt hep-ph/0507256}}].

\end{thebibliography}\endgroup
\end{document}